%% file: PaperJHEPVersion.tex
\documentclass[a4paper,12pt]{article}
\pdfoutput=1
\usepackage{geometry}
\usepackage{amsmath}
\usepackage{amsfonts}
\usepackage{amssymb}
\usepackage{mathtools}
\usepackage{ytableau}
\usepackage{verbatim}
\usepackage[hidelinks]{hyperref}
\usepackage{color}
\usepackage{graphicx}
\usepackage{tikz}
\usetikzlibrary{decorations.pathreplacing}
\usepackage{multirow}

\numberwithin{equation}{section}

\newcommand{\bea}{\begin{eqnarray}}
\newcommand{\eea}{\end{eqnarray}}
\def\cN{ \mathcal{N} }
\def\cP{ \mathcal{P} }
\def\mR{ \mathbb{R} }
\def\mC{ \mathbb{C} }

\begin{document}

\begin{flushright}
QMUL-PH-17-22\\
\end{flushright}

\bigskip

\begin{center}

{\Large\bf BPS operators in $\mathcal{N}=4$  $SO(N)$   super Yang-Mills  theory:
  plethysms, dominoes  and words
}

\medskip 


\bigskip

{
Christopher Lewis-Brown $^{a,*}$ and Sanjaye Ramgoolam$^{a,b,\dag}  $}

\bigskip
$^{a}${\em School of Physics and Astronomy} , {\em  Centre for Research in String Theory}\\
{\em Queen Mary University of London, London E1 4NS, United Kingdom }\\
\medskip
$^{b}${\em  School of Physics and Mandelstam Institute for Theoretical Physics,} \\   
{\em University of Witwatersrand, Wits, 2050, South Africa} \\
\medskip
E-mails:  $^{*}$ c.h.lewis-brown@qmul.ac.uk,
\quad $^{\dag}$s.ramgoolam@qmul.ac.uk

\begin{abstract}

\noindent Permutations and associated algebras allow the construction  of half and quarter BPS operators in maximally supersymmetric Yang Mills theory with $U(N)$, $SO(N)$ and $Sp(N) $ gauge groups. The construction leads to bases for the operators, labelled by Young diagrams and associated group theory data, which have been shown to be  orthogonal under the inner product defined by the free field two-point functions. In this paper, we study in detail the orientifold projection map between the Young diagram basis for $U(N)$ theories and the Young diagram basis for  $ SO(N)$  (and $ Sp(N)$) half-BPS operators. We  find a simple connection between this map and the plethystic refinement of the Littlewood Richardson coefficients which  couple triples of Young diagrams where two of them are identical. This plethystic refinement is known to be computable using an algorithm based on domino tilings of Young diagrams.  We discuss the domino combinatorics of the orientifold projection map in terms of giant graviton branes.  The permutation construction of $SO(N)$ operators is used to find  large N generating functions for both the half and quarter BPS sectors. The structure of these generating functions is elucidated using the combinatorics of words, organised according to the number of periods. Aperiodic words in the $U(N)$ theory are Lyndon words and an $SO(N)$ analogue of Lyndon words is defined using a minimally periodic condition. We calculate the normalization factor for the orthogonal basis of Young diagram operators in the baryonic sector of $SO(N)$.

\end{abstract}

\end{center}


\newpage 

\setcounter{tocdepth}{2}
\tableofcontents

\section{Introduction}

	\ytableausetup{centertableaux}
		
	The AdS/CFT correspondence \cite{Maldacena1998,Gubser1998,Witten1998}  allows us to use the physics of $\mathcal{N}=4$ $U(N)$ super Yang-Mills to find new properties of type IIB string theory on $AdS_5 \times S^5$. Following the association of sub-determinant operators with single giant graviton branes in $S^5$ \cite{BBNS01}, the construction of the Young diagram basis for the half-BPS sector of  $U(N)$ gauge theory \cite{Corley2002}, which diagonalizes the CFT 2-point functions,  has led to an explicit correspondence between the half-BPS operators and non-perturbative states in the AdS space. These states include general giant gravitons \cite{MST2000,GMT2000,HHI2000} as well as LLM geometries arising from the backreaction of giant gravitons on the space-time \cite{LLM}. The finite $N$ cut-off in the diagrams is dual to the stringy exclusion principle \cite{Maldacena:1998bw}. Progress towards Young diagram bases for multi-matrix systems was made in \cite{BHLN02,BBFH04,DSS07I}, motivated by the study of open strings attached to giant gravitons. The diagonalisation property  for two-point functions of Young diagram bases in multi-matrix systems  was proved in \cite{Kimura:2007wy,Brown:2007xh,Bhattacharyya:2008rb, Bhattacharyya:2008xy,Brown:2008ij}. The $U(N)$ Young diagram basis was extended in \cite{Pasukonis2013,Mattioli2014,Mattioli:2016gyl} to more general quiver gauge theories. 
		
	In a different direction, the dual description of different gauge groups was considered in \cite{Witten1998a}. A $\mathbb{Z}_2$ orientifold operation, acting as a $ \mathbb{Z}_2$ orbifold in space-time,  takes the $S^5$ factor of $AdS_5 \times S^5$ to $ \mathbb{R} P^5$. This operation can produce, depending on topological factors, a theory with either an orthogonal or symplectic gauge group. Following earlier work by \cite{AABF02}, recently a Young diagram  basis of the half-BPS sector was found for these groups and was used to compute exact correlators for elements of this basis, and for certain traces \cite{Caputa2013,Caputa2013a}. These considerations were extended to the free field quarter-BPS sector in \cite{Kemp2014,Kemp1406}. These results involve significant extensions of the $U(N)$ story, where wreath products of symmetric groups play an interesting role. These wreath products are described in more detail in section 2.

	In this paper we revisit the construction and counting of operators in the half and quarter-BPS sectors of $\mathcal{N}=4$ SYM with $SO(N)$ and $Sp(N)$ gauge groups. In the orthogonal case, these operators are of two types, which we call mesonic and baryonic.  Essentially, the mesonic operators are constructed by contracting indices of matrices with Kronecker $\delta$ invariants common to $SO(N)$ and $U(N)$ theories (the $Sp(N)$ equivalent is the symplectic form). The baryonic operators are constructed using $\varepsilon $ invariants which are specific to $SO(N)$ theories ($Sp(N)$ theories also contain such operators, but they are linearly dependent on the mesonic ones). These two types of operators are described in more detail in Section 3. A natural question is: how are the mesonic Young diagram operators of $U(N)$ theories related to the mesonic Young diagram operators of the $SO(N)$ (or $Sp(N)$ theories)?  From the physical perspective, the relation is given by the orientifold map. So the question we are asking can be posed as: how does the orientifold projection operation of string theory act on the Young diagram operators of the $U(N)$ half-BPS sector (and quarter-BPS sector) to produce the Young diagram operators of the $SO(N)$ theory? Surprisingly, we find that this question, in the case of the half-BPS sector, has a simple and elegant answer in terms of a classic concept in the combinatorics of Young diagrams, called plethysms of Young diagrams. 
		
	Consider a  Young diagram $t$ with $m$ boxes and a positive integer $k$.  There is a  representation $V_t$ of $U(N) $ corresponding to $t$. We take $N$ to be large here, more precisely $ N \geq m k $.   Now consider the tensor product $ V_t^{ \otimes k } $. This is a representation of $U(N)$ under the diagonal action where the group element $U \in U(N)$ acts as  $ U \otimes U \otimes \cdots U$. 		This diagonal action of $ U(N)$ commutes with the $ S_k$ permutation group acting on $V_t^{ \otimes k }$ by permuting the different factors of  the tensor product. So we can decompose $ V_{t}^{ \otimes k } $ according to irreps of $ U(N) \times S_k$ which correspond to pair  $ ( R , \Lambda ) $ where $R$ is a Young diagram with $k m $ boxes and $ \Lambda $ is  a Young diagram with $k$ boxes. The multiplicity of $(R , \Lambda)$, denoted $ \mathcal{P} ( t , \Lambda , R )$ is known as a plethysm coefficient. They were defined by D. E. Littlewood \cite{PlethysmL} and remain the subject of important questions in combinatorics \cite{Stanley1999}. The sum over $\Lambda $ of $ \cP ( t , \Lambda , R )$ can be expressed in terms of Littlewood-Richardson coefficients. For the case where $k =2$,  the Young diagram $ \Lambda $ can be either the symmetric with a row of length $2$, denoted as $\Lambda = [2]$, or it can be anti-symmetric, denoted as $ \Lambda = [1,1]$ for two rows of length $1$. The sum $ \cP (  t , [2] , R )  + \cP (  t , [1,1] , R )  $ is a Littlewood-Richardson coefficient: the number of times $ R$ appears in $V_{ t }^{ \otimes 2 }$ when this is decomposed into irreps of the diagonal $U(N)$. Thus $ \cP (  t , [2] , R )  $ and $ \cP (  t , [ 1,1] , R ) $  are plethystic refinements of the Littlewood-Richardson coefficients. It turns out that the orientifold projection map can be expressed in terms of the plethysm coefficients $ \cP ( t , [2] , R  ) $ and $ \cP ( t ,  [1,1] , R  )$. A combinatorial rule for finding these coefficients was given in \cite{Carre1995}, refining the Littlewood-Richardson rule by replacing the standard Littlewood-Richardson tableaux with Yamanouchi domino tableaux. The derivation of this connection between the orientifold operation of string theory and the plethysm coefficients is the first main result of this paper.
		 
	It has been recognized for a while that connections between the combinatorics of words and the classification of gauge invariant operators form an important pillar of gauge-string duality \cite{Sundborg2000,Polyakov2001,BDHO07}. This has seen the application of Polya  theory in the study of the thermodynamics of $N=4$ SYM theory. Another aspect of word combinatorics,  underlying the structure of counting functions for gauge invariants in  general quiver gauge theories, was highlighted in \cite{Mattioli2014}.  These were free monoids of  words, which are sets of words obtained by multiplying a few generating letters in arbitrary order without assuming any commutativity of the multiplication. Words can be composed by concatenation, giving rise to a monoid structure. There is no explicit mention of cyclicity in these free monoids, which makes it a little surprising that these have anything to do with counting gauge invariants, which are traces of one kind or another. Nevertheless the counting of gauge invariants built from two matrices $X, Y $ at large $N$ has an interesting connection with the free monoid generated by two letters $x, y$.  The key to understanding this relation is to think about the organisation of traces of two letters according to the number of periods. For example Tr$( XYXY )$ has $2$ periods, while Tr$(X^2Y^2)$ has only $1$. Figure \ref{figure: U(N) diagram} gives a glimpse of the importance of aperiodic single traces. They lie at the apex of this diagram. To the right of the diagram, are multi-traces of aperiodic traces. To the left, are single traces of all periodicities. At the bottom of the diagram is the full space of gauge-invariant operators made from two matrices. As we explain in section \ref{section: structure of the space}, aperiodic multi-traces (appearing in the right box of the figure) are in one to one correspondence with the free monoid generated by two letters $x,y$. This builds on known results concerning Lyndon words \cite{Lyndon54}, which play an important role in the field of ``combinatorics on words'',  an area with diverse applications in mathematics \cite{wiki-words,Lothaire1983combinatorics}. As we  further explain, there are counting functions for each of these boxes, and there are relations between them, involving the plethystic exponential and the M{\"o}bius map. The plethystic exponential has been emphasised and its applications in many problems of counting chiral operators in supersymmetric gauge theories have been developed in \cite{Benvenuti2006} and subsequent literature. 

	The second main result of this paper is to develop the analogous picture for the counting of $SO(N)$ gauge invariant operators made from two matrices, i.e. the free field quarter-BPS sector of $N=4$ SYM. This involves defining an analogue of the notion of Lyndon words, appropriate for mesonic operators in the large $N$ limit of $SO(N)$ groups, which we call {\it orthogonal Lyndon words}. These are defined by a minimally periodic condition which replaces the aperiodic condition on traces for $U(N)$ theories. Figure \ref{figure: SO(N) diagram} shows the $SO(N)$ analogue: the minimally periodic words are at the apex, and there are maps leading to the counting of all the multi-traces. As we will see, the minimally periodic words can have either one or two periods. Consequently, there is a different organization of $SO(N)$ two-matrix multi-traces which respects the number of periods. This is shown in Figure \ref{figure: SO(N) periodicity diagram}.
	
	Alongside demonstrating the structure of the space of two-matrix multi-traces, the diagrams in figures \ref{figure: U(N) diagram}, \ref{figure: SO(N) diagram} and \ref{figure: SO(N) periodicity diagram} give different forms for the generating function of the quarter-BPS sector. For the $U(N)$ theory this function is already well known \cite{Bhattacharyya:2008rb}, and can also be written as the sum of squares of Littlewood-Richardson coefficients. The $SO(N)$ generating function has been given previously \cite{Kemp2014} as a linear sum of Littlewood-Richardson coefficients, but to the best of our knowledge, the explicit expression we derive here is a new mathematical result, of interest to mathematicians \cite{Willenbring07} as well as physicists.

	The paper is organised as follows. Section \ref{section: notation} gives a brief summary of our notation and conventions. In particular it introduces the wreath product group $S_n[S_2]$, which is one of the key mathematical structures in analysing $SO(N)$ invariants. In section \ref{section: Construction and counting of operators} we explicitly construct a Fourier basis for the quarter-BPS sector. This is done by studying the group invariances of the different methods of contracting the $SO(N)$ indices and then using Young diagrams to build operators that align with these invariances. The two $SO(N)$ invariant tensors $\delta_{ij}$ and $\varepsilon_{i_1 \ldots i_N}$ lead to two types of operators, which we call mesonic and baryonic. We also review the quarter-BPS sector of the $U(N)$ theory.
	
	In section \ref{section: Z2 quotient} we specialise these operators to the half-BPS sector and look in detail at how Young diagram operators in the $U(N)$ theory behave when projected to the $SO(N)$ theory. This leads to surprising connections with the combinatorics of domino tableaux.
	
	In sections \ref{section:U(N) results} and \ref{section: Generating function at infinite N}, we study the vector spaces spanned by $U(N)$ traces (multi-traces of two arbitrary complex matrices) and $SO(N)$ traces (multi-traces of two anti-symmetric complex matrices) respectively in the large $N$ limit. In particular we look at how the structure of these spaces reflects the factorisation of multi-traces into single traces and the classification of traces by the number of periods/repetitions. We also study how this structure is exhibited in the respective Hilbert series. This leads to many different counting formulae, and an associated list of number sequences is given in Appendix \ref{section: sequences}. 
	
	Having studied the half-BPS projection in section \ref{section: Z2 quotient}, we proceed to the quarter-BPS case in section \ref{section: quarter-BPS projection}. Much of the difficulty here is in finding a suitable labelling set for generic 2-matrix multi-traces. For the $U(N)$ theory this set is provided by partitions labelled by Lyndon words, while for $SO(N)$, the partitions are labelled by orthogonal Lyndon words. The projection coefficients can be expressed as a sum over these labelling sets, where the summand is given in terms of restricted characters. Finding a combinatoric interpretation of the projection coefficients, generalising the domino combinatorics of the half-BPS case, is an interesting problem for the future.
	
	Correlators of the mesonic operators have been studied before in \cite{Kemp2014}. In section \ref{section: correlators} we complete the quarter-BPS picture by giving the correlator of baryonic operators. We give two methods for this calculation. The first uses Schur-Weyl duality to connect the baryonic calculation with the mesonic result. The second is given in appendix \ref{section: alternative baryonic correlator}, and goes via a nice generalisation of the result \eqref{mesonic Jucys-Murphy result}.
	
	It is well known that invariants of the symplectic and orthogonal groups are connected. In section \ref{section: Symplectic gauge group} we repeat all of the above, but with a symplectic gauge group instead of an orthogonal one. All the results follow along similar lines, and in many cases are entirely identical.
	
	Finally, section \ref{section: Outlook} gives a brief overview of questions arising from and  future research directions related to the results of this paper.

		\section{Notation and conventions}
		\label{section: notation}
		
		In this paper, we will make extensive use of the symmetric group $S_n$, made up of permutations of $n$ distinct objects, usually taken to be $ \{ 1 , 2 , \cdots  , n \} $. Permutations $ \sigma \in S_n$ are maps  $ \sigma : \{ 1 , 2 , \cdots , n \} \mapsto \{  1, 2, \cdots , n \} $ which are one to one and invertible. The image of $ i $ is denoted $ \sigma (i ) $ and the product is defined by $( \sigma \tau ) (i) = \tau ( \sigma (i))$. The group algebra $ \mC ( S_n ) $ is the space of linear combinations $ \sum_{ \sigma \in S_n } a_{ \sigma } \sigma $ where $ a_{ \sigma } \in \mC $.  The product on this space is inherited from the product on the group. We define $(-1)^\sigma = $ sgn$(\sigma)$ to be the sign of the permutation $\sigma$.
		
		Conjugacy classes in $S_n$ are labelled by partitions $p$ of $n$, for which we use the standard notation $p \vdash n$. We write partitions in two distinct ways, depending on which is more suitable for the situation. Firstly, we write $p = [\lambda_1, \lambda_2, \ldots ]$, where the $\lambda_i$ are just the components of $p$ in (weakly) decreasing order. Secondly, we use $p = (1^{p_1}, 2^{p_2}, 3^{p_3}, \ldots )$, where $p_i$ is the multiplicity of $i$ as a component of $p$. So $p_1$ is the number of $\lambda$s equal to 1, $p_2$ is the number of $\lambda$s equal to 2, and similarly for $p_3,p_4$ etc.  When speaking about  partitions in general (so for example when considering the set of all partitions), we normally use multiplicities, but when giving specific examples of partitions, we will typically work with components. To avoid confusion between the two notations, we will use Greek letters for the components of a partition and Latin ones for the multiplicities. Since we use the multiplicities more often for named partitions, we will generally give partitions Latin names ($p$, $q$, etc).
		
		As an example of the two notations, the partition $p = [5,3,3,2,1] \vdash 14$ has components $\lambda_1 = 5$, \ldots, $\lambda_5 = 1$ and multiplicities $p_1 = p_2 = p_5 = 1$ and $p_3 = 2$.
		
		We denote the sum of a partition $p$ by
		\begin{equation}
		|p| = \sum_i \lambda_i = \sum_i i p_i = n
		\nonumber
		\end{equation}
		and the number of components by
		\begin{equation}
		l(p) = \# \text{ of non-zero } \lambda_i = \sum_i p_i
		\nonumber
		\end{equation}
		Since all permutations of cycle type $p$ have the same sign, we define the sign of a partition to be the sign of any permutation with that cycle type. Explicitly
		\begin{equation}
		(-1)^p = \prod_i (-1)^{\lambda_i + 1} = \prod_{i \text{ even}} (-1)^{p_i}
		\nonumber
		\end{equation}
		Given two partitions $p \vdash n$ and $q \vdash m$, there are two ways of `adding' them together to create a partition of $n+m$. Firstly, we can add the components together, which we denote by $p+q$. So given $q = [\mu_1, \mu_2, \ldots ]$, we have $p+q = [\lambda_1+\mu_1 , \lambda_2 + \mu_2, \ldots ]$. Secondly, we can add the multiplicities together, which we denote by $p \cup q$. So $p \cup q = (1^{p_1+q_1}, 2^{p_2 + q_2}, \ldots )$. Intuitively, $+$ corresponds to concatenating Young diagram left to right, while $\cup$ concatenates them top to bottom. This notation was used in \cite{Macdonald1995}.
		
		It will be useful to define the partitions $2p = p + p$, $3p = p + p + p$ and so on. In terms of components and multiplicities, $kp = [k \lambda_1 , k \lambda_2 , \ldots ] = (k^{p_1}, (2k)^{p_2}, \ldots )$.
		
		An important quantity for $p \vdash n$ is given by
		\begin{equation}
		z_p = \prod_i i^{p_i} p_i !
		\label{z_p}
		\end{equation}
		For $\sigma \in S_n$ of cycle type $p$, $z_p$ gives the size of the centraliser of $\sigma$, that is the subgroup of $S_n$ that commutes with $\sigma$. Using the orbit-stabiliser theorem \cite{cameron1999permutation} then tells us that the size of the conjugacy class (number of elements in $S_n$ with cycle type $p$) is $\frac{n!}{z_p}$.
		
		The definition \eqref{z_p} interacts nicely with the definition of $kp$
		\begin{equation}
		z_{kp} = k^{l(p)} z_p
		\label{z_kp}
		\end{equation}
		As is standard, we define the number of distinct partitions of $n$ to be $p(n)$.
		
		For a partition $p \vdash n$, we denote the conjugate (transposed) partition by $p^c$. The operations $\cup$ and $+$ are conjugate to each other:
		\begin{equation}
		(p+q)^c = p^c \cup q^c
		\label{conjugation of partition addition}
		\end{equation}
		The irreducible representations (irreps) of $S_n$ are also labelled by partitions. For various purposes, it is useful to think of these visually in terms of Young diagrams. These are arrangements of boxes such that the number of boxes in each row corresponds to the components of the partition. So for example the partition $R = [4,4,2]$ has the Young diagram
		\begin{equation}
		R = \ydiagram{4,4,2}
		\nonumber
		\end{equation}		
		We will use the terms Young diagram and partition interchangeably, and we denote them in the same way, $R \vdash n$. 
		
		It is well known that $S_n$ representations are real and that the representation space can be given an inner product so as to make them orthogonal. We denote the dimension of $R$ by $d_R$.
		
		The matrix representatives of group or group algebra elements are denoted by $D^R ( \sigma )$. These matrices satisfy the orthogonality relations
		\begin{equation}
		\sum_{\sigma \in S_n} D^R_{ij} (\sigma) D^S_{kl} (\sigma^{-1}) = \frac{n!}{d_R} \delta^{RS} \delta_{il} \delta_{jk}
		\label{orthogonality of matrix elements}
		\end{equation}
		For each irrep $R$ we can define a projector in $\mathbb{C}(S_n)$
		\begin{equation}
		P_R = \frac{d_R}{n!} \sum_{\sigma \in S_n} \chi_R ( \sigma ) \sigma
		\label{projector}
		\end{equation}
		These satisfy the multiplication identity
		\begin{equation}
		P_R P_S = \delta_{RS} P_R
		\nonumber
		\end{equation}
		They are represented by the identity matrix in the corresponding irrep and the zero matrix in all other irreps
		\begin{equation}
		D^S (P_R) = \delta_{RS}
		\nonumber
		\end{equation}
		Therefore in any direct sum representation $P_R$ projects to the $R$ subspace (or subspaces).
		
		The character of a permutation in an irrep $R$ depends only on its cycle type, so taking $\sigma \in S_n$ to be of cycle type $p$ we have
		\begin{equation}
		\chi_R ( p ) = \chi_R (\sigma)
		\nonumber
		\end{equation}		
		The definition of $z_p$ \eqref{z_p} allows us to neatly write down the orthogonality relations for characters. They are
		\begin{equation}
		\sum_{p \vdash n} \frac{1}{z_p} \chi_R (p) \chi_S (p) = \delta_{RS} \qquad \qquad \sum_{R \vdash n} \chi_R (p) \chi_R (q) = z_p \delta_{pq}
		\label{character orthogonality}
		\end{equation}
		For a Young diagram $R \vdash n$, the irrep $R^c$ is isomorphic to the tensor product of $R$ with the sign (anti-symmetric) representation, so
		\begin{equation}
		\chi_{R^c} (p) = (-1)^p \chi_R (p)
		\end{equation}

		\subsection{The wreath product $S_n[S_2]$}
		\label{section: Sp[S2]}
		
\begin{figure}
			\centering
			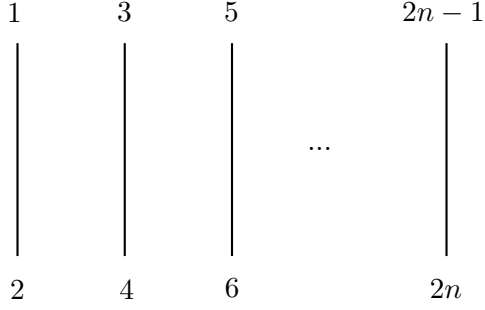
			\caption{The set on which $S_n[S_2]$ acts. A group element can permute the $n$ pairs, while switching or not switching each individual pair}
			\label{Sn[S2]}
		\end{figure}

		We will have particular use for a certain subgroup of $S_{2n}$, called $S_n[S_2]$. This can be thought of as the permutations of $n$ pairs of objects. Each pair can be individually switched, and the $n$ pairs can be permuted among themselves, so we have $|S_n[S_2]|=2^n n!$. By labelling the $2n$ objects as $\{1,2\}, \{3,4\}, \ldots ,\{2n-1,2n\}$, where the brackets denote the pairings, we see that $S_n[S_2]$ is a subgroup of $S_{2n}$ as claimed. It is simple to check that it is the centraliser of the permutation $(1,2)(3,4) \ldots (2n-1,2n)$. Figure \ref{Sn[S2]} shows the set on which $S_n[S_2]$ acts.		
		
		More formally, $S_n[S_2]$ is defined as the wreath product of $S_n$ with $S_2$, or equivalently as the semi-direct product of $S_n$ with $(S_2)^n$, where the $S_n$ acts on $(S_2)^n$ by permutation of the factors.
		
		As with \eqref{projector}, we can define projection operators onto irreps of $S_n[S_2]$. For $[r]$ an irrep (we use square brackets to denote that this is an irrep of $S_n[S_2]$ rather than $S_{2n}$) we define
		\begin{equation}
		P_{[r]} = \frac{d_{[r]}}{2^n n!} \sum_{\sigma \in S_n[S_2]} \chi_{[r]} (\sigma) \sigma 
		\nonumber
		\end{equation}
		There are two one-dimensional irreps of $S_n[S_2]$ that we will use. The trivial (symmetric) representation takes $\sigma$ to 1, and the anti-symmetric (sign) representation takes $\sigma$ to $(-1)^\sigma$, which is defined by considering $\sigma \in S_n[S_2] \leq S_{2n}$. We denote these two representations by $[S]_n$ and $[A]_n$ respectively. We will sometimes write these without the subscript when it is clear which $n$ we are referring to. The projectors of $[S]_n$ and $[A]_n$ are given by
		\begin{equation}
		P_{[S]_n} = \frac{1}{2^n n!} \sum_{\sigma \in S_n[S_2]} \sigma \qquad \qquad P_{[A]_n} = \frac{1}{2^n n!} \sum_{\sigma \in S_n[S_2]} (-1)^\sigma \sigma
		\label{Sn[S2] projectors} 
		\end{equation} 
		
		\subsection{Tensor space}
		\label{section: tensor space}
		
		Let $V$ be the (complex) carrier space for the $N$-dimensional fundamental representation of $U(N)$ and $SO(N)$. Then $V$ has two distinct inner products corresponding to the two gauge groups. The $U(N)$ inner product is a hermitian form, and therefore the dual space to $V$ (with this inner product) is $V^*$, which is defined as the conjugate space to $V$. These two spaces are non-isomorphic. The $SO(N)$ inner product is a symmetric form, and therefore the dual space is just $V$ itself. Hence $U(N)$ (or $\mathfrak{u}(N)$) matrices $U$ acting on $V$ have indices $U^i_j$ while $SO(N)$ (or $\mathfrak{so}(N)$) matrices $O$ acting on $V$ can be given indices $O^{ij}$. In general, upstairs indices will correspond to objects in $V$ while downstairs indices will correspond to objects in the conjugate space $V^*$.
		
		We will be constructing invariants from multiple copies of two matrices, $X$ and $Y$. Depending on context, these will belong to $\mathfrak{u}(N)$ or $\mathfrak{so}(N)$. In any particular instance, it should be clear which is the case from the index structure. For compactness we will use the notation
		\begin{align}
		\left( X^{\otimes n} Y^{\otimes m} \right)^I_J & = X^{i_1}_{j_1} X^{i_2}_{j_2} \ldots X^{i_n}_{j_n} Y^{i_{n+1}}_{j_{n+1}} Y^{i_{n+2}}_{j_{n+2}} \ldots Y^{i_{n+m}}_{j_{n+m}} \nonumber \\
		\left( X^{\otimes n}  Y^{\otimes m} \right)^I & = X^{i_1 i_2} \ldots X^{i_{2n-1} i_{2n}} Y^{i_{2n+1} i_{2n+2}} \ldots Y^{i_{2n+2m-1} i_{2n+2m}}
		\nonumber
		\end{align}
		for $\mathfrak{u}(N)$ and $\mathfrak{so}(N)$ respectively.
		
		$S_n$ acts on $V^{\otimes n}$ by permutation of the factors. Explicitly, on a pure product state we have
		\begin{equation}
		\sigma \left( v_1 \otimes v_2 \otimes \ldots \otimes v_n \right) = v_{\sigma (1)} \otimes v_{\sigma(2)} \otimes \ldots \otimes v_{\sigma(n)} 
		\nonumber
		\end{equation}
		and the action is extended linearly to the whole of $V^{\otimes n}$. In index notation, it is simple to check that this corresponds to
		\begin{equation}
		\sigma^I_J = \sigma^{i_1 i_2 \ldots i_n}_{j_1 j_2 \ldots j_n} = \delta^{i_1}_{j_{\sigma(1)}} \delta^{i_2}_{j_{\sigma(2)}} \ldots \delta^{i_n}_{j_{\sigma(n)}}
		\label{permutations acting on tensor space}
		\end{equation}
		This definition can then be extended linearly to the symmetric algebra $\mathbb{C}(S_{n})$. Note that since we use the convention $(\sigma \tau) (i) = \tau( \sigma (i))$, we have $\sigma^I_J \tau^J_K = (\sigma \tau)^I_K$.
		
		For $U(N)$, tracing these permutations with the tensor product of $X$ over $V^{\otimes n}$ provides a nice way of of generating multi-traces of $X$. Consider $\sigma \in S_n$ with just a single cycle, e.g.  $\sigma = (1,2,3, \ldots ,n)$. Then
		\begin{equation}
		\text{Tr} \left( \sigma X^{\otimes n} \right) = \sigma^I_J \left( X^{\otimes n} \right)^J_I = X^{i_1}_{i_2} X^{i_2}_{i_3} \ldots X^{i_{n-1}}_{i_n} X^{i_n}_{i_1} = \left( X^n \right)^i_i = \text{Tr} X^n
		\nonumber
		\end{equation}
		By doing this for each cycle, we see that if $\sigma$ has cycle type $p$, we have
		\begin{equation}
		\text{Tr} \left( \sigma X^{\otimes n} \right) = \prod_i \left( \text{Tr} X^i \right)^{p_i}
		\label{U(N) permutations to traces}
		\end{equation}
		The equivalent statement for $SO(N)$ is
		\begin{equation}
		X^{i_1 i_{\sigma(1)}} X^{i_2 i_{\sigma(2)}} \ldots X^{i_n i_{\sigma(n)}} = \prod_i \left( \text{Tr} X^i \right)^{p_i}
		\label{SO(N) permutations to traces}
		\end{equation}
		Using \eqref{permutations acting on tensor space}, we can let permutations in $S_{2n}$ act on $\left( X^{\otimes n} \right)^I$. The relation between this action and the $S_n$ permutations in \eqref{SO(N) permutations to traces} is explained in section \ref{section: SO(N) permutations to traces}.
		
		Note that in \eqref{U(N) permutations to traces} we have used an unadorned trace to mean traces over two different spaces. On the left-hand side the trace is over $V^{\otimes n}$ while on the right it is over $V$. In this paper we consider traces over various different vector spaces, so wherever there is the potential for confusion we will add a subscript of the appropriate vector space (e.g. $\text{Tr}_R$ for a trace over the irrep $R$). Traces over $V$ and $V^{\otimes n}$ will generally be left unadorned.
		
		Since $S_n[S_2]$ is a subgroup of $S_{2n}$ it acts on $V^{\otimes 2n}$. The properties of this action are easiest to see if we label the indices slightly differently. Consider $A \in V^{\otimes 2n}$ with the indices labelled as follows
		\begin{equation}
		A^I = A^{i_{1,1} i_{1,2} i_{2,1} i_{2,2} \ldots i_{n,1} i_{n,2}}
		\nonumber
		\end{equation}
		Then the $S_n$ part of $S_n[S_2]$ act on the first index ($j$ in $i_{j,k}$) while the $n$ copies of $S_2$ acts on the second index ($k$). Therefore if $M$ is a symmetric (anti-symmetric) matrix, $\left( M^{\otimes n} \right)^I$ will be invariant (anti-invariant) under the action of $S_n[S_2]$. More formally, it will transform in the $[S]_n$ ($[A]_n$) representation of $S_n[S_2]$.

		\section{Construction and counting of  $SO(N)$ quarter-BPS operators}
		\label{section: Construction and counting of operators}
		
		$\mathcal{N}=4$ super Yang-Mills contains 3 complex (or 6 real) scalar fields in the adjoint of the Lie algebra of the gauge group. For $\mathfrak{u}(N)$, the adjoint is the real vector space of anti-Hermitian matrices, so the complex scalar fields are arbitrary complex matrices. In contrast, $\mathfrak{so}(N)$ contains anti-symmetric matrices, which is a linear condition under complexification, and therefore the orthogonal complex scalar fields are anti-symmetric complex matrices. The quarter-BPS sector of the theory consists of gauge-invariant combinations of two of these three fields, while the half-BPS sectors uses just one.
		
		With a $U(N)$ gauge group, both these sectors have been studied extensively for infinite $N$ and at finite $N$. A basis for the half-BPS sector was constructed in \cite{Corley2002} and used to find exact correlators of giant gravitons in the AdS dual theory. This basis was labelled by Young diagrams, and has two important properties that we will look to emulate in the $SO(N)$ theory. Firstly it is orthogonal under the two-point function, and secondly it allows a simple description of the finite $N$ cut-off. Going from the infinite $N$ theory to the finite $N$ corresponds to the vanishing of those operators whose Young diagrams have more than $N$ rows.
		
		This basis was extended to the quarter-BPS sector, via various methods, in \cite{Kimura:2007wy, Brown:2007xh, Bhattacharyya:2008rb, Bhattacharyya:2008xy, Brown:2008ij}. Similarly to the half-BPS basis, these bases are labelled by Young diagrams, are orthogonal under the two-point function, and have a nice description of the finite $N$ cut-off.
		
		Considering an $SO(N)$ gauge group instead, there is a similar story of bases labelled by Young diagrams. The half-BPS sector was studied in \cite{Caputa2013,Caputa2013a}, and the quarter-BPS in \cite{Kemp2014,Kemp1406}. In this section we construct the same operators as found in these papers. Our perspective is focused on the description of gauge invariant operators in terms of permutations and their equivalences, where the key features of the construction are described as consequences of these equivalences. Properties of correlators are developed at a second stage of the discussion (see section \ref{section: correlators}). This perspective is  close to that of \cite{Brown:2007xh,Brown:2008ij} and also, in the context where $SO(N)$ appears as a flavour group, in \cite{Kimura:2016bzo}.

		We start by reviewing the approach taken in \cite{Caputa2013,Caputa2013a} and how that differs from the arguments presented here. We then construct our basis, and in the process find the classification into mesonic and baryonic operators. As a by-product of constructing the basis, we obtain an expression for the counting of gauge-invariant operators. A quick review of the $U(N)$ quarter-BPS operators allows us to make some nice connections between the permutation algebras relevant for the construction of $SO(N)$ and $U(N)$ operators.

		\subsection{Half-BPS sector}
		
		In \cite{Caputa2013}, the starting point was the Wick contractions of the scalar field $X^{ij}$. The authors noticed that these could be described by the projector onto the symmetric representation of $S_n[S_2]$.
		\begin{equation}
		\left\langle \left( X^{\otimes n} \right)^I \left( X^{\otimes n} \right)_J \right\rangle = 2^n n! \left( P_{[S]_n} \right)^I_J
		\nonumber
		\end{equation}
		where $X_{ij}$ is the conjugate matrix $\left( X^* \right)^{ij}$. This allowed them to construct operators $\left( T_R \right)^I = \left( P_R \right)^I_J \left( X^{\otimes n} \right)^J$ labelled by $R \vdash 2n$ that diagonalised the Wick contractions. Finally they contracted the indices to create gauge-invariant operators that (by construction) diagonalised the inner product.
		\begin{align}
		\mathcal{O}_R = \left( T_R \right)^{i_1 i_2 i_2 i_1 i_3 i_4 i_4 i_3 \ldots i_{n-1} i_n i_n i_{n-1}}
		\label{Robert's operators}
		\end{align}
		Although the $T_R$ were labelled by any Young diagram $R \vdash 2n$, this contraction was found to be identically zero for any $R$ not constructed from 2$\times$2 blocks (note that this implies $n$ is even).
		
		The $\mathcal{O}_R$ (for $R$ constructed from 2$\times$2 blocks) form a basis of the space of gauge-invariant operators. There are $p \left( \frac{n}{2} \right)$ such $R$. The generating function for these numbers is known, and is given by
		\begin{equation}
		\prod_{n=1}^\infty \frac{1}{(1-x^{2n})}
		\nonumber
		\end{equation}
		To compare with the approach developed here, we rewrite \eqref{Robert's operators} in a different form
		\begin{align}
		\mathcal{O} = C_I \beta^I_J \left( X^{\otimes n} \right)^J
		\label{Robert's operators 2}
		\end{align}
		where $C_I$, which we call a contractor, is given by
		\begin{equation}
		C_I = \delta_{i_1 i_2} \delta_{i_3 i_4} \ldots \delta_{i_{2n-1} i_{2n}}
		\nonumber
		\end{equation}
		and $\beta \in \mathbb{C}(S_{2n})$ is given by
		\begin{equation}
		\beta = \rho P_R \qquad , \qquad \rho = (1,3)(5,7) \ldots (2n-3,2n-1)
		\label{index rearrangement}
		\end{equation}
		The permutation $\rho$ is needed to change the contraction pattern of $C_I$, which contracts indices in the pairs $(1,2),(3,4), \ldots ,(2n-1,2n)$, to the contraction pattern in \eqref{Robert's operators}, which has pairs $(1,4),(2,3),(5,8),(6,7), \ldots ,(2n-3,2n),(2n-2,2n-1)$.
		
		We will approach the problem in the opposite direction to the above, and also generalise it to the 2-matrix setting. We start with the expression \eqref{Robert's operators 2}, but with $\beta \in \mathbb{C}(S_{2n})$ arbitrary and study the invariance properties of $\mathcal{O}$ as a function of $\beta$. Since $C_I$ is invariant under $S_n[S_2]$ permutations and $X^{\otimes n}$ is anti-invariant under $S_n[S_2]$ permutations, $\mathcal{O}$ remains unchanged under the transformation
		\begin{equation}
		\beta \mapsto (-1)^\gamma \alpha \beta \gamma^{-1} \qquad , \qquad \alpha, \gamma \in S_n[S_2]
		\label{half-bps invariance 0}
		\end{equation}
		We consider the subspace of $\mathbb{C}( S_{2n} )$ defined by those elements which are invariant under \eqref{half-bps invariance 0} and derive a basis for this subspace, labelled by the same Young diagrams as seen in \eqref{Robert's operators}. The basis of operators then follows by contracting the subspace basis, and the operators we define differ from \eqref{Robert's operators} only by a factor.

		\subsection{Quarter-BPS set-up}
		\label{section: Set-up}
		
		In the quarter-BPS sector we consider operators constructed from 2 complex anti-symmetric matrices $X^{ij}$ and $Y^{ij}$. The most general gauge-invariant operator constructed from $n$ copies of $X$ and $m$ copies of $Y$ is:
		\begin{align}
		\mathcal{O} & = C_{i_1 i_2 \ldots i_{2n} j_1 j_2 \ldots j_{2m}}  X^{i_1 i_2} X^{i_3 i_4} \ldots X^{i_{2n-1} i_{2n}} Y^{j_1 j_2} Y^{j_3 j_4} \ldots Y^{j_{2m-1} j_{2m}} \nonumber \\
		& = C_{I} \left( X^{\otimes n} Y^{\otimes m} \right)^I 
		\label{general SO(N) operator}
		\end{align}
		where $C_I$ is constructed from $SO(N)$ invariant tensors. We have two such tensors to choose between, namely $\delta_{ij}$ and $\varepsilon_{i_1 i_2 \ldots i_N}$, and their tensor products. Since two $\varepsilon$s can be expressed as a sum of ($N$-fold tensor products of) $\delta$s, there are two linearly independent possibilities for $C_I$. Either it is made of $n+m$ $\delta$s or (if $N$ is even) $n+m-\frac{N}{2}$ $\delta$s and an $\varepsilon$. In analogy with $SU(N)$ terminology, we call these mesonic and baryonic operators respectively. 
		
		Among mesonic (or baryonic) operators, there are many different ways to arrange the indices on the $n+m$ $\delta$s. However, all the different arrangements are related by permutations. We saw an example of this already in \eqref{index rearrangement} where we introduced the permutation $\rho$ to change the contraction from one index arrangement to another.
		
\begin{figure}
			\centering
			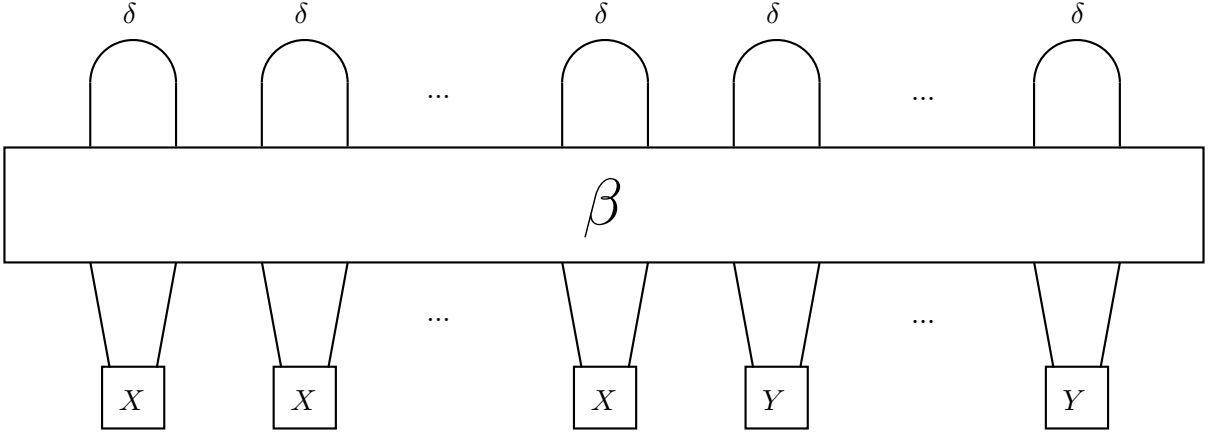
			\caption{A diagrammatic representation of the index contraction in a mesonic operator, where each line represents an index. There are $n$ $X$s and $m$ $Y$s, and $\beta \in \mathbb{C}(S_{2n+2m})$.}
			\label{figure: Mesonic contractor picture}
		\end{figure}

\begin{figure}
			\centering
			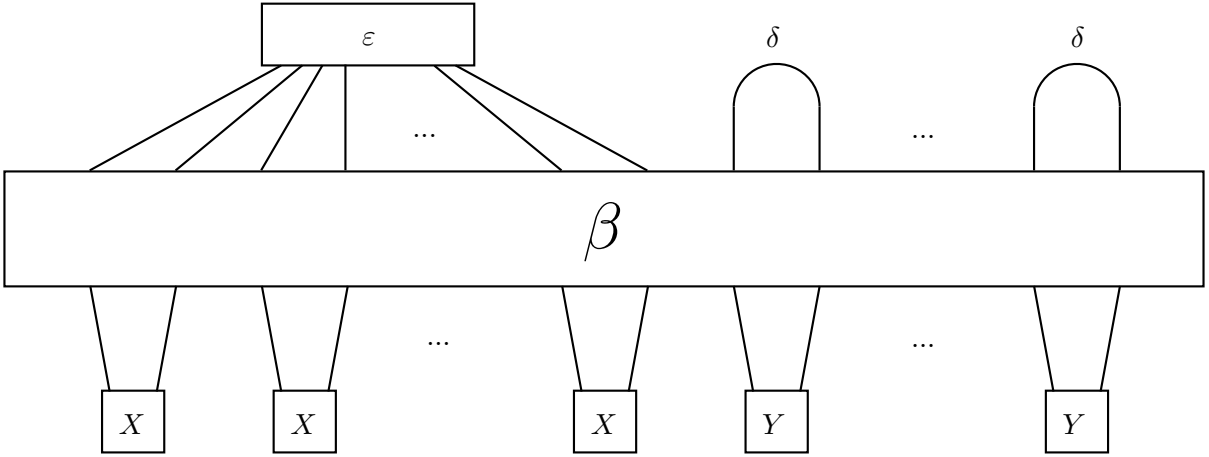
			\caption{A diagrammatic representation of the index contraction in a baryonic operator. The $\varepsilon$ vertex has $N$ legs and there are $n$ $X$s, $m$ $Y$s and $q$ $\left(  = n + m - \frac{N}{2} \right)$ $\delta$s. For convenience, this diagram shows $N=2n$, but in general this does not have to be the case.}
			\label{figure: Baryonic contractor picture}
		\end{figure}

		A mesonic contractor could be composed of a linear combination of all different index arrangements. Using permutations we can absorb all of these into a single element $\beta \in \mathbb{C}(S_{2n+2m})$ and a contractor with the standard index arrangement (defined below). The exact same process applies for the baryonic operators. We call the contractors with the standard index arrangement $C^{(\delta)}$ and $C^{(\varepsilon)}$ respectively. Explicitly, the mesonic operators are:
		\begin{align}
		\mathcal{O}^{(\delta)}_{\beta} & = C^{(\delta)}_I \beta^I_J \left( X^{\otimes n} Y^{\otimes m} \right)^J \nonumber \\ 
		& = \left( \delta_{i_1 i_2} \delta_{i_3 i_4} \ldots \delta_{i_{2n+2m-1} i_{2n+2m}} \right) \beta^{i_1 \ \ldots \ i_{2n+2m}}_{j_1 \ldots j_{2n} k_1 \ldots k_{2m}} \nonumber \\
		& \hspace{6.5cm}  ( X^{j_1 j_2} \ldots X^{j_{2n-1} j_{2n}} ) ( Y^{k_1 k_2} \ldots Y^{k_{2m-1} k_{2m}} ) 
		\label{mesonic operator}
		\end{align}
		Figure \ref{figure: Mesonic contractor picture} shows a diagrammatic representation of this contraction. The baryonic operators are:
		\begin{align}
		\mathcal{O}^{(\varepsilon)}_{\beta} & = C^{(\varepsilon)}_I \beta^I_J \left( X^{\otimes n} Y^{\otimes m} \right)^J \nonumber \\ & = \left( \varepsilon_{i_1 \ldots i_N} \delta_{i_{N+1} i_{N+2}}  \ldots \delta_{i_{2n+2m-1} i_{2n+2m}} \right) \beta^{i_1 \ \ldots \ i_{2n+2m}}_{j_1 \ldots j_{2n} k_1 \ldots k_{2m}} \nonumber \\
		& \hspace{6.5cm} ( X^{j_1 j_2} \ldots X^{j_{2n-1} j_{2n}} ) ( Y^{k_1 k_2} \ldots Y^{k_{2m-1} k_{2m}} ) 
		\label{baryonic operator}
		\end{align}
		Figure \ref{figure: Baryonic contractor picture} shows this contraction.
			
		The most general gauge-invariant operator is then a sum of \eqref{mesonic operator} and \eqref{baryonic operator} (clearly $\beta$ will in general differ between the two). We now look in more detail at the two types of operator.

		\subsection{Mesonic operators}
		\label{section: Mesonic operators}
		
		\subsubsection{Mesonic operators as multi-traces}
		\label{section: mesonic = multi-trace}
		
		Examine the definition \eqref{mesonic operator} with $\beta = \sigma \in S_{2n+2m}$ (so we consider a single permutation rather than a linear combination). We have
		\begin{align}
		\mathcal{O}_\sigma = \delta_{i_1 i_2} \delta_{i_3 i_4} \ldots \delta_{i_{2n+2m-1} i_{2n+2m}} \delta^{i_1}_{j_{\sigma(1)}} \delta^{i_2}_{j_{\sigma(2)}} \ldots \delta^{i_{2n+2m}}_{j_{\sigma(2n+2m)}}  X^{j_1 j_2} \ldots Y^{j_{2n+2m-1} j_{2n+2m}}  
		\nonumber
		\end{align}
		By evaluating the $\delta$s and then rearranging and renaming indices this becomes
		\begin{equation}
		\mathcal{O}_\sigma = \pm X^{i_1 i_{\tau(1)}} \ldots X^{i_n i_{\tau(n)}} Y^{i_{n+1} i_{\tau(n+1)}} \ldots Y^{i_{n+m} i_{\tau(n+m)}}
		\label{mesonic = multi-trace}
		\end{equation}
		where $\tau \in S_{n+m}$ is a permutation related to $\sigma$ in a non-trivial way that we will study in detail in section \ref{section: 2-matrix permutations to traces}. The $\pm$ arises from the anti-symmetry of $X$ and $Y$, since during the rearranging of indices to arrive at this expression, some of the $X$s and $Y$s may be transposed.
		
		We can see that \eqref{mesonic = multi-trace} is a multi-trace of $X$ and $Y$ (the one-matrix version is given in \eqref{SO(N) permutations to traces}). The more general form \eqref{mesonic operator} is therefore a linear combination of multi-traces. Clearly any multi-trace can be generated from \eqref{mesonic = multi-trace} given the appropriate $\tau$, and we prove in \eqref{SO(N) to U(N) contraction} that any $\tau \in S_{n+m}$ can be induced from the appropriate $\sigma$. Therefore we can generate all multi-traces of $X$ and $Y$ from the formula \eqref{mesonic operator}.
		
		Thus the space of mesonic operators is exactly the space of multi-traces. An obvious basis is therefore just the distinct multi-traces. In the large $N$ limit, this basis is orthogonal under the two-point function. However at finite $N$ this is no longer the case, and we will find an exactly orthogonal basis at all $N$.

		\subsubsection{Construction and counting}
		
		Since $X$ and $Y$ are anti-symmetric, $X^{\otimes n} \otimes Y^{\otimes m}$ is anti-invariant under $S_n[S_2] \times S_m[S_2]$ permutations, while the contractor $C^{(\delta)}$ is invariant under $S_{n+m}[S_2]$. Combining this anti-invariance and invariance means that $\mathcal{O}_{\beta}$ is invariant under
		\begin{equation}
		\beta \mapsto \left( -1 \right)^{\gamma} \alpha \beta \gamma^{-1} \qquad  \qquad \alpha \in S_{n+m}[S_2] \ , \ \gamma \in S_n[S_2] \times S_m[S_2]
		\label{SO(N) group algebra invariance}
		\end{equation}
		We have used $\gamma^{-1}$ rather than $\gamma$ so that this forms an action of the direct product group $S_{n+m}[S_2] \times \left( S_n[S_2] \times S_m[S_2] \right)$, but the statement would have been entirely equivalent had we used just $\gamma$ instead. For simplicity, we will use $\gamma$ for the remainder of this section. Using this invariance and the fact that $\mathcal{O}_\beta$ is linear in $\beta$, we have
		\begin{align}
		\mathcal{O}_\beta & = \frac{1}{2^{(n+m)}(n+m)!} \sum_{\alpha \in S_{n+m}[S_2]} \frac{1}{2^n n! 2^m m!} \sum_{\gamma \in S_n[S_2] \times S_m[S_2]} \mathcal{O}_{ \text{sgn}(\gamma) \alpha \beta \gamma} \nonumber \\
		& = \mathcal{O}_{\bar{\beta}} 
		\label{Mesonic operator labelled by invariant algebra element}
		\end{align}
		where
		\begin{equation}
		\bar{\beta} = \frac{1}{2^{2n+2m} (n+m)! n! m!} \sum_{\alpha \in S_{n+m}[S_2]} \sum_{\gamma \in S_n[S_2] \times S_m[S_2]} (-1)^\gamma \alpha \beta \gamma
		\label{averaging beta}
		\end{equation}
		The elements which are invariant under \eqref{SO(N) group algebra invariance} form a subspace of $\mathbb{C}(S_{2n+2m})$ that we call $\mathcal{A}^{SO}_{n,m}$. Since $\bar{\beta}$ is invariant under this transformation, we see that without loss of generality, mesonic operators are labelled by $\mathcal{A}^{SO}_{n,m}$ rather than the full algebra $\mathbb{C}(S_{2n+2m})$. Note that we have put $SO$ in the superscript rather than $SO(N)$ because this subspace depends only on the invariance \eqref{SO(N) group algebra invariance} and is therefore independent of $N$.
		
		It is well known that the group algebra $\mathbb{C}(S_n)$ is isomorphic to the algebra of complex functions on the group (with multiplication defined by convolution). Explicitly, given a function $f : S_n \rightarrow \mathbb{C}$, we can define a corresponding algebra element $\beta = \sum_{\sigma \in S_n} f ( \sigma ) \sigma$. Conversely, given an algebra element $\beta$, the coefficients in its linear expansion give us a function $f$.
		
		Under this isomorphism, $\mathcal{A}^{SO}_{n,m}$ maps to a subspace of the full space of functions. This subspace consists of functions satisfying the equivalent (anti-)invariance to \eqref{SO(N) group algebra invariance}, given by
		\begin{equation}
		f(\sigma) = (-1)^{\gamma} f(\alpha \sigma \gamma)
		\nonumber
		\end{equation}
		We call this space $\mathcal{F}_{n,m}$.
		
		To find a basis of gauge-invariant operators we want to find a basis for $\mathcal{A}^{SO}_{n,m}$. Since this is isomorphic to $\mathcal{F}_{n,m}$, we can equivalently find a basis for the function space.
		
		Given an arbitrary function $f$ on the group, we can project it to $\mathcal{F}_{n,m}$ by averaging over the double coset, just as we did in \eqref{averaging beta}. This mapping from an arbitrary function to an (anti-)invariant one is surjective since if $f$ is already (anti-)invariant, we have $\bar{f} = f$. Thus we can produce a spanning set for $\mathcal{F}_{n,m}$ by averaging a basis of ordinary functions.
		
		An obvious choice for a basis would be the functions $ \left\{ f_\sigma : \sigma \in S_{2n+2m} \right\}$, where $f_\sigma$ evaluates to $1$ on $\sigma$ and $0$ on any other permutation. These would lead to the basis of multi-traces that we already considered at the end of section \ref{section: mesonic = multi-trace}. Since we want an exactly orthogonal basis for the $\mathcal{F}_{n,m}$, we instead consider the matrix elements of the irreducible representations
		\begin{equation}
		D^T_{IJ} ( \sigma ) = \langle I, T | D^T ( \sigma ) | J,T \rangle
		\label{matrix elements of irreps}
		\end{equation}
		where $T \vdash 2n+2m$ is a Young diagram, and $\{ |T,I \rangle \}$, $\{ |T,J \rangle \}$ are two bases, not necessarily the same, for the carrier space of $T$. Clearly $I$ and $J$ must take $d_T$ different values each, but these values may have more structure than just the numbers $1,2, \ldots ,d_T$. Averaging \eqref{matrix elements of irreps} gives
		\begin{align*}
		\bar{D}^T_{IJ} ( \sigma ) & =  \frac{1}{ 2^{2n+2m} (n+m)! n!m! } \sum_{\alpha \in S_{n+m}[S_2]} \sum_{\gamma \in S_n[S_2] \times S_m[S_2]} (-1)^{\gamma} D^T_{IJ} ( \alpha \sigma \gamma) \\
		& = D^T_{IJ} \left( P_{[S]_{n+m}} \sigma P_{[A]_n \otimes [A]_m} \right) \\
		& = D^T_{IJ} \left( P_{[S]_{n+m}} \sigma \left[ P_{[A]_n} \otimes \mathbb{I}_{2m} \right]  \left[ \mathbb{I}_{2n} \otimes P_{[A]_m} \right] \right)
		\end{align*}
		where we have used the projectors defined in \eqref{Sn[S2] projectors}.

		Now we see that a convenient choice for the bases $\{|T,I \rangle \}$ and $\{ |T,J \rangle \}$ will be ones that align nicely with the projectors $P_{[S]_{n+m}}$ and $P_{[A]_n \otimes [A]_m}$ respectively. This is most easily done by aligning the bases with arbitrary representations of $S_{n+m}[S_2]$ and $S_n[S_2] \times S_m[S_2]$.
		
		By restricting the representation $T$ of $S_{2n+2m}$ to $S_{n+m}[S_2]$, we get the decomposition
		\[
		T = \bigoplus_{[t]} \nu_{T, [t]} [t]
		\]
		Where $[t]$ runs over the irreps of $S_{n+m}[S_2]$ and $[t]$ occurs with multiplicity $\nu_{T, [t]}$. So the new basis is
		\[
		\left| T, [t], \mu_{T, [t]}, I_{[t]} \right\rangle
		\]
		Where $1 \leq \mu_{T, [t]} \leq \nu_{T, [t]}$ counts which copy of $[t]$ we are in and $1 \leq I_{[t]} \leq $dim$[t]$ indexes the basis vectors of $[t]$.
		
		For $S_n[S_2] \times S_m[S_2]$, we first decompose the representation $T$ of $S_{2n+2m}$ into a direct sum of representations of $S_{2n} \times S_{2m}$
		\[
		T = \bigoplus_{R,S} g_{R,S;T} R \otimes S
		\]
		where $R$ runs over Young diagrams with $2n$ boxes, $S$ runs over Young diagrams with $2m$ boxes and the $g_{R,S;T}$ are the Littlewood-Richardson coefficients. Then $R$ and $S$ can be decomposed into $S_n[S_2]$ and $S_m[S_2]$ representations as we did above with $T$. Doing this, we get the basis
		\begin{align*}
		& \left| T, R, S, \lambda, [r], \mu_{R,[r]}, I_{[r]}, [s], \mu_{S,[s]}, I_{[s]} \right\rangle = \left| R, \lambda, [r], \mu_{R,[r]}, I_{[r]} \right\rangle \otimes \left| S, \lambda, [s], \mu_{S,[s]}, I_{[s]} \right\rangle
		\end{align*}
		where $1 \leq \lambda \leq g_{R,S;T}$ indexes which copy of $R \otimes S$ we are in inside $T$. More formally, $\lambda$ indexes the basis vectors of the multiplicity space for $R \otimes S$ inside $T$. Choosing this basis requires the use of an algebra defined in section \ref{section: U(N) construction}, so we delay reviewing this until section \ref{section: basis of multiplicity space}.
		
		Therefore we can construct a spanning set for $\mathcal{F}_{n,m}$ by considering the matrix elements of
		\[
		D^T \left( P_{[S]_{n+m}} \sigma  P_{[A]_n \otimes [A]_m} \right)
		\]
		with respect to the $S_{n+m}[S_2]$ basis on the left and the $S_n[S_2] \times S_m[S_2]$ basis on the right. Explicitly, they are
		\begin{align}
		& F_{T,[t],\mu_{T,[t]},I_{[t]},R,S,\lambda,[r],\mu_{R,[r]},I_{[r]}, [s], \mu_{S,[s]},I_{[s]}} (\sigma) \nonumber \\
		& \qquad = \left\langle T, [t], \mu_{T, [t]}, I_{[t]} \right| D^T \left( P_{[S]_{n+m}} \sigma  P_{[A]_n \otimes [A]_m} \right)  \nonumber \\
		& \hspace{6.5cm}  \Big( \left| R, \lambda, [r], \mu_{R,[r]}, I_{[r]} \right\rangle \otimes \left| S, \lambda, [s], \mu_{S,[s]}, I_{[s]} \right\rangle \Big)
		\label{anti-invariant functions}
		\end{align}
		It is a result from the representation theory of the wreath product \cite[Chapter VII.2]{Macdonald1995} that when an irrep $R$ of $S_{2n}$ is restricted to be a representation of $S_n[S_2]$, the completely symmetric representation of $S_n[S_2]$ appears in the direct sum decomposition if and only if $R$ has an even number of boxes in each row, and then it appears with multiplicity 1. By transposing the Young diagrams, the completely anti-symmetric representation of $S_n[S_2]$ appears in the direct sum decomposition if and only if $R$ has an even number of boxes in each column, and then it appears with multiplicity 1. Therefore, using the adapted bases, we can write the representatives of the projectors as
		\begin{align*}
		D^T \left( P_{[S]_{n+m}} \right) & = 
		\begin{cases}
		| T, [S] \rangle \langle T, [S] | & \text{if $T$ has rows of even length} \\
		0 & \text{otherwise}
		\end{cases}\\
		D^T \left( P_{[A]_n \otimes [A]_m} \right) & = \sum_{\substack{R \vdash 2n \\ S \vdash 2m \\ R,S \text{ have columns} \\ \text{of even length}}} \sum_{\lambda = 1}^{g_{R,S;T}} | R, \lambda, [A] \rangle \langle R, \lambda, [A] | \otimes | S, \lambda, [A] \rangle \langle S, \lambda, [A] |
		\end{align*}
		Where we have dropped the $\mu_{T,[S]_{n+m}}$ and $I_{[S]_{n+m}}$ indices because the multiplicity and dimension of $[S]_{n+m}$ are 1 (similarly for $[A]_n$ and $[A]_m$). Now we see that the vast majority of the functions \eqref{anti-invariant functions} vanish. They are only non-zero when
		\begin{align*}
		& T \vdash 2n+2m \text{ has even row lengths} \\
		& [t] = [S]_{n+m} \\
		& R \vdash 2n \text{ and } S \vdash 2m \text{ have even column lengths} \\
		& [r] = [A]_n \\
		& [s] = [A]_m 
		\end{align*}
		As we already noticed, the $\mu$ multiplicities and $I$ dimension indices have become trivial, so we can drop those. Since $[t],[r]$ and $[s]$ are predetermined, we can also drop them. Therefore the non-zero functions are indexed only by $T,R,S, \lambda$, with $1 \leq \lambda \leq g_{R,S;T}$
		\begin{equation}
		F_{T,R,S, \lambda} (\sigma) = \frac{d_T}{(2n+2m)!} \langle T, [S] | D^T (\sigma ) \Big( | R, \lambda, [A] \rangle \otimes | S, \lambda, [A] \rangle \Big)
		\label{mesonic invariant functions}
		\end{equation}
		where $T,R,S$ satisfy the appropriate conditions. Note we have introduced a normalisation factor for later convenience.
		
		Since $F_{T,R,S,\lambda}$ are different matrix elements of different irreps of $S_{2n+2m}$, they are linearly independent in the full algebra of functions, and therefore they are linearly independent in $\mathcal{F}_{n,m}$. Hence they form a basis.
		
		There is an ambiguity in \eqref{mesonic invariant functions}, stemming from the choice of basis vectors. We chose $|T,[S] \rangle$ to be the invariant basis vector, but we could equally well have chosen $|T,[S] \rangle' =-|T,[S] \rangle$ (since representations of $S_{2n+2m}$ are real, we cannot have a complex phase) and followed the exact same process. There is a similar ambiguity in the vectors $|R,\lambda,[A] \rangle$ and $|S,\lambda,[A] \rangle$. Since we are concerned with the function $F$, it is only the total sign of the matrix element that needs to be determined. This can be done by selecting a permutation $\sigma$ and setting its matrix element to be positive (or negative). Note that this requires the matrix element to be non-zero. $F_{T,R,S,\lambda}$ is then determined on the whole of $S_{2n+2m}$. Unfortunately finding a permutation with guaranteed non-zero matrix element for a given $T,R,S,\lambda$ is not easy, so for now we leave $F_{T,R,S,\lambda}$ defined up to a sign.
		
		From the basis \eqref{mesonic invariant functions} for $\mathcal{F}_{n,m}$, we get a basis for $\mathcal{A}^{SO}_{n,m}$
		\begin{align}
		\beta_{T,R,S,\lambda} & = \sum_{\sigma \in S_{2n+2m}} F^{T,R,S,\lambda}(\sigma) \sigma \nonumber \\
		& = \frac{d_T}{(2n+2m)!} \sum_{\sigma \in S_{2n+2m}} \langle T, [S] | D^T (\sigma ) \Big( | R, \lambda, [A] \rangle \otimes | S, \lambda, [A] \rangle \Big) \sigma
		\label{mesonic algebra elements}
		\end{align}
		There is a caveat before we proceed to a basis of gauge-invariant operators. Since $N$ is finite, an $SO(N)$ vector index  can only take a maximum of $N$ distinct values. The construction of Young diagram representations in $V^{\otimes 2n+2m}$ involves anti-symmetrising down each of the columns. Thus if $T$ has more than $N$ rows ($l(T) > N$), the associated operator will vanish. This extra condition gives a subspace of $\mathcal{A}^{SO}_{n,m}$ (similarly for $\mathcal{F}_{n,m}$) relevant for constructing operators at finite $N$, although the space remains unchanged if $N \geq n+m$. Incorporating the restriction $l(T) \leq N$ in addition to those already in place on $R,S$ and $T$, the basis of operators is
		\begin{align}
		\mathcal{O}_{T,R,S,\lambda} & = \frac{d_T}{(2n+2m)!} \sum_{\sigma \in S_{2n+2m}} \hspace{-5pt} \langle T, [S] | D^T (\sigma ) \Big( | R, \lambda, [A] \rangle \otimes | S, \lambda, [A] \rangle \Big) \nonumber \\ 
		& \hspace{8cm} C^{(\delta)}_I \sigma^I_J \left( X^{\otimes n} Y^{\otimes m} \right)^J
		\label{mesonic operators definition}
		\end{align}
		In this section we have only proved that these operators span the space of (mesonic) gauge-invariant observables. In section \ref{section: correlators} we prove they are orthogonal under the two point function, and therefore they are also linearly independent. Thus they do form a basis, as claimed.
		 
		The operators \eqref{mesonic operators definition} were presented in \cite{Kemp2014}. The normalisation there differs from \eqref{mesonic operators definition} by a factor of $\frac{(2n+2m)!}{d_T (2n)! (2m)!}$.
		
		From the labelling in \eqref{mesonic operators definition}, we know the number of linearly independent mesonic operators for $n$ $X$ fields and $m$ $Y$ fields is
		\begin{equation}
		N^{SO(N); \delta}_{n,m} = \sum_{ \substack{
				R \vdash 2n \text{ with even column lengths} \\
				S \vdash 2m \text{ with even column lengths} \\
				T \vdash 2n+2m \text{ with even row lengths} \\
				l(T) \leq N } }
		g_{R,S;T}
		\label{2-matrix delta counting}
		\end{equation}
		This counting of operators (and the baryonic counting \eqref{2-matrix epsilon counting}) can be obtained directly from group integral formulae for the generating function of the quarter-BPS sector. This calculation is given explicitly in \cite{Kemp2014}.
		
		We now check that \eqref{mesonic operators definition} and \eqref{2-matrix delta counting} agree with the half-BPS results presented in \cite{Caputa2013}. To reduce the quarter-BPS objects to half-BPS ones, we set $m=0$.
		
		Since $S \vdash 2m$, it must now be the empty Young diagram. Therefore $R$ and $T$ must be the same, and the Littlewood-Richardson coefficient $g_{R,S;T}$ is just 1, so $\lambda$ is fixed as well. Hence the half-BPS operators are labelled only by $T$, which must be a Young diagram with even column and row lengths. Thus it must be constructed from 2$\times$2 blocks, just as found in \cite{Caputa2013}. Since $T \vdash 2n$, this can only occur when $n$ is even. 
		
		Define $t \vdash \frac{n}{2}$ to be the Young diagram defined by replacing each $2 \times 2$ block in $T$ with a single box. More formally, we have $T = 2t \cup 2t$. Since $t$ is unconstrained as a partition of $\frac{n}{2}$, there are $p \left( \frac{n}{2} \right)$ half-BPS operators when $n$ is even, and none when $n$ is odd.
		
		Explicitly, by setting $m=0$ in \eqref{mesonic operators definition}, the half-BPS operators are
		\begin{equation}
		\mathcal{O}_T = \frac{d_T}{(2n)!} \sum_{\sigma \in S_{2n}} \langle T, [S] | D^T (\sigma )| T, [A] \rangle  C^{(\delta)}_I \sigma^I_J \left( X^{\otimes n} \right)^J
		\label{half-BPS mesonic operator}
		\end{equation}
		It can be shown that the operators in \cite{Caputa2013} differ from this by a factor of
		\begin{equation}
		\frac{1}{d_T} \langle T, [S] | D^T ( \rho ) | T , [A] \rangle
		\nonumber
		\end{equation}
		where $\rho = (1,3)(5,7) \ldots (2n-3,2n-1)$ is the same permutation we saw in \eqref{index rearrangement}.
		
		This matrix element was calculated in \cite{Ivanov1999}, and is given by
		\begin{equation}
		\langle T, [S] | D^T ( \rho ) | T , [A] \rangle = \frac{ d_t }{2^{\frac{n}{2}} n! } \sqrt{ \frac{(2n)!}{d_T} }
		\label{special case of Ivanov's formula}
		\end{equation}
		This is just a special case of the full result in \cite{Ivanov1999}, which gives this matrix element for any permutation $\sigma \in S_{2n}$. We need to develop several ideas before we can present this, and it is given in \eqref{Ivanov's matrix element}.

		\subsubsection{Resolving sign ambiguity}
		\label{section: resolving ambiguity}
		
		We noted earlier that \eqref{mesonic invariant functions} contained an ambiguity in the choice of basis vectors that meant the functions $F_{T,R,S,\lambda}$ (and therefore $\mathcal{O}_{T,R,S,\lambda}$) were only defined up to a minus sign. The equality \eqref{special case of Ivanov's formula} allows us to resolve this ambiguity in the half-BPS sector. Since this matrix element is always non-zero, for any $T$, we can choose it to be positive (and have already done so implicitly in \eqref{special case of Ivanov's formula}).
		
		When plugged into \eqref{mesonic operator}, the permutation $\rho$ produces the multi-trace $\left( \text{Tr} X^2 \right)^{\frac{n}{2}}$, so this gives us another way of determining the sign of $\mathcal{O}_T$. Rather than decreeing that the matrix element be positive, we could instead choose the coefficient of $\left( \text{Tr} X^2 \right)^{\frac{n}{2}}$ in $\mathcal{O}_T$ to be positive. These two methods are entirely equivalent. However, the latter allows us to more easily express a proposed resolution for the quarter-BPS ambiguity. Note that this second approach relies on all multi-traces being linearly independent, so that the coefficient of $\left( \text{Tr} X^2 \right)^{\frac{n}{2}}$ is uniquely defined. This is true provided $n+m \leq N$.
		
		To resolve the sign ambiguity in the quarter-BPS case, we can use a similar technique. Simply choose an ordering for the multi-traces of order $n,m$. Then if the coefficient of the first multi-trace is non-zero, set it to be positive. If the first coefficient is zero, use the second, and so on.
		
		In the second appendix of \cite{Kemp2014}, Kemp gives explicit expressions for $\mathcal{O}_{T,R,S,\lambda}$ (up to constants of proportionality) for $n=m=1$, $n=m=2$ and $n=3,m=1$. He finds that the coefficients for $\text{Tr} XY$, $\left( \text{Tr} XY \right)^2$ and $\text{Tr} XY \text{Tr} X^2$ are non-zero for all the appropriate $T,R,S$. Therefore at these low orders, only a single multi-trace is needed, independent of $T,R,S,\lambda$ (the $\lambda$ index is trivial for low orders). In general this may not be possible and instead two or more traces will be required. Numerical experiments at higher orders are difficult because of the size of the permutation groups involved, and we leave these to the future.
		
		At even higher orders we can have $g_{R,S;T} > 1$ (the first case of this occurs at $n+m=11$), so the $\lambda$ index becomes important. At this point the exact method of choosing the basis for the Littlewood-Richardson multiplicity space becomes relevant, and we would have to be more specific about how this is done (see section \ref{section: basis of multiplicity space}).

		\subsection{Baryonic operators}
		\label{section: baryonic operators}
		
		The construction of the baryonic operators follows in exactly the same way as the mesonic case, just with a different group invariance. The contractor $C^{(\varepsilon)}$ involves a single $\varepsilon$ and $q = n+m - \frac{N}{2}$ $\delta$s, so the symmetry transformations on the left are now controlled by the group $S_N \times S_{q}[S_2]$, where $C^{(\varepsilon)}$ is anti-invariant under the $S_N$ factor and invariant under the $S_{q}[S_2]$ part. The invariances on the right are unchanged compared to the mesonic case, so $\beta$ in \eqref{baryonic operator} is invariant under
		\begin{equation}
		\beta \mapsto (-1)^{\alpha_1} (-1)^\gamma ( \alpha_1 \alpha_2 ) \beta \gamma^{-1} \qquad \qquad (\alpha_1 , \alpha_2) \in S_N \times S_q[S_2] \ \ , \ \ \gamma \in S_n[S_2] \times S_m[S_2] \vspace{2pt}
		\label{baryonic invariance}
		\end{equation}
		Elements of $\mathbb{C}(S_{2n+2m})$ that are invariant under \eqref{baryonic invariance} define a subspace that we call $\mathcal{A}^{SO(N);\varepsilon}_{n,m}$. Note that this space, unlike $\mathcal{A}^{SO}_{n,m}$, depends on $N$, and so we include $N$ in the superscript. As for the mesonic case, we have an equivalent space of invariant functions on the group which we call $\mathcal{F}^{N;\varepsilon}_{n,m}$.	Running through the same argument, we find a basis for $\mathcal{F}^{N;\varepsilon}_{n,m}$	
		\begin{equation*}
		F^{N;\varepsilon}_{P, T, \mu, R, S, \lambda} ( \sigma ) = \frac{d_P}{(2n+2m)!} \Big( \langle [1^N], \mu | \otimes \langle T, \mu,  [S] | \Big) D^P ( \sigma  ) \Big( | R, \lambda, [A] \rangle \otimes | S, \lambda, [A] \rangle \Big)
		\end{equation*}
		With the constraints
		\begin{gather*}
		P \text{ is a Young diagram with } 2n+2m \text{ boxes} \\
		T \text{ is Young diagram with } 2q = 2n+2m - N \text{ boxes and even row lengths} \\
		\mu \text{ is multiplicity index between } 1 \text{ and } g_{[1^N], T;P} \\
		R \text{ is a Young diagram with } 2n \text{ boxes and even column lengths} \\
		S \text{ is a Young diagram with } 2m \text{ boxes and even column lengths} \\
		\lambda \text{ is a multiplicity index between } 1 \text{ and } g_{R, S;P}
		\end{gather*}
		where $[1^N]$ is the Young diagram with $N$ rows, each consisting of a single box, i.e. it is the completely anti-symmetric representation of $S_N$. We see that $P$ is a Young diagram found in the tensor product $[1^N] \otimes T$, so imposing the constraint that $P$ must have at most $N$ rows means it must be formed of a single column of $N$ boxes with the Young diagram $T$ attached to the right of that column. Using the notation defined in section \ref{section: notation}, this is $[1^N] + T$. For example if $N= 6$ and
		\[
		T = \ydiagram{4,4,2}
		\]
		Then we must have
		\[
		P = [1^N] + T = \ydiagram{5,5,3,1,1,1}
		\]
		Note that the Littlewood-Richardson coefficient for the triple $ \left( [1^N], T, [1^N]+T \right)$ is just one, so in the basis of gauge-invariant operators we can drop both $P$ and $\mu$. For convenience we also drop the square brackets from $[1^N]$.
		
		As in the mesonic case, this restriction defines a subspace of $\mathcal{A}^{SO(N);\varepsilon}_{n,m}$ that contributes to making operators.
		
		So the bases of the subspaces of $\mathcal{F}^{N;\varepsilon}_{n,m}$ and $\mathcal{A}^{SO(N);\varepsilon}_{n,m}$ relevant for operator construction are
		\begin{gather}
		F^{N;\varepsilon}_{T,R,S,\lambda} (\sigma)  = \frac{d_{1^N+T}}{(2n+2m)!} \Big( \langle 1^N | \otimes \langle T, [S] | \Big) D^{1^N+T}(\sigma) \Big( | R, \lambda, [A] \rangle \otimes | S, \lambda, [A] \rangle \Big) 
		\label{baryonic function basis} \\
		\beta^{N;\varepsilon}_{T,R,S,\lambda}  = \frac{d_{1^N+T}}{(2n+2m)!} \sum_{\sigma \in S_{2n+2m}} \! \! \! \! \Big( \langle 1^N | \! \otimes \langle T, [S] | \Big) D^{1^N \! +T}(\sigma) \Big( | R, \lambda, [A] \rangle \! \otimes | S, \lambda, [A] \rangle \Big) \sigma
		\label{baryonic algebra element}		
		\end{gather}
		and the operators themselves are given by
		\begin{gather}
		\mathcal{O}^{\varepsilon}_{T,R,S,\lambda}  = \frac{d_{1^N+T}}{(2n+2m)!} \sum_{\sigma \in S_{2n+2m}} \Big( \langle 1^N | \! \otimes \langle T, [S] | \Big) D^{1^N \! + T}(\sigma) \Big( | R, \lambda, [A] \rangle \otimes | S, \lambda, [A] \rangle \Big) \nonumber \\  \qquad \qquad \qquad \qquad \qquad \qquad \qquad \qquad \qquad \qquad \qquad C^{(\varepsilon)}_I \sigma^I_J  \left( X^{\otimes n} Y^{\otimes m} \right)^J
		\label{baryonic operators definition}
		\end{gather}
		From the labelling set of the above, the number of linearly independent baryonic operators using $n$ $X$ fields and $m$ $Y$ fields is
		\begin{equation}
		N^{SO(N); \varepsilon}_{n,m} = \sum_{\substack{
				R \vdash 2n \text{ with even column lengths} \\
				S \vdash 2m \text{ with even column lengths} \\
				T \vdash 2n+2m - N \text{ with even row lengths} \\
				l(T) \leq N}}
		g_{R,S;1^N+T}
		\label{2-matrix epsilon counting}
		\end{equation}
		Setting $m=0$, we see that the gauge-invariant operators are labelled by $T$ only, where $T \vdash 2q = 2n - N$ has an even number of boxes in each row and column. This confirms conjectures made in \cite{Caputa2013a}. Similarly to the mesonic case, the operators \eqref{baryonic operators definition} differ from those defined in \cite{Caputa2013a} by a factor of
		\begin{equation}
		\frac{1}{d_{1^N+T}} \Big( \langle 1^N | \otimes \langle T, [S] | \Big) D^{1^N+T}(\rho) | 1^N + T, [A] \rangle
		\nonumber
		\end{equation}
		where $\rho = (N+1,N+3)(N+5,N+7) \ldots (2n-3,2n-1)$. This matrix element is harder to evaluate than the mesonic equivalent, and we have not managed to find a simpler expression.
		
		As with the mesonic case, there is a sign ambiguity in \eqref{baryonic function basis}, \eqref{baryonic algebra element} and \eqref{baryonic operators definition} relating to the choice of basis vectors. It is more difficult to make suggestions for a resolution, as the baryonic operators are intrinsically a finite $N$ object, and therefore different multi-traces have linear dependencies among themselves. This means that coefficients of multi-traces are not necessarily well-defined, and so we cannot base our positivity condition on the coefficient of a particular multi-trace as we did in section \ref{section: resolving ambiguity}.

		\subsection{$U(N)$ construction}
		\label{section: U(N) construction}

		\begin{figure}
			\centering
			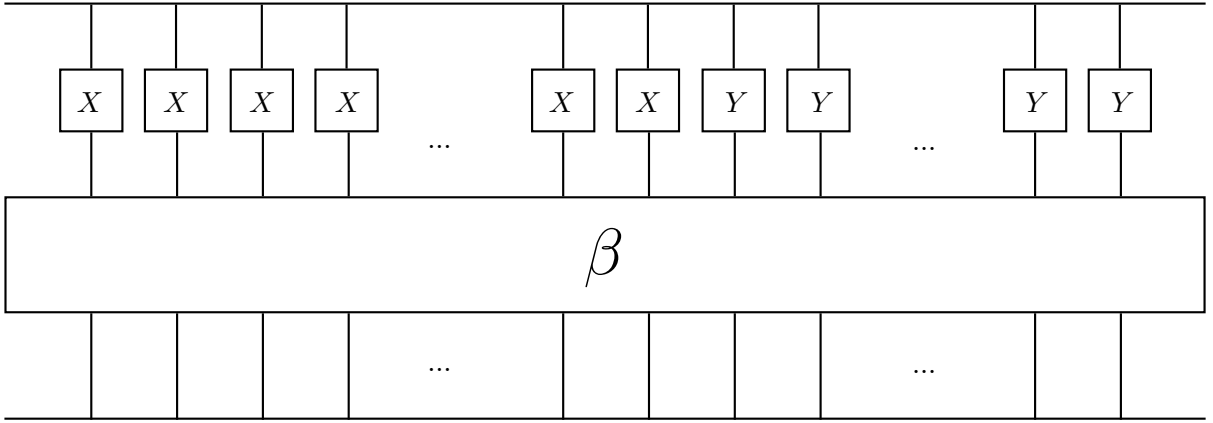
			\caption{Diagrammatic representation of the $U(N)$ contraction. Each vertical line represents an index, while the horizontal lines at the top and bottom indicate that we have traced over these indices.}
			\label{figure: U(N) contractor picture}
		\end{figure}
		
		We briefly review the construction of the quarter-BPS operators in the $U(N)$ gauge theory. The basis we use was first constructed in \cite{Bhattacharyya:2008rb, Bhattacharyya:2008xy} following earlier work in \cite{BHLN02,BBFH04}. The notation we use was developed in \cite{EHS08}.
		
		As explained in section \ref{section: tensor space}, the complex scalar fields are now generic matrices, and have index structure $X^i_j$ and $Y^i_j$. We start with an expression for the most general gauge-invariant operator
		\begin{equation}
		\mathcal{O}^{U(N)} = \beta^J_I \left( X^{\otimes n} Y^{\otimes m} \right)^I_J = \beta^{j_1 j_2 \ldots j_{n+m}}_{i_1 i_2 \ldots i_{n+m}} X^{i_1}_{j_1} \ldots X^{i_n}_{j_n} Y^{i_{n+1}}_{j_{n+1}} \ldots Y^{i_{n+m}}_{j_{n+m}}
		\label{general U(N) operator}
		\end{equation}
		where $\beta^I_J$ is constructed from $n+m$ copies of the only $U(N)$ invariant tensor, $\delta^i_j$. Since this is the only available tensor, $\beta^I_J$ must be formed of $N$-fold tensor products of $\delta^i_j$. In general it will be a linear combination of the different possible index arrangements. Noting the definition \eqref{permutations acting on tensor space}, each term in this linear combination is just a permutation acting on the tensor space, and therefore $\beta^I_J$ is an element of $\mathbb{C}(S_{n+m})$. Figure \ref{figure: U(N) contractor picture} gives a diagrammatic description of the index contraction in \eqref{general U(N) operator}.
		
		Following the example of section \ref{section: Mesonic operators}, we look at the group invariances of \eqref{general U(N) operator}. In this case $\mathcal{O}^{U(N)}$ is invariant under the transformation
		\begin{equation}
		\beta \mapsto \alpha \beta \alpha^{-1} \qquad \qquad \alpha \in S_n \times S_m
		\label{U(N) group algebra invariance}
		\end{equation}
		Elements of $\mathbb{C}(S_{n+m})$ invariant under \eqref{U(N) group algebra invariance} define a sub-algebra that we call $\mathcal{A}_{n,m}^{U}$. A similar process to that in section \ref{section: Mesonic operators} leads us to a basis for $\mathcal{A}_{n,m}^{U}$, given by
		\begin{equation}
		\beta^{U(N)}_{T,R,S,\mu,\nu} = \frac{d_T}{(n+m)!} \sum_{\sigma \in S_{n+m}} \text{Tr}_T \left[ P^T_{R,S;\mu \rightarrow \nu} D^T (\sigma) \right] \sigma
		\nonumber	
		\end{equation}
		where the labels are $T \vdash n+m$, $R \vdash n$, $S \vdash m$ and $1 \leq \mu, \nu \leq g_{R,S;T}$ and the $P^T_{R,S;\mu \rightarrow \nu}$ are defined as operators on the representation space of $T$ that take a vector in the $\mu$th copy of $R \otimes S$ to the equivalent vector in the $\nu$th copy (and act as 0 on all other vectors). We can write these explicitly by introducing orthonormal bases $\{ | R, I \rangle \}, \{ | S, J \rangle \}$ for $R$ and $S$ ($1 \leq I \leq d_R, 1 \leq J \leq d_S$). Using these, we denote the bases for the $\mu$th and $\nu$th copies of $R \otimes S$ by
		\begin{equation}
		\Big\{ | R, \mu, I \rangle \otimes | S,\mu,J \rangle \Big\} \qquad \qquad \Big\{ |R,\nu,I \rangle \otimes | S,\nu,J \rangle \Big\}
		\nonumber
		\end{equation}
		Then we can write
		\begin{equation}
		P^T_{R,S;\mu \rightarrow \nu} = \sum_{I,J} \Big( | R,\nu, I \rangle\otimes | S,\nu,J \rangle \Big) \Big(  \langle R,\mu,I | \otimes \langle S,\mu,J | \Big)
		\nonumber
		\end{equation}
		It is simple to show that this is independent of the basis used. Note that these satisfy
		\begin{equation}
		P^T_{R,S;\mu \rightarrow \nu} P^T_{R',S';\mu' \rightarrow \nu'} = \delta_{R R'} \delta_{S S'} \delta_{\mu \nu'} P^T_{R,S;\mu' \rightarrow \nu}
		\nonumber
		\end{equation}
		The trace 
		\begin{equation*}
		\text{Tr}_T \left[ P^T_{R,S;\mu \rightarrow \nu} D^T(\sigma) \right]
		\end{equation*}
		is called the restricted character of $\sigma$, and reduces to the standard character if we set one of $n,m=0$.
		
		At finite $N$ ($N < n + m$), only a subspace of $\mathcal{A}^U_{n,m}$ is relevant for constructing operators. This is spanned by those basis elements with $l(T) \leq N$. Adding this restriction to the existing conditions on $R,S,T,\mu, \nu$, the corresponding basis for operators is
		\begin{align}
		\mathcal{O}^{U(N)}_{T,R,S,\mu,\nu} & = \frac{d_T}{(n+m)!} \sum_{\sigma \in S_{n+m}} \text{Tr}_T \left[ P^T_{R,S;\mu \rightarrow \nu} D^T (\sigma) \right] \sigma^I_J \left( X^{\otimes n} Y^{\otimes m} \right)^J_I \nonumber \\
		& = \frac{d_T}{(n+m)!} \sum_{\sigma \in S_{n+m}} \text{Tr}_T \left[ P^T_{R,S;\mu \rightarrow \nu} D^T (\sigma) \right] \text{Tr} \left( \sigma X^{\otimes n} Y^{\otimes m} \right)
		\label{U(N) operators}
		\end{align}
		From the labelling in \eqref{U(N) operators} we can see that the counting of quarter-BPS operators in the $U(N)$ gauge theory is given by
		\begin{equation}
		N^{U(N)}_{n,m} = \sum_{\substack{
				R \vdash n \\
				S \vdash m \\
				T \vdash n+m \\
				l(T) \leq N}}
		g_{R,S;T}^2
		\label{U(N) 2 matrix counting}
		\end{equation}
		If we set $m=0$ to reduce to the half-BPS case, as studied in \cite{Corley2002}, the operators are labelled only by a Young diagram $R \vdash n$, and are given by
		\begin{align}
		\mathcal{O}^{U(N)}_R & = \frac{d_R}{n!} \sum_{\sigma \in S_n} \chi_R ( \sigma ) \sigma^I_J \left( X^{\otimes n} \right)^J_I \nonumber \\
		& = \text{Tr} \left( P_R X^{\otimes n} \right)
		\label{U(N) half-BPS operators}
		\end{align}

		\subsection{Basis of Littlewood-Richardson multiplicity space}
		\label{section: basis of multiplicity space}
		
		In sections \ref{section: Mesonic operators} and \ref{section: baryonic operators} we consider the Littlewood-Richardson decomposition of a $S_{n+m}$ representation $T$ into $S_n \times S_m$ representations $R \otimes S$. These come with a multiplicity given by the Littlewood-Richardson coefficients $g_{R,S;T}$. More formally, we have
		\begin{equation}
		V_T = \bigoplus_{\substack{R \vdash n \\ S \vdash m}} V_R \otimes V_S \otimes V^{mult}_{R,S;T} \nonumber
		\end{equation}
		where $V^{mult}_{R,S;T}$ is the multiplicity space and has dimension $g_{R,S;T}$. 
		
		Now consider the algebra $\mathcal{A}^U_{n,m}$ defined in section \ref{section: U(N) construction}. For each pair $R \vdash n, S \vdash m$, we define a sub-algebra by projecting onto the $R \otimes S$ representation of $S_n \times S_m$. Explicitly,
		\begin{equation}
		\mathcal{A}^U_{R,S} = P_{R \otimes S} \mathcal{A}^U_{n,m} = \operatorname{Span}\left\{ \beta^{U(N)}_{T,R,S,\mu,\nu} : 1 \leq \mu , \nu \leq g_{R,S;T} \right\}
		\nonumber
		\end{equation}
		Since $\mathcal{A}^U_{R,S}$ is a sub-algebra of $\mathbb{C}(S_{n+m})$, it acts on $V_T$. The projection onto $R \otimes S$ means it acts only on the $R,S$ subspace and annihilates all others.
		As it is invariant under \eqref{U(N) group algebra invariance}, and $P_{R \otimes S}$ is a central element of $\mathbb{C}(S_n \times S_m)$, it commutes with $S_n$ and $S_m$, and therefore acts proportional to the identity operator on $V_{R} \otimes V_{S}$. Therefore $\mathcal{A}^U_{R,S}$ acts purely on the multiplicity space $V_{R,S;T}^{mult}$.
		
		One can then use the behaviour of vectors in $V_{R,S;T}^{mult}$ under $\mathcal{A}^U_{R,S}$ to choose an orthogonal basis. Simply choose a maximal commuting set of operators, and use the eigenbasis. This is the standard procedure in many representation theory contexts, including the Young basis for representations of $S_n$, for which the Jucys-Murphy elements form a maximal commuting set (see appendix \ref{section: Jucys-Murphy elements}).
		
		For a more complete description of how one chooses these operators, or the maximal commuting sub-algebra they span, see \cite{Mattioli:2016eyp}.

		\subsection{$SO(N)$ states ($\mathcal{A}^{SO}_{n,m}$) as a module over $U(N)$ states ($\mathcal{A}^U_{2n,2m}$)}
		
		For the $SO(N)$ and $U(N)$ gauge theories, the gauge-invariant operators arise from sub-algebras $\mathcal{A}^{SO}_{n,m}$ and $\mathcal{A}^{U}_{n,m}$ of $\mathbb{C}(S_{2n+2m})$ and $\mathbb{C}(S_{n+m})$, defined by the respective invariance properties \eqref{SO(N) group algebra invariance} and \eqref{U(N) group algebra invariance}. Both sub-algebras, inherit multiplication from the full symmetric group algebra, and in this section we study this structure. Multi-matrix operators, and specifically their classification, complexity  and correlators, have been discussed from the perspective of these algebras in \cite{Mattioli:2016eyp} (for a related discussion see \cite{Kimura:2017vpp}). 
		
		For the $SO(N)$ case, the minus sign in \eqref{SO(N) group algebra invariance} means multiplication is identically zero. The $U(N)$ case is more interesting. Using the definition \eqref{U(N) operators}, we have
		\begin{align}
		& \beta^{U(N)}_{T,R,S,\mu,\nu} \beta^{U(N)}_{T',R',S',\mu',\nu'} \nonumber \\ 
		& \qquad = \frac{d_T d_{T'}}{(n+m)!^2} \sum_{\sigma, \rho \in S_{n+m}} \text{Tr}_T \left[ P^T_{R,S;\mu \rightarrow \nu} D^T (\sigma) \right] \text{Tr}_{T'} \left[ P^{T'}_{R',S';\mu' \rightarrow \nu'} D^{T'} (\rho) \right] \sigma \rho 
		\nonumber \\
		& \qquad = \frac{d_T d_{T'}}{(n+m)!^2} \sum_{\sigma, \rho \in S_{n+m}} \text{Tr}_T \left[ P^T_{R,S;\mu \rightarrow \nu} D^T (\sigma \rho^{-1}) \right] \text{Tr}_{T'} \left[ P^{T'}_{R',S';\mu' \rightarrow \nu'} D^{T'} (\rho) \right] \sigma 
		\nonumber \\
		& \qquad = \frac{d_T}{(n+m)!} \delta_{TT'} \sum_{\sigma \in S_{n+m}} \text{Tr}_T \left[ P^T_{R',S';\mu' \rightarrow \nu'} P^T_{R,S;\mu \rightarrow \nu} D^T (\sigma) \right] \sigma
		\nonumber \\
		& \qquad = \delta_{TT'} \delta_{RR'} \delta_{SS'} \delta_{\nu \mu'} \beta^{U(N)}_{T,R,S,\mu, \nu'}
		\label{U(N) algebra elements multiplication property}
		\end{align}
		where in going from the second line to the third we have used the orthogonality of matrix elements, \eqref{orthogonality of matrix elements}. 
	
		From \eqref{U(N) algebra elements multiplication property} we see that we can represent the $\beta_{T,R,S,\mu,\nu}^{U(N)}$ as block diagonal matrices. There is a block for each trio $( T,R,S)$, of size $g_{R,S;T}$. Explicitly, the representative of $\beta_{T,R,S,\mu,\nu}^{U(N)}$ is the matrix containing only zeroes in each block except the $(T,R,S)$ block, in which there is a single 1 in the $(\mu,\nu)$th position.
		
		A basis for $\mathcal{A}^{SO}_{n,m}$ is given by \eqref{mesonic algebra elements}. Although they give 0 when multiplied with each other, we can multiply them on the right by elements of $\mathcal{A}^{U}_{2n,2m}$
		\begin{align}
		& \beta^{SO(N);\delta}_{T,R,S,\lambda} \beta^{U(N)}_{T',R',S',\mu,\nu} \nonumber  \\
		& \qquad = \frac{d_T d_{T'}}{ (2n+2m)!^2} \sum_{\sigma , \rho \in S_{2n+2m}} \langle T,[S] | D^T (\sigma) \left( |R,\lambda,[A] \rangle \otimes |S,\lambda, [A] \rangle \right) \nonumber \\ 
		& \hspace{9.5cm}
		\text{Tr}_{T'} \left[ P^{T'}_{R',S';\mu \rightarrow \nu} D^{T'} (\rho) \right] \sigma \rho
		\nonumber \\
		& \qquad = \frac{d_T d_{T'}}{(2n+2m)!^2} \sum_{\sigma , \rho \in S_{2n+2m}} \hspace{-0.5cm} \langle T,[S] | D^T (\sigma \rho^{-1}) \Big( |R,\lambda,[A] \rangle \otimes |S,\lambda, [A] \rangle \Big) 
		\nonumber \\
		& \hspace{9.5cm}
		\text{Tr}_{T'} \left[ P^{T'}_{R',S';\mu \rightarrow \nu} D^{T'} (\rho) \right] \sigma 
		\nonumber \\
		& \qquad = \frac{\delta_{TT'} d_T}{(2n+2m)!}  \sum_{\sigma \in S_{2n+2m}}  \langle T,[S]| D^T (\sigma) P^T_{R',S';\mu \rightarrow \nu} \Big( | R,\lambda,[A] \rangle \otimes | S,\lambda,[A] \rangle \Big) \sigma 
		\nonumber \\
		& \qquad =  \frac{\delta_{TT'} \delta_{RR'} \delta_{SS'} \delta_{\mu \lambda} d_T}{(2n+2m)!} \sum_{\sigma \in S_{2n+2m}} \langle T,[S] | D^T (\sigma) \Big( | R,\nu,[A] \rangle \otimes |S,\nu,[A] \rangle \Big) \sigma \nonumber \\
		& \qquad = \delta_{TT'} \delta_{RR'} \delta_{SS'} \delta_{\mu \lambda} \beta^{SO(N);\delta}_{T,R,S,\nu}
		\label{module multiplication}
		\end{align}
		So $\mathcal{A}^{SO}_{n,m}$ forms a right-module over $\mathcal{A}^{U}_{2n,2m}$. Thinking of the $U(N)$ elements as block diagonal matrices, $\beta^{SO(N);\delta}_{T,R,S,\lambda}$ form row vectors with zero entries in all sections except that corresponding to $(T,R,S)$, in which it has a single 1 at the $\lambda$th position.
		
		This gives a nice interpretation of the $SO(N)$ counting \eqref{2-matrix delta counting} and its $U(N)$ equivalent \eqref{U(N) 2 matrix counting}. The $U(N)$ counting contains squares of Littlewood-Richardson coefficients because it is composed of block diagonal matrices of size $g_{R,S;T}$, while the $SO(N)$ counting contains Littlewood-Richardson coefficients to the first power because it lies in the fundamental (of a subset of the blocks) of the $U(N)$ algebra.
		
		Similarly, $\mathcal{A}^{SO(N);\varepsilon}_{n,m}$ also forms a module of $\mathcal{A}^{U}_{2n,2m}$. The algebra elements are given by \eqref{baryonic algebra element} and they satisfy
		\begin{equation}
		\beta^{SO(N);\varepsilon}_{T,R,S,\lambda} \beta^{U(N)}_{T',R',S',\mu,\nu} = \delta_{1^N+T,T'} \delta_{RR'} \delta_{SS'} \delta_{\lambda \mu} \beta^{SO(N);\varepsilon}_{T,R,S,\nu}
		\nonumber
		\end{equation}		
		So the baryonic elements lie in the fundamental of a different set of blocks.

		\section{The orientifold quotient from $U(N)$ to $SO(N)$ in the half-BPS sector :   plethysms, dominoes and branes  }
		\label{section: Z2 quotient}
		
		In \cite{Witten1998a}, the $SO(N)$ (and $Sp(N)$) gauge theory was considered as the dual of type IIB string theory on $AdS_5 \times \mathbb{R} P^5$. This string theory was obtained from the standard $AdS_5 \times S^5$ theory by performing an orientifold operation on the $S^5$ factor. Depending on topological considerations, the orientifold quotient can lead to either a $SO(N)$ or a $Sp(N)$ dual theory. We now study this quotient in the half-BPS sector from the field theory point of view. The distinction between the orthogonal and symplectic quotient is much less subtle here, we either put the scalar field $X$ in the adjoint of $\mathfrak{so}(N)$ or $\mathfrak{sp}(N)$. Here we only study the orthogonal quotient, and leave the symplectic case until section \ref{section: Symplectic gauge group}.
		
		In \cite{Corley2002} a basis of the half-BPS sector of the $U(N)$ theory was discussed, while in \cite{Caputa2013} an equivalent was found for the $SO(N)$ theory. When we perform the orientifold quotient on an arbitrary $U(N)$ state, it becomes a linear combination of the $SO(N)$ basis. This section focuses on finding the coefficients in this expansion. They have a surprising interpretation in terms of plethysms of Young diagrams, which in turn are related to the combinatorics of domino tableaux.
		
		Explicitly, the quotient takes the matrix $X$, unconstrained in the $U(N)$ theory, and makes it anti-symmetric.
		
		This section concerns only the mesonic sector of the $SO(N)$ theory, since the $U(N)$ operators are all multi-traces, and the $\mathbb{Z}_2$ quotient takes multi-traces to multi-traces. The baryonic operators do not arise from the quotient in this way.

		\subsection{An example: $n=4$}
		\label{section: n=4 example}
		
		As an example of the quotient we look at the case $n=4$, so the $U(N)$ diagrams have 4 boxes while the $SO(N)$ diagrams have 8. Using the definitions \eqref{half-BPS mesonic operator} and \eqref{U(N) half-BPS operators}, the operators are
		\ytableausetup{boxsize=5pt}
		\begin{align*}
		\mathcal{O}^{U(N)}_{\ydiagram{4}} & = \hspace{9pt} \frac{1}{4} \text{Tr} X^4 & + \frac{1}{8} \left( \text{Tr}X^2 \right)^2 & + \frac{1}{4} \left( \text{Tr}X^2 \right) \left( \text{Tr}X \right)^2 & + \frac{1}{3} \left( \text{Tr} X^3 \right) \left( \text{Tr}X \right) & + \frac{1}{24} \left( \text{Tr}X \right)^4 \\
		\mathcal{O}^{U(N)}_{\ydiagram{3,1}} & = - \frac{3}{4} \text{Tr} X^4 & - \frac{3}{8} \left( \text{Tr}X^2 \right)^2 &  + \frac{3}{4}\left( \text{Tr}X^2 \right) \left( \text{Tr}X \right)^2 & & + \frac{3}{8} \left( \text{Tr}X \right)^4 \\
		\mathcal{O}^{U(N)}_{\ydiagram{2,2}} & = & \frac{1}{2} \left( \text{Tr}X^2 \right)^2 & & - \frac{2}{3} \left( \text{Tr} X^3 \right) \left( \text{Tr}X \right) & + \frac{1}{6} \left( \text{Tr}X \right)^4 \\
		\mathcal{O}^{U(N)}_{\ydiagram{2,1,1}} & = \hspace{9pt} \frac{3}{4} \text{Tr} X^4 & - \frac{3}{8} \left( \text{Tr}X^2 \right)^2 & - \frac{3}{4}\left( \text{Tr}X^2 \right) \left( \text{Tr}X \right)^2 & & + \frac{3}{8} \left( \text{Tr}X \right)^4 \\
		\mathcal{O}^{U(N)}_{\ydiagram{1,1,1,1}} & = \hspace{2pt} \underset{\text{Survive the } \mathbb{Z}_2 \text{ quotient}}{\underbrace{\phantom{- \frac{1}{4} \text{Tr} X^4  + \frac{1}{8} \left( \text{Tr}X^2 \right)^2 }}} \hspace{-120pt}
		- \frac{1}{4} \text{Tr} X^4 &  + \frac{1}{8} \left( \text{Tr}X^2 \right)^2 &
		\hspace{6pt} \underset{\text{Annihilated by the } \mathbb{Z}_2 \text{ quotient}}{\underbrace{\phantom{- \frac{1}{4} \left( \text{Tr}X^2 \right) \left( \text{Tr}X \right)^2 + \frac{1}{3} \left( \text{Tr} X^3 \right) \left( \text{Tr}X \right) + \frac{1}{24} \left( \text{Tr}X \right)^4}}} \hspace{-267pt}
		- \frac{1}{4} \left( \text{Tr}X^2 \right) \left( \text{Tr}X \right)^2 & + \frac{1}{3} \left( \text{Tr} X^3 \right) \left( \text{Tr}X \right) & + \frac{1}{24} \left( \text{Tr}X \right)^4
		\end{align*}
		\begin{align*}
		\mathcal{O}^{SO(N)}_{\ydiagram{4,4}} & = \frac{4}{\sqrt{5}} \text{Tr} X^4 + \frac{2}{\sqrt{5}} \left( \text{Tr}X^2 \right)^2 \\
		\mathcal{O}^{SO(N)}_{\ydiagram{2,2,2,2}} & = - \frac{4}{\sqrt{5}} \text{Tr} X^4 + \frac{2}{\sqrt{5}} \left( \text{Tr}X^2 \right)^2
		\end{align*}		
		We can see that
		\begin{align}
		\mathcal{O}^{U(N)}_{\ydiagram{4}} & \overset{\mathbb{Z}_2}{\longrightarrow} \frac{\sqrt{5}}{16}	\mathcal{O}^{SO(N)}_{\ydiagram{4,4}} \label{projection of [4]} \\
		\mathcal{O}^{U(N)}_{\ydiagram{3,1}} & \overset{\mathbb{Z}_2}{\longrightarrow}  - \frac{3 \sqrt{5}}{16} \mathcal{O}^{SO(N)}_{\ydiagram{4,4}} \label{projection of [3,1]} \\
		\mathcal{O}^{U(N)}_{\ydiagram{2,2}} & \overset{\mathbb{Z}_2}{\longrightarrow} \frac{\sqrt{5}}{8} \left( \mathcal{O}^{SO(N)}_{\ydiagram{4,4}} + \mathcal{O}^{SO(N)}_{\ydiagram{2,2,2,2}} \right) \label{projection of [2,2]} \\
		\mathcal{O}^{U(N)}_{\ydiagram{2,1,1}} & \overset{\mathbb{Z}_2}{\longrightarrow} - \frac{3 \sqrt{5}}{16} \mathcal{O}^{SO(N)}_{\ydiagram{2,2,2,2}} \label{projection of [2,1,1]} \\
		\mathcal{O}^{U(N)}_{\ydiagram{1,1,1,1}} & \overset{\mathbb{Z}_2}{\longrightarrow} \frac{\sqrt{5}}{16} \mathcal{O}^{SO(N)}_{\ydiagram{2,2,2,2}} \label{projection of [1,1,1,1]}
		\end{align}
		In order to perform this projection for larger $n$, we will need to have an expression for the operators in terms of multi-traces. Since the operators are defined by sums over $S_n$ and $S_{2n}$, this leads us to study how permutations produce multi-traces.


		\subsection{From permutations to traces}
		\label{section: half-BPS permutations to traces}

		In \eqref{general SO(N) operator} and \eqref{general U(N) operator} we saw two different ways of contracting the indices of the scalar fields using the action of permutations on the tensor space. The former is more general: it allows the construction of arbitrary multi-traces of $X, X^T, Y$ and $Y^T$ (as well as Pfaffian type objects if we use the baryonic contractor) and allows us to encode the anti-symmetry of $X$ and $Y$ into an invariance of $\mathbb{C}(S_{2n+2m})$. However in the $U(N)$ theory a trace made from $X$ and $X^T$ is not gauge invariant, so we instead use the simpler formulation \eqref{general U(N) operator}, which only admits multi-traces of $X$ and $Y$. Clearly all $SO(N)$ multi-traces can be constructed using either approach. We now investigate how different permutations lead to different traces, and the relation between the two contraction types, in the simpler case when we only have $X$ matrices and no $Y$s. The full two-matrix version is studied in section \ref{section: 2-matrix permutations to traces}.
		
		Once we have relations between permutations and traces, we will use these to turn the sums over permutations in \eqref{half-BPS mesonic operator} and \eqref{U(N) half-BPS operators} into sums over traces (or more accurately their labelling sets, which will be partitions). To do this, we will also need to know the size of the set of permutations that lead to a particular multi-trace. These sets are the orbits of group actions on $S_n$ and $S_{2n}$, so we devote much of the coming section to studying these actions.

		\subsubsection{$U(N)$}
		
		$U(N)$ multi-traces of order $n$ are indexed by partitions of $n$. For a partition $p = [1^{p_1}, 2^{p_2}, \ldots ]$, the corresponding trace is
		\begin{equation}
		\prod_i \left( \text{Tr} X^i \right)^{p_i}
		\label{multi-trace}
		\end{equation}
		This is related to permutations in $S_n$ by \eqref{U(N) permutations to traces}. The set of permutations producing \eqref{multi-trace} is just the conjugacy class labelled by $p$. To convert sums over permutations into sums over partitions, we will need the size of the conjugacy class, which can be found using the orbit-stabiliser theorem. With the conjugation action, the stabiliser of a permutation $\sigma$ is just all elements which commute with it.
		
		For a single cycle, the centraliser is just the cyclic group generated by the cycle. We think of these as a rotation group, since when they conjugate the cycle they cyclically rotate the elements. For multiple cycles, we have the direct product of these individual rotation groups, and then additionally a permutation group factor (incorporated via a semi-direct product) arising from permuting multiple cycles of the same length. Explicitly, for a permutation of cycle type $p$ the stabiliser group is isomorphic to
		\begin{equation}
		\bigtimes_i \Big( S_{p_i} \ltimes \left( \mathbb{Z}_i \right)^{p_i} \Big) = \bigtimes_i S_{p_i} \left[ \mathbb{Z}_i \right]
		\label{U(N) stabiliser}
		\end{equation}
		where the notation of the right denotes the wreath product of $S_{p_i}$ with $\mathbb{Z}_i$, as seen in section \ref{section: Sp[S2]}. From the explicit form above, we see the stabiliser has size $z_p$, defined in \eqref{z_p}. Applying the orbit-stabiliser theorem, the size of the conjugacy classes is
		\begin{equation}
		\frac{n!}{z_p}
		\label{size of conjugacy class}
		\end{equation}

		\subsubsection{$SO(N)$}
		\label{section: SO(N) permutations to traces}
		
\begin{figure}
			\centering
			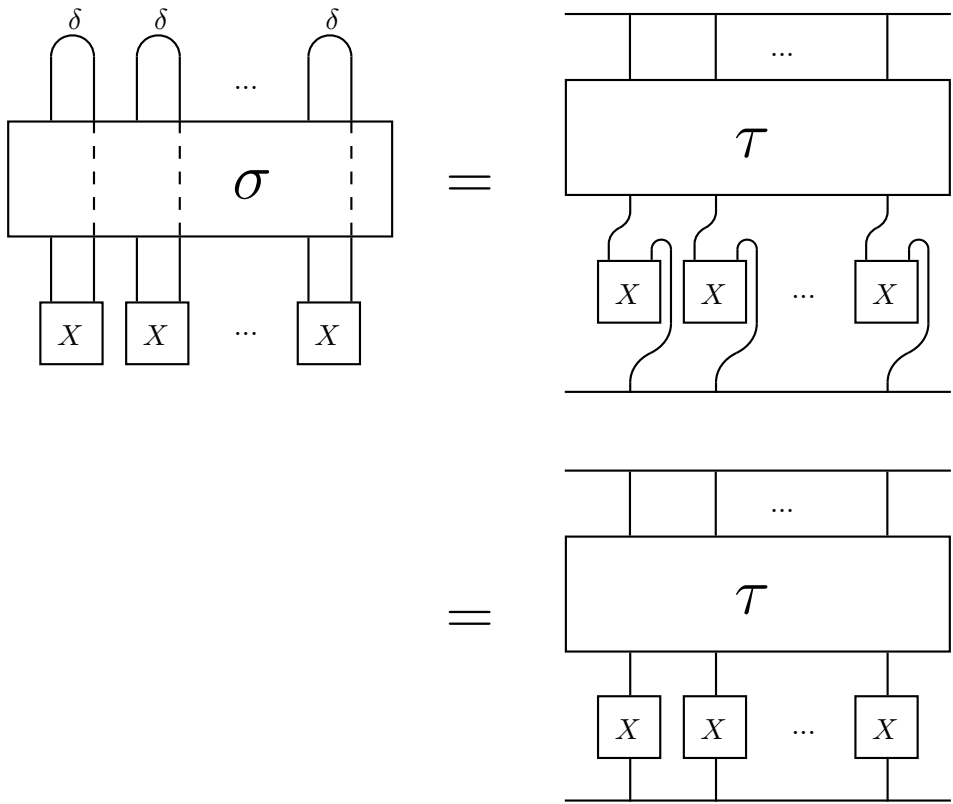
			\caption{A diagrammatic version of \eqref{Trace from SO(N) contraction}. The dotted lines represent the fact that $\sigma$ fixes all even numbers. The first row keeps the index positions in $X$ constant, while the second  b+reaks our index conventions and uses the index structure $X^i_{\ j} = X^{ij}$ to illustrate that using $\sigma \in S_n^{(odd)}$ has changed the $SO(N)$ type contraction into the $U(N)$ type contraction (see figures \ref{figure: Mesonic contractor picture} and \ref{figure: U(N) contractor picture})}
			\label{figure: SO(N) simplified contraction}
		\end{figure}

		For $SO(N)$, $X$ is anti-symmetric, so Tr$X = $Tr$X^3 = $Tr$X^5 \ldots = 0$, and hence we only consider $n$ even. Since the odd single traces vanish, the non-zero multi-traces are indexed by a partition $q \vdash \frac{n}{2}$. The trace corresponding to $q$ is
		\begin{equation}
		\prod_i \left( \text{Tr}X^{2i} \right)^{q_i}
		\label{SO(N) multi-trace}
		\end{equation}		
		The relation to permutations in $S_{2n}$ is more complicated. Taking $m=0$ in \eqref{SO(N) group algebra invariance} we see that for $\beta \in \mathbb{C}(S_{2n})$ the $SO(N)$ contraction is invariant under
		\begin{equation}
		\beta \mapsto (-1)^\gamma \alpha \beta \gamma^{-1} \qquad \qquad \alpha, \gamma \in S_n[S_2]
		\label{half-bps invariance}
		\end{equation}
		We wish to consider individual permutations, so we look at just the group theory part of this action. In particular we are interested in the orbits of $\sigma \in S_{2n}$ under the action
		\begin{equation}
		\sigma \mapsto \alpha \sigma \gamma^{-1} \qquad \qquad \alpha, \gamma \in S_n[S_2]
		\label{hal-bps invariance without minus sign}
		\end{equation}
		These orbits are called the double cosets of $S_{2n}$ over $S_n[S_2]$ and have been well studied. It was shown in \cite[Chapter VII.2]{Macdonald1995} that the double cosets are indexed by partitions of $n$. For a partition $p \vdash n$, we choose the double coset representative to be any permutation $\sigma \in S_{2n}$ that fixes $\{ 2, 4, 6, \ldots , 2n \}$ and acts with cycle type $p$ on the set $\{ 1, 3, 5, \ldots , 2n-1 \}$. Clearly $\sigma$ is a member of the subgroup $S^{(odd)}_n \leq S_{2n}$ defined by acting only on $\{ 1, 3, 5, \ldots , 2n-1 \}$ (for later convenience, we also define $S_n^{(even)}$ in the analogous way). When thinking of $\sigma$ just as an element of $S_n \equiv S_n^{(odd)}$, we call it $\tau$. Using this notation, we have
		\begin{align}
		C^{(\delta)}_I \sigma^I_J \left( X^{\otimes n} \right)^J & = \delta_{i_1 j_1} \delta_{i_2 j_2} \ldots \delta_{i_n j_n} \sigma^{i_1 j_1 i_2 j_2 \ldots i_n j_n}_{k_1 l_1 k_2 l_2 \ldots k_n l_n} X^{k_1 l_1} X^{k_2 l_2} \ldots X^{k_n l_n} \nonumber \\
		& = \delta_{i_1 j_1} \ldots \delta_{i_n j_n} \tau^{i_1 i_2 \ldots i_n}_{k_1 k_2 \ldots k_n} \delta^{j_1}_{l_1} \delta^{j_2}_{l_2} \ldots \delta^{j_n}_{l_n} X^{k_1 l_1} X^{k_2 l_2} \ldots X^{k_n l_n} \nonumber \\
		& = \tau^{i_1 i_2 \ldots i_n}_{k_1 k_2 \ldots k_n} X^{k_1 i_1} X^{k_2 i_2} \ldots X^{k_n i_n} \nonumber \\
		& = X^{k_1 k_{\tau(1)}} X^{k_2 k_{\tau(2)}} \ldots X^{k_n k_{\tau(n)}} \nonumber \\
		& = \prod_i \left( \text{Tr} X^i \right)^{p_i}
		\label{Trace from SO(N) contraction}
		\end{align}
		where for the last line we recall \eqref{SO(N) permutations to traces}. Figure \ref{figure: SO(N) simplified contraction} shows a diagrammatic expression of this equality (excluding the last line). Since odd order single traces vanish, this tells us that if the partition $p$ has any odd components, the $SO(N)$ contraction of any member of the corresponding double coset will vanish, as the trace is zero. We are only interested in those double cosets whose partition $p$ has only even components. From the invariance \eqref{half-bps invariance} we know that permutations in the same double coset will produce the same trace up to a sign.
		
		We now give another characterisation, in more group theoretic language, of the double cosets on which the contraction vanishes. The crucial characteristic is whether we can use the transformation \eqref{half-bps invariance} to take $\sigma \mapsto -\sigma$. If we can do so, then we have
		\begin{equation}
		C_I^{(\delta)} \sigma^I_J \left( X^{\otimes n} \right)^J = - C_I^{(\delta)} \sigma^I_J \left( X^{\otimes n} \right)^J = 0
		\label{vanishing contraction}
		\end{equation}
		and similarly
		\begin{equation}
		\langle T, [S] | D^T (\sigma )| T, [A] \rangle = - \langle T, [S] | D^T (\sigma )| T, [A] \rangle = 0
		\label{vanishing matrix element}
		\end{equation}
		Clearly this occurs if we can find $\alpha, \gamma \in S_n[S_2]$ such that $\alpha \sigma \gamma^{-1} = \sigma$ and $\gamma$ odd. Consider $Stab(\sigma)$, the stabiliser group of $\sigma$. This is the subgroup of $S_n[S_2] \times S_n[S_2]$ defined by
		\begin{equation}
		Stab(\sigma) = \left\{ (\alpha, \gamma) : \alpha \sigma \gamma^{-1} = \sigma \right\}
		\nonumber
		\end{equation}
		Those elements with $\gamma$ even define a subgroup of $Stab(\sigma)$ 
		\begin{equation}
		G(\sigma) = \left\{ ( \alpha, \gamma ) : \alpha \sigma \gamma^{-1} = \sigma , (-1)^\gamma = 1 \right\}
		\nonumber
		\end{equation}
		Note that since $\alpha \sigma \gamma^{-1} = \sigma$, $\alpha$ and $\gamma$ must have the same sign. We could therefore have defined $G(\sigma)$ with $\alpha$ even instead of $\gamma$. This means the analysis done here, and in particular the split into even and odd double cosets (defined below), applies to the symplectic case, where the invariance and anti-invariance have switched sides (see section \ref{section: Symplectic gauge group}).
		
		There are two possibilities for $G(\sigma)$. Firstly, we could have $G(\sigma) = Stab(\sigma)$. In this case, \eqref{vanishing contraction} and \eqref{vanishing matrix element} do not hold and we cannot conclude anything further.
		
		Secondly, suppose $G(\sigma) \neq Stab(\sigma)$. Therefore there exists a pair $(\hat{\alpha}, \hat{\gamma}) \in Stab(\sigma)$ with $\hat{\gamma}$ odd. It is easy to prove that the coset $(\hat{\alpha}, \hat{\gamma})G(\sigma)$ is the set of permutations with an odd right hand factor
		\begin{equation}
		(\hat{\alpha}, \hat{\gamma})G(\sigma) = \left\{ ( \alpha, \gamma ) : \alpha \sigma \gamma^{-1} = \sigma , (-1)^\gamma = -1 \right\}
		\nonumber
		\end{equation}
		Since $\gamma$ must be even or odd, this implies $G(\sigma) \cup (\hat{\alpha}, \hat{\gamma})G(\sigma) = Stab(\sigma)$. So in this case $G(\sigma)$ makes up exactly half of $Stab(\sigma)$.
		
		Suppose we take $\sigma, \tau$ to be in the same double coset. Then we have $\tau = \alpha \sigma \gamma^{-1}$ for some $\alpha, \gamma \in S_n[S_2]$. The stabiliser of $\tau$ is given by $(\alpha,\gamma) Stab(\sigma) (\alpha,\gamma)^{-1}$, and therefore $G(\tau)$ is the same size as $G(\sigma)$. Thus the behaviour of $G(\sigma)$ (whether it is the whole stabiliser or half of it) is a property of the double coset. If a double coset has $G(\sigma) = Stab(\sigma)$, we call it an even double coset, while if $G(\sigma) \neq Stab(\sigma)$ we call it an odd double coset.
		
		From \eqref{vanishing contraction} and \eqref{vanishing matrix element} we know that odd double cosets have vanishing contraction and matrix element. From \eqref{Trace from SO(N) contraction}, we also know that a double coset has vanishing contraction if the corresponding partition has one or more odd component. We now prove that these two conditions are equivalent.
		
		To do this we study the stabiliser of a coset representative in more detail. Take $\sigma \in S_n^{(odd)}$ of cycle type $p \vdash n$ (when discussing the cycle type of $\sigma$, we will always ignore the $n$ 1-cycles arising from the fixed even numbers). This is a representative of the double coset labelled by $p$. We want to find $(\alpha, \gamma) \in S_n[S_2] \times S_n[S_2]$ such that
		\begin{equation}
		\alpha \sigma \gamma^{-1} = \sigma
		\nonumber
		\end{equation}
		Note that this is equivalent to 
		\begin{equation}
		\alpha = \sigma \gamma \sigma^{-1}
		\nonumber
		\end{equation}
		Therefore rather than searching for the pair $(\alpha, \gamma)$, we instead look for $\gamma \in S_n[S_2]$ such that
		\begin{equation}
		\sigma \gamma \sigma^{-1} \in S_n[S_2]
		\label{half-bps stabiliser condition}
		\end{equation}
		It is simple to show that $\gamma$ satisfying \eqref{half-bps stabiliser condition} form a subgroup of $S_n[S_2]$. This subgroup is isomorphic to $Stab ( \sigma )$ via the bijection
		\begin{equation}
		\gamma \qquad \qquad \longleftrightarrow \qquad \qquad ( \sigma \gamma \sigma^{-1}, \gamma )
		\nonumber
		\end{equation}
		Technically, $Stab(\sigma)$ is a subgroup of $S_n[S_2] \times S_n[S_2]$, but for convenience we will refer to the subgroup of $S_n[S_2]$ defined by \eqref{half-bps stabiliser condition} as $Stab(\sigma)$. It will be clear from context which we are talking about, and since the two are isomorphic it makes sense to identify them.
		
		Under this bijection, $G(\sigma)$ maps to the subgroup of $S_n[S_2]$ defined by \eqref{half-bps stabiliser condition} with the additional condition that $\gamma$ is an even permutation. Again, we will refer to this subgroup of $S_n[S_2]$ as $G(\sigma)$.
		
		It is clear from \eqref{half-bps stabiliser condition} that any $\gamma \in S_n[S_2]$ that commutes with $\sigma$ will automatically be in $Stab(\sigma)$.	Consider a permutation $\tau^{(odd)} \in S_n^{(odd)}$. There is an equivalent permutation $\tau^{(even)} \in S_n^{(even)}$ which acts exactly the same as $\tau^{(odd)}$ but permutes the even numbers instead of the odd ones. Then the combination $\tau = \tau^{(odd)} \tau^{(even)}$ lies in $S_n[S_2]$. This embedding of $S_n$ into $S_n[S_2]$ (the diagonal subgroup of $S_n^{(odd)} \times S_n^{(even)}$) is exactly the $S_n$ subgroup in the definition of $S_n[S_2]$, since it moves the $n$ pairs around without any swaps. If we take $\tau^{(odd)}$ to be a member of the centraliser of $\sigma$ in $S_n^{(odd)}$, then $\tau$ will commute with $\sigma$, and hence $\tau \in Stab(\sigma)$. Therefore the centraliser of $\sigma$ in $S_n^{(odd)}$, see \eqref{U(N) stabiliser}, is a subgroup of $Stab(\sigma)$. In particular, we have a rotation subgroup corresponding to each cycle. Note that all $\tau$ produced in this way are even permutations, and are therefore in $G(\sigma)$.
		
		In addition to these rotations, there are reflection-type elements for each cycle. For example, if we consider $ \sigma = (1,3,5,..,2n-1)$ with just a single cycle, one possible reflection is $\gamma = (1,2n)(3,2n-2) \ldots (2n-3,4)(2n-1,2)$. One can check that both $\gamma$ and $\sigma \gamma \sigma^{-1}$ are in $S_n[S_2]$, and therefore $\gamma$ is a member of $Stab(\sigma)$. The generator of the rotations is $\tau = (1,3,5, \ldots ,2n-1)(2,4,6, \ldots ,2n)$, and we see that $\gamma \tau \gamma^{-1} = \tau^{-1}$. This (along with $\tau^n = \gamma^2 = 1$) is the defining relation between the generators of the dihedral group $D_n$, and therefore $Stab(\sigma)$ contains a dihedral subgroup. 
		
		Note that the sign of $\gamma$ is $(-1)^n$, so $\gamma \in G(\sigma)$ if and only if $n$, the length of the cycle, is even. In particular, $G(\sigma) \neq Stab(\sigma)$ when $n$ is odd.
		
		Recall that $S_n[S_2]$ is the centraliser of the permutation $\pi = (1,2) (3,4) \ldots (2n-1,2n)$, and therefore $\gamma \in S_n[S_2]$ if and only if
		\begin{equation}
		\gamma = \pi \gamma \pi
		\label{conjugation by pi}
		\end{equation}
		Defining $\pi' = \sigma^{-1} \pi \sigma = (2,3) (4,5) \ldots (2n-2,2n-1) (2n,1)$, \eqref{half-bps stabiliser condition} is equivalent to
		\begin{equation}
		\gamma = \pi' \gamma \pi'
		\label{conjugation by pi'}
		\end{equation}
		Therefore $\gamma \in Stab(\sigma)$ if and only if $\gamma$ is invariant under conjugation by $\pi$ and $\pi'$.
		
		Now suppose $\gamma \in Stab(\sigma)$ with $\gamma(1) = i$ for some $1 \leq i \leq 2n$. Then \eqref{conjugation by pi} implies that $\gamma(2) = \left( \pi \gamma \pi \right) (2) = \pi(i)$. Plugging this into \eqref{conjugation by pi'} implies $\gamma(3) = \pi' ( \pi (i))$. We can now repeat to find $\gamma(4) = \pi ( \pi' ( \pi ( i ) ) )$ and so on. Therefore the value of $\gamma(1)$ determines $\gamma$ completely, and hence $Stab(\sigma)$ can have at most $2n$ members. Since we already know it contains $D_n$, of size $2n$, we must have $Stab(\sigma) = D_n$.
		
		If $\sigma$ has multiple cycles, the above can be repeated for each one. Therefore the stabiliser contains a direct product of dihedral groups. There are permutation group factors arising from permuting cycles of the same length, just as in the $U(N)$ case \eqref{U(N) stabiliser}. Explicitly, for $\sigma$ in the double coset labelled by $p \vdash n$, we have
		\begin{equation}
		Stab(\sigma) \cong \bigtimes_i \Big( S_{p_i} \ltimes \left( D_i \right)^{p_i} \Big) = \bigtimes_i S_{p_i} \left[ D_i \right]
		\label{SO(N) stabiliser}
		\end{equation}
		From the construction of the stabiliser group, we see that $G(\sigma) \neq Stab(\sigma)$ exactly when there is one or more cycle of odd length in $\sigma$. Since $\sigma$ is of cycle type $p$, this corresponds exactly to $p$ containing one or more odd component, as claimed.
		
		The dihedral group $D_k$ is defined as the symmetry group of a $k$-gon, made up of $k$ rotations and $k$ reflections. It therefore has size $\left| D_k \right| = 2k$. Hence the size of the $Stab(\sigma)$ is given by
		\begin{equation}
		\prod_i (2i)^{p_i} (p_i)! = z_{2p}
		\nonumber
		\end{equation}
		where the partition $2p$ is defined in section \ref{section: notation}. It has components that are double those of $p$. The factor of two appears here because the cyclic group in \eqref{U(N) stabiliser} has been replaced by a dihedral group in \eqref{SO(N) stabiliser}.
			
		Applying the orbit-stabiliser theorem, the size of a double coset is
		\begin{equation}
		\frac{\left| S_n[S_2] \times S_n[S_2] \right|}{|\text{stabiliser}|} = \frac{2^{2n} (n!)^2}{z_{2p}}
		\label{size of double coset 1}
		\end{equation}		
		The even double cosets are of the form $p = 2q$, where $q \vdash \frac{n}{2}$. In terms of $q$, the size of an even double coset is
		\begin{equation}
		\frac{2^{2n} (n!)^2}{z_{4q}}
		\label{size of double coset 2}
		\end{equation}
		The above proof of this result was intended to provide intuition in preparation for the more complicated two-matrix version in section \ref{section: 2-matrix permutations to traces}. It is a well known result, and a more rigorous treatment can be found in \cite[Chapter VII.2]{Macdonald1995}.

		\subsection{Projection coefficients}
		\label{section: projection coefficients}
		
		Let $R \vdash n$ ($n$ even, otherwise all $U(N)$ operators project to 0) index a $U(N)$ operator, then we know that
		\begin{equation}
		\mathcal{O}^{U(N)}_R \overset{\mathbb{Z}_2}{\longrightarrow} \sum_{T} \alpha^T_R \mathcal{O}^{SO(N)}_T
		\label{Z2 projection of U(N) operator}
		\end{equation}
		where $T$ runs over the set of Young diagrams with $2n$ boxes made from $2 \times 2$ blocks. When defining the coefficients $\alpha^T_R$, we take $N$ to be infinite (equivalently, $N \geq n$) so that all the $\mathcal{O}^{SO(N)}_T$ are linearly independent. This means the $\alpha^T_R$ are defined uniquely and are independent of $N$.
		
		The single matrix $U(N)$ and $SO(N)$ operators are defined in \eqref{U(N) half-BPS operators} and \eqref{half-BPS mesonic operator} respectively, and both contain sums over permutations. Using the results of the previous section, we can re-express these as sums over partitions of $n$ and $\frac{n}{2}$. In the $U(N)$ case this is simply summing over conjugacy classes, and is possible as the summand of \eqref{U(N) half-BPS operators} is invariant under conjugation of $\sigma$. Similarly, the summand of \eqref{half-BPS mesonic operator} is invariant under pre- or post- multiplication of $\sigma$ by $S_n[S_2]$, and so we can reduce the sum to one over the double cosets. Using \eqref{size of conjugacy class} and \eqref{size of double coset 2} for the sizes of the conjugacy classes and double cosets respectively and recalling from section \ref{section: half-BPS permutations to traces} how permutations get contracted with tensor products of $X$ to produce traces, we have
		\begin{align}
		\mathcal{O}^{U(N)}_R & = d_R \sum_{p \vdash n} \frac{ \chi_R (p) }{z_p} \prod_i \left( \text{Tr} X^i \right)^{p_i}
		\label{U(N) operator from partitions} \\
		\mathcal{O}_T^{SO(N)} & = \frac{ d_T 2^{2n} \left( n! \right)^2}{(2n)!} \sum_{q \vdash \frac{n}{2}} \frac{1}{z_{4q}} \langle T, [S] | D^T \left( \sigma_{2q} \right) | T, [A] \rangle \prod_i \left( \text{Tr} X^{2i} \right)^{q_i}		
		\label{SO(N) operator from partitions}
		\end{align}
		where $\sigma_{2q} \in S_n^{(odd)}$ is of cycle type $2q$. 
		
		The matrix element $\langle T,[S] | D^T \left( \sigma_{2q} \right) | T, [A] \rangle$ was calculated in \cite{Ivanov1999}. A different representative of the double coset was used there, but it is simple to show that this differs by left multiplication only from an element of the form $\sigma_{2q}$, and therefore the matrix element is the same. Explicitly
		\begin{equation}
		\langle T,[S]| D^T \left( \sigma_{2q} \right) | T, [A] \rangle = \frac{2^{l(q)}}{2^n n!} \sqrt{\frac{(2n)!}{d_T}} \chi_t( q )
		\label{Ivanov's matrix element}
		\end{equation}
		where $t \vdash \frac{n}{2}$ is the Young diagram defined by taking each $2 \times 2$ block in $T$ and replacing it with a single square, so that $T = 2t \cup 2t$.
		
		Recalling \eqref{z_kp}, we find that
		\begin{equation}
		\mathcal{O}^{SO(N)}_T = 2^n n! \sqrt{\frac{d_T}{(2n)!}}  \sum_{q \vdash \frac{n}{2}} \frac{1}{z_{2q}} \chi_t (q) \prod_i \left( \text{Tr} X^{2i} \right)^{q_i}
		\label{SO(N) half-bps operator}
		\end{equation}
		We can use the character orthogonality relations \eqref{character orthogonality} to invert this, and find the multi-trace in terms of $\mathcal{O}_T^{SO(N)}$. We see that for a partition $p \vdash \frac{n}{2}$
		\begin{equation}
		\prod_i \left( \text{Tr} X^{2i} \right)^{p_i} = \frac{2^{l(p)}}{2^n n!} \sum_{t \vdash \frac{n}{2}} \sqrt{\frac{(2n)!}{d_T}} \chi_t (p) \mathcal{O}_T^{SO(N)}
		\label{multi-traces from SO(N) operators}
		\end{equation}
		Now we consider the projection of the $U(N)$ operators to the $SO(N)$ theory. This sets Tr$X^i = 0$ if $i$ odd, so the sum is restricted to only run over partitions with even parts. Reparameterising \eqref{U(N) operator from partitions} in terms of partitions of $\frac{n}{2}$, we get
		\begin{equation}
		\mathcal{O}_R^{U(N)} \overset{\mathbb{Z}_2}{\longrightarrow} d_R \sum_{p \vdash \frac{n}{2}} \frac{1}{z_{2p}} \chi_R (2p) \prod_i \left( \text{Tr} X^{2i} \right)^{p_i}
		\label{quotient of U(N) operator}
		\end{equation}
		Substituting in \eqref{multi-traces from SO(N) operators} gives
		\begin{equation}
		\mathcal{O}^{U(N)}_R \overset{\mathbb{Z}_2}{\longrightarrow} d_R \sum_{t \vdash \frac{n}{2}} \frac{1}{2^n n!} \sqrt{ \frac{(2n)!}{d_T}} \sum_{p \vdash \frac{n}{2}} \frac{1}{z_p} \chi_R (2p) \chi_t (p) \mathcal{O}^{SO(N)}_T
		\nonumber
		\end{equation}
		Then by comparison with \eqref{Z2 projection of U(N) operator}, we find
		\begin{equation}
		\alpha^T_R = \frac{d_R}{2^n n!} \sqrt{ \frac{(2n)!}{d_T}} \sum_{p \vdash \frac{n}{2}} \frac{1}{z_p} \chi_R (2p) \chi_t (p)
		\label{projection coefficients}
		\end{equation}
		The sum in this expression is particularly interesting, so we drop the normalisation factor to get
		\begin{equation}
		\bar{\alpha}^T_R = \sum_{p \vdash \frac{n}{2}} \frac{1}{z_p} \chi_R \left( 2p \right) \chi_t ( p )
		\label{reduced projection coefficients 1st definition}
		\end{equation}
		which we call reduced projection coefficients. Note that if we normalise $O_R^{U(N)}$ and $O_T^{SO(N)}$ to have identical two-point functions in the leading large $N$ limit (see \eqref{U(N) two-point function} and the $m=0$ simplification of \eqref{mesonic correlator}), the projection coefficients are exactly $\bar{\alpha}^T_R$.
		
		We give some low $n$ examples of $\bar{\alpha}^T_R$ in tables \ref{table: n=6 reduced projection coefficients} and \ref{table: n=8 reduced projection coefficients}, calculated in GAP using the above formula. The $n=4$ coefficients can be read off from (\ref{projection of [4]},~\ref{projection of [3,1]},~\ref{projection of [2,2]},~\ref{projection of [2,1,1]},~\ref{projection of [1,1,1,1]}). We see that the reduced projection coefficients are integers, and further numerical exploration gives many nice relations between the different coefficients. Since they are integers, we hope to find some combinatoric interpretation that will shed light on these relations. We find two combinatoric rules involving domino tableaux of shape $R$ and $T$ respectively, given in \eqref{combinatorial rule for difference} and \eqref{combinatorial rule for individual plethysms}.
		
		Note that one pattern we see in tables \ref{table: n=6 reduced projection coefficients} and \ref{table: n=8 reduced projection coefficients} that does not generalise is $\left| \bar{\alpha}^T_R \right| \leq 1$. The first coefficient that breaks this pattern has is found at $n=12$, with $R = [4,4,2,2]$ and $t = [3,2,1]$. For this $R,t$ we have $\bar{\alpha}^T_R = 2$.
		
		We now manipulate \eqref{reduced projection coefficients 1st definition} into a suitable form to relate the coefficients to the combinatorics of domino tableaux.
		
		\begin{table}
			\begin{center}
				\begin{tabular}{c | c | c | c}
					& [3] & [2,1] & [1,1,1] \\ \hline
					[6] & 1 & 0 & 0 \\ \hline
					[5,1] & -1 & 0 & 0 \\ \hline
					[4,2] & 1 & 1 & 0 \\ \hline
					[4,1,1] & 0 & -1 & 0 \\ \hline
					[3,3] & -1 & -1 & 0 \\ \hline
					[3,2,1] & 0 & 0 & 0 \\ \hline
					[3,1,1,1] & 0 & 1 & 0 \\ \hline
					[2,2,2] & 0 & 1 & 1 \\ \hline
					[2,2,1,1] & 0 & -1 & -1 \\  \hline
					[2,1,1,1,1] & 0 & 0 & 1 \\ \hline
					[1,1,1,1,1,1] & 0 & 0 & -1 \\
				\end{tabular}
			\end{center}
			\caption{Reduced projection coefficients $\bar{\alpha}^T_R$ at $n=6$. The leftmost column indexes $R \vdash n$ while the top row indexes $t \vdash \frac{n}{2}$. The $T$ in $\bar{\alpha}^T_R$ is constructed from $t$ by replacing each individual square in the Young diagram with a $2 \times 2$ block.}
			\label{table: n=6 reduced projection coefficients}
		\end{table}
		
		\begin{table}
			\begin{center}
				\begin{tabular}{c | c | c | c | c | c}
					& [4] & [3,1] & [2,2] & [2,1,1] & [1,1,1,1]  \\ \hline
					[8] & 1 & 0 & 0 & 0 & 0 \\ \hline
					[7,1] & -1 & 0 & 0 & 0 & 0 \\ \hline
					[6,2] & 1 & 1 & 0 & 0 & 0 \\ \hline
					[6,1,1] & 0 & -1 & 0 & 0 & 0 \\ \hline
					[5,3] & -1 & -1 & 0 & 0 & 0 \\ \hline
					[5,2,1] & 0 & 0 & 0 & 0 & 0 \\ \hline
					[5,1,1,1] & 0 & 1 & 0 & 0 & 0 \\ \hline
					[4,4] & 1 & 1 & 1 & 0 & 0 \\ \hline
					[4,3,1] & 0 & 0 & -1 & 0 & 0 \\ \hline
					[4,2,2] & 0 & 1 & 1 & 1 & 0 \\ \hline
					[4,2,1,1] & 0 & -1 & 0 & -1 & 0 \\ \hline
					[4,1,1,1,1] & 0 & 0 & 0 & 1 & 0 \\ \hline
					[3,3,2] & 0 & -1 & 0 & -1 & 0 \\ \hline
					[3,3,1,1] & 0 & 1 & 1 & 1 & 0 \\ \hline
					[3,2,2,1] & 0 & 0 & -1 & 0 & 0 \\ \hline
					[3,2,1,1,1] & 0 & 0 & 0 & 0 & 0 \\ \hline
					[3,1,1,1,1,1] & 0 & 0 & 0 & -1 & 0 \\ \hline
					[2,2,2,2] & 0 & 0 & 1 & 1 & 1 \\ \hline
					[2,2,2,1,1] & 0 & 0 & 0 & -1 & -1 \\ \hline
					[2,2,1,1,1,1] & 0 & 0 & 0 & 1 & 1 \\ \hline
					[2,1,1,1,1,1,1] & 0 & 0 & 0 & 0 & -1 \\ \hline
					[1,1,1,1,1,1,1,1] & 0 & 0 & 0 & 0 & 1 \\
				\end{tabular}
			\end{center}
			\caption{Reduced projection coefficients $\bar{\alpha}^T_R$ at $n=8$. The leftmost column indexes $R \vdash n$ while the top row indexes $t \vdash \frac{n}{2}$.The $T$ in $\bar{\alpha}^T_R$ is constructed from $t$ by replacing each individual square in the Young diagram with a $2 \times 2$ block.}
			\label{table: n=8 reduced projection coefficients}
		\end{table}

		Using \eqref{size of conjugacy class}, we see we can replace the sum over partitions of $\frac{n}{2}$ by a sum over $S_{\frac{n}{2}}$.
		\begin{equation}
		\bar{\alpha}^T_R = \frac{1}{\left( \frac{n}{2} \right)!} \sum_{\sigma \in S_{\frac{n}{2}}} \chi_R \left( \sigma \tau \right) \chi_t ( \sigma )
		\label{reduced projection coefficients from permutations}
		\end{equation}
		Where $\tau \in S_n$ is defined by
		\begin{equation}
		\tau = \left( 1, 1 + \frac{n}{2} \right) \left( 2, 2 + \frac{n}{2} \right) \ldots \left( \frac{n}{2} , n \right)
		\label{tau definition}
		\end{equation}
		and we have embedded $S_\frac{n}{2}$ in $S_n$ by having it act on $\{ 1,2, \ldots ,\frac{n}{2} \}$. It is then easy to check that if $\sigma \in S_{\frac{n}{2}}$ has cycle type $p$, $\sigma \tau$ will have cycle type $2p$.
		
		The sum over $S_\frac{n}{2}$ in \eqref{reduced projection coefficients from permutations} is proportional to the projector onto irrep $t$ of $S_\frac{n}{2}$, defined in \eqref{projector}. So we have
		\begin{equation}
		\bar{\alpha}^T_R = \frac{1}{d_t} \text{Tr}_R \left( P_t \tau \right)
		\nonumber
		\end{equation}
		From the definition \eqref{tau definition}, we see that $\tau$ switches the sets $\{ 1,2, \ldots , \frac{n}{2} \}$ and $\left\{ \frac{n}{2} + 1, \frac{n}{2} + 2, \ldots ,n \right\}$. So for $\sigma \in S_\frac{n}{2}$, conjugating by $\tau$ takes $\sigma$ to the equivalent element of a different embedding of $S_\frac{n}{2}$, namely that defined by acting on $\{ \frac{n}{2} + 1, \frac{n}{2} + 2, \ldots ,n \}$. Therefore conjugating $P_t$ by $\tau$ gives the projector onto the $t$ irrep of this different embedding of $S_\frac{n}{2}$. We call this $\hat{P}_t$.
		
		Then using properties of projectors and traces
		\begin{align*}
		\bar{\alpha}^T_R & = \frac{1}{d_t} \text{Tr}_R \left( P_t P_t \tau \right) \\
		& = \frac{1}{d_t} \text{Tr}_R \left( P_t \hat{P}_t \tau \right) \\
		& = \frac{1}{d_t} \text{Tr}_R \left( P_{t \otimes t} \tau \right)
		\end{align*}
		where $P_{t \otimes t} = P_t \hat{P}_t = \hat{P}_t P_t$ is the projector onto the irrep $t \otimes t$ of $S_\frac{n}{2} \times S_\frac{n}{2}$. From this expression we can see that $\bar{\alpha}^T_R$ is related to the Littlewood-Richardson coefficient $g_{t,t;R}$, since this keeps track of the number of distinct copies of $t \otimes t$ contained in $R$. One immediate consequence is that if $g_{t,t;R}=0$, we must have $\bar{\alpha}^T_R = 0$.
		
		To go further in evaluating $\bar{\alpha}^T_R$, we use Schur-Weyl duality to change the trace from one over an irrep of $S_n$ to one over an irrep of $U(N)$. 
		
		Let $V$ be the vector space for the fundamental of $U(N)$. Then by the standard rules of tensor product representations, $V^{\otimes n}$ carries a representation of $U(N)$. We can also define an action of $S_n$ by permutation of the tensor factors. Schur-Weyl duality states that these two actions commute and that the tensor product space can be decomposed as
		\begin{equation}
		V^{\otimes n} = \bigoplus_{\substack{R \vdash n \\ l(R) \leq N}} V_R^{U(N)} \otimes V_R^{S_n}
		\label{Schur-Weyl duality}
		\end{equation}
		where $V_R^{U(N)}$ and $V_R^{S_n}$ are the representation spaces for the irreps of $U(N)$ and $S_n$ labelled by $R$. We denote the dimensions of the $U(N)$ representations by $d_R^{U(N)}$ and keep the notation $d_R$ for the $S_n$ representations. On the right hand side of this identification, $U(N)$ acts only on the $U(N)$ tensor factor, and similarly for $S_n$. Note the restriction $l(R) \leq N$. This means the following arguments only apply when $l(R) \leq N$. However, since $\alpha^T_R$ are independent of $N$, the conclusion \eqref{reduced projection coefficients} holds true for all $N$.
		
		The structure \eqref{Schur-Weyl duality} means that traces over $V^{\otimes n}$ can be decomposed into traces over $U(N)$ and $S_n$ irreps
		\begin{equation}
		\text{Tr}_{V^{\otimes n}} \left( \sigma U \right) = \sum_{R \vdash n} \left( \text{Tr}_{V_R^{U(N)}} U \right) \left( \text{Tr}_{V_R^{S_n}} \sigma \right) \qquad \qquad \sigma \in S_n , \ \  U \in U(N)
		\nonumber
		\end{equation}
		In direct analogy to the projector \eqref{projector} we can define an operator that projects onto the $R$ irrep of $U(N)$. Since $U(N)$ is a compact Lie group, the sum is replaced by an integral over the Haar measure (normalised so that the volume of the group is 1).
		\begin{equation}
		P_R^{U(N)} = \int \text{d}U \ \chi_R^{U(N)} \left( U^{-1} \right) U
		\nonumber
		\end{equation}
		We can use this to express $\bar{\alpha}^T_R$ as a trace over the whole of $V^{\otimes n}$
		\begin{align*}
		\bar{\alpha}^T_R & = \frac{1}{d_t} \text{Tr}_{V_R^{S_n}} \left( P_{t \otimes t} \, \tau \right) \\
		& = \frac{1}{d_R^{U(N)} \, d_t} \text{Tr}_{V_R^{U(N)} \otimes V_R^{S_n}} \left( P_{t \otimes t} \, \tau \right) \\
		& = \frac{1}{d_R^{U(N)} \, d_t} \text{Tr}_{V^{\otimes n}} \left( P_R^{U(N)} P_{t \otimes t} \, \tau \right)
		\end{align*}
		To reduce this to a trace over a $U(N)$ representation, we now decompose $V^{\otimes n}$ in a way that will allow us to use $P_{t \otimes t}$ in the same manner that $P_R^{U(N)}$ was used above.
		
		Trivially, we have $V^{\otimes n} = V^{\otimes \frac{n}{2}} \otimes V^{\otimes \frac{n}{2}}$, so we can do a Schur-Weyl decomposition on each of the two factors
		\begin{align}
		V^{\otimes n} & = \left( \bigoplus_{r \vdash \frac{n}{2}} V_r^{U(N)} \otimes V_r^{S_{n\over 2 } } \right) \otimes \left( \bigoplus_{t \vdash \frac{n}{2}} V_t^{U(N)} \otimes V_t^{S_{n\over 2 } } \right) \nonumber \\
		& = \bigoplus_{r,t \vdash \frac{n}{2}} V_r^{U(N)} \otimes V_t^{U(N)} \otimes V_r^{S_{n \over 2 } } \otimes V_t^{S_{n\over 2 } }
		\label{halved schur-weyl decomposition}
		\end{align}
		The permutation $\tau$ acts on $V^{\otimes \frac{n}{2}} \otimes V^{\otimes \frac{n}{2}}$ by exchanging the two factors
		\begin{equation}
		\tau \left( u, v \right) = \left( v, u \right) \qquad \qquad u,v \in V^{\otimes \frac{n}{2}}
		\nonumber
		\end{equation}
		After decomposing the two copies of $V^{\otimes \frac{n}{2}}$, \eqref{halved schur-weyl decomposition}, we see that for $r \neq t$, $\tau$ exchanges the spaces labelled by $(r,t)$ and $(t,r)$. However, on the spaces with $r=t$, $\tau$ splits into a tensor product operator
		\begin{equation}
		\tau = \tau^{U(N)} \otimes \tau^{S_n}
		\nonumber
		\end{equation}
		where $\tau^{U(N)}$ acts on $V_t^{U(N)} \otimes V_t^{U(N)}$ and $\tau^{S_n}$ acts on $V_t^{S_{n\over 2 } } \otimes V_t^{S_{n \over 2 } }$, both by exchanging the factors. This allows us to go further in evaluating $\bar{\alpha}^T_R$
		\begin{align*}
		\bar{\alpha}^T_R & = \frac{1}{d_R^{U(N)} \, d_t} \text{Tr}_{V^{\otimes n}} \left(  P_{t \otimes t} P_R^{U(N)} \, \tau \right) \\
		& =  \frac{1}{d_R^{U(N)} \, d_t} \text{Tr}_{V_t^{U(N)} \otimes V_t^{U(N)} \otimes V_t^{S_{n\over 2 } } \otimes V_t^{S_{ n \over 2 } }} \left( P_R^{U(N)} \tau \right) \\
		& = \frac{1}{d_R^{U(N)} \, d_t} \text{Tr}_{V_t^{U(N)} \otimes V_t^{U(N)}} \left( P_R^{U(N)} \tau^{U(N)} \right) \text{Tr}_{V_t^{S_{n\over 2 } } \otimes V_t^{S_{n\over 2 } }} \left( \tau^{S_n} \right)
		\end{align*}
		We can split $V_t^{S_{n\over 2 } } \otimes V_t^{S_{n\over 2 } }$ into its symmetric part and its anti-symmetric part, on which $\tau^{S_n}$ acts as 1 and $-1$ respectively. This gives us
		\begin{align*}
		\text{Tr}_{V_t^{S_{ n\over 2 } } \otimes V_t^{S_{ n\over 2 } }} \left( \tau^{S_n} \right) & = \text{Dim} \left[ S^2 \left( V_t^{S_{ n \over 2 } } \right) \right] - \text{Dim} \left[ \Lambda^2 \left( V_t^{S_{ n\over 2 } } \right) \right] \\
		& = \frac{d_t \left( d_t + 1 \right)}{2} - \frac{d_t \left( d_t - 1 \right)}{2} \\
		& = d_t
		\end{align*}
		We can apply the same process to $V_t^{U(N)} \otimes V_t^{U(N)}$, giving
		\begin{equation}
		\bar{\alpha}^T_R = \frac{1}{d_R^{U(N)}} \left[ \text{Tr}_{S^2\left( V_t^{U(N)} \right)} \left( P_R^{U(N)} \right) -  \text{Tr}_{\Lambda^2\left( V_t^{U(N)} \right)} \left( P_R^{U(N)} \right) \right]
		\nonumber
		\end{equation}
		Each of the two terms is just the multiplicity of the $R$ irrep of $U(N)$ in $S^2 \left( V_t^{U(N)} \right)$ and $\Lambda^2 \left( V_t^{U(N)} \right)$ respectively. So we have
		\begin{align}
		\bar{\alpha}^T_R & = \text{Mult} \left[ R, S^2 \left( V_t^{U(N)} \right) \right] - \text{Mult} \left[ R, \Lambda^2 \left( V_t^{U(N)} \right) \right] \nonumber \\
		& = \mathcal{P}(t,[2],R) - \mathcal{P}(t,[1,1],R)
		\label{reduced projection coefficients}
		\end{align}
		where the plethysm coefficients $\mathcal{P}(t,\Lambda,R)$ were defined in the introduction.
		
		The Littlewood-Richardson coefficient is
		\begin{align}
		g_{t,t;R} & = \text{Mult} \left( R, V_t^{U(N)} \otimes V_t^{U(N)} \right) \nonumber \\
		& = \mathcal{P}(t,[2],R) + \mathcal{P}(t,[1,1],R)
		\label{LRcoefficient = sum of plethysms}
		\end{align}
		so again we see that $g_{t,t;R} = 0$ is a sufficient condition for $\alpha^T_R = 0$. Additionally, this shows that the parity of $g_{t,t;R}$ is the same as the parity of $\bar{\alpha}^T_R$.
		
		The plethysm coefficients $\mathcal{P}(t,[2],R)$ and $\mathcal{P}(t,[1,1],R)$  were the subject of the paper \cite{Carre1995}. They present two combinatorial rules, the first gives the difference $\mathcal{P}(t,[2],R) - \mathcal{P}(t,[1,1],R) = \bar{\alpha}^T_R$ directly, while the second gives the two plethysm coefficients individually. Both rules involve Yamanouchi domino tableaux, which we now define.

		\subsection{Domino tableaux and combinatorics of plethysms}
		\label{section: domino tableaux}

		A domino tiling of shape $R \vdash n$ ($n$ even) is a tiling of the shape $R$ with $2 \times 1$ or $1 \times 2$ rectangles, which are called dominoes. A domino tableau is a tiling where each domino contains a positive integer, such that the numbers increase weakly along the rows and strictly down the columns. Note that each domino occupies 2 rows and 1 column (or 2 columns and 1 row), and the integers contained within the dominoes must be correctly ordered in both rows (columns).
		
		Each column in a domino tableau defines a word by reading the numbers in the column from bottom to top, where horizontal dominoes, which span two columns, only contribute to the leftmost column. The column reading of the tableau is then defined by concatenating these words, starting on the left and ending on the right.
		
		A Yamanouchi word is a word on the alphabet of positive integers such that each suffix contains at least as many 1s as 2s, at least as many 2s as 3s, and, more generally, at least as many $i$s as $i+1$s for every $i$. A Yamanouchi domino tableau is a domino tableau for which the column reading is a Yamanouhci word.
		
		For a given Yamanouchi domino tableau, let the number of integers $i$ in the tableau be given by $\lambda_i$. We define the evaluation of the tableau to be $\lambda = [\lambda_1,\lambda_2, \ldots ]$. Clearly $\sum_i \lambda_i = \frac{n}{2}$, and the Yamanouchi condition ensures that $\lambda$ is a partition of $\frac{n}{2}$, i.e. the $\lambda_i$ are weakly decreasing.
		
		As an example of the above definitions, figure \ref{figure: Yamanouchi domino tableaux} gives the ten Yamanouchi domino tableaux of shape $[4,4,3,3,1,1]$ along with their evaluations.
				
		\begin{figure}		
		\begin{gather*}
		\begin{tikzpicture}
		\draw (0,0) -- (0,9em) ;
		\draw (1.5em,0) -- (1.5em,9em) ;
		\draw (3em,3em) -- (3em,9em) ;
		\draw (4.5em,3em) -- (4.5em,9em) ;
		\draw (6em,6em) -- (6em,9em) ;
		\draw (0,0) -- (1.5em,0) ;
		\draw (0,3em) -- (4.5em,3em) ;
		\draw (0,6em) -- (6em,6em) ;
		\draw (0,9em) -- (6em,9em) ;
		\node at (0.75em,1.5em) {3} ;
		\node at (0.75em,4.5em) {2} ;
		\node at (2.25em,4.5em) {2} ;
		\node at (3.75em,4.5em) {2} ;
		\node at (0.75em,7.5em) {1} ;
		\node at (2.25em,7.5em) {1} ;
		\node at (3.75em,7.5em) {1} ;
		\node at (5.25em,7.5em) {1} ;
		\node at (3em,-1.5em) {$[4,3,1]$} ;
		\end{tikzpicture}
		\hspace{1.2em}
		\begin{tikzpicture}
		\draw (0,0) -- (0,9em) ;
		\draw (1.5em,0) -- (1.5em,9em) ;
		\draw (3em,6em) -- (3em,9em) ;
		\draw (4.5em,3em) -- (4.5em,9em) ;
		\draw (6em,6em) -- (6em,9em) ;
		\draw (0,0) -- (1.5em,0) ;
		\draw (0,3em) -- (4.5em,3em) ;
		\draw (1.5em,4.5em) -- (4.5em,4.5em) ;
		\draw (0,6em) -- (6em,6em) ;
		\draw (0,9em) -- (6em,9em) ;
		\node at (0.75em,1.5em) {3} ;
		\node at (0.75em,4.5em) {2} ;
		\node at (3em,3.75em) {3} ;
		\node at (3em,5.25em) {2} ;
		\node at (0.75em,7.5em) {1} ;
		\node at (2.25em,7.5em) {1} ;
		\node at (3.75em,7.5em) {1} ;
		\node at (5.25em,7.5em) {1} ;
		\node at (3em,-1.5em) {$[4,2,2]$} ;
		\end{tikzpicture}
		\hspace{1.2em}
		\begin{tikzpicture}
		\draw (0,0) -- (0,9em) ;
		\draw (1.5em,0) -- (1.5em,9em) ;
		\draw (3em,6em) -- (3em,9em) ;
		\draw (4.5em,3em) -- (4.5em,9em) ;
		\draw (6em,6em) -- (6em,9em) ;
		\draw (0,0) -- (1.5em,0) ;
		\draw (0,3em) -- (4.5em,3em) ;
		\draw (1.5em,4.5em) -- (4.5em,4.5em) ;
		\draw (0,6em) -- (6em,6em) ;
		\draw (0,9em) -- (6em,9em) ;
		\node at (0.75em,1.5em) {4} ;
		\node at (0.75em,4.5em) {2} ;
		\node at (3em,3.75em) {3} ;
		\node at (3em,5.25em) {2} ;
		\node at (0.75em,7.5em) {1} ;
		\node at (2.25em,7.5em) {1} ;
		\node at (3.75em,7.5em) {1} ;
		\node at (5.25em,7.5em) {1} ;
		\node at (3em,-1.5em) {$[4,2,1,1]$} ;
		\end{tikzpicture}
		\hspace{1.2em}
		\begin{tikzpicture}
		\draw (0,0) -- (0,9em) ;
		\draw (1.5em,0) -- (1.5em,9em) ;
		\draw (3em,3em) -- (3em,9em) ;
		\draw (4.5em,3em) -- (4.5em,6em) ;
		\draw (6em,6em) -- (6em,9em) ;
		\draw (0,0) -- (1.5em,0) ;
		\draw (0,3em) -- (4.5em,3em) ;
		\draw (0,6em) -- (6em,6em) ;
		\draw (3em,7.5em) -- (6em,7.5em) ;
		\draw (0,9em) -- (6em,9em) ;
		\node at (0.75em,1.5em) {3} ;
		\node at (0.75em,4.5em) {2} ;
		\node at (2.25em,4.5em) {2} ;
		\node at (3.75em,4.5em) {3} ;
		\node at (0.75em,7.5em) {1} ;
		\node at (2.25em,7.5em) {1} ;
		\node at (4.5em,6.75em) {2} ;
		\node at (4.5em,8.25em) {1} ;
		\node at (3em,-1.5em) {$[3,3,2]$} ;
		\end{tikzpicture}
		\hspace{1.2em}
		\begin{tikzpicture}
		\draw (0,0) -- (0,9em) ;
		\draw (1.5em,0) -- (1.5em,9em) ;
		\draw (3em,3em) -- (3em,9em) ;
		\draw (4.5em,3em) -- (4.5em,6em) ;
		\draw (6em,6em) -- (6em,9em) ;
		\draw (0,0) -- (1.5em,0) ;
		\draw (0,3em) -- (4.5em,3em) ;
		\draw (0,6em) -- (6em,6em) ;
		\draw (3em,7.5em) -- (6em,7.5em) ;
		\draw (0,9em) -- (6em,9em) ;
		\node at (0.75em,1.5em) {4} ;
		\node at (0.75em,4.5em) {2} ;
		\node at (2.25em,4.5em) {2} ;
		\node at (3.75em,4.5em) {3} ;
		\node at (0.75em,7.5em) {1} ;
		\node at (2.25em,7.5em) {1} ;
		\node at (4.5em,6.75em) {2} ;
		\node at (4.5em,8.25em) {1} ;
		\node at (3em,-1.5em) {$[3,3,1,1]$} ;
		\end{tikzpicture}
		\\
		\begin{tikzpicture}
		\draw (0,0) -- (0,9em) ;
		\draw (1.5em,0) -- (1.5em,9em) ;
		\draw (3em,6em) -- (3em,9em) ;
		\draw (4.5em,3em) -- (4.5em,6em) ;
		\draw (6em,6em) -- (6em,9em) ;
		\draw (0,0) -- (1.5em,0) ;
		\draw (0,3em) -- (4.5em,3em) ;
		\draw (1.5em,4.5em) -- (4.5em,4.5em) ;
		\draw (0,6em) -- (6em,6em) ;
		\draw (3em,7.5em) -- (6em,7.5em) ;
		\draw (0,9em) -- (6em,9em) ;
		\node at (0.75em,1.5em) {3} ;
		\node at (0.75em,4.5em) {2} ;
		\node at (3em,5.25em) {3} ;
		\node at (3em,3.75em) {4} ;
		\node at (0.75em,7.5em) {1} ;
		\node at (2.25em,7.5em) {1} ;
		\node at (4.5em,6.75em) {2} ;
		\node at (4.5em,8.25em) {1} ;
		\node at (3em,-1.5em) {$[3,2,2,1]$} ;
		\end{tikzpicture}
		\hspace{1.2em}
		\begin{tikzpicture}
		\draw (0,0) -- (0,9em) ;
		\draw (1.5em,0) -- (1.5em,3em) ;
		\draw (1.5em,6em) -- (1.5em,9em) ;
		\draw (3em,3em) -- (3em,9em) ;
		\draw (4.5em,3em) -- (4.5em,6em) ;
		\draw (6em,6em) -- (6em,9em) ;
		\draw (0,0) -- (1.5em,0) ;
		\draw (0,3em) -- (4.5em,3em) ;
		\draw (0,4.5em) -- (3em,4.5em) ;
		\draw (0,6em) -- (6em,6em) ;
		\draw (3em,7.5em) -- (6em,7.5em) ;
		\draw (0,9em) -- (6em,9em) ;
		\node at (0.75em,1.5em) {4} ;
		\node at (1.5em,3.75em) {3} ;
		\node at (1.5em,5.25em) {2} ;
		\node at (3.75em,4.5em) {3} ;
		\node at (0.75em,7.5em) {1} ;
		\node at (2.25em,7.5em) {1} ;
		\node at (4.5em,6.75em) {2} ;
		\node at (4.5em,8.25em) {1} ;
		\node at (3em,-1.5em) {$[3,2,2,1]$} ;
		\end{tikzpicture}
		\hspace{1.2em}
		\begin{tikzpicture}
		\draw (0,0) -- (0,9em) ;
		\draw (1.5em,0) -- (1.5em,9em) ;
		\draw (3em,6em) -- (3em,9em) ;
		\draw (4.5em,3em) -- (4.5em,6em) ;
		\draw (6em,6em) -- (6em,9em) ;
		\draw (0,0) -- (1.5em,0) ;
		\draw (0,3em) -- (4.5em,3em) ;
		\draw (1.5em,4.5em) -- (4.5em,4.5em) ;
		\draw (0,6em) -- (6em,6em) ;
		\draw (3em,7.5em) -- (6em,7.5em) ;
		\draw (0,9em) -- (6em,9em) ;
		\node at (0.75em,1.5em) {5} ;
		\node at (0.75em,4.5em) {2} ;
		\node at (3em,5.25em) {3} ;
		\node at (3em,3.75em) {4} ;
		\node at (0.75em,7.5em) {1} ;
		\node at (2.25em,7.5em) {1} ;
		\node at (4.5em,6.75em) {2} ;
		\node at (4.5em,8.25em) {1} ;
		\node at (3em,-1.5em) {$[3,2,1,1,1]$} ;
		\end{tikzpicture}
		\hspace{1.2em}
		\begin{tikzpicture}
		\draw (0,0) -- (0,9em) ;
		\draw (1.5em,0) -- (1.5em,6em) ;
		\draw (3em,6em) -- (3em,9em) ;
		\draw (4.5em,3em) -- (4.5em,6em) ;
		\draw (6em,6em) -- (6em,9em) ;
		\draw (0,0) -- (1.5em,0) ;
		\draw (0,3em) -- (4.5em,3em) ;
		\draw (1.5em,4.5em) -- (4.5em,4.5em) ;
		\draw (0,6em) -- (6em,6em) ;
		\draw (0,7.5em) -- (6em,7.5em) ;
		\draw (0,9em) -- (6em,9em) ;
		\node at (0.75em,1.5em) {4} ;
		\node at (0.75em,4.5em) {3} ;
		\node at (3em,5.25em) {3} ;
		\node at (3em,3.75em) {4} ;
		\node at (1.5em,6.75em) {2} ;
		\node at (1.5em,8.25em) {1} ;
		\node at (4.5em,6.75em) {2} ;
		\node at (4.5em,8.25em) {1} ;
		\node at (3em,-1.5em) {$[2,2,2,2]$} ;
		\end{tikzpicture}
		\hspace{1.2em}
		\begin{tikzpicture}
		\draw (0,0) -- (0,9em) ;
		\draw (1.5em,0) -- (1.5em,6em) ;
		\draw (3em,6em) -- (3em,9em) ;
		\draw (4.5em,3em) -- (4.5em,6em) ;
		\draw (6em,6em) -- (6em,9em) ;
		\draw (0,0) -- (1.5em,0) ;
		\draw (0,3em) -- (4.5em,3em) ;
		\draw (1.5em,4.5em) -- (4.5em,4.5em) ;
		\draw (0,6em) -- (6em,6em) ;
		\draw (0,7.5em) -- (6em,7.5em) ;
		\draw (0,9em) -- (6em,9em) ;
		\node at (0.75em,1.5em) {5} ;
		\node at (0.75em,4.5em) {3} ;
		\node at (3em,5.25em) {3} ;
		\node at (3em,3.75em) {4} ;
		\node at (1.5em,6.75em) {2} ;
		\node at (1.5em,8.25em) {1} ;
		\node at (4.5em,6.75em) {2} ;
		\node at (4.5em,8.25em) {1} ;
		\node at (3em,-1.5em) {$[2,2,2,1,1]$} ;
		\end{tikzpicture}
		\end{gather*}
		\caption{The possible Yamanouchi domino tableaux of shape [4,4,3,3,1,1]. The evaluation of each tableau is given beneath.}
		\label{figure: Yamanouchi domino tableaux}
		\end{figure}
		
		A key property of a domino tiling is the number of horizontal or vertical dominoes. Take $R \vdash n$, with components $R_1, R_2, \ldots , R_k$. Assume that $R$ admits a domino tiling, and let $r$ be such a tiling. Then define $h_i(r)$ to be the number of horizontal dominoes in row $i$ of $r$, $v_i(r)$ be the number of vertical dominoes with their uppermost box in row $i$, and $h(r)$ and $v(r)$ be the total number of horizontal and vertical dominoes. Then we have
		\begin{gather*}
		R_1 = 2 h_1(r) + v_1(r) \\ 
		R_2 = 2 h_2(r) + v_1(r) + v_2(r) \\ 
		R_3 = 2 h_3(r) + v_2(r) + v_3(r) \\ 
		\vdots \\
		R_{k-1} = 2 h_{k-1}(r) + v_{k-2}(r) + v_{k-1}(r) \\
		R_k = 2 h_k(r) + v_{k-1}(r)
		\end{gather*}
		Therefore
		\begin{equation}
		(-1)^{R_1 + R_3 + \ldots } = (-1)^{2\left( h_1(r) + h_3(r) + \ldots \right) + v_1(r) + v_2(r) + \ldots + v_{k-1}(r)} = (-1)^{v(r)} 
		\end{equation}
		Crucially, if a domino tiling of shape $R$ exists, the parity of $v(r)$ (and therefore the parity of $h(r)$) depends only on $R$, and not on how the dominoes are arranged. In light of this, we define $\varepsilon_2 (R)$, the \emph{2-sign} of $R$, to be $(-1)^{v(r)}$ if $R$ admits a domino tiling, and 0 otherwise.
		
		This allows us to give the first combinatorial rule, proved in \cite{Carre1995}, for finding $\bar{\alpha}^T_R$. Defining $D^R_\lambda$ to be the number of Yamanouchi domino tableau of shape $R$ and evaluation $\lambda$, we have
		\begin{equation}
		\bar{\alpha}^T_R = \mathcal{P}(t,[2],R) - \mathcal{P}(t,[1,1],R) = \varepsilon_2 (R) D^R_t
		\label{combinatorial rule for difference}
		\end{equation}
		Note this means the sign of the non-zero $\bar{\alpha}^T_R$ depends only on $R$ and not $T$, since $D^R_t \geq 0$. This can be seen in tables \ref{table: n=6 reduced projection coefficients} and \ref{table: n=8 reduced projection coefficients}, where each row consists only of zeroes and positive numbers, or zeroes and negative numbers.
		
		For the second rule, consider $T \vdash 2n$, constructed from $2 \times 2$ blocks. Clearly we can tile $T$ with dominoes by putting 2 horizontal dominoes in each $2 \times 2$ block. Therefore in any domino tableau of $T$, there must be an even number of horizontal (and vertical) dominoes. We split the domino tableau of shape $T$ into two classes, based on the number of pairs of horizontal dominoes. If a tableau has an even number of pairs, we say it has spin $1$, while if it has an odd number of pairs it has spin $-1$. So for $T$ of this type, we define $D^T_{+,R}$ and $D^T_{-,R}$ to be the number of Yamanouchi domino tableaux of evaluation $R$ and positive and negative spin respectively. The second combinatorial rule, which gives the two plethysm coefficients individually, is
		\begin{align}
		\mathcal{P} \left( t, [2],R \right) & = D^T_{+,R} & \mathcal{P} \left( t, [1,1],R \right) & = D^T_{-,R}
		\label{combinatorial rule for individual plethysms}
		\end{align}
		This leads to a second expression for \eqref{reduced projection coefficients}
		\begin{equation}
		\bar{\alpha}^T_R = D^T_{+,R} - D^T_{-,R}
		\end{equation}		
		Note that $D^T_{+,R} + D^T_{-,R} = D^T_R$, so from \eqref{LRcoefficient = sum of plethysms} we have
		\begin{equation}
		g_{t,t;R} = D^T_R \nonumber
		\end{equation}
		The two combinatoric methods of finding $\bar{\alpha}^T_R$ are independent of each other. For example if we take $R=[3,2,1]$, then there are no domino tableau of shape $R$, so \eqref{combinatorial rule for difference} gives 0 trivially. However if we look at Yamanouchi domino tableau of shape $T = [4,4,2,2]$ (corresponding to $t = [2,1]$) and evaluation $R$, we find two such tableaux, one contributing to each of the two plehtysm coefficients. These two tableaux are 
		\begin{gather}
		\begin{gathered}
		\begin{tikzpicture}	
		\draw (0,0) -- (0,6em) ;
		\draw (1.5em,3em) -- (1.5em,6em) ;
		\draw (3em, 0) -- (3em, 6em) ;
		\draw (6em, 3em) -- (6em, 6em) ;
		\draw (0,0) -- (3em,0) ;
		\draw(0,1.5em) -- (3em, 1.5em) ;
		\draw (0, 3em) -- (6em, 3em) ;
		\draw (3em, 4.5em) -- (6em, 4.5em) ;
		\draw (0, 6em) -- (6em, 6em) ;
		\node at (1.5em,0.75em) {3} ;
		\node at (1.5em,2.25em) {2} ;
		\node at (0.75em,4.5em) {1} ;
		\node at (2.25em,4.5em) {1} ;
		\node at (4.5em,5.25em) {1} ;
		\node at (4.5em,3.75em) {2} ;
		\end{tikzpicture}	
		\end{gathered}
		\qquad \qquad
		\begin{gathered}
		\begin{tikzpicture}	
		\draw (0,0) -- (0,6em) ;
		\draw (1.5em,0) -- (1.5em,6em) ;
		\draw (3em, 0) -- (3em, 6em) ;
		\draw (6em, 3em) -- (6em, 6em) ;
		\draw (0,0) -- (3em,0) ;
		\draw (0, 3em) -- (6em, 3em) ;
		\draw (3em, 4.5em) -- (6em, 4.5em) ;
		\draw (0, 6em) -- (6em, 6em) ;
		\node at (0.75em,1.5em) {2} ;
		\node at (2.25em,1.5em) {3} ;
		\node at (0.75em,4.5em) {1} ;
		\node at (2.25em,4.5em) {1} ;
		\node at (4.5em,5.25em) {1} ;
		\node at (4.5em,3.75em) {2} ;
		\end{tikzpicture}	
		\end{gathered}
		\label{two domino tableaux}
		\end{gather}
		We see that the first tableau has spin $+1$ while the second has spin $-1$. Using \eqref{combinatorial rule for individual plethysms}, we get $\mathcal{P}(t,[2],R) = \mathcal{P}(t,[1,1],R)= 1$, and therefore $\bar{\alpha}^T_R = 0$ as claimed.
		
		The two tableaux in \eqref{two domino tableaux} can also be interpreted with the roles of $T$ and $R$ switched. If we take $R = [4,2,2]$ and $t = [3,2,1]$ then these tableaux contribute to $D^R_t$, and by \eqref{combinatorial rule for difference} we find $\bar{\alpha}^T_R = 2$. This is the lowest $n$ example of a reduced projection coefficient taking a value with modulus greater than 1.
		
		For most Young diagrams, it is hard to be more explicit that the two rules \eqref{combinatorial rule for difference} and \eqref{combinatorial rule for individual plethysms}. However for two large families, namely Hook diagrams and `staircases + dominoes', we can evaluate these rules in general and provide explicit formulae for the projection coefficients.
		
		In tables \ref{table: n=6 reduced projection coefficients} and \ref{table: n=8 reduced projection coefficients} we gave some low $n$ ($n=6,8$) examples of $\bar{\alpha}^T_R$, calculated using \eqref{reduced projection coefficients 1st definition}. In addition, the $n=4$ coefficients can be read off (\ref{projection of [4]},~\ref{projection of [3,1]},~\ref{projection of [2,2]},~\ref{projection of [2,1,1]},~\ref{projection of [1,1,1,1]}).  We have then checked these tables against the two combinatorial rules \eqref{combinatorial rule for difference} and \eqref{combinatorial rule for individual plethysms}. In all cases the results match.

	\subsection{Hook diagrams}
	\label{section: hook diagrams}
	
	For $R$ a hook diagram, we can use the rules \eqref{combinatorial rule for difference} and \eqref{combinatorial rule for individual plethysms} to find $\alpha^T_R$ explicitly. Consider
	\ytableausetup{boxsize=normal}
	\begin{gather*}
	\begin{gathered}
	\begin{tikzpicture}
	\draw [decorate,decoration={brace,amplitude=10pt},xshift=-2pt,line width=0.8pt]
	(0,0) -- (0,7.5em) node [midway,xshift=-1.25cm] {$R = \quad c$} ;
	\draw [decorate,decoration={brace,amplitude=10pt},yshift=2pt,line width=0.8pt]
	(0,7.5em) -- (7.5em,7.5em) node [midway,yshift=+0.6cm] {$r$} ;
	\draw (0,0) rectangle (1.5em,1.5em) ;
	\draw (0,3em) rectangle (1.5em,4.5em) ;
	\draw (0,4.5em) rectangle (1.5em,6em) ;
	\draw (0,6em) rectangle (1.5em,7.5em) ;
	\draw (1.5em,6em) rectangle (3em,7.5em) ;
	\draw (3em,6em) rectangle (4.5em,7.5em) ;
	\draw (6em,6em) rectangle (7.5em,7.5em) ;
	\node at (0.75em,2.5em) {$\vdots$} ;
	\node at (5.35em,6.75em) {$\ldots$} ;
	\end{tikzpicture}
	\end{gathered}	
	\end{gather*}
	where $n = r+c-1$. For $g_{t,t;R}$ to be non-zero, $t$ must be contained within $R$, and therefore it must also be a hook diagram. By considering possible Littlewood-Richardson tableaux, we see that there is only one possible hook $t$ (and corresponding $T$) for which $g_{t,t;R} \neq 0$, namely
	\begin{gather*}
	\begin{gathered}
	\begin{tikzpicture}
	\draw [decorate,decoration={brace,amplitude=10pt},xshift=-2pt,line width=0.8pt]
	(0,0) -- (0,7.5em) node [midway,xshift=-1.35cm] {$t = \quad \left\lceil \frac{c}{2} \right\rceil$} ;
	\draw [decorate,decoration={brace,amplitude=10pt},yshift=2pt,line width=0.8pt]
	(0,7.5em) -- (7.5em,7.5em) node [midway,yshift=+0.7cm] {$\left\lceil \frac{r}{2} \right\rceil$} ;
	\draw (0,0) rectangle (1.5em,1.5em) ;
	\draw (0,3em) rectangle (1.5em,4.5em) ;
	\draw (0,4.5em) rectangle (1.5em,6em) ;
	\draw (0,6em) rectangle (1.5em,7.5em) ;
	\draw (1.5em,6em) rectangle (3em,7.5em) ;
	\draw (3em,6em) rectangle (4.5em,7.5em) ;
	\draw (6em,6em) rectangle (7.5em,7.5em) ;
	\node at (0.75em,2.5em) {$\vdots$} ;
	\node at (5.35em,6.75em) {$\ldots$} ;
	\end{tikzpicture}
	\qquad \qquad 
	\begin{tikzpicture}
	\draw [decorate,decoration={brace,amplitude=10pt},xshift=-2pt,line width=0.8pt]
	(0,0) -- (0,7.5em) node [midway,xshift=-1.5cm] {$T = \quad 2 \left\lceil \frac{c}{2} \right\rceil$} ;
	\draw [decorate,decoration={brace,amplitude=10pt},yshift=2pt,line width=0.8pt]
	(0,7.5em) -- (7.5em,7.5em) node [midway,yshift=+0.7cm] {$2 \left\lceil \frac{r}{2} \right\rceil$} ;
	\draw (0,0) rectangle (3em,1.5em) ;
	\draw (1.5em,0) -- (1.5em,1.5em) ;
	\draw (0,3em) -- (0,7.5em) ;
	\draw (1.5em,3em) -- (1.5em,7.5em) ;
	\draw (3em,3em) -- (3em,7.5em) ;
	\draw (4.5em,4.5em) -- (4.5em,7.5em) ;
	\draw (0,3em) -- (3em,3em) ;
	\draw (0,4.5em) -- (4.5em,4.5em) ;
	\draw (0,6em) -- (4.5em,6em) ;
	\draw (0,7.5em) -- (4.5em,7.5em) ;
	\draw (6em,4.5em) rectangle (7.5em,7.5em) ;
	\draw (6em,6em) -- (7.5em,6em) ;
	\node at (0.75em,2.5em) {$\vdots$} ;
	\node at (2.25em,2.5em) {$\vdots$} ;
	\node at (5.35em,6.75em) {$\ldots$} ;
	\node at (5.35em,5.25em) {$\ldots$} ;
	\end{tikzpicture}
	\end{gathered}
	\end{gather*}
	The dimension of a Young diagram representation of $S_n$ is given by the hook length formula.
	\begin{equation}
	d_R = \frac{n!}{H_R}
	\label{hook length formula}
	\end{equation}
	where $H_R$ is defined to be the product of the hook lengths of each box in $R$. Therefore we can rewrite the normalisation in \eqref{projection coefficients} as
	\begin{equation}
	\alpha^T_R = \frac{1}{2^n} \frac{\sqrt{H_T}}{H_R} \bar{\alpha}^T_R
	\nonumber
	\end{equation}
	It is simple to find $H_T$ and $H_R$
	\begin{align*}
	H_R & = n (r-1)! (c-1)! \\
	H_T & = \left[ n \left( \left\lceil r \right\rceil_2 - 2 \right)! \left( \left\lceil c \right\rceil_2 - 2 \right)! \right]^2 (n+1)(n-1) \left( \left\lceil r \right\rceil_2 -1 \right) \left( \left\lceil c \right\rceil_2 -1 \right)
	\end{align*}
	where we have defined $\left\lceil r \right\rceil_2 = 2 \left\lceil \frac{r}{2} \right\rceil$ to be $r$ rounded up to the nearest multiple of 2. Then splitting into cases for $r$ even and odd (and recalling that $c=n+1-r$, so $c$ and $r$ have opposite parity)
	\begin{equation}
	\frac{\sqrt{H_T}}{H_R} = \sqrt{n^2-1} \begin{cases}
	\sqrt{\frac{c}{r-1}} & r \text{ even} \\
	\sqrt{\frac{r}{c-1}} & r \text{ odd}
	\end{cases}
	\nonumber
	\end{equation}
	To evaluate $\bar{\alpha}^T_R$ we can use either \eqref{combinatorial rule for difference} and \eqref{combinatorial rule for individual plethysms}. Clearly both methods give the same answer, so we give the tableaux for both as an example of their use. In either case we only have one tableau to consider.
	
	The form of the tableaux depend on the parity of $r$. We look at $r$ even first. In this case the relevant tableaux are
	\begin{gather*}
	\begin{gathered}
	\begin{tikzpicture}
	\draw [decorate,decoration={brace,amplitude=10pt},xshift=-2pt,line width=0.8pt]
	(0,0) -- (0,12em) node [midway,xshift=-0.6cm] {$c$} ;
	\draw [decorate,decoration={brace,amplitude=10pt},yshift=2pt,line width=0.8pt]
	(0,12em) -- (10.5em,12em) node [midway,yshift=+0.6cm] {$r$} ;
	\draw (0,0) rectangle (1.5em,3em) ;
	\draw (0,4.5em) rectangle (1.5em,7.5em) ;
	\draw (0,7.5em) rectangle (1.5em,10.5em) ;
	\draw (0,10.5em) rectangle (3em,12em) ;
	\draw (3em,10.5em) rectangle (6em,12em) ;
	\draw (7.5em,10.5em) rectangle (10.5em,12em) ;
	\node at (0.75em,1.5em) {$\frac{c+1}{2}$} ;
	\node at (0.75em,4em) {$\vdots$} ;
	\node at (0.75em,6em) {3} ;
	\node at (0.75em,9em) {2} ;
	\node at (1.5em,11.25em) {1} ;
	\node at (4.5em,11.25em) {1} ;
	\node at (6.85em,11.25em) {$\ldots$} ;
	\node at (9em,11.25em) {1} ;
	\end{tikzpicture}
	\end{gathered}
	\qquad \qquad \qquad
	\begin{gathered}
	\begin{tikzpicture}
	\draw [decorate,decoration={brace,amplitude=10pt},xshift=-2pt,line width=0.8pt]
	(0,0) -- (0,9em) node [midway,xshift=-0.9cm] {$c+1$} ;
	\draw [decorate,decoration={brace,amplitude=10pt},yshift=2pt,line width=0.8pt]
	(0,9em) -- (7.5em,9em) node [midway,yshift=+0.6cm] {$r$} ;	
	\draw (0,0) -- (0,1.5em) -- (3em, 1.5em) -- (3em,0) -- (0,0) ;
	\draw (0,3em) -- (0,9em) -- (4.5em,9em) -- (4.5em,6em) -- (3em,6em) -- (3em,3em) -- (0,3em) ;
	\draw (6em,6em) -- (6em, 9em) -- (7.5em, 9em) -- (7.5em,6em) -- (6em,6em) ;
	\draw (0,4.5em) -- (3em,4.5em) ;
	\draw (0,6em) -- (3em,6em) ;
	\draw (1.5em,6em) -- (1.5em,9em) ;
	\draw (3em,6em) -- (3em,9em) ;
	\node at (1.5em,0.75em) {$c$} ;
	\node at (1.5em,2.5em) {$\vdots$} ;
	\node at (1.5em,3.75em) {$3$} ;
	\node at (1.5em,5.25em) {$2$} ;
	\node at (0.75em,7.5em) {$1$} ;
	\node at (2.25em,7.5em) {$1$} ;
	\node at (3.75em,7.5em) {$1$} ;
	\node at (5.35em,7.5em) {$\ldots$} ;
	\node at (6.75em,7.5em) {$1$} ;
	\end{tikzpicture}
	\end{gathered}
	\end{gather*}
	The tableau on the left has $\frac{c-1}{2}$ vertical dominoes, so its 2-sign is $\varepsilon(R) = (-1)^{\frac{c-1}{2}}$. The tableau on the right has $c-1$ horizontal dominoes, so its spin is $(-1)^\frac{c-1}{2}$. So with either method we find $\bar{\alpha}^T_R = (-1)^{\frac{c-1}{2}}$. 
	
	For $r$ odd we have
	\begin{gather*}
	\begin{gathered}
	\begin{tikzpicture}
	\draw [decorate,decoration={brace,amplitude=10pt},xshift=-2pt,line width=0.8pt]
	(0,0) -- (0,10.5em) node [midway,xshift=-0.6cm] {$c$} ;
	\draw [decorate,decoration={brace,amplitude=10pt},yshift=2pt,line width=0.8pt]
	(0,10.5em) -- (9em,10.5em) node [midway,yshift=+0.6cm] {$r$} ;
	\draw (0,0) rectangle (1.5em,3em) ;
	\draw (0,4.5em) rectangle (1.5em,7.5em) ;
	\draw (0,7.5em) rectangle (1.5em,10.5em) ;
	\draw (1.5em,9em) rectangle (4.5em,10.5em) ;
	\draw (6em,9em) rectangle (9em,10.5em) ;
	\node at (0.75em,1.5em) {$\frac{c}{2}$} ;
	\node at (0.75em,4em) {$\vdots$} ;
	\node at (0.75em,6em) {2} ;
	\node at (0.75em,9em) {1} ;
	\node at (3em, 9.75em) {1} ;
	\node at (5.35em,9.75em) {$\ldots$} ;
	\node at (7.5em,9.75em) {1} ;
	\end{tikzpicture}
	\end{gathered}
	\qquad \qquad \qquad
	\begin{gathered}
	\begin{tikzpicture}
	\draw [decorate,decoration={brace,amplitude=10pt},xshift=-2pt,line width=0.8pt]
	(0,0) -- (0,9em) node [midway,xshift=-0.6cm] {$c$} ;
	\draw [decorate,decoration={brace,amplitude=10pt},yshift=2pt,line width=0.8pt]
	(0,9em) -- (10.5em,9em) node [midway,yshift=+0.6cm] {$r+1$} ;	
	\draw (0,0) -- (0,1.5em) -- (3em, 1.5em) -- (3em,0) -- (0,0) ;
	\draw (0,3em) -- (0,9em) -- (4.5em,9em) -- (4.5em,6em) -- (3em,6em) -- (3em,3em) -- (0,3em) ;
	\draw (6em,6em) -- (6em, 9em) -- (10.5em, 9em) -- (10.5em,6em) -- (6em,6em) ;
	\draw (0,4.5em) -- (3em,4.5em) ;
	\draw (0,6em) -- (3em,6em) ;
	\draw (1.5em,6em) -- (1.5em,9em) ;
	\draw (3em,6em) -- (3em,9em) ;
	\draw (7.5em,6em) -- (7.5em,9em) ;
	\draw (7.5em,7.5em) -- (10.5em,7.5em) ;
	\node at (1.5em,0.75em) {$c$} ;
	\node at (1.5em,2.5em) {$\vdots$} ;
	\node at (1.5em,3.75em) {4} ;
	\node at (1.5em,5.25em) {3} ;
	\node at (0.75em,7.5em) {1} ;
	\node at (2.25em,7.5em) {1} ;
	\node at (3.75em,7.5em) {1} ;
	\node at (5.35em,7.5em) {$\ldots$} ;
	\node at (6.75em,7.5em) {1} ;
	\node at (9em,8.25em) {1} ;
	\node at (9em,6.75em) {2} ;
	\end{tikzpicture}	
	\end{gathered}
	\end{gather*}
	The tableau on the left has $\frac{c}{2}$ vertical dominoes, so its 2-sign is $\varepsilon (R) = (-1)^{\frac{c}{2}}$, while the tableau on the right has $c$ horizontal dominoes, and therefore the spin is $(-1)^{\frac{c}{2}}$. Either way $\bar{\alpha}^T_R = (-1)^{\frac{c}{2}}$. 
	
	Putting together the results for $r$ even and odd, we get
	\begin{equation}
	\alpha^T_R = \frac{\sqrt{n^2-1}}{2^n} (-1)^{\left\lfloor \frac{c}{2} \right\rfloor}
	\begin{cases}
	\sqrt{\frac{c}{r-1}} & r \text{ even} \\
	\sqrt{\frac{r}{c-1}} & r \text{ odd}
	\end{cases}
	\end{equation}
	One can check that this formula agrees with the coefficients in (\ref{projection of [4]},~\ref{projection of [3,1]},~\ref{projection of [2,2]},~\ref{projection of [2,1,1]},~\ref{projection of [1,1,1,1]}) and tables \ref{table: n=6 reduced projection coefficients} and \ref{table: n=8 reduced projection coefficients}.

		\subsection{Vanishing coefficients}
		\label{section: vanishing coefficients}
		
		From equation \eqref{combinatorial rule for difference} we see that there is a family of $R$ for which the projection coefficient $\alpha^T_R$ vanishes for all $T$, or equivalently, a family for which $\mathcal{O}^{U(N)}_R$ vanishes under the $\mathbb{Z}_2$ projection. These $R$ are characterised by not admitting a domino tiling. We already saw, at the end of section \ref{section: domino tableaux}, that $R = [3,2,1]$ has this property.
		
		As is standard when considering whether a shape can be tiled by dominoes, we colour $R$ in a chessboard pattern, starting with a white square in the top left. Let $\Delta_R$ be the number of white squares in $R$ minus the number of black squares. Then clearly if $\Delta_R \neq 0$, $R$ can't be tiled by dominoes and all the $\alpha^T_R$ vanish. We now prove that the converse holds; if $\Delta_R = 0$ then $R$ can be tiled by dominoes.
		
		The key point in the proof is to note that in any Young diagram except staircase diagrams (those of the form $[k, k-1, k-2, ... , 2,1]$), a domino can be removed to obtain a smaller Young diagram.\footnote{The authors would like to thank the referee for pointing this out.} To see this, note that a vertical domino can be removed if two rows have the same length, while a horizontal domino can be removed if consecutive rows differ by two or more.
		
		Staircase diagrams can be assembled by adding diagonals of increasing length. Each diagonal alternates in colour, and it is simple to see that the staircase of height $k$ has $\Delta = \frac{k+1}{2}$ if $k$ is odd and $\Delta = - \frac{k}{2}$ if $k$ is even. In particular a Young diagram with $\Delta = 0$ cannot be a staircase, and therefore it must be possible to remove a domino and obtain a smaller Young diagram (provided the original is not the empty diagram).
		
		Therefore given any $R$ with $\Delta_R = 0$, we can remove a domino to obtain $\bar{R}$, a Young diagram with two fewer boxes. Clearly $\bar{R}$ also has $\Delta_{\bar{R}} = 0$, and therefore we can repeat this process, inductively removing another domino each time to obtain a still smaller Young diagram. Since $R$ has finite size, this process terminates when we reach the empty Young diagram. Hence there is sequence of domino removals that takes $R$ to the empty Young diagram, which is exactly a domino tiling of $R$. We therefore have a complete, simple characterisation of the $R$ which vanish under the $\mathbb{Z}_2$ projection, namely $\Delta_R \neq 0$.
		
		A similar process allows us to give a domino description of Young diagrams $R$ with $\Delta_R \neq 0$. If $\Delta_R$ is positive, then $R$ can be constructed by adding dominoes to the staircase of height $k = 2 \Delta - 1$. If $\Delta_R$ is negative, $R$ is obtained by adding dominoes to the staircase of height $k = -2 \Delta$.

		We can give an another characterisation of the vanishing $R$ using the Murnaghan-Nakayama rule (for a mathematical description of the rule see \cite{james1984representation}, and see \cite{Berenstein2017} for a free fermion description). Consider the character $\chi_R \left( p \right)$, where $R \vdash n$ and $p$ is the partition $\left[ 2^\frac{n}{2} \right]$. For this $p$, border strip tableaux are just standard domino tableaux (standard means that each positive integer from 1 to $\frac{n}{2}$ appears once in the tableau) of shape $R$, and the sign of the border strip tableaux is given by the parity of the number of vertical dominoes. Since this parity depends only on $R$ (it is just the 2-sign of $R$), all of the border strip tableaux contribute to $\chi_R \left( p \right)$ with the same sign. Therefore
		\begin{equation}
		\chi_R \left( p \right) = \varepsilon(R) \left( \text{ \# of standard domino tableau of shape } R \right)
		\nonumber
		\end{equation}
		Clearly $R$ admits a domino tiling if and only if it admits a standard domino tableau. Therefore $\alpha^T_R$ vanishes for all $T$ if and only if $\chi_R ( p )$ vanishes.
		
		One can confirm these results in tables \ref{table: n=6 reduced projection coefficients} and \ref{table: n=8 reduced projection coefficients} and our numerical experiments match up to $n=16$.

		\subsection{Conjugation}
		
		
		To examine how $\alpha^T_R$ behaves under conjugation of its arguments, we return to the formula \eqref{projection coefficients}. For $S$ an arbitrary Young diagram, we have $S^c = $sgn$\otimes S$, so $d_{S^c} = d_S$ and $\chi_{S^c} (p) = (-1)^p \chi_S (p)$. Therefore the summand in \eqref{projection coefficients} changes by a factor of $(-1)^p (-1)^{2p}$. By definition we have $(-1)^p = (-1)^{p_2 + p_4 + p_6 + \ldots }$. The doubled partition $2p$ has $(-1)^{2p} = (-1)^{p_1 + p_2 + p_3 + \ldots }$. The product is
		\begin{equation}
		(-1)^p (-1)^{2p} = (-1)^{p_1 + 2 p_2 + 3 p_3 + \ldots } = (-1)^{|p|} = (-1)^{\frac{n}{2}}
		\nonumber
		\end{equation}
		Therefore
		\begin{equation}
		\alpha^{T^c}_{R^c} = (-1)^{\frac{n}{2}} \alpha^T_R
		\label{conjugation of projection coefficients}
		\end{equation}
		By comparing the coefficients in (\ref{projection of [4]},~\ref{projection of [3,1]},~\ref{projection of [2,2]},~\ref{projection of [2,1,1]},~\ref{projection of [1,1,1,1]}), one can see this relation for $n=4$. Similarly, tables \ref{table: n=6 reduced projection coefficients} and \ref{table: n=8 reduced projection coefficients} exhibit the relation at $n=6$ and $8$ respectively.

		\subsection{Semi-classical giant graviton regimes and brane interpretation of domino algorithm   }\label{sec:branesdominoes}
		
		Half-BPS operators labelled by  Young diagrams  $R$ with a single column of length comparable to $N$ are dual to single giant gravitons which are $ S^3$ expanding in $S^5$. Multiple column Young diagrams with a number of columns of order $1$ and column lengths comparable to $N$, are dual to multi-giants having $S^3$ expanding in the $AdS_5$. It is instructive to consider the domino algorithm for $ \bar \alpha_R^T$ in these regimes and develop a heuristic interpretation in terms of branes and orientifolds. 
		
		A natural first postulate is that the analogous picture for the connection between branes and rows or columns of the Young diagram works for $t$ in the $SO(N) $ theory. A single column $t$, with length comparable to $N$, is a single giant graviton with a large $S^3$ world-volume in the directions inside $ \mR P^5$  of $ AdS^5 \times \mR P^5$. Multiple long columns correspond to multi-giants of this type. A single long row with length of order $N$ corresponds to a single giant, with large spatial  world-volume in $AdS^5$. Multiple long rows correspond to multiple giants of this type. Note that among the giants which are large in the $ \mR P^5$ we also have those with worldvolume $ \mR P^3$ \cite{Mukhi:2005cv} corresponding to baryonic operators involving the $ \epsilon$-invariant. Since our focus here is on the  projection to mesonic operators, these will not be part of the  discussion that follows here.  
		
		One simple qualitative property which can be anticipated from the  brane interpretation of $R$ and $t$ is an inequality constraining the number of rows/columns in $t$ by the number of rows/columns in $R$. A brane state which survives the orientifold projection can consist of pairs of branes in spacetime related as mirror images under the  $\mathbb{Z}_2$ action. At the other extreme, we can have a single brane state which is $\mathbb{Z}_2$ invariant. Focusing on a regime of long-rowed Young diagrams corresponding to AdS giants, the number of rows, equivalently the length of the first column, is the number of giant gravitons. Allowing for the two types of $\mathbb{Z}_2$ invariant states, we expect 
		\bea\label{ineqc1}  
		{ c_1( R ) \over 2 } \le  		   c_1 ( t ) \le c_1 ( R ) 
		\eea
		Similar reasoning in the regime of sphere giants suggests 
		\bea\label{ineqc2} 
		{ r_1( R ) \over 2 } \le  		   r_1 ( t ) \le r_1 ( R ) 
		\eea		
		The last inequality is easy to derive from the domino algorithm. To maximise $ r_1 ( t ) $, we need to maximise the number of  dominoes in the first row of $R$. This maximum is achieved if all the dominoes involved in the first row are vertical (this requires that $r_2(R) = r_1 (R)$), and leads to $r_1(t) = r_1(R)$. The minimum is achieved when all the dominoes in the first row of $R$ are horizontal. This leads to $ r_1(t) = r_1(R)/2$. This prove the inequality. By applying the conjugation property of the projection coefficients, \eqref{conjugation of projection coefficients}, we obtain (\ref{ineqc2}).

		We can also formulate a more detailed brane interpretation of the domino algorithm. For a single column Young diagram $R$, a domino tiling  exists only if the length of the first column is even. Single giant gravitons with $L$ units of angular momentum can be usefully thought of as composites of $L$ quanta. Pairs of quanta are invariant under  the orientifold action, consistent with the fact that only single column Young diagrams of even length survive the projection. The projection of these single column Young diagrams $R$ are single column Young diagrams $t$, which should therefore also be interpreted as single giants in the orientifold theory. Similarly the quanta of angular momentum forming a single long row (AdS giant) are paired by the domino algorithm 
		into $ \mathbb{Z}_2 $ invariant pairs, resulting in a single giant in the quotient. 
		
		Now consider 2-row Young diagrams with row lengths $(r_1 , r_2) $, in the regime where $r_1, r_2 $ are comparable to $N$ and their difference is also comparable to $N$, e.g. $  (r_1 , r_2 ) = (2N , N )$. Consider a domino tiling with a number $s_1 < r_2$ of vertical dominoes, with the remaining boxes $ (r_1 - s_1 , r_2 - s_1 )$ occupied by horizontal dominoes. This results in $ t = ( ( r_1 +s_1 ) /2 , ( r_2 - s_2 ) /2 )$. The vertical dominoes stretch across boxes in the first and second row, which can be viewed as quanta constituting the two branes described by the Young diagram $R$. The horizontal dominoes are constituents of the same brane. A horizontal domino in row one or two of $R $ contributes  a box to the first or second row of $t$.  The vertical dominoes, even though they span row one and two of $R$,  contribute to the first row of $t$. The domino combinatorics thus encodes, in a precise way, a recombination of  angular momentum quanta between the two branes of angular momenta $r_1, r_2 $ described by $R$, which accompanies	the orientifold procedure. For multi-row Young diagrams, the domino algorithm pairs quanta of angular momentum in adjacent rows, equivalently adjacent giant gravitons in the LLM plane. An analogous discussion holds for multi-column states, where horizontally tiled dominoes pair quanta from distinct giants and vertically tiled dominoes pair quanta within a giant worldvolume. 
		
		It would be interesting to deduce  connections between the brane interpretation of the orientifold  projection coefficients discussed heuristically above, 
		from more general frameworks for brane dynamics in the presence of orientifolds, as developed for example in \cite{Gukov:1999yn,Hanany:2000fq}. In the AdS/CFT context, a useful discussion of orientifolds is in \cite{Mukhi:2005cv}.

		\subsection{Inverse projection coefficients and $U(N)$ correlators of $SO(N)$ operator} \label{section: U(N) correlators of SO(N) operators}
		
		In section \ref{section: half-BPS permutations to traces}, we saw that the $U(N)$ half-BPS sector is spanned by multi-traces of the form \eqref{multi-trace}, while the $SO(N)$ half-BPS sector is spanned by multi-traces of the form \eqref{SO(N) multi-trace}. Therefore one can consider half-BPS sector of the $SO(N)$ theory as a subspace of the equivalent in the $U(N)$ theory.  
		
		This leads to the question, what does the $U(N)$ inner product look like on this subspace? Clearly the $SO(N)$ theory has its own inner product (studied in detail in section \ref{section: correlators}), but this is a different pairing which will have a different structure. In this paper, we have made extensive use of permutations as a way to describe bases of gauge-invariant operators in different theories ($U(N), SO(N), Sp(N)$). They give us a uniform way of talking about operators in different gauge theories, namely about how the indices of matrices $X, Y$ are contracted without being specific about whether these are generic matrices in the Lie algebra $\mathfrak{u}(N)$, anti-symmetric matrices in $\mathfrak{so}(N)$, or matrices in $\mathfrak{sp}(N)$. These different theories, via AdS/CFT duality, correspond to different string theory backgrounds. In this sense, permutations are background independent structures, while the pairings we put on them are theory-dependent. Here we will see that exploring the $U(N)$ inner product which survive the projection to $SO(N)$ has interesting relations to an appropriately defined inverse of the plethysm coefficients we encountered earlier. 
		
		Consider the $SO(N)$ operators \eqref{SO(N) half-bps operator}, but where $X$ is an arbitrary complex matrix rather than anti-symmetric. We can express this as a sum of $U(N)$ operators
		\begin{equation}
		\mathcal{O}_T^{SO(N)} = \sum_{R \vdash n} \beta^R_T \mathcal{O}^{U(N)}_R
		\label{inverse projection coefficient definition}
		\end{equation}	
		If we consider taking the $\mathbb{Z}_2$ quotient of this expression, clearly the left hand side remains unchanged, while we can can evaluate the right hand side using definition \eqref{Z2 projection of U(N) operator}. This leads to
		\begin{equation}
		\mathcal{O}_T^{SO(N)} = \sum_{R \vdash n} \sum_{T'} \beta^R_T \alpha^{T'}_R \mathcal{O}^{SO(N)}_{T'}
		\nonumber
		\end{equation}
		Since this holds for all $T$, we have
		\begin{equation}
		\sum_{R \vdash n} \beta^R_T \alpha^{T'}_R = \delta_T^{T'}
		\label{inverse projection coefficient relation}
		\end{equation}
		so we call $\beta^R_T$ inverse projection coefficients. Clearly they are not true inverses to $\alpha^T_R$, since $R$ has more degrees of freedom than $T$, and so summing over $T$ will not lead to $\delta^R_{R'}$, as one would expect for true inverses. For the same reason, the relation \eqref{inverse projection coefficient relation} does not uniquely define the $\beta^R_T$ (note that they are well defined by \eqref{inverse projection coefficient definition}).
		
		To find $\beta^R_T$, we can use the orthogonality relation \eqref{character orthogonality} to invert \eqref{U(N) operator from partitions} to give traces in terms of Young diagram operators
		\begin{equation}
		\prod_i \left( \text{Tr} X^i \right)^{p_i} = \sum_{R \vdash n} \frac{1}{d_R} \chi_R (p) \mathcal{O}_R^{U(N)}
		\nonumber
		\end{equation}
		Plugging this into \eqref{SO(N) half-bps operator}, we have
		\begin{equation}
		\mathcal{O}_T^{SO(N)} = 2^n n! \sqrt{ \frac{d_T}{(2n)!} } \sum_{p \vdash \frac{n}{2}} \frac{1}{z_{2p}} \chi_t (p) \sum_{R \vdash n} \frac{1}{d_R} \chi_R (2p) \mathcal{O}_R^{U(N)}
		\nonumber
		\end{equation}
		and therefore
		\begin{equation}
		\beta^R_T = \frac{2^n n!}{d_R} \sqrt{ \frac{d_T}{(2n)!} } \sum_{p \vdash \frac{n}{2}} \frac{1}{z_{2p}} \chi_R (2p) \chi_t (p)
		\label{inverse projection coefficient}
		\end{equation}
		Notice the similarities between \eqref{inverse projection coefficient} and \eqref{projection coefficients}. The normalisation factor is upside down and we have an extra factor of $2^{l(p)}$ inside the sum over partitions (turning $z_p$ into $z_{2p}$). We have not managed to find a combinatoric interpretation of $\beta^R_T$.
		
		We can now give the $U(N)$ correlators of the $SO(N)$ operators. The correlators of $\mathcal{O}^{U(N)}_R$ were given in \cite{Corley2002} (their operators differ in normalisation by a factor of $d_R$) and are given by
		\begin{equation}
		\left\langle \mathcal{O}^{U(N)}_R \overline{\mathcal{O}}^{U(N)}_S \right\rangle = \delta_{RS} d_R n! d_R^{U(N)}
		\label{U(N) two-point function}
		\end{equation}
		where $d_R^{U(N)}$ is the dimension of the $U(N)$ representation labelled by $R$.
		
		It is then simple to show that
		\begin{equation}
		\left\langle \mathcal{O}_T^{SO(N)} \overline{\mathcal{O}}^{SO(N)}_{T'} \right\rangle = \sum_{R \vdash n} \beta^R_T \beta^R_{T'} d_R n! d_R^{U(N)}
		\nonumber
		\end{equation}
		So the $SO(N)$ orthogonal basis operators are not orthogonal under the $U(N)$ inner product, even at large $N$.

		\section{Periodicities of traces and integer-graded word combinatorics in $U(N)$ quarter-BPS counting.  }
		\label{section:U(N) results}

		The counting of quarter-BPS operators in the free limit $ \cN=4$ SYM for $U(N)$ gauge group (at large $N$) was given in terms of an infinite product generating function in \cite{BDHO07}.
		\begin{equation}
		F_{U(N)}(x,y) = \prod_{k=1}^\infty \frac{1}{1-x^k-y^k}
		\label{U(N) multi trace generating function}
		\end{equation}
		The factors are obtained from the substitutions $ ( x , y ) \rightarrow ( x^i , y^i )$ in what we will call the {\it root function } $ ( 1 - x - y )^{ -1}$. In \cite{Mattioli2014}, an interpretation of the root function in terms of word counting was given and this interpretation was extended to the generating function for the counting of gauge-invariants  in free quiver gauge theories with $U(N)$ gauge groups, derived in \cite{Pasukonis2013}. This combinatorics of gauge invariants is closely related to 
		 paths on graphs, which have interesting number theoretic aspects studied recently \cite{GR2017}. 
		 
		 Consider, for the 2-matrix case, the root function 
		\[
		\frac{1}{1-x-y}
		\]
		The coefficient of $x^n y^m$ is $\binom{n+m}{n}$, which counts the number of different ways of ordering $n$ $x$s and $m$ $y$s, or equivalently the number of different words that can be made from $n$ $\hat x$s and $m$ $\hat y$s, in the space of words generated freely by two generators $ \hat x , \hat y$. This space of words form a monoid, where the product is given by concatenation. In this paper, we will consider the implications of interpretaing the whole infinite product  $ F_{U(N)}(x,y)$ in terms of words.  The coefficient of $x^n y^m$ in $(1-x^2-y^2)^{-1}$ counts the number of words formed from $n$ $\hat x$s and $m$ $\hat y$s, but now the letters have weight 2. We denote the $\hat x$s and $\hat y$s with weight one by $\hat x_1 $ and $\hat y_1$ and those with weight 2 by $\hat x_2$ and $\hat y_2$. Multiplying the two generating functions then counts words made from all four available letters, where the weight 1 letters commute with weight 2 letters. So the coefficient of $x^n y^m$ in 
		\[
		\frac{1}{(1-x-y)(1-x^2-y^2)}
		\]
		counts words constructed from $n_1$ $\hat x_1$s, $m_1$ $\hat y_1$s, $n_2$ $\hat x_2$s and $m_2$ $\hat y_2$s such that $n_1 + 2n_2 = n$ and $m_1 + 2m_2 = m$ and the $(\hat x_1,\hat y_1)$s always precede the $(\hat x_2,\hat y_2)$s. Equivalently, we can take  the $ \hat x_1 , \hat y_1$ to commute with the $ \hat x_2, \hat y_2$. Repeating this process, we see that $F_{U(N)}(x,y)$ counts words constructed from $\hat x$s and $\hat y$s of all weights (i.e. $\hat x_k, \hat y_k$ with $k$ any positive integer), where within each level, $\hat x_k$ and $\hat y_k$ are non-commutative, but different levels commute with each other. We will refer to this kind of word counting problem as an integrally-graded word combinatorics. A natural problem is to give a bijection between the words in this counting and the traces of two matrices $ X,Y$ in the large $N$ limit. In this section, we will describe such a bijection. The analogous results for gauge invariants in $SO(N)$ gauge theory will be developed in section \ref{section: Generating function at infinite N}.

		\subsection{Structure of the space of $U(N)$ gauge-invariant functions of two matrices }
		\label{section: structure of the space}
				
		First we consider the global structure of the set of multi-traces, as well as how this structure is reflected in \eqref{U(N) multi trace generating function}. We find it is simplest to express this in the language of vector spaces, so we consider $T$, the space spanned by the $U(N)$ multi-traces.
				
		The generating function \eqref{U(N) multi trace generating function} is then the Hilbert series of $T$, where $T$ is graded by how many $X$s and $Y$s appear in each multi-trace. More explicitly, we can split $T$ into a direct sum of subspaces $T_{(n,m)}$ spanned by those multi-traces composed of $n$ $X$s and $m$ $Y$s. Then the Hilbert series is defined by
		\begin{equation}
		H_T(x,y) = \sum_{n,m} x^n y^m \text{dim} T_{(n,m)}
		\nonumber
		\end{equation}
		When studying the structure of $T$, we will need to keep track of how this interacts with the grading. 
				
		Note that we use the term `Hilbert Series' only with reference to graded vector spaces. When the vector space also has the structure of an algebra, the Hilbert series imparts information about the relations between the generating elements of the algebra. While many of the vector spaces we consider do have an algebra structure, we will not focus on this aspect.
				
		To describe the factorisation of multi-traces into single traces, we divide the full space $T$ into subspaces $T_r$ spanned by multi-traces formed from $r$ single traces.
		\begin{equation}
		T = \bigoplus_{r=0}^\infty T_r
		\nonumber
		\end{equation}
		Then $T_0$ is the one-dimensional space spanned by 1, thought of as the trivial multi-trace (the multi-trace containing no single traces). We define $T_{ST}$ to be the space spanned by the single traces, so that $T_1$ is just $T_{ST}$. $T_2$ contains multi-traces with two single traces in their factorisation. Initially we might think this space is simply $T_{ST} \otimes T_{ST}$, but this is not quite right. In this space there is a distinction between $t_1 \otimes t_2$ and $t_2 \otimes t_1$, but given the two traces $t_1$ and $t_2$, clearly there is a unique multi-trace formed from their product. Instead we have $T_2 = \operatorname{Sym}^2 \left( T_{ST} \right)$, defined to be the symmetric part of $T_{ST} \otimes T_{ST}$. Similarly, $T_r = \operatorname{Sym}^r \left( T_{ST} \right)$, defined to be the completely symmetric part of $\left( T_{ST} \right)^{\otimes r}$. So we have
		\begin{align*}
		T & = \mathbb{C} \oplus T_{ST} \oplus \operatorname{Sym}^2\left( T_{ST} \right) \oplus \ldots \\
		& = \bigoplus_{r=0}^\infty \operatorname{Sym}^r\left( T_{ST} \right) \\
		& \coloneqq \operatorname{Sym}\left( T_{ST} \right)
		\end{align*}
		Clearly $T_{ST}$ is graded by how many $X$s and $Y$s appear in a single trace, and so has its own Hilbert series, which is the generating function for the counting of single traces. The counting of single traces is obviously related to the counting of multi-traces. This relation is made explicit in the plethystic exponential. Given the generating function
		\begin{equation}
		f(x,y) = \sum_{n,m} A_{n,m} x^n y^m
		\nonumber
		\end{equation} 
		for the single traces, the generating function for the multi-traces is given by
		\begin{align}
		F(x,y) = \operatorname{PExp} (f)(x,y) & = \exp \left( \sum_{k=1}^\infty \frac{ f ( x^k , y^k ) }{k} \right) \nonumber \\ 
		& = \prod_{ n,m } \frac{1}{(1-x^n y^m)^{A_{n,m}}}
		\label{plethystic exponential}
		\end{align}
		Note that this diverges if $ f(0,0) = A_{0,0} \neq 0$. This is expected, since a single trace operator of weight 0 would lead to an infinity of multi-trace operators of weight 0. Clearly there is no single trace operator containing no matrices, and hence this is not a problem.
		
		For an explanation of why the plethystic exponential takes the single trace counting to the multi-trace counting, and for more details on the interesting properties of the plethystic exponential, see \cite{Benvenuti2006,Feng2007}. 
		
		The plethystic exponential can be inverted, up to the arbitrary constant already discussed, using the plethystic logarithm
		\begin{equation}
		f(x,y) = PLog (F)(x,y) = \sum_{k=1}^\infty \frac{\mu(k)}{k} \log F(x^k,y^k)
		\label{plethystic logarithm}
		\end{equation}
		where $\mu$ is the M{\"o}bius function defined in \eqref{mobius mu function}. The proof that these two are inverses of each other comes from the identity \eqref{mobius function identity}. See appendix \ref{section: mobius inversion} for a more detailed description of the useful properties of the M{\"o}bius function.
				
		So the Hilbert series for $T$ and $T_{ST}$ are related by
		\begin{equation}
		H_{T} = \operatorname{PExp} \left( H_{T_{ST}} \right) \qquad \qquad H_{T_{ST}} = PLog \left( H_T \right)
		\label{plethystic exponential and sym}
		\end{equation}
		Now we look at the structure of $T_{ST}$. A single trace can be written as Tr$( \ldots )^k$, where the interior of the brackets is an aperiodic matrix word, and $k$ is the number of periods. So for example Tr$XY$ has 1 period while Tr$XYXY = $Tr$(XY)^2$ has 2. Clearly the number of periods and the aperiodic matrix word (which is only defined up to cyclic rotations) identify the trace. Therefore we have
		\begin{equation}
		T_{ST} = K \otimes T_{ST}^{(1)}
		\nonumber
		\end{equation}
		where $T_{ST}^{(1)}$ is spanned by the aperiodic single traces and $K$ is spanned by the positive integers. Consider an element $k \otimes w$, where $w$ is an aperiodic single trace of weight $(n,m)$ (i.e. contains $n$ $X$s and $m$ $Y$s), then the weight of $k \otimes w$ is $(kn,km)$. So the two tensor factors interact non-trivially with respect to the weightings. Taking account of this, the Hilbert series of $T_{ST}$ and $T_{ST}^{(1)}$ are related by
		\begin{equation}
		H_{T_{ST}}(x,y) = \sum_{k=1}^\infty H_{T_{ST}^{(1)}} (x^k,y^k)
		\label{hilbert series mobius transform}
		\end{equation}
		where the $k$th term in the sum corresponds to the subspace $k \otimes T_{ST}^{(1)}$ of $T_{ST} = K \otimes T_{ST}^{(1)}$. Defining the coefficients of the two Hilbert series by
		\begin{equation}
		H_{T_{ST}} (x,y) = \sum_{n,m} A_{n,m} x^n y^m \qquad \qquad H_{T_{ST}^{(1)}} (x,y) = \sum_{n,m} a_{n,m} x^n y^m
		\label{coefficients of Hilbert series}
		\end{equation}
		the relation \eqref{hilbert series mobius transform} becomes
		\begin{equation}
		A_{n,m} = \sum_{d|n,m} a_{\frac{n}{d}, \frac{m}{d}}
		\label{A in terms of a}
		\end{equation}
		where $d | n,m$ means $d$ is a divisor of both $n$ and $m$.
		
		We can invert this relation using the M{\"o}bius inversion formula \eqref{multi variate mobius inversion formula} to get
		\begin{equation}
		a_{n,m} = \sum_{d|n,m} \mu(d) A_{\frac{n}{d}, \frac{m}{d}}
		\label{mobius transform on coefficients}
		\end{equation}
		In terms of the Hilbert series, this becomes
		\begin{equation}
		H_{T_{ST}^{(1)}} (x,y) = \sum_{k=1}^\infty \mu(k) H_{T_{ST}} (x^k, y^k)
		\label{mobius transform on function}
		\end{equation}
		We call $H_{T_{ST}^{(1)}}$ the M{\"o}bius transform of $H_{T_{ST}}$
		\begin{equation}
		H_{T_{ST}^{(1)}} = \mathcal{M} \left( H_{T_{ST}} \right) \qquad \qquad H_{T_{ST}} = \mathcal{M}^{-1} \left( H_{T_{ST}^{(1)}} \right)
		\nonumber
		\end{equation}
		In full, $T$ can be decomposed as
		\begin{equation}
		T = \operatorname{Sym} \left( K \otimes T_{ST}^{(1)} \right)
		\label{structure of U(N) space}
		\end{equation}
		and the corresponding decomposition in the generating function is
		\begin{align}
		H_T & = \operatorname{PExp} \left[ \mathcal{M}^{-1} \left( H_{T_{ST}^{(1)}} \right) \right] \nonumber \\
		& = \operatorname{PExp} \left[ \sum_{k=1}^\infty H_{T_{ST}^{(1)}}(x^k,y^k) \right] \nonumber \\
		& = \prod_{k=1}^\infty \operatorname{PExp} \left[ H_{T_{ST}^{(1)}} \right] (x^k, y^k)
		\label{vector space structure to generating function structure}
		\end{align}
		So far, we have split the multi-traces into single traces, and then decomposed the single traces by the number of periods. We could have done this the other way round. A multi-trace can be split into factors, where each factor is a multi-trace with a specified number of peroids. We can then decompose these factors into single traces with the specified number of periods. Doing things in this order gives the structure
		\begin{equation}
		T =  \left[ T^{(1)} \right]^{\otimes K} : = \left[ \operatorname{Sym} \left( T_{ST}^{(1)} \right) \right]^{\otimes K}
		\label{structure of trace vector space 2}
		\end{equation}
		where by $V^{\otimes K}$, we mean
		\begin{equation}
		V^{\otimes K} = V_1 \otimes V_2 \otimes V_3 \otimes \ldots =  \bigotimes_{k=1}^\infty V_k
		\nonumber
		\end{equation}
		and each $V_k$ is a copy of $V$ but with all weights multiplied by $k$. The Hilbert series of $V^{\otimes K}$ is then given by
		\begin{equation}
		H_{V^{\otimes K}} (x,y) = \prod_{k=1}^\infty H_V (x^k, y^k)
		\nonumber
		\end{equation}
		Just as for the sum \eqref{hilbert series mobius transform}, we can invert this
		\begin{equation}
		H_V (x,y) = \prod_{k=1}^\infty H_{V^{\otimes K}} (x^k, y^k)^{\mu (k)}
		\nonumber
		\end{equation}
		The proof of this inversion relies on the multiplicative version of the M{\"o}bius inversion formula, \eqref{multiplicative mobius inversion}. We say $H_{V}$ is the multiplicative M{\"o}bius transform of $H_{V^{\otimes K}}$
		\begin{equation}
		H_V = \mathcal{M}_{mult} \left( H_{V^{\otimes K}} \right) \qquad \qquad H_{V^{\otimes K}} = \mathcal{M}_{mult}^{-1} \left( H_V \right)
		\nonumber
		\end{equation}
		So the generating function version of \eqref{structure of trace vector space 2} is
		\begin{align}
		H_T & = \mathcal{M}_{mult}^{-1} \left( H_{T^{(1)}} \right) = \mathcal{M}_{mult}^{-1} \left[ \operatorname{PExp}  \left( H_{T_{ST}^{(1)}} \right) \right] \nonumber \\
		H_{T}(x,y) & = \prod_{k=1}^\infty H_{T^{(1)}} (x^k, y^k) = \prod_{k=1}^\infty \operatorname{PExp} \left[ H_{T_{ST}^{(1)}} \right] (x^k, y^k) \label{U(N) hilbert series relations}
		\end{align}
		which matches \eqref{vector space structure to generating function structure}. So we see that the (not immediately obvious) result
		\begin{equation}
		\operatorname{Sym} \left( K \otimes V \right) = \left( \operatorname{Sym} V \right)^{\otimes K}
		\nonumber
		\end{equation}
		corresponds to the trivial result
		\begin{equation}
		\operatorname{PExp} \left( \sum_{k=1}^\infty H_V \right) = \prod_{k=1}^\infty \operatorname{PExp} \left( H_V \right)
		\nonumber
		\end{equation}	
		Comparing \eqref{U(N) hilbert series relations} with \eqref{U(N) multi trace generating function} we see that $H_{T^{(1)}}$ is what we called the root function. Additionally, we find the root function is not the most fundamental object. It is the plethystic exponential of $H_{T_{ST}^{(1)}}$, and we should think of this Hilbert series as the fundamental object of interest. It would be interesting to see whether this additional structure of the root function has an analogue in the general quiver theory explored in \cite{Mattioli2014}.
			
\begin{figure}
			\centering
			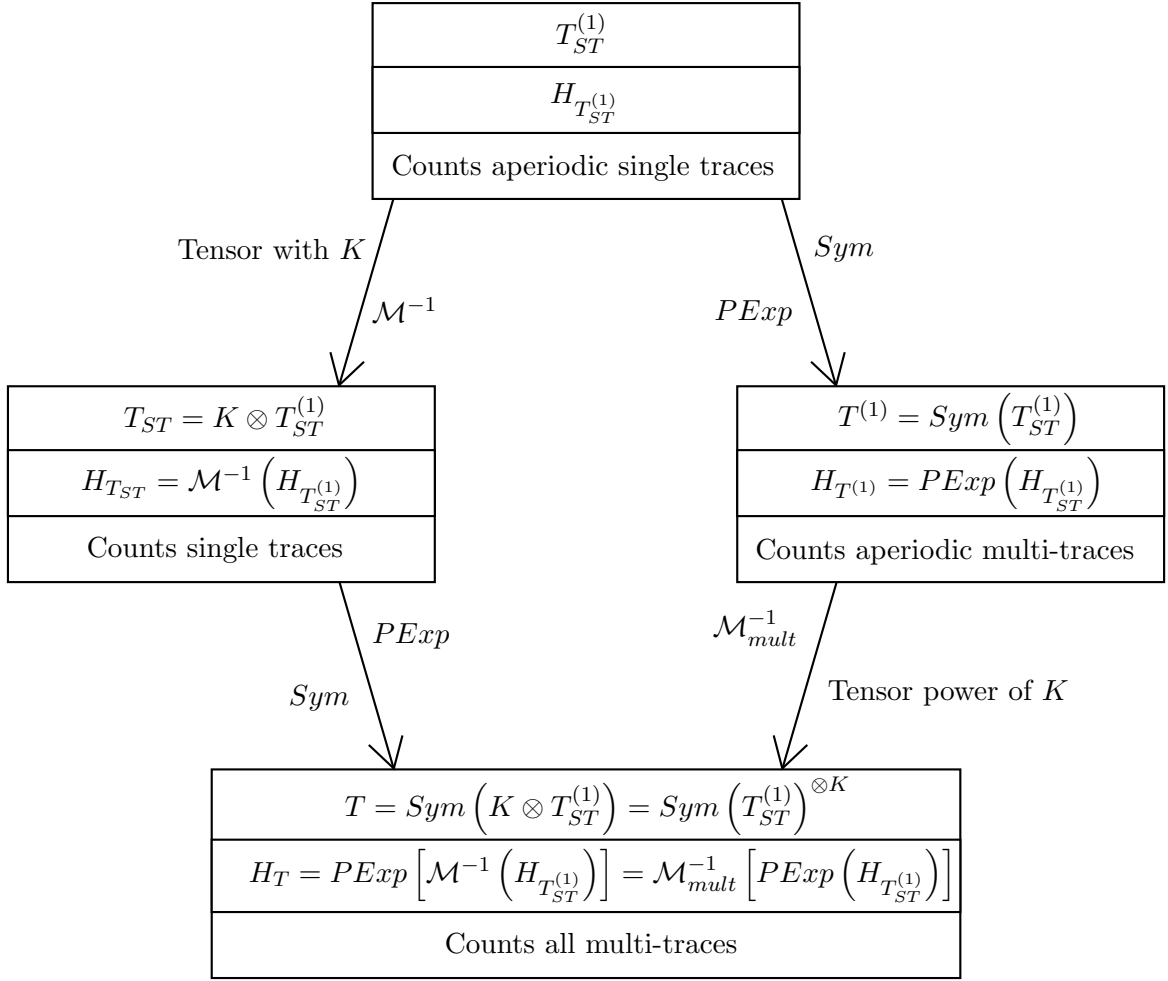
			\caption{Diagram summarising the structure of $T$, the space of $U(N)$ multi-traces, and its relation to $T_{ST}^{(1)}$, the space of $U(N)$ aperiodic single traces}
			\label{figure: U(N) diagram}
		\end{figure}

		The structure described above, both for the vector spaces and their associated Hilbert series, is summarised in figure \ref{figure: U(N) diagram}.

		\subsection{Explicit Hilbert series (generating functions)}
		\label{section: hilbert series}
		
		In the above (and summarised in figure \ref{figure: U(N) diagram}), we explained the relations between the Hilbert series associated to $T$, $T_{ST}$, $T^{(1)}$ and $T_{ST}^{(1)}$. Since these relations are invertible, we can find all the Hilbert series from just one of them. We know $H_T$ counts $U(N)$ multi-traces, as does \eqref{U(N) multi trace generating function}, so we have
		\begin{equation}
		H_T (x,y) = F_{U(N)} (x,y) = \prod_{k=1}^\infty \frac{1}{1-x^k-y^k}
		\label{U(N) multi-traces}
		\end{equation}
		Comparing with \eqref{U(N) hilbert series relations}, we see that
		\begin{equation}
		H_{T^{(1)}} (x,y) = \frac{1}{1-x-y}
		\label{U(N) aperiodic multi-traces}
		\end{equation}
		which counts aperiodic multi-traces. This allows us to interpret the product in \eqref{U(N) multi-traces}. The factor $(1-x-y)^{-1}$ counts multi-traces constructed only from aperiodic single traces. Similarly the factor $(1-x^k-y^k)^{-1}$ counts multi-traces constructed only from single traces with $k$ periods.

		Applying the plethystic logarithm to \eqref{U(N) multi-traces} and \eqref{U(N) aperiodic multi-traces} gives
		\begin{align}
		H_{T_{ST}} (x,y) & = \sum_{l=1}^\infty \frac{\mu(l)}{l} \log \left( \prod_{k=1}^\infty \frac{1}{1-x^{kl} - y^{kl}} \right) 
		\nonumber \\
		& = - \sum_{k,l=1}^\infty \frac{\mu(l)}{l} \log \left( 1-x^{kl} -y^{kl} \right)
		\nonumber\\
		& = - \sum_{d=1}^\infty \log \left( 1-x^d - y^d \right) \sum_{l|d} \frac{\mu(l)}{l} 
		\nonumber \\
		& = - \sum_{d=1}^\infty \frac{\phi(d)}{d} \log \left( 1 - x^d - y^d \right)
		\label{U(N) single traces} \\
		H_{T_{ST}^{(1)}} (x,y) & = - \sum_{d=1}^\infty \frac{\mu(d)}{d} \log \left( 1 - x^d - y^d \right)
		\label{U(N) aperiodic single traces}
		\end{align}
		where in the first calculation we have changed variables from $(k,l : 1 \leq k,l \leq \infty)$ to $(d,l : 1 \leq l \leq \infty , l | d)$ by setting $d = kl$. We have also used the identity \eqref{phi mu identity}, and $\phi (d)$ is the Euler totient function defined in \eqref{phi identity}.
		
		These two series count single traces and aperiodic single traces respectively. Later it will be important to have explicit formulae for the counting of these traces. 
		
		Expanding the logarithm in \eqref{U(N) single traces}, we get
		\begin{equation}
		H_{T_{ST}} (x,y) = \sum_{d,k=1}^{\infty} \frac{\phi(d)}{dk} (x^d + y^d)^k = \sum_{d,k=1}^\infty \frac{\phi(d)}{dk} \sum_{r=0}^k \binom{k}{r} x^{dr} y^{d(k-r)}
		\nonumber
		\end{equation}
		We wish to find an expression for the coefficient of $x^n y^m$, so reparameterise $r$ and $k$ in terms of $n = dr$ and $m = d(k-r)$
		\begin{equation}
		H_{T_{ST}} (x,y) = \sum_{n,m} x^n y^m \frac{1}{n+m} \sum_{d| n,m} \phi(d) \binom{\frac{n+m}{d}}{\frac{n}{d}}
		\label{binomial expansion for A_n,m}
		\end{equation}
		where the sum excludes $n = m = 0$. Similarly we find
		\begin{equation}
		H_{T_{ST}^{(1)}} (x,y) = \sum_{n,m} x^n y^m \frac{1}{n+m} \sum_{d| n,m} \mu(d) \binom{\frac{n+m}{d}}{\frac{n}{d}}
		\nonumber
		\end{equation}
		Comparing with \eqref{coefficients of Hilbert series}, we see that, for $(n,m) \neq (0,0)$
		\begin{align}
		A_{n,m} & = \frac{1}{n+m} \sum_{d|n,m} \phi(d) \binom{\frac{n+m}{d}}{\frac{n}{d}}
		\label{U(N) single trace counting} \\
		a_{n,m} & = \frac{1}{n+m} \sum_{d | n,m} \mu(d) \binom{\frac{n+m}{d}}{\frac{n}{d}}	
		\label{U(N) aperiodic single trace counting}	
		\end{align}
		and $A_{0,0} = a_{0,0} = 0$. The counting interpretation of these sequences is as follows: $a_{n,m}$ is the number of aperiodic single traces that can be constructed from $n$ $X$s and $m$ $Y$s, while $A_{n,m}$ is the number of single traces (with any number of periods) that can be constructed from $n$ $X$s and $m$ $Y$s. Tables of values for these sequences are given in appendix \ref{section: sequences}. Note that they are related by \eqref{A in terms of a} and \eqref{mobius transform on coefficients}.

		\subsection{Bijection between words and traces}
		\label{section: bijection between words and traces}

		Since $F_{U(N)}(x,y)$ counts both multi-traces and words, we hope to find some kind of natural bijection between the two. Before we can describe the bijection, we need to define Lyndon words. For simplicity, we will use the alphabet $\{ 0,1 \}$ in the definition, and then replace this with $\{ \hat{x}, \hat{y} \}$ when constructing the bijection.
				
		A Lyndon word is an aperiodic word which is smallest among cyclic rotations of its letters. For example the word 000101 is aperiodic and is smaller than its cyclic rotations 001010, 010100, 101000, 010001,100010, and is therefore a Lyndon word. The Lyndon words of length $\leq 5$ are
		\begin{gather*}
		0 \ , \ 1 \\ 01 \\ 001 \ , \ 011 \\ 0001 \ , \ 0011 \ , \ 0111 \\ 00001 \ , \ 00011 \ , \ 00101 \ , \ 00111 \ , \ 01011 \ , \ 01111
		\end{gather*}
		The usefulness of Lyndon words comes from the Chen-Fox-Lyndon theorem \cite[Theorem, 5.1.5]{Lothaire1983combinatorics} which states that all words can be uniquely factorised as a sequence of `non-increasing' Lyndon words. 
				
		Before going further, we must define the ordering on Lyndon words (and indeed all other words), so that `non-increasing' makes sense. With the binary alphabet, this is particularly easy. View the strings as being the binary expansions of numbers between 0 and 1. Then the ordering we want (called the lexicographic ordering) is just the same as the ordinary ordering of numbers between 0 and 1. If two words would form the same number (for example 01, 010, 0100, etc), then the longer word is larger. This last provision gives the set of words a total ordering, but is not needed for Lyndon words as they cannot end in a 0 (with the exception of 0 itself).
				
		We provide some factorisations as an example
		\begin{align*}
		100101 & = 1 \circ 00101 \\
		110010 & = 1 \circ 1 \circ 001 \circ 0 \\
		011010 & = 011 \circ 01 \circ 0
		\end{align*}
		where we have used $\circ$ as the binary operation in the free monoid on 0 and 1. Note that we require the restriction to non-increasing sequences of Lyndon words, otherwise for example we could also factorise the first word as $1 \circ 001 \circ 01$, or even $1 \circ 0 \circ 0 \circ 1 \circ 0 \circ 1$.
				
		We need to consider words constructed not just from $\hat{x}_1, \hat{y}_1$, but also $\hat{x}_2, \hat{y}_2, \hat{x}_3, \hat{y}_3, \ldots$. To deal with this we consider the set of Lyndon words for each level. The factorisation of a multi-level word then consists of the factorisation of its level one component, the factorisation of its level two component, and so on.
				
		In section \ref{section: structure of the space} we saw the structure of the vector space of traces. Clearly for a bijection to exist between traces and words, the vector space of words must also have the same structure. Define $W$ to be the space spanned by the multi-level words. As argued in section \ref{section: structure of the space}, the factorisation of words into (multi-level) Lyndon words corresponds to
		\begin{equation}
		W = \operatorname{Sym} \left( W_{LW} \right)
		\nonumber
		\end{equation}
		where $W_{LW}$ is the space spanned by the Lyndon words of all levels. Clearly a Lyndon word is identified by its level and an un-levelled Lyndon word. As before, this corresponds to
		\begin{equation}
		W_{LW} = K \otimes W_{LW}^{(1)}
		\nonumber
		\end{equation}
		where the weight of a levelled Lyndon word $k \otimes l$ is given by $k$ times the weight of the un-levelled Lyndon word $l$. This is exactly the structure we saw in $T$.
				
		So to find a bijection between the bases of $W$ and $T$ (i.e. between words and traces), we only need to find a bijection between $W_{LW}^{(1)}$ and $T_{ST}^{(1)}$. Intuitively, what we have done is matched the two factorisations (words into Lyndon words and multi-traces into single traces) and the two level structures (periodicities and word level). Therefore we only need to find a bijection between level 1 Lyndon words and aperiodic single traces to find a bijection between all multi-level words and all multi-traces.
				
		The final ingredient is now clear. An aperiodic trace is equivalent to an aperiodic word constructed from $X$ and $Y$, up to cyclic rotations. In particular we can choose a representative from the orbit of cyclic rotations as that which is smallest (where the ordering is as defined earlier with $X$ replaced by 0 and $Y$ by 1). Then the aperiodic word, by definition, is just a Lyndon word on the two letters $X$ and $Y$. Replacing those letters with $\hat{x}_1$ and $\hat{y}_1$ gives us a bijection.

		\subsection{$SO(2,1)$ representation}
		\label{section: so21}
		
		The structure found in section \ref{section: structure of the space} carries a representation of the algebra $\mathfrak{so} (2,1)$. Let $\mathbf{e}_k$ ($k=1,2,3, \ldots$) be the basis vectors for $K$. The generators for $\mathfrak{so}(2,1)$ are $J_+, J_-, J_3$. We define their action on $K$ by
		\begin{align*}
		J_+ \mathbf{e}_k & = k \, \mathbf{e}_{k+1} \\
		J_3 \mathbf{e}_k & = k \, \mathbf{e}_k \\
		J_- \mathbf{e}_k & = \begin{cases}
		k \mathbf{e}_{k-1} & k > 1 \\
		0 & k = 1
		\end{cases}
		\end{align*}
		The commutation relations for these are
		\begin{align*}
		[ J_3, J_+] & = J_+ \\
		[ J_3, J_- ] & = - J_- \\
		[ J_+, J_- ] & = -2 J_3
		\end{align*}
		Which are indeed the commutation relations for $\mathfrak{so}(2,1)$.
		
		Using the standard rules of tensor product representations, we can use this to show $T_{ST} = K \otimes T_{ST}^{(1)}$ carries a representation of $\mathfrak{so}(2,1)$, where $T_{ST}^{(1)}$ is given the trivial representation. 
		
		Let $V$ be the carrier space for an arbitrary representation of $\mathfrak{so}(2,1)$. We note that $\operatorname{Sym}^r \left( V \right)$ is an invariant subspace of $V^{\otimes r}$ with the standard tensor product representation. Therefore $\operatorname{Sym}^r \left( V \right)$ is also the carrier space for a representation of $\mathfrak{so}(2,1)$. Using this fact, it is easy to see that $T= \operatorname{Sym} \left( T_{ST} \right)$ carries a representation of $\mathfrak{so}(2,1)$.
		
		It will be interesting to investigate whether this $\mathfrak{so}(2,1)$ can be interpreted geometrically in terms of spectrum generating algebras (SGAs)  in the dual space-time, in the context of gauge-string duality for the zero coupling quarter BPS sector.  SGAs of the form $SO(p,1)$ were discussed in the context of AdS/CFT in \cite{BGM99}.

			\section{$SO(N)$ generating functions at infinite $N$ and $SO(N)$ analogues of Lyndon words }
			\label{section: Generating function at infinite N}
			
			In section \ref{section:U(N) results} we investigated the structure of the space of $U(N)$ gauge-invariant functions of two matrices in the large $N$ limit. In particular we looked at the level structure corresponding to the number of periods in a trace, and the factorisation arising from the decomposition of multi-traces into their single trace constituents. These two processes were reflected in the Hilbert series by the inverse M{\"o}bius transform $\mathcal{M}^{-1}$, defined in \eqref{hilbert series mobius transform}, and the plethystic exponential. This structure, for both vector space and Hilbert series, is summarised in figure \ref{figure: U(N) diagram}. Furthermore, we found a bijection between $U(N)$ aperiodic single traces and Lyndon words, which generalised to a bijection between the full space of multi-traces and a levelled word monoid.
			
			In this section we find the analogue picture for the $SO(N)$ theory. Let $\widetilde{T}$ be the space of $SO(N)$ gauge-invariant functions of two matrices in the large $N$ limit (note this means there are no baryonic operators). We find that the structure exhibited in figure \ref{figure: U(N) diagram} also applies to $\widetilde{T}$, with the aperiodic single traces of the $U(N)$ theory being replaced by minimally periodic traces in the $SO(N)$. These correspond to a transformed set of Lyndon words that we call \emph{orthogonal Lyndon words}. As suggested by the change in name, the minimally periodic traces can have one or two periods. This leads to an alternate structure of $\widetilde{T}$ which respects the absolute number of periods, rather than the number of repetitions of the minimally periodic units. The two different structures of $\widetilde{T}$ are summarised in figures \ref{figure: SO(N) diagram} and \ref{figure: SO(N) periodicity diagram}.
			
			These two structures give relations between the Hilbert series for the relevant vectors spaces. However, unlike the $U(N)$ case, we do not already have the Hilbert series for $\widetilde{T}$. In appendix \ref{section: alternative SO(N) generating function} we present an argument that derives it directly from the formula \eqref{2-matrix delta counting}. Here we will give a shorter, more direct approach to finding the function that gives more insight into its structure. This generating function is of interest to mathematicians \cite{Willenbring07}, and we believe that our explicit evaluation of it is a new mathematical result.

		\subsection{Structure of the space of $SO(N)$ multi-traces}
		\label{Different expressions for the SO(N) generating function}

		As in section \ref{section:U(N) results}, we will consider various different vector spaces in addition to $\widetilde{T}$. In general, those relating to $SO(N)$ traces will have a tilde on top, whereas those primarily to do with $U(N)$ objects will not. Some vector spaces we define will be relevant to both, so the divide is not a sharp one. Similarly to the notation used in section \ref{section: structure of the space}, we use superscripts in brackets to refer to a space with a specified number of periods, and subscripts to add extra information on the type of traces being considered.
		
		To get from the $U(N)$ theory to the $SO(N)$ theory, we replace the generic complex matrices of the $U(N)$ theory with the anti-symmetric complex matrices of the $SO(N)$ theory. This is a $\mathbb{Z}_2$ quotient on the space of traces. 
		
		We examine the effect of the $\mathbb{Z}_2$ quotient by looking at an arbitrary $U(N)$ single trace. It is specified by $k$, the number of periods, and an aperiodic matrix word $W$. Since a trace is invariant under transposition, we have
		\begin{equation}
		\text{Tr}W^k = \text{Tr} \left( W^T \right)^k
		\nonumber
		\end{equation}
		After the quotient, $X$ and $Y$ are related to their transposes, so this relation reduces the number of independent single traces. The transpose reverses the matrix word - we call the reversed word $W^{(r)}$ - and introduces a factor of $(-1)^{k \, l(W)}$, where $l(W)$ is the length of $W$.
		\begin{equation}
		\text{Tr} W^k = (-1)^{k \, l(W)} \text{Tr} \left( W^{(r)} \right)^k
		\label{relation between traces from antisymmetry}
		\end{equation}
		There are now two sets of two possibilities: either $W$ and $W^{(r)}$ are the same (up to cyclic rotations), or they are not, and $k \, l(W)$ is either even or odd.
		
		If $W \neq W^{(r)}$, then \eqref{relation between traces from antisymmetry} tells us that two distinct traces that were previously unrelated are no longer independent. This is true whether $k \, l(W)$ is even or odd.
		
		If $W = W^{(r)}$, then \eqref{relation between traces from antisymmetry} does depend on whether $k \, l(W)$ is even or odd. If it is even, then \eqref{relation between traces from antisymmetry} is trivial, and gives us no new information. If it is odd, then \eqref{relation between traces from antisymmetry} implies that the trace vanishes. So for example, Tr$X$, Tr$Y^3$, Tr$X^2Y$ and Tr$(X^4Y)^5$ all vanish.
		
		To encode this structure into the $U(N)$ vector space $T$ we split $T_{ST}^{(1)}$ into three distinct subspaces
		\begin{equation}
		T_{ST}^{(1)} = T_{ST;inv;even}^{(1)} \oplus T_{ST;inv;odd}^{(1)} \oplus T_{ST;var}^{(1)} 
		\label{decomposition into variant, invariant even and invariant odd}
		\end{equation}
		The first space is spanned by those traces of even length with $W = W^{(r)}$ (`inv' stands for invariant); the second space is spanned by traces of odd length with $W = W^{(r)}$; the third space is spanned by traces of any length with $W \neq W^{(r)}$ (`var' stands for variant). From previous arguments, $T_{ST;inv;even}^{(1)}$ is unchanged under the $\mathbb{Z}_2$ quotient. The other two spaces are more complex.
		
		We saw that for reversal-invariant $W$ of odd length, the determining factor between whether the trace vanishes or not is whether $k$ is odd or even respectively. If $k$ is even, $T_{ST;inv;odd}^{(1)}$ is unchanged by the quotient, while if $k$ is odd, it vanishes. So we have
		\begin{equation}
		K \otimes T_{ST;inv;odd}^{(1)} \underset{\mathbb{Z}_2}{\longrightarrow}  K_{even}  \otimes T_{ST;inv;odd}^{(1)}
		\label{Z2 projection of odd invariant traces}
		\end{equation}
		where $K_{even}$ is the space spanned by the even integers. Clearly this is isomorphic to $K$, but we cannot just replace $K_{even}$ with $K$ as then we lose information about the weight of a given trace. Formally, they are isomorphic as vector spaces but not as graded vector spaces. However, we can recover $K$ as a tensor factor of the graded vector space if we double the weight of the space $T_{ST;inv;odd}$ to make up for halving the weight of the $K_{even}$ factor. So we have
		\begin{equation}
		K_{even} \otimes T_{ST;inv;odd}^{(1)} = K \otimes \left( T_{ST;inv;odd}^{(2)} \right)
		\label{Z2 projection of subset of single traces}
		\end{equation}
		Effectively what we have done here is say rather than consider $X$ (or $Y$, $X^2Y$, $X^4Y$, \ldots ) as the aperiodic word identifying the trace, instead we consider $X^2$ (or $Y^2$, $(X^2Y)^2$, $(X^4Y)^2$, \ldots  ) as the `aperiodic word'. Since these are the lowest order at which the aperiodic words appear, we call the doubled versions minimally periodic words.
		
\begin{figure}
			\centering
			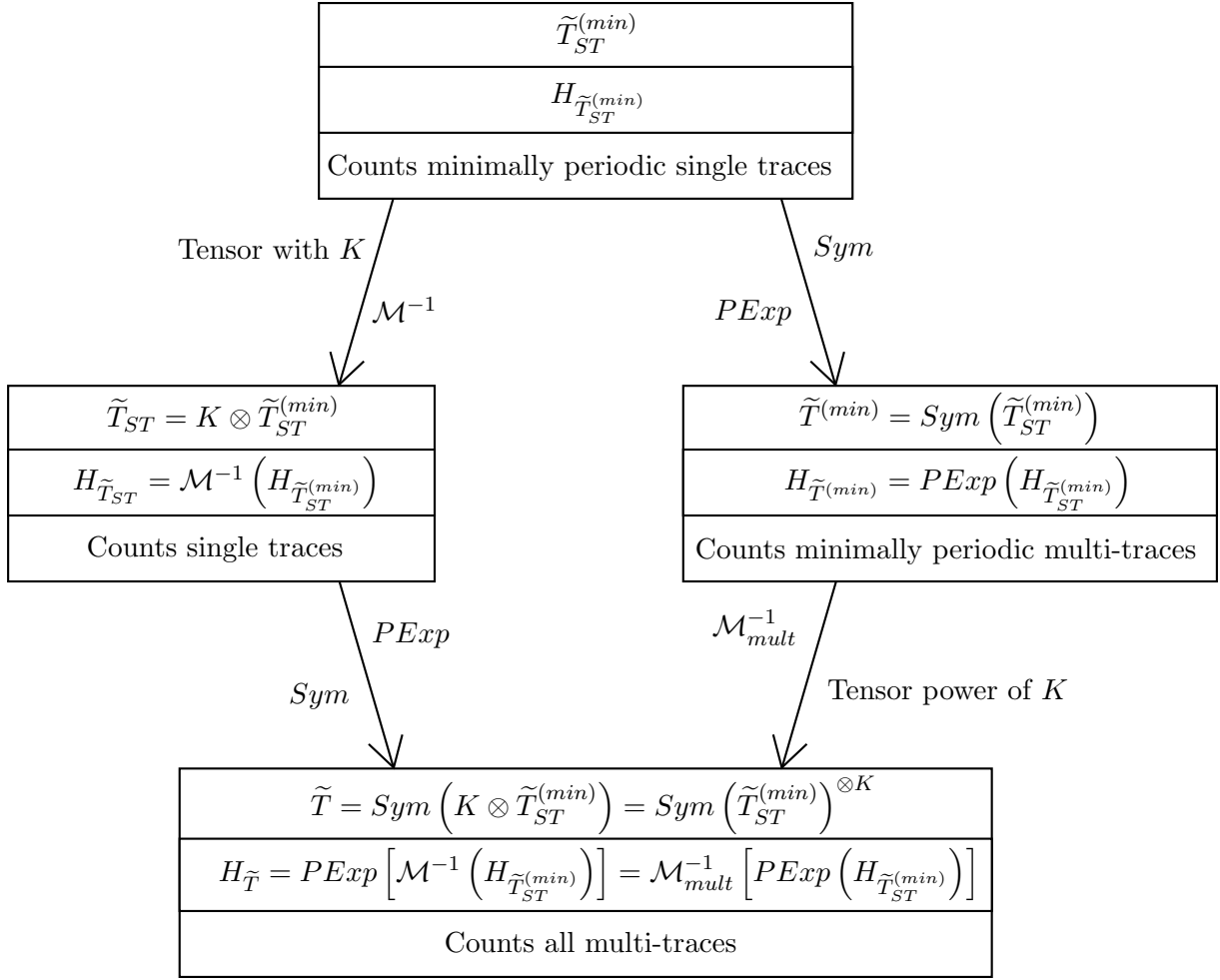
			\caption{Diagram summarising the structure of $\widetilde{T}$, the space of $SO(N)$ multi-traces, and its relation to $\widetilde{T}_{ST}^{(min)}$, the space of $SO(N)$ minimally periodic single traces.}
			\label{figure: SO(N) diagram}
		\end{figure}

		Finally we consider $T_{ST;var}^{(1)}$. It is spanned by aperiodic matrix words (up to cyclic rotations) which change under reversal. So we can split the spanning set into orbits (of size 2) under reversal. Then defining $\widetilde{T}_{ST;var}^{(1)}$ to be the space spanned by these orbits, we have
		\begin{equation}
		T_{ST;var}^{(1)} \underset{\mathbb{Z}_2}{\longrightarrow} \widetilde{T}_{ST;var}^{(1)}
		\label{Z2 projection of variant traces}
		\end{equation}
		It will be useful later to note that $T_{ST;var}^{(1)}$ is just two copies of $\widetilde{T}_{ST;var}^{(1)}$
		\begin{equation}
		T_{ST;var}^{(1)} = \widetilde{T}_{ST;var}^{(1)} \oplus \widetilde{T}_{ST;var}^{(1)}
		\label{variant space halves after projection}
		\end{equation}
		In full, the $\mathbb{Z}_2$ quotient of $T_{ST}$ is
		\begin{equation}
		T_{ST} = K \otimes T_{ST}^{(1)} \underset{\mathbb{Z}_2}{\longrightarrow} \widetilde{T}_{ST} = K \otimes \widetilde{T}_{ST}^{(min)} = K \otimes \left( T_{ST;inv;even}^{(1)} \oplus T_{ST;inv;odd}^{(2)} \oplus \widetilde{T}_{ST;var}^{(1)} \right)
		\label{Z2 projection of aperiodic single traces}
		\end{equation}
		where the `min' superscript refers to the words being minimally periodic, as opposed to aperiodic. Extrapolating to the full space of multi-traces
		\begin{align}
		T = \operatorname{Sym} \left( K \otimes T_{ST}^{(1)} \right) & \underset{\mathbb{Z}_2}{\longrightarrow} \widetilde{T} =  \operatorname{Sym} \left( K \otimes \widetilde{T}_{ST}^{(min)} \right)
		\label{Z2 projection of multi-traces}
		\end{align}
		We see this has the same structure as \eqref{structure of U(N) space}, but with a base space $\widetilde{T}_{ST}^{(min)}$. This allows us to reproduce figure \ref{figure: U(N) diagram}, but with the new base space, shown in figure \ref{figure: SO(N) diagram}.
		
		Furthermore, we saw in section \ref{section: so21} that the structure \eqref{structure of U(N) space} allowed $T$ to carry a representation of $\mathfrak{so}(2,1)$. By the same argument, $\widetilde{T}$ will also carry such a representation.
		
		In section \ref{section: bijection between words and traces} we saw that Lyndon words on $x$ and $y$ give a good description of the spanning set for $T_{ST}^{(1)}$. The definition of $\widetilde{T}_{ST}^{(min)}$ (implicit in \eqref{Z2 projection of aperiodic single traces}) allows us to give it a similar description in terms of `orthogonal' Lyndon words. 
		
		These orthogonal Lyndon words fall into one of three categories, depending on whether the corresponding trace comes from the spanning set for $T_{ST;inv;even}^{(1)}$, $T_{ST;inv;odd}^{(2)}$ or $\widetilde{T}_{ST;var}^{(1)}$. We say the orthogonal Lyndon words are of types 1A,1B or 2 respectively. Type 1A orthogonal Lyndon words are normal Lyndon words of even length that are invariant under reversal (up to cyclic rotations). A type 1B word is the square of a normal Lyndon word of odd length that is invariant under reversal. A type 2 word is the first (lexicographically) of a pair of normal Lyndon words that transform into each other when reversed. The lowest order examples of the three types are shown in table \ref{table: examples of orthogonal lyndon words}.
		
		\begin{table}
			\begin{center}
				\begin{tabular}{c | c }
					& $x y$ \\
					Type 1A & $ x^3 y $ , $ x^2 y^2$ , $ x y^3 $ \\
					& $ x^5 y$, $x^4 y^2$, $x^3 y x y$, $x^3 y^3$, $x^2 y^4$, $x y x y^3$, $x y^5$ \\ \hline
					& $x^2$, $y^2$ \\
					Type 1B & $x^2 y x^2 y$, $x y^2 x y^2$ \\
					& $x^4 y x^4 y$, $x^3 y^2 x^3 y^2$, $x^2 y x y x^2 y x y$, $x^2 y^3 x^2 y^3$, $x y x y^2 x y x y^2$, $x y^4 x y^4$ \\ \hline
					& $x^2 y x y^2$ \\
					Type 2 & $x^3 y x y^2, x^2 y x y^3$ \\
					& $x^3 y x^2 y^2$, $x^4 y x y^2$, $x^3 y x y^3$, $x^2 y x y x y^2$, $x^2 y x y^4$, $x^2 y^2 x y^3$
				\end{tabular}
			\end{center}
			\caption{Lowest order examples of the three distinct types of `orthogonal' Lyndon words}
			\label{table: examples of orthogonal lyndon words}
		\end{table}

		\subsection{Periodicity structure}
		\label{section: Periodicity structure}
		
		We now briefly return to the description of the $U(N)$ single trace space $T_{ST}$. We had
		\begin{equation}
		T_{ST} = K \otimes T_{ST}^{(1)} = \left( 1 \otimes T_{ST}^{(1)} \right) \oplus \left(  2 \otimes T_{ST}^{(1)} \right) \oplus \left( 3 \otimes T_{ST}^{(1)} \right) \oplus \ldots
		\nonumber
		\end{equation}
		We know that $\widetilde{T}_{ST}$ also has this structure, but there is a difference in interpretation. The subspace $k \otimes T_{ST}^{(1)}$ of $T_{ST}$ corresponds to the traces with $k$ periods, whereas the subspace $k \otimes \widetilde{T}_{ST}^{(min)}$ of $\widetilde{T}_{ST}$ does not, instead it contains traces with $k$ repetitions of the minimally periodic words. Since these words can contain two periods (if they are of type 1B), $k \otimes \widetilde{T}_{ST}^{(min)}$ contains traces with $k$ or $2k$ periods. We now decompose $\widetilde{T}_{ST}$ into subspaces corresponding to the number of periods rather than the number of repetitions. Explicitly, we want to find $V_k$ such that
		\begin{equation}
		\widetilde{T}_{ST} = \left( 1 \otimes V_1 \right) \oplus \left(  2 \otimes V_2 \right) \oplus \left( 3 \otimes V_3 \right) \oplus \ldots
		\nonumber
		\end{equation}
		and $k \otimes V_k$ is the vector space of single traces with $k$ periods.
		
		We saw in \eqref{Z2 projection of odd invariant traces} that for odd length, reversal invariant aperiodic matrix words, only the even periodicities survive the $\mathbb{Z}_2$ projection. For all other aperiodic matrix words, there is no distinction between even and odd periodicities. Therefore $V_k$ will depend only on whether $k$ is even or odd. From the discussions in section \ref{section: single traces and plethystic exponential}, we can write down the appropriate vector spaces. They are
		\begin{align}
		\widetilde{T}_{ST}^{(odd)} & = T_{ST;inv;even}^{(1)} \oplus \widetilde{T}_{ST;var}^{(1)} \label{SO(N) odd periodicity vector space} \\
		\widetilde{T}_{ST}^{(even)} & = T_{ST;inv}^{(1)} \oplus \widetilde{T}_{ST;var}^{(1)} \nonumber \\ & = T_{ST;inv;even}^{(1)} \oplus T_{ST;inv;odd}^{(1)} \oplus \widetilde{T}_{ST;var}^{(1)} \label{SO(N) even periodicity vector space}
		\end{align}
		Note that the odd and even superscripts refer to periodicities, while the odd and even subscripts refer to the length of the aperiodic trace/matrix word. Splitting $K = K_{odd} \oplus K_{even}$ in the obvious way, we have
		\begin{equation}
		\widetilde{T}_{ST} = \left( K_{odd} \otimes \widetilde{T}_{ST}^{(odd)} \right) \oplus \left( K_{even} \otimes \widetilde{T}_{ST}^{(even)} \right)
		\nonumber
		\end{equation}
		Now the combination of $K_{odd}$ and $K_{even}$ keeps track of the true periodicities of the traces.
		
\begin{figure}
			\centering
			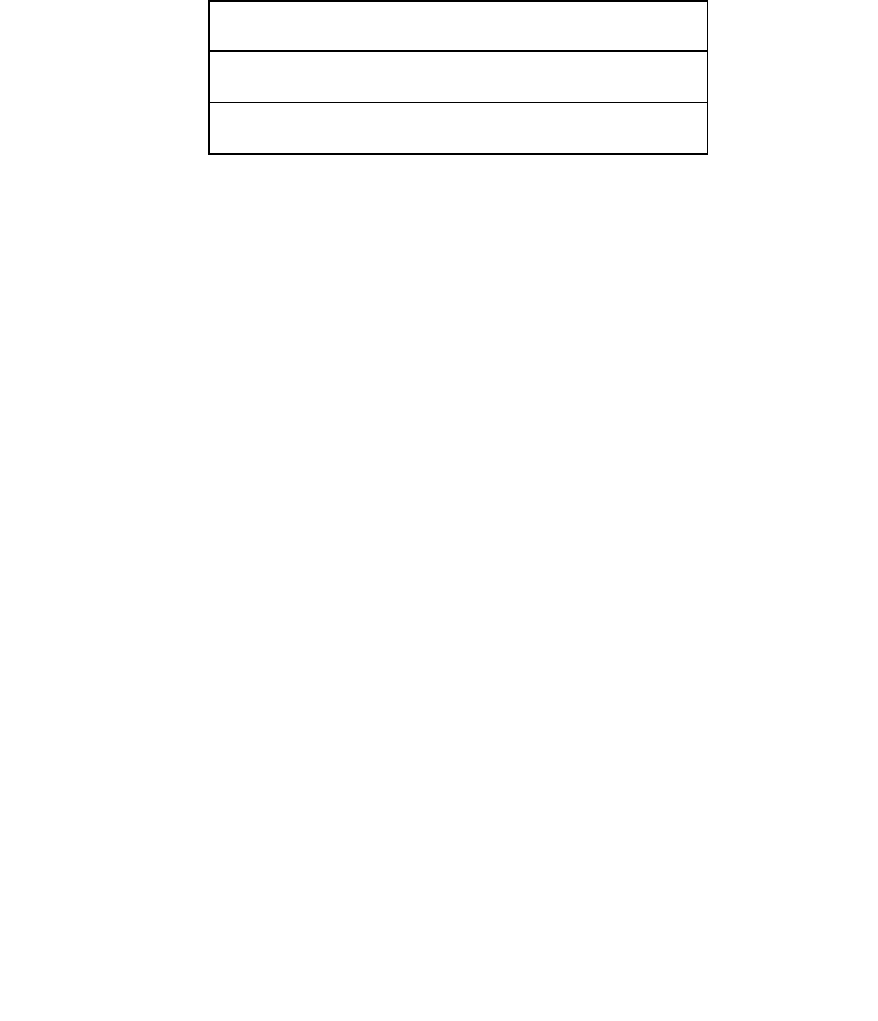
			\caption{Diagram summarising the structure of $\widetilde{T}$, the space of $SO(N)$ multi-traces, and its relation to $\widetilde{T}_{ST}^{(odd)}$, the space of $SO(N)$ single traces with a specified odd number of periods, and $\widetilde{T}_{ST}^{(even)}$, the space of $SO(N)$ single traces with a specified even number of periods.}
			\label{figure: SO(N) periodicity diagram}
		\end{figure}

		Doing a analysis of the Hilbert series associated with these vector spaces, similar to that done in section \ref{section: structure of the space}, we arrive at the relations shown in figure \ref{figure: SO(N) periodicity diagram}. The transformations $\mathcal{S}$ and $\mathcal{S}_{mult}$  are defined by
		\begin{align*}
		\mathcal{S} \left[ f,g \right] (x,y) = \sum_{k \text{ odd}} f (x^k, y^k) + \sum_{k \text{ even}} g(x^k,y^k) \\
		\mathcal{S}_{mult} \left[ f,g \right] (x,y) = \left( \prod_{k \text{ odd}} f(x^k, y^k) \right) \left( \prod_{k \text{ even}} g(x^k, y^k) \right)
		\end{align*}
		Note that $\mathcal{S}$, while being similar to $\mathcal{M}^{-1}$, has a distinct disadvantage to it's analogue, namely it is not invertible. Given $S[f,g]$, there are multiple $f,g$ which would produce the same $\mathcal{S}$. This means we cannot instantly find the Hilbert series for $\widetilde{T}_{ST}^{(odd)}$ and $\widetilde{T}_{ST}^{(even)}$ just from the Hilbert series for $\widetilde{T}$. Instead we need to investigate the structures \eqref{SO(N) odd periodicity vector space} and \eqref{SO(N) even periodicity vector space}.
		
		In order to do this, we introduce names for the coefficients of various Hilbert series. These are shown in table \ref{table: Hilbert series coefficients}, along with a description of which set of traces these coefficients count. Tables of values are given in appendix \ref{section: sequences}. Note that since the coefficients listed all count single traces, they all vanish when $n=m=0$. Therefore in the later explicit expressions for these sequences, we implicitly set the $n=m=0$ term to be 0.
		
		\begin{table}
			\centering
			\renewcommand{\arraystretch}{1.3}
			\begin{tabular}{c|c|c}
				Hilbert series & Vector space & Counting interpretation \\
				coefficients & & \\ \hline
				$b_{n,m}$ & $\widetilde{T}_{ST}^{(min)}$ & minimally periodic $SO(N)$ single traces \\ \hline
				$B_{n,m}$ & $\widetilde{T}_{ST}$ & all $SO(N)$ single traces \\ \hline
				\multirow{2}{*}{$a_{n,m}^{inv}$} & \multirow{2}{*}{$T_{ST;inv}^{(1)}$} & aperiodic, reversal invariant\\
				& & $U(N)$ single traces \\ \hline				
				\multirow{2}{*}{$a_{n,m}^{var}$} & \multirow{2}{*}{$\widetilde{T}^{(1)}_{ST;var}$} & aperiodic pairs of $U(N)$ single traces \\
				& & that reverse into each other \\ \hline
				$A_{n,m}^{inv}$ & $T_{ST;inv} = K \otimes T_{ST;inv}^{(1)}$ & all reversal invariant $U(N)$ single traces \\ \hline
				\multirow{2}{*}{$A_{n,m}^{var}$} & \multirow{2}{*}{$\widetilde{T}_{ST;var} = K \otimes \widetilde{T}_{ST;var}^{(1)} $} &  all pairs of $U(N)$ single traces \\
				& & that reverse into each other \\ \hline
				\multirow{2}{*}{$b_{n,m}^{(odd)}$} & \multirow{2}{*}{$\widetilde{T}_{ST}^{(odd)}$} & $SO(N)$ single traces with a \\
				& & specified odd number of periods  \\ \hline
				\multirow{2}{*}{$b_{n,m}^{(even)}$} & \multirow{2}{*}{$\widetilde{T}_{ST}^{(even)}$} & $SO(N)$ single traces with a \\
				& & specified even number of periods
			\end{tabular}
			\caption{Definition of various single trace counting sequences. Formally, they are defined as the coefficients of Hilbert series for certain vector spaces. We also give the counting interpretation.}
			\label{table: Hilbert series coefficients}
		\end{table}
		
		Recall that $a_{n,m}$ are the coefficients in the Hilbert series for $T_{ST}^{(1)}$. Then from definition \eqref{decomposition into variant, invariant even and invariant odd}, and recalling \eqref{variant space halves after projection}, we have
		\begin{equation}
		a_{n,m} = a_{n,m}^{inv} + 2 a_{n,m}^{var}
		\nonumber
		\end{equation}
		The lower case sequences count aperiodic single traces, while the upper case ones count single traces of all periodicities. This leads to relations \eqref{A in terms of a} and \eqref{mobius transform on coefficients} between the $a$s and $A$s (although shown only for the undecorated versions, this is also true for both superscripts). Using these, we have
		\begin{equation}
		A_{n,m} = A_{n,m}^{inv} + 2 A_{n,m}^{var}
		\nonumber
		\end{equation}
		From the definitions \eqref{SO(N) odd periodicity vector space} and \eqref{SO(N) even periodicity vector space}, we also get
		\begin{align}
		b_{n,m}^{(even)} & = a_{n,m}^{var} + a_{n,m}^{inv} \nonumber \\
		& = \frac{1}{2} \left[ a_{n,m} + a_{n,m}^{inv} \right]
		\label{b_{n,m;even}} \\
		b_{n,m}^{(odd)} & = \begin{cases}
		a_{n,m}^{var} + a_{n,m}^{inv} & n+m \text{ even} \\
		a_{n,m}^{var} & n+m \text{ odd} 
		\end{cases} \nonumber \\
		& = \frac{1}{2} \left[ a_{n,m} + (-1)^{n+m} a_{n,m}^{inv} \right]
		\label{b_{n,m;odd}}
		\end{align}
		So to find the desired Hilbert series, we first need to find the generating function for the $a_{n,m}^{inv}$, or equivalently the $A_{n,m}^{inv}$, since they are related by \eqref{A in terms of a} and \eqref{mobius transform on coefficients}.
		
		In \eqref{decomposition into variant, invariant even and invariant odd} we decomposed $T_{ST}^{(1)}$ into subspaces that were invariant or variant under reversal. We now do the same to $T_{ST}$.
		\begin{equation}
		T_{ST} = T_{ST;inv} \oplus T_{ST;var} = \left( T_{ST;inv;odd} \oplus T_{ST;inv;even} \right) \oplus T_{ST;var}
		\nonumber
		\end{equation}
		where the odd and even parts refer to the length of the entire single trace, not (as before) the length of the aperiodic matrix word which, along with the number of periods, defined the single trace. We have
		\begin{equation}
		T_{ST;inv} = K \otimes T_{ST;inv}^{(1)} \qquad \qquad T_{ST;var} = K \otimes T_{ST;var}^{(1)}
		\nonumber
		\end{equation}
		but the split into odd and even parts does not respect the $K$ tensor product. Instead, we have
		\begin{align}
		T_{ST;inv;even} & = \left( K \otimes T_{ST;inv;even}^{(1)} \right) \oplus \left( K_{even} \otimes T_{ST;inv;odd}^{(1)} \right)
		\label{invariant even traces} \\
		T_{ST;inv;odd} & = K_{odd} \otimes T_{ST;inv;odd}^{(1)}
		\label{invariant odd traces}
		\end{align}
		By repeating the earlier analysis, or by comparing \eqref{invariant even traces} and \eqref{invariant odd traces} with \eqref{Z2 projection of odd invariant traces}, \eqref{Z2 projection of subset of single traces} and \eqref{Z2 projection of variant traces}, we see that under the $\mathbb{Z}_2$ quotient, $T_{ST;inv;odd}$ disappears, $T_{ST;inv;even}$ is unchanged, and $T_{ST;var}$ is `halved' to $\widetilde{T}_{ST;var}$ as before. So looking at all single traces, rather than just aperiodic single traces, we have
		\begin{equation}
		T_{ST} \underset{\mathbb{Z}_2}{\longrightarrow} \widetilde{T}_{ST} = T_{ST;inv;even} \oplus \widetilde{T}_{ST;var}
		\label{Z2 projection on single traces}
		\end{equation}
		The coefficients of $H_{\widetilde{T}_{ST}}$ are $B_{n,m}$, and therefore
		\begin{align}
		B_{n,m} & = \begin{cases}
		A^{var}_{n,m} + A^{inv}_{n,m} & n+m \text{ even} \\
		A^{var}_{n,m} & n+m \text{ odd}
		\end{cases} \nonumber \\ 
		& = \frac{1}{2} \left[ A_{n,m} + (-1)^{n+m} A^{inv}_{n,m} \right]
		\label{SO(N) single trace counting 2nd form}
		\end{align}
		We previously found a formula for $A_{n,m}$, \eqref{U(N) single trace counting}, and in the next section we find an expression for $B_{n,m}$, \eqref{SO(N) single trace counting}. Comparing these with \eqref{SO(N) single trace counting 2nd form} allows us to find $A_{n,m}^{inv}$. Since $a_{n,m}^{inv}$ are related to $A_{n,m}^{inv}$ via the M{\"o}bius transform, we can then use \eqref{b_{n,m;even}} and \eqref{b_{n,m;odd}} to find the Hilbert series for  $\widetilde{T}_{ST}^{(even)}$ and $\widetilde{T}_{ST}^{(odd)}$.

		\subsection{Derivation of Hilbert series}
		\label{section: single traces and plethystic exponential}
		
		The previous two sections have found the structure of $\widetilde{T}$ and how this structure is made manifest in the Hilbert series $H_{\widetilde{T}}$. We now find the various related Hilbert series explicitly.
		
		As explained in section \ref{section: hilbert series}, any of the Hilbert series in figure \ref{figure: SO(N) diagram} determines all others, and from the argument in the previous section, we know finding the $B_{n,m}$ (or equivalently $H_{\widetilde{T}_{ST}}$) will give all the series in figure \ref{figure: SO(N) periodicity diagram}. It is therefore sufficient to find just the series $H_{\widetilde{T}_{ST}}$.
		
		To find the $B_{n,m}$, think about the matrix words contained inside the traces. These words are constructed from $n$ $X$s and $m$ $Y$s. In the $U(N)$ gauge theory, they are equivalent up to cyclic rotations only, but in the $SO(N)$ gauge theory, we also have to consider the effect of transposition. As seen in \eqref{relation between traces from antisymmetry}, this reverses the word and also multiplies by a factor of $(-1)^M$ ($M=n+m$ is the total number of matrices in the trace). The cyclic rotations and the reversal act as $D_M$ on the matrix word. Note that we already encountered a dihedral group in section \ref{section: SO(N) permutations to traces}, where it arose in the stabiliser group of an individual cycle. Since single cycles correspond to single traces, this is the same dihedral group as appears here.
		
		The factors of $-1$ mean this dihedral group action on its own is not sufficient to describe the effect of the anti-symmetry. Rather than consider $D_M$ acting on the set of words, we consider it acting on the vector space spanned by the set of words. Explicitly, let $V$ be the vector space spanned by two vectors, $e_X$ and $e_Y$. Then a basis for $V^{\otimes M}$ is clearly labelled by the set of words of length $M$ constructed from $X$ and $Y$. Let $\sigma$ be the generator of rotations in $D_M$ and $\tau$ the reflection/transposition. Then they act as
		\begin{align}
		\sigma \left[ e_{i_1} \otimes e_{i_2} \otimes \ldots \otimes e_{i_M} \right] & = e_{i_2} \otimes e_{i_3} \otimes \ldots \otimes e_{i_M} \otimes e_{i_{1}} \nonumber \\
		\tau \left[ e_{i_1} \otimes e_{i_2} \otimes \ldots \otimes e_{i_M} \right] & = (-1)^M e_{i_M} \otimes \ldots \otimes e_{i_2} \otimes e_{i_1} \label{D_M action}
		\end{align}
		where $i_j \in \{ X,Y \}$. This action of the dihedral group on the space of matrix words was considered in \cite{Bae:2017fcs}. 
		
		To get the vector space spanned by traces of anti-symmetric matrices, we project down to those states which are invariant under this action of $D_M$. This is done using the projector
		\[
		P = \frac{1}{2M} \sum_{ \rho \in D_M } \rho = \frac{1}{2M} \sum_{i=1}^M \sigma^i ( 1 + \tau )
		\]
		At this point, we've only sorted the words by their length, as opposed to how many $X$s and $Y$s they contain. To do this more refined sorting, we define an operator $Q$ on $V$ by
		\[
		Q e_X = x e_X \qquad \qquad Q e_Y = y e_Y
		\]
		Let $\widehat{Q} = Q \otimes \ldots \otimes Q$ be the equivalent on $V^{\otimes M}$. Then words constructed from $n$ $X$s and $m$ $Y$s have eigenvalue $x^n y^m$ under $\widehat{Q}$, so it is the action of $P$ on the eigenspaces of $\widehat{Q}$ that we are interested in (clearly the number of $X$s and $Y$s in a word is not changed by the action of $D_M$, and therefore $\widehat{Q}$ and $P$ commute, or equivalently, P acts on each of the eigenspaces of $\widehat{Q}$ independently of the other eigenspaces). The $B_{n,m}$ are then the dimension of the projected eigenspaces. To find these we consider
		\begin{equation}
		\text{ Tr } ( \widehat{ Q } P ) = \sum_{ n + m = M } x^n y^m B_{n,m}
		\label{level M generating function 1} 
		\end{equation}
		This can be calculated explicitly by noting that if we take $\rho \in S_M$ to be a permutation on the factors of $V^{\otimes M}$ (note that if we forget about the factor of $(-1)^M$ in \eqref{D_M action} briefly, our elements of $D_M$ are permutations), we have
		\begin{align}
		\text{ Tr } ( \rho \widehat{Q} )  & = (\text{ Tr } Q )^{ c_1 ( \rho ) } (\text{ Tr } Q^2 )^{ c_2 ( \rho ) } ( \text{ Tr } Q^3 )^{ c_3 ( \rho ) } \ldots. \nonumber \\
		& = ( x + y )^{ c_1 ( \rho ) } ( x^2 + y^2 )^{ c_2 ( \rho ) } ( x^3 + y^3 )^{ c_3 ( \rho ) } \ldots \nonumber
		\end{align}
		where $c_i(\rho)$ is the number of $i$-cycles in the cycle decomposition of $\rho$. We have already seen the above statement in \eqref{U(N) permutations to traces}. Now restoring the $(-1)^M$ and using \eqref{level M generating function 1}, we have
		\begin{align}
		\sum_{ n + m = M } x^n y^m B_{n,m} & = \frac{1}{2M} \left[ \sum_{ i = 1 }^M ( x + y )^{ c_1 ( \sigma^i ) } ( x^2 + y^2 )^{ c_2 ( \sigma^i ) } ( x^3 + y^3 )^{ c_3 ( \sigma^i ) } \ldots \right. \nonumber \\
		& \qquad + \left. (-1)^M \sum_{ i = 1 }^M ( x + y )^{ c_1 ( \sigma^i \tau ) } ( x^2 + y^2 )^{ c_2 ( \sigma^i \tau ) } ( x^3 + y^3 )^{ c_3 ( \sigma^i \tau ) } \ldots \right]
		\label{level M generating function 2}
		\end{align}
		We can evaluate this using the cycle index polynomial of $D_M$. For a subgroup $H$ of the symmetric group $S_M$, the cycle index polynomial of $H$ is defined to be
		\begin{align}
		Z^H ( t_1 , t_2 , \ldots ) & = \frac{1}{|H|} \sum_{ \rho \in H } t_1^{ c_1 ( \rho ) } t_2^{ c_2 ( \rho ) } t_3^{ c_3 ( \rho ) } \ldots \nonumber \\
		& = \sum_{ p \vdash M } Z^H_p \prod_i t_i^{ p_i }
		\label{cycle polynomial definition}
		\end{align}
		where $Z^H_p$ is the number of elements of $H$ with cycle type $p$ divided by $|H|$. 
		
		Were it not for the $(-1)^M$ in \eqref{level M generating function 2}, we could just replace $t_i$ with $x^i + y^i$ in $Z^{ D_M }$ to get the order $M$ part of the generating function for the $B_{n,m}$. As it is, we need to know slightly more about the structure of the $Z^{D_M}$ polynomials. Fortunately they are well known
		\begin{align}
		Z^{D_M} (t_1, t_2, \ldots ) = 
		\frac{1}{2M} \sum_{d|M} \phi(d) t_d^{\frac{M}{d}} + 
		\begin{cases} 
		\frac{1}{2} t_1 t_2^{ \frac{ M - 1 }{ 2 }} & M \text{ odd } \\
		\frac{1}{4} t_2^{ \frac{ M - 2 }{ 2 }} \left( t_1^2  + t_2 \right) & M \text{ even (and } \geq 2)
		\end{cases} \nonumber
		\end{align}
		where $\phi ( d )$ is the Euler totient function defined in \eqref{phi identity}. The first part of the polynomials is just half the cycle index polynomial of the cyclic group $C_M$. This corresponds to the rotations in $D_M$. The second part is the reflections, where the differences between $M$ odd and even come from the fact that odd-sided polygons only have one type of line of symmetry, those going through a vertex and bisecting the opposite side; while even-sided polygons have two types of lines of symmetry, those going through pairs of opposite vertices and those bisecting pairs of opposite lines. Now we know which part of $Z^{D_M}$ comes from reflections, we can see that \eqref{level M generating function 2} is
		\begin{align}
		\sum_{n+m=M} x^n y^m B_{n,m} & = \frac{1}{2M} \sum_{d|M} \phi(d) (x^d + y^d)^{\frac{M}{d}}
		\nonumber  \\
		&  + \begin{cases}
		- \frac{1}{2} (x+y) (x^2+y^2)^{\frac{M-1}{2}} & M \text{ odd} \\
		\frac{1}{4} (x^2+y^2)^{\frac{M-2}{2}} \left[ (x+y)^2  + (x^2+y^2) \right] & M \text{ even}
		\end{cases}
		\label{order M generating function}
		\end{align}
		To find $B_{n,m}$ explicitly we binomially expand the above. The first half of the expression was already expanded in \eqref{binomial expansion for A_n,m}, and is just (half) the order $M$ generating function for the $A_{n,m}$, so we focus on the second half. For $M$ odd, we have
		\begin{equation}
		(x+y) \left( x^2 + y^2 \right)^{\frac{M-1}{2}} = \sum_{r=0}^{\frac{M-1}{2}} \binom{\frac{M-1}{2}}{r} \left( x^{2r+1} y^{M-2r-1} + x^{2r} y^{M-2r} \right)
		\label{M odd expansion}
		\end{equation}
		and
		\begin{align}
		(x^2 + xy + y^2) \left( x^2 + y^2 \right)^{\frac{M-2}{2}} = \sum_{r=0}^{\frac{M-2}{2}} \binom{\frac{M-2}{2}}{r} ( x^{2r+2} y^{M-2r-2} & + x^{2r+1} y^{M-2r-1} \nonumber \\ & + x^{2r} y^{M-2r} )
		\label{M even expansion}
		\end{align}
		for $M$ even.
		
		Consider the coefficient of $x^n y^m$ if both $n$ and $m$ are even. Two of the three terms in \eqref{M even expansion} can contribute. Provided $n,m \geq 2$, we get contributions from $r = \frac{n}{2}, \frac{n}{2}-1$. This leads to the coefficient
		\begin{equation}
		\binom{\frac{n+m}{2} - 1}{\frac{n}{2}} + \binom{\frac{n+m}{2} - 1}{\frac{n}{2} - 1} = \binom{\frac{n}{2} + \frac{m}{2}}{\frac{n}{2}}
		\label{n,m even coefficient}
		\end{equation}
		Checking the cases where $n=0$ or $m=0$, we get 1 as a coefficient, which agrees with \eqref{n,m even coefficient}.
		
		Performing similar analyses for the other possible parity combinations leads to the coefficients
		\begin{align*}
		\binom{\frac{n}{2} + \frac{m-1}{2}}{\frac{n}{2}} & & n \text{ even}, m \text{ odd} \\
		\binom{\frac{n-1}{2} + \frac{m}{2}}{\frac{n-1}{2}} & & n \text{ odd}, m \text{ even} \\
		\binom{\frac{n-1}{2} + \frac{m-1}{2}}{\frac{n-1}{2}} & & n \text{ odd}, m \text{ odd}
		\end{align*}
		All four cases can be summarised by the coefficient
		\begin{equation*}
		\binom{\lfloor \frac{n}{2} \rfloor + \lfloor \frac{m}{2} \rfloor}{\lfloor \frac{n}{2} \rfloor}
		\end{equation*}
		Taking account of the signs and factors of a half in \eqref{order M generating function}, we have
		\begin{align}
		B_{n,m} & = \frac{1}{2} A_{n,m} + \frac{(-1)^{n+m}}{2}   \binom{\lfloor \frac{n}{2} \rfloor + \lfloor \frac{m}{2} \rfloor}{\lfloor \frac{n}{2} \rfloor} \nonumber  \\
		& = \frac{1}{2n+2m} \sum_{d|n,m} \phi(d) \binom{\frac{n+m}{d}}{\frac{n}{d}} + \frac{(-1)^{n+m}}{2}   \binom{\lfloor \frac{n}{2} \rfloor + \lfloor \frac{m}{2} \rfloor}{\lfloor \frac{n}{2} \rfloor}
		\label{SO(N) single trace counting}
		\end{align}
		To find the full generating function for $B_{n,m}$, we sum \eqref{order M generating function} from $M=1$ to $\infty$. We already know how to sum the first half of this expression from \eqref{U(N) single traces} (it is just the generating function for $A_{n,m}$), and the second half is simple to evaluate directly. Explicitly, we get
		\begin{align}
		f_{SO(N)}(x,y) = H_{\widetilde{T}_{ST}} (x,y) = \frac{1}{2}  & \left[ - \sum_{d=1}^\infty \frac{\phi(d)}{d}  \log(1-x^d-y^d) \right. \nonumber \\ & \qquad \qquad \left. + \frac{x^2+xy+y^2-x-y}{1-x^2-y^2} \right]
		\label{SO(N) single trace generating function}
		\end{align}
		We can now take the plethystic exponential, given in \eqref{plethystic exponential}, to get the multi-trace generating function
		\begin{equation}
		F_{SO(N)}(x,y) = H_{\widetilde{T}}(x,y) = \prod_{k=1}^\infty \frac{1}{\sqrt{1-x^k-y^k}} \text{exp} \left[ \frac{x^{2k} + x^k y^k + y^{2k} - x^k -y^k}{2k (1 - x^{2k} - y^{2k})} \right]
		\label{SO(N) multi trace generating function}
		\end{equation}
		where to evaluate the infinite products/sums we have used a change of variables similar to those in \eqref{U(N) aperiodic single traces} and \eqref{binomial expansion for A_n,m} as well as the identity \eqref{phi identity}. After completion of this paper we became aware of \cite{Yokoyama2016}, which gives a similar counting formula in the context of $SO(N)$ superconformal indices.
		
		
		As a sanity check, we can set $y=0$ and check that we recover the generating function for single matrix operators found in \cite{Caputa2013a}. Using \eqref{plethystic exponential}, this gives us
		\begin{equation}
		F_{SO(N)}(x,0) = \prod_{n=1}^\infty \frac{1}{(1-x^n)^{B_{n,0}}}
		\label{single matrix generating function}
		\end{equation}
		Setting $m=0$ in \eqref{SO(N) single trace counting} and using \eqref{phi identity} we get
		\begin{equation}
		B_{n,0} = \frac{1}{2} \left( 1 + (-1)^n \right) = \begin{cases}
		1 & n \text{ even} \\
		0 & n \text{ odd}
		\end{cases}
		\nonumber
		\end{equation}
		Plugging this into \eqref{single matrix generating function} gives us
		\begin{equation}
		F_{SO(N)}(x,0) = \prod_{n=1}^\infty \frac{1}{1-x^{2n}}
		\nonumber
		\end{equation}
		which matches the result found in \cite{Caputa2013a}.
		
		We can now use the relations given  in figure \ref{figure: SO(N) diagram} to find $H_{\widetilde{T}^{(min)}}$ and $H_{\widetilde{T}_{ST}^{(min)}}$.  Taking the M\"obius transform (see \eqref{mobius transform on function}) of \eqref{SO(N) single trace generating function} gives
		\begin{equation}
		H_{\widetilde{T}_{ST}^{(min)}} (x,y) = \frac{1}{2} \sum_{d=1}^\infty \mu(d) \left[ -\frac{1}{d} \log (1-x^d - y^d) + \frac{x^{2d} +x^d y^d + y^{2d} - x^d - y^d}{1 - x^{2d} - y^{2d}} \right]
		\label{SO(N) minimally periodic trace generating function}
		\end{equation}
		where we have used the identity \eqref{phi mu identity 2}. Expanding to find the coefficients gives
		\begin{equation}
		b_{n,m} = \frac{1}{2} \sum_{d | n,m} \mu (d) \left[ \frac{1}{n+m}  \binom{\frac{n+m}{d}}{\frac{n}{d}} + (-1)^{\frac{n+m}{d}} \binom{ \lfloor \frac{n}{2d} \rfloor + \lfloor \frac{m}{2d} \rfloor}{\lfloor \frac{n}{2d} \rfloor} \right]
		\nonumber
		\end{equation}
		Taking the plethystic exponential of \eqref{SO(N) minimally periodic trace generating function}, we get
		\begin{equation*}
		H_{\widetilde{T}^{(min)}} (x,y) = \frac{1}{\sqrt{1-x-y}} \prod_{k=1}^\infty \text{exp} \left[ \frac{1}{2k} \frac{x^{2k} + x^k y^k + y^{2k} - x^k - y^k}{1 - x^{2k} - y^{2k}} \sum_{d|k} d \mu(d) \right]
		\end{equation*}
		where we have used the identity \eqref{mobius function identity}.
		
		The numbers appearing in the exponential here, $c_k = \sum_{d|k} d \mu (d)$, form an interesting mathematical sequence. It is sequence A023900 in the OEIS \cite{oeis}, and has the alternative expression
		\begin{equation*}
		c_k = \prod_{\substack{p | k \\ p \text{ prime}}} (1-p)
		\end{equation*}
		To find the Hilbert series for the vector spaces shown in figure \ref{figure: SO(N) periodicity diagram}, we first compare \eqref{SO(N) single trace counting 2nd form} with \eqref{SO(N) single trace counting} and \eqref{U(N) single trace counting} to find
		\begin{equation}
		A^{inv}_{n,m} = \binom{\lfloor \frac{n}{2} \rfloor + \lfloor \frac{m}{2} \rfloor}{\lfloor \frac{n}{2} \rfloor}
		\nonumber
		\end{equation}
		Summing over $n,m$ gives the Hilbert series for $T_{ST;inv}$
		\begin{equation}
		H_{T_{ST;inv}}(x,y) = \frac{(1+x)(1+y)}{1-x^2-y^2} - 1 = \frac{x^2 + x y + y^2 + x + y}{1-x^2-y^2}
		\label{generating function for invariant traces}
		\end{equation}
		where the $-1$ comes from setting $A_{0,0}^{inv} = 0$. Note that we already saw this, up to a change in sign of $x$ and $y$, as the second half of the generating function \eqref{SO(N) single trace generating function}.
		
		Since $A_{n,m}^{inv}$ and $a_{n,m}^{inv}$ are related by a M{\"o}bius transform, we have
		\begin{equation}
		H_{T^{(1)}_{ST;inv}} (x,y) = \mathcal{M} \left( H_{T_{ST;inv}} \right) (x,y) = \sum_{d=1}^\infty \mu(d) \frac{x^{2d} + x^d y^d + y^{2d} + x^d + y^d}{1 - x^{2d} - y^{2d}}
		\label{generating function for aperiodic invaraint traces}
		\end{equation}
		Then using the Hilbert series equivalents of the formulae \eqref{b_{n,m;even}} and \eqref{b_{n,m;odd}}, we have
		\begin{align*}
		H_{\widetilde{T}_{ST}^{(odd)}}(x,y) & = \frac{1}{2} \left[ H_{T_{ST}^{(1)}}(x,y) + H_{T^{(1)}_{ST;inv}} (-x,-y) \right] \\
		& = \frac{1}{2} \sum_{d=1}^\infty \mu(d)  \left[ - \frac{1}{d} \log (1-x^d - y^d) \right. \\ & \hspace{150pt} \left. + \frac{x^{2d}+x^d y^d + y^{2d} + (-x)^d + (-y)^d}{1 - x^{2d} - y^{2d}}   \right] \\
		H_{\widetilde{T}_{ST}^{(even)}} (x,y) & = \frac{1}{2} \left[ H_{T_{ST}^{(1)}}(x,y) + H_{T^{(1)}_{ST;inv}} (x,y) \right] \\
		& = \frac{1}{2} \sum_{d=1}^\infty \mu(d)  \left[- \frac{1}{d} \log (1-x^d - y^d) + \frac{x^{2d}+x^d y^d + y^{2d} + x^d + y^d}{1 - x^{2d} - y^{2d}}   \right]
		\end{align*}
		Note the similarities between these series, which count single traces with odd and even numbers of periods, and the minimally periodic version \eqref{SO(N) minimally periodic trace generating function}. The only difference between the three series is in the sign of the last two terms.
		
		From these three Hilbert series we can derive explicit expressions for the coefficients $a_{n,m}^{inv}, b_{n,m}^{(odd)}$ and $b_{n,m}^{(even)}$. These are given in appendix \ref{section: sequences}.
		
		Finally, taking the plethystic exponential gives
		\begin{align*}
		H_{\widetilde{T}^{(odd)}}(x,y) & = \frac{1}{\sqrt{1-x-y}} \prod_{k=1}^\infty \hspace{-2pt} \text{exp} \hspace{-2pt} \left[ \sum_{d|k} \frac{d\mu(d)}{2k} \frac{x^{2k} + x^k y^k + y^{2k} + (-1)^d (x^k + y^k)}{1- x^{2k} - y^{2k}} \right] \\
		H_{\widetilde{T}^{(even)}}(x,y) & = \frac{1}{\sqrt{1-x-y}} \prod_{k=1}^\infty \text{exp} \left[  \frac{x^{2k} + x^k y^k + y^{2k} + x^k + y^k}{2k(1- x^{2k} - y^{2k})} \sum_{d|k} d\mu(d) \right] 
		\end{align*}
		where we have used the identity \eqref{mobius function identity}. This gives us all the Hilbert series featured in figure \ref{figure: SO(N) periodicity diagram}.

		\section{The orientifold quotient in the quarter-BPS sector}
		\label{section: quarter-BPS projection}
		
		In section \ref{section: Z2 quotient} we looked at the orientifold quotient that takes the $U(N)$ theory to the $SO(N)$ theory in the half-BPS sector. This map had remarkable connections to plethysms of Young diagrams and the combinatorics of domino tableaux. The key result that enabled us to link these together was the matrix element \eqref{Ivanov's matrix element}, proved in \cite{Ivanov1999}.
		
		We now generalise to the quarter-BPS sector. As with the half-BPS sector, the quotient simply takes the matrices $X$ and $Y$ and makes them anti-symmetric. Since the $U(N)$ operators are multi-traces, their quotient must also be multi-traces, and therefore (similarly to half-BPS) the baryonic operators do not feature. However, as demonstrated in sections \ref{section: structure of the space} and \ref{Different expressions for the SO(N) generating function}, the set of two-matrix traces is significantly more complicated than the one-matrix version. Our first task is to give a labelling set for generic multi-traces in both the $U(N)$ and $SO(N)$ theories. These make use of Lyndon words and orthogonal Lyndon words respectively.
		
		After establishing a notation for generic multi-traces, we investigate how an individual $U(N)$ multi-trace behaves under the quotient. This is more complicated than the half-BPS case since two distinct $U(N)$ multi-traces can now give the same (non-zero) $SO(N)$ multi-trace. For example Tr$X^2YXY^2$ and Tr$X^2 Y^2 X Y$ are distinct in the $U(N)$ theory but when $X$ and $Y$ are made anti-symmetric, they reduce to the same object.
		
		Both the $U(N)$ basis \eqref{U(N) operators} and the mesonic $SO(N)$ basis \eqref{mesonic operators definition} are defined in terms of sums over permutations. Following the route in \ref{section: half-BPS permutations to traces}, we investigate how these reduce to sums over (the labelling sets of) multi-traces. To do this, we study the group action which leaves the mesonic contraction invariant, \eqref{SO(N) group algebra invariance}, and in particular we find the stabiliser group for a representative of a double coset.
		
		We then put all the pieces together to find the coefficients involved in the quotient of a $U(N)$ operator to a linear combination of $SO(N)$ operators. Unfortunately we have not found an analogue of \eqref{Ivanov's matrix element}, so the simplifications of the half-BPS do not occur here and we have not been able to find a combinatoric interpretation of our results.

		\subsection{Labelling of traces}
		\label{section: labelling of traces}
		
		A $U(N)$ single trace is described by a Lyndon word $w$ and the number of periods, while a multi-trace is defined by a collection of these single traces. Consider a generic $U(N)$ multi-trace, and let the number of constituent single traces with Lyndon word $w$ and number of periods $i$ be $p_{w,i}$, then the multi-trace can be written
		\begin{equation}
		T^{U(N)}_{\mathcal{P}} = \prod_{w,i}  \left( \text{Tr} W^i \right)^{p_{w,i}}
		\nonumber
		\end{equation}
		where $W$ is the matrix word equivalent of the Lyndon word $w$.	This trace is characterised by the set of numbers $\{ p_{w,i} \}$. A convenient way to package these numbers is to define a partition $p_w$ for each Lyndon word
		\begin{equation}
		p_w = (1^{p_{w,1}}, 2^{p_{w,2}}, \ldots )
		\nonumber
		\end{equation}
		Then the label for a $U(N)$ multi-trace is
		\begin{equation}
		\mathcal{P} = \left\{ p_w : w \text{ a Lyndon word} \right\} = \left\{ p_x, p_y , p_{xy}, p_{x^2y}, p_{x y^2}, \ldots \right\}
		\nonumber
		\end{equation}	
		Define $l_x (w)$, $l_y(w)$ and $l(w)$ be the number of $x$s, the number of $y$s and the total length of $w$ respectively. Then clearly $l(w) = l_x(w) + l_y(w)$, and the number of $X$s and $Y$s in a multi-trace is
		\begin{equation}
		n = \sum_{w} l_x(w) |p_w| \qquad \qquad m = \sum_w l_y (w) |p_w|
		\nonumber
		\end{equation}
		We summarise this with $\mathcal{P} \Vdash (n,m)$.
		
		As an example of this labelling, table \ref{table: example trace labelling} lists the 10 different $\mathcal{P} \Vdash (2,2)$ and their associated multi-traces.

		\begin{table}
			\begin{center}
				\begin{tabular}{c | c }
					$\mathcal{P}$ & $T^{U(N)}_{\mathcal{P}}$ \\ \hline
					$p_x = [1,1] \ , \ p_y = [1,1]$ & $\left( \text{Tr} X \right)^2 \left( \text{Tr} Y \right)^2$ \\
					$p_x = [1,1] \ , \ p_y = [2]$ & $ \left( \text{Tr} X \right)^2 \left( \text{Tr} Y^2 \right)$ \\
					$p_x = [2] \ , \ p_y = [1,1]$ & $ \left( \text{Tr} X^2 \right) \left( \text{Tr} Y \right)^2$ \\
					$ p_x = [1] \ , \ p_y = [1] \ , \ p_{xy} = [1]$ & $ \left( \text{Tr} X \right) \left( \text{Tr} XY \right) \left( \text{Tr} Y \right) $ \\
					$ p_x = [2] \ , \ p_y = [2] $ & $ \left( \text{Tr} X^2 \right) \left( \text{Tr} Y^2 \right) $ \\
					$ p_{xy} = [1,1]$ & $ \left( \text{Tr} XY \right)^2 $ \\
					$ p_{x^2y} = [1] \ , \ p_y = [1] $ & $ \left( \text{Tr} X^2 Y \right) \left( \text{Tr} Y \right)$ \\
					$ p_{xy^2} = [1] \ , \ p_x = [1] $ & $ \left( \text{Tr} X \right) \left( \text{Tr} X Y^2 \right) $ \\
					$ p_{xy} = [2] $ & $ \text{Tr} (XY)^2 $ \\
					$p_{x^2 y^2} = [1] $ & $ \text{Tr} X^2 Y^2$
				\end{tabular}
			\end{center}
			\caption{The 10 different $U(N)$ multi-traces at $n=m=2$ along with their labels. Any constituent partitions of $\mathcal{P}$ that are not explicitly listed are set to zero.}
			\label{table: example trace labelling}
		\end{table}
		
		A $SO(N)$ single trace is described by an orthogonal Lyndon word $\tilde{w}$ (as defined in section \ref{Different expressions for the SO(N) generating function}) and the number of repetitions $i$ (note this is not the number of periods). Consider a $SO(N)$ multi-trace, and let $p_{\tilde{w},i}$ be the number of constituent single traces with orthogonal Lyndon word $\tilde{w}$ and number of repetitions $i$. Then we have a partition $p_{\tilde{w}} = (1^{p_{\tilde{w},1}}, 2^{p_{\tilde{w},2}}, \ldots )$ for each orthogonal Lyndon word and we denote the combination by
		\begin{align*}
		\widetilde{\mathcal{P}} & = \left\{ p_{\tilde{w}} : \tilde{w} \text{ an orthogonal Lyndon word} \right\} \\
		& = \left\{ p_{x^2}, p_{xy}, p_{y^2}, p_{x^3 y}, p_{x^2 y^2}, p_{xy^3}, \ldots , p_{x^2yxy^2}, \ldots \right\}
		\end{align*}
		The multi-trace corresponding to $\widetilde{\mathcal{P}}$ is
		\begin{equation}
		T^{SO(N)}_{\widetilde{\mathcal{P}}} = \prod_{\tilde{w}, i} \left( \text{Tr} \widetilde{W}^i \right)^{p_{\tilde{w},i}}
		\label{SO(N) multi-trace convention}
		\end{equation}
		where $\widetilde{W}$ is the matrix word corresponding to the orthogonal Lyndon word $\tilde{w}$. As for the normal Lyndon words, let $l_x(\tilde{w})$, $l_y(\tilde{w})$ and $l(\tilde{w})$ be the number of $x$s, number of $y$s and total length of $\tilde{w}$ respectively. Then
		\begin{equation}
		n = \sum_{\tilde{w}} l_x(\tilde{w}) |p_{\tilde{w}}| \qquad \qquad m = \sum_{\tilde{w}} l_y (\tilde{w}) |p_{\tilde{w}}|
		\nonumber
		\end{equation}
		We use the same notation $\widetilde{\mathcal{P}} \ \Vdash \ (n,m)$ as for the $U(N)$ traces. It will always be clear whether we are referring to a $SO(N)$ or $U(N)$ trace.
		
		We give the 9 different $\widetilde{\mathcal{P}} \Vdash (3,3)$ in table \ref{table: example SO(N) trace labelling}.
		
		\begin{table}
			\begin{center}
				\begin{tabular}{c | c }
					$\widetilde{\mathcal{P}}$ & $T^{SO(N)}_{\widetilde{\mathcal{P}}}$ \\ \hline
					$p_{x^2} = [1] \ , \ p_{xy} = [1] \ , \ p_{y^2} = [1] $ & $\left( \text{Tr} X^2 \right) \left( \text{Tr} XY \right) \left( \text{Tr} Y^2 \right) $ \\
					$p_{x y} = [1,1,1]$ & $ \left( \text{Tr} XY \right)^3 $ \\
					$p_{x^3 y} = [1] \ , \ p_{y^2} = [1]$ & $ \left( \text{Tr} X^3Y \right) \left( \text{Tr} Y^2 \right)$ \\
					$ p_{x^2 y^2} = [1] \ , \ p_{xy} = [1] $ & $ \left( \text{Tr} X^2 Y^2 \right) \left( \text{Tr} XY \right) $ \\
					$ p_{x y^3} = [1] \ , \ p_{x^2} = [1] $ & $ \left( \text{Tr} X Y^3 \right) \left( \text{Tr} X^2 \right) $ \\
					$ p_{xy} = [2,1]$ & $ \text{Tr} \left( XY \right)^2 \left( \text{Tr} XY \right) $ \\
					$ p_{x^3 y^3} = [1] $ & $ \text{Tr} X^3 Y^3 $ \\
					$ p_{x^2 y x y^2} = [1] $ & $ \text{Tr} X^2 Y X Y^2 $ \\
					$ p_{xy} = [3] $ & $ \text{Tr} (XY)^3 $
				\end{tabular}
			\end{center}
			\caption{The 9 different $SO(N)$ multi-traces at $n=m=3$ along with their labels. Any constituent partitions of $\widetilde{\mathcal{P}}$ that are not explicitly listed are set to zero.}
			\label{table: example SO(N) trace labelling}
		\end{table}
		
		It will also be helpful to consider traces of symmetric matrices $X$ and $Y$. A single trace will be labelled by a Lyndon word up to reversal, $\bar{w}$, and the number of periods. This means $\bar{w}$ can be split into two types; either it is a Lyndon word that is invariant under reversal (type 1), or it is the first (lexicographically) of a pair of Lyndon words that transform into each other under reversal (type 2). This differs from the $SO(N)$ case (anti-symmetric matrices) in that there is no distinction between odd and even length words. We define $p_{\bar{w}}$, $\bar{\mathcal{P}}$, $\bar{W}$, $l_x (\bar{w})$, $l_y(\bar{w})$, $l(\bar{w})$ and $\Vdash$ in an analogous way to the $U(N)$ and $SO(N)$ traces.

		\subsection{Projection of a trace}
		
		Consider a $U(N)$ multi-trace $T^{U(N)}_{\mathcal{P}}$ and project it to the $SO(N)$ theory by turning each of the $X$s and $Y$s into anti-symmetric matrices. If any of the constituent single traces vanish when $X$ and $Y$ are anti-symmetric, then clearly the projection is zero. In section \ref{Different expressions for the SO(N) generating function} we studied how the single traces behave under the projection. In the language of this section, $T_{\mathcal{P}}^{U(N)}$ will vanish if $p_{w,i} \neq 0$ for a pair $(w,i)$ such that $i$ is odd and $w$ is reversal-invariant and of odd length. Equivalently, if $\mathcal{P}$ contains a partition $p_w$ (where $w$ is reversal-invariant and of odd length) which has an odd component.
		
		For the remaining $\mathcal{P}$, $T^{U(N)}_{\mathcal{P}}$ projects to a non-zero $SO(N)$ multi-trace, whose constituent single traces fall into 4 categories. They are (powers of) type 1A orthogonal Lyndon words, type 1B words, type 2 words or the reversal of type 2 words. To turn the trace into the form $T^{SO(N)}_{\widetilde{\mathcal{Q}}}$ for some $\widetilde{\mathcal{Q}}$ we transpose the traces in the last category. For a single trace with Lyndon word $w = \tilde{w}^{(r)}$ and number of periods $i$ (in this case the number of periods and repetitions match), this introduces a factor of $(-1)^{i \, l(w)}$, as shown in \eqref{relation between traces from antisymmetry}. Multiplying up all the sign factors from the constituent single traces gives
		\begin{equation}
		\text{OrthSign} ( \mathcal{P} ) 
		=  \prod_{\substack{w,i \\ \text{ where } w = \tilde{w}^{(r)} \text{ for } \\ \tilde{w} \text{ an orthogonal Lyndon } \\ \text{ word of type 2}}} (-1)^{i \, l(w) p_{w,i}}
		=  \prod_{\substack{w \\ \text{ where } w = \tilde{w}^{(r)} \text{ for } \\ \tilde{w} \text{ an orthogonal Lyndon } \\ \text{ word of type 2}}} (-1)^{l(w) | p_w |}
		\nonumber
		\end{equation}
		So for those $T^{U(N)}_{\mathcal{P}}$ which don't vanish under the projection, we have
		\begin{equation}
		T^{U(N)}_{\mathcal{P}} \overset{\mathbb{Z}_2}{\longrightarrow} \text{OrthSign}(\mathcal{P}) T^{SO(N)}_{\text{Orth} (\mathcal{P})}
		\label{quarter-BPS trace projection}
		\end{equation}
		where Orth$(\mathcal{P})$ is the $SO(N)$ multi-partition composed of $q_{\tilde{w},i}$, defined by
		\begin{equation}
		q_{\tilde{w},i} =
		\begin{cases}
		p_{\tilde{w},i} & \tilde{w} \text{ of type 1A} \\
		p_{\sqrt{\tilde{w}},2i} & \tilde{w} \text{ of type 1B} \\
		p_{\tilde{w},i} + p_{\tilde{w}^{(r)},i} & \tilde{w} \text{ of type 2}
		\end{cases}
		\nonumber
		\end{equation}
		and for $\tilde{w}$ of type 1B, we define $\sqrt{\tilde{w}}$ to be the Lyndon word that, when repeated, gives $\tilde{w}$. So for example $\sqrt{x^2} = x$ and $\sqrt{x^2 y x^2 y} = x^2 y$. 
		
		We give a few examples of the full projection in table \ref{table: example projections}.

		\begin{table}
			\begin{center}
				\begin{tabular}{c | c }
				$U(N)$ multi-trace & Image after projection \\ \hline
				$\left( \text{Tr} X^2 Y X Y^2 \right)^2 $ & $\left( \text{Tr} X^2 Y X Y^2 \right)^2$ \\
				$\left( \text{Tr} X^2 Y X Y^2 \right) \left( \text{Tr} X^2 Y^2 X Y \right)$ & $\left( \text{Tr} X^2 Y X Y^2 \right)^2$ \\
				$\left( \text{Tr} X^2 Y^2 X Y \right)^2 $ & $\left( \text{Tr} X^2 Y X Y^2 \right)^2$ \\
				$ \left( \text{Tr} X^3 Y X Y^2 \right)^2 $ & $ \left( \text{Tr} X^3 Y X Y^2 \right)^2 $ \\
				$\left( \text{Tr} X^3 Y X Y^2 \right) \left( \text{Tr} X^3 Y^2 X Y \right)$ & $ - \left( \text{Tr} X^3 Y X Y^2 \right)^2$ \\
				$ \left( \text{Tr} X^3 Y^2 X Y \right)^2 $ & $ \left( \text{Tr} X^3 Y X Y^2 \right)^2 $
				\end{tabular}
			\end{center}
			\caption{Examples of projections of individual $U(N)$ multi-traces}
			\label{table: example projections}
		\end{table}

		\subsection{From permutations to traces}
		\label{section: 2-matrix permutations to traces}
		
		\subsubsection{$U(N)$}
		\label{section: U(N) quarter-bps permutations to traces}
		
		Permutations $\sigma \in S_{n+m}$ produce multi-traces via the formula \eqref{general U(N) operator}. As in the half-BPS case, each cycle in the permutation corresponds to a single trace, but we can now have two cycles of the same length producing different traces. We give a few examples in the first column of table \eqref{table: example permutations to traces}.
		
		\begin{table}
		\begin{center}
			\begin{tabular}{c | c | c}
				$U(N)$ Permutation & $SO(N)$ Permutation & Corresponding trace \\ \hline
				$(1,2,n+1,n+2)$ & $(1,3,2n+1, 2n+3)$ & Tr$X^2 Y^2$ \\
				$(1,n+1,2,n+2)$ & $(1,2n+1, 3, 2n+3)$ & Tr$(XY)^2$ \\
				$(1,2,3,n+1,4,n+2)$ & $(1,3,5,2n+1,7,2n+3)$ & Tr$X^3YXY$ \\
				$(1,2,n+1,3,4,n+2)$ & $(1,3,2n+1,5,7,2n+3)$ & Tr$\left( X^2 Y \right)^2$
			\end{tabular}
		\end{center}
		\caption{Examples of multi-traces and the permutations which produce them via the $SO(N)$ contraction \eqref{mesonic operator} and the $U(N)$ contraction \eqref{general U(N) operator}.}
		\label{table: example permutations to traces}
		\end{table}
		
		These examples make it clear how a permutation produces a trace. Writing out the permutation in cycle notation, a number in $\{ 1,2, \ldots ,n\}$ corresponds to an $X$ while a number in $\{ n+1,n+2, \ldots ,n+m\}$ corresponds to a $Y$. From \eqref{U(N) group algebra invariance}, we see that the set of permutations producing the same multi-trace is no longer a standard conjugacy class, but is instead the orbit under conjugation by $S_n \times S_m$.
		
		Each different conjugacy class produces a different multi-trace, and conversely each multi-trace corresponds to a conjugacy class. Therefore the labelling set for the conjugacy classes is exactly the same as that for the traces, given by the $\mathcal{P}$ defined in section \ref{section: labelling of traces}.
		
		The size of these conjugacy classes is found using the orbit-stabiliser theorem. Take $\sigma$ to be a representative member of the conjugacy class labelled by $\mathcal{P}$. The stabiliser of $\sigma$ is composed of the elements of $S_n \times S_m$ that commute with $\sigma$. As in the half-BPS, each cycle has a rotation subgroup attached to it. However, conjugation by $S_n \times S_m$ rather than by $S_{n+m}$ means we can only rotate the numbers $1,2, \ldots ,n$ amongst themselves (and similarly for $n+1, n+2, \ldots , n+m$). Therefore for a single cycle labelled by Lyndon word $w$ and number of repetitions $i$ (remember cycles correspond to single traces), the rotation group has size $i$ (rather than $i l(w)$, which is the length of the cycle). As in the half-BPS case, different cycles with the same labels can be permuted, and therefore the stabiliser is given by
		\begin{equation}
		Stab(\sigma) \cong \bigtimes_{w,i} S_{p_{w,i}} \left[ \mathbb{Z}_i \right]
		\nonumber
		\end{equation}
		which has size
		\begin{equation}
		Z_{\mathcal{P}} = \prod_{w,i} i^{p_{w,i}} \left( p_{w,i} \right)! = \prod_w z_{p_w}
		\nonumber
		\end{equation}
		So by the orbit-stabiliser theorem, the size of $S_n \times S_m$ conjugacy classes is
		\begin{equation}
		\frac{n! m!}{Z_{\mathcal{P}}}
		\nonumber
		\end{equation}

		\subsubsection{$SO(N)$}
		
		Permutations $\sigma \in S_{2n+2m}$, or elements of $\mathbb{C}(S_{2n+2m})$, produce traces via the formula \eqref{mesonic operator}. This contraction is invariant under the algebra transformation \eqref{SO(N) group algebra invariance}. Following the route we took in section \ref{section: SO(N) permutations to traces}, we study the orbits of $\sigma \in S_{2n+2m}$ under the action
		\begin{equation}
		\sigma \mapsto \alpha \sigma \gamma^{-1} \qquad \qquad \alpha \in S_{n+m}[S_2] \ \ , \ \  \gamma \in S_n[S_2] \times S_m[S_2]
		\label{SO(N) invariance without minus sign}
		\end{equation}
		These orbits are called double cosets, but in contrast to the half-BPS case, the groups are different on the left and right. If we took $X$ and $Y$ to be symmetric matrices in \eqref{mesonic operator}, then it would be invariant under \eqref{SO(N) invariance without minus sign}. Therefore the orbits under this action correspond to multi-traces of symmetric matrices. Thus the $\bar{\mathcal{P}}$, defined in section \ref{section: labelling of traces}, form the labelling set for the orbits.
		
		We can repeat the steps in \eqref{Trace from SO(N) contraction} and figure \ref{figure: SO(N) simplified contraction}, but including copies of both $X$ and $Y$, to show that if $\sigma \in S_{n+m}^{(odd)} \leq S_{2n+2m}$, the $SO(N)$-style contraction reduces to the $U(N)$-style contraction. Explicitly, let $\tau \in S_{n+m}$ be the equivalent permutation to $\sigma$, then
		\begin{equation}
		C_I^{(\delta)} \sigma^I_J \left( X^{\otimes n} Y^{\otimes m} \right)^J = X^{k_1 k_{\tau(1)}} \ldots X^{k_n k_{\tau(n)}} Y^{k_{n+1} k_{\tau(n+1)}} \ldots Y^{k_{n+m} k_{\tau(n+m)}}
		\label{SO(N) to U(N) contraction}
		\end{equation}
		So by comparison with the $U(N)$, if we write out $\sigma \in S_{n+m}^{(odd)}$ in cycle notation, an odd number in $\{ 1,3,5, \ldots ,2n-1\}$ will correspond to an $X$, while an odd number in $\{ 2n+1, 2n+3, \ldots , 2n+2m-1\}$ will correspond to a $Y$. Examples are given in the second column of table \ref{table: example permutations to traces}. From \eqref{SO(N) to U(N) contraction}, we can see that any multi-trace can be produced by permutations in $S_{n+m}^{(odd)}$, and therefore we can take the double coset representatives to be in $S_{n+m}^{(odd)}$.
		
		As with their half-BPS equivalents, the double cosets can be split into two categories, odd and even, depending on whether the stabilisers of a representative element can have an odd permutation in their right hand factor. By analogous reasoning to \eqref{vanishing contraction} and \eqref{vanishing matrix element}, the odd double cosets produce vanishing traces and matrix elements. Just as in the half-BPS case, the even double cosets are those which produce non-zero traces, so we expect (and prove below) that they will be labelled by $\widetilde{\mathcal{P}}$ (defined in section \ref{section: labelling of traces}).
		
		With the half-BPS even/odd double cosets we were able to characterise them more simply by whether the corresponding partition had all even components or not. We now find the corresponding characterisation for the quarter-BPS double cosets, for which purpose we study the detailed structure of the stabiliser.
		
		Take $\sigma \in S_{n+m}^{(odd)}$ to be a representative of the double coset labelled by $\bar{\mathcal{P}}$. The pair $(\alpha, \gamma ) \in S_{n+m}[S_2] \times \left( S_n[S_2] \times S_m[S_2] \right)$ is in the stabiliser of $\sigma$ if $\alpha \sigma \gamma^{-1} = \sigma$. This is equivalent to $\alpha = \sigma \gamma \sigma^{-1}$, so we look for $\gamma \in S_n[S_2] \times S_m[S_2]$ such that
		\begin{equation}
		\sigma \gamma \sigma^{-1} \in S_{n+m}[S_2]
		\label{stabiliser condition}
		\end{equation}
		Similarly to the half-BPS case, this is trivially true if $\gamma$ commutes with $\sigma$. By following the same argument as given in section \ref{section: SO(N) permutations to traces}, we can embed $S_n \times S_m$ into $S_{2n+2m}$ in such a way that the conjugation (by $S_n^{(odd)} \times S_m^{(odd)}$) stabiliser of $\sigma$ in $S_{n+m}^{(odd)}$ is a subgroup of the $SO(N)$ stabiliser. This is exactly the group we already found in section \ref{section: U(N) quarter-bps permutations to traces}. Note that the form of this embedding means that all members of this subgroup are even in $S_{2n+2m}$.
		
		This tells us that for each individual cycle of type $(\bar{w},i)$ we have a corresponding rotation group of order $i$. Just as for half-BPS, the $SO(N)$ stabiliser differs from the $U(N)$ one in that it has reflections as well these rotations. However, unlike the half-BPS case, this does not occur for all cycles. To see why, we explain how these reflections are constructed by giving examples and then explain the general case.
		
		If we take $c = (1,2n+1, 3 , 2n+3 , 5 , 2n+5)$ (this is labelled by $\bar{w} = xy$ and $i=3$), a reflection is given by $\gamma = (1,2) ( 2n+1 , 2n+6 )( 3 , 6 ) ( 2n+3 , 2n+4 ) ( 5 , 4 ) ( 2n+5 , 2n+2 )$. Given $c = ( 1 , 3 , 2n+1 , 5 , 7 , 2n+3 )$ (labelled by $\bar{w}=x^2y$ and $i=2$), a reflection is $\gamma = ( 1 , 8 ) ( 3 , 6 ) ( 2n+1 , 2n+2 ) ( 5 , 4 ) ( 7 , 2 ) ( 2n+3 , 2n+4 )$.
		
		In general, for a cycle $c \in S_{n+m}^{(odd)}$ labelled by $\bar{w}$ and $i$, the reflections $\gamma$ can be constructed from $i \, l(\bar{w})$ transpositions, each consisting of one odd and one even number. Order the transpositions so that the odd numbers appear in the same order as they do in $c$. Then the even numbers should appear in the reverse order. This produces a $\gamma \in S_{n+m}[S_2]$ satisfying \eqref{stabiliser condition}, but for $\gamma \in Stab(c)$ we need $\gamma \in S_n[S_2] \times S_m[S_2]$. This can only be done if it is possible to match the even and odd numbers so that each transposition only consists of numbers $\leq 2n$ or $\geq 2n+1$. Since the ordering of the numbers is governed by the word $\bar{w}$, this can only be done if $\bar{w}$ is invariant under reversal (up to cyclic rotations).
		
		Therefore $\bar{w}$ of type 1 do have a reflection symmetry in their stabiliser group, while $\bar{w}$ of type 2 do not. The sign of the reflections is given by $(-1)^{i \, l(\bar{w})}$, so if $\sigma$ contains a cycle labelled by an odd length $\bar{w}$ of type 1 with an odd number of repetitions, $\sigma$ represents an odd double coset. In terms of $\bar{\mathcal{P}}$, the double coset is odd if any of the constituent partitions $p_{\bar{w}}$, for reversal-invariant (type 1) $\bar{w}$ of odd length, has an odd component.
		
		Therefore for even double cosets, the partitions $p_{\bar{w}}$ (type 1 $\bar{w}$ of odd length) must have even components, and we can therefore give a more streamlined parameterisation by setting	$p_{\bar{w}} = 2 p_{\bar{w} \bar{w}}$ for $p_{\bar{w} \bar{w}} \vdash \frac{1}{2} | p_{\bar{w}}|$. Note that $\bar{w} \bar{w}$ is exactly a type 1B orthogonal Lyndon word. Replacing all such constituent partitions of $\bar{\mathcal{P}}$ with their `halved' counterparts, we see that the even double cosets are indeed labelled by $\widetilde{\mathcal{P}}$.
		
		Consider an even double coset $\widetilde{\mathcal{P}}$ and a representative member $\sigma$. The contraction \eqref{mesonic operator} will produce the trace $T^{SO(N)}_{ \widetilde{ \mathcal{P}}}$, up to a possible sign. We call $\sigma$ a positive or negative representative depending on this sign.
		
		From the above analysis of the stabiliser, we see that for $\sigma$ a representative for the double coset $\bar{\mathcal{P}}$, we have
		\begin{equation}
		Stab(\sigma) \cong \left( \bigtimes_{\substack{\bar{w} \text{ of type 1} \\ i}} S_{p_{\bar{w},i}} \left[ D_i \right] \right) \times \left( \bigtimes_{\substack{\bar{w} \text{ of type 2} \\ i}} S_{p_{\bar{w},i}} \left[ \mathbb{Z}_i \right] \right)
		\nonumber
		\end{equation}
		which has size
		\begin{align*}
		\bar{Z}_{\bar{\mathcal{P}}} & = 
		\left(  \prod_{\bar{w} \text{ of type 1}} z_{2 p_{\bar{w}}} \right) 
		\left(  \prod_{\substack{\bar{w} \text{ of type 2}}} z_{p_{\bar{w}}} \right) 
		\end{align*}
		Since we are only interested in even double cosets, we can re-express these for $\widetilde{\mathcal{P}}$
		\begin{equation}
		Stab(\sigma) \cong \left( \bigtimes_{\substack{\tilde{w} \text{ of type 1A} \\ i}} S_{p_{\tilde{w},i}} \left[ D_i \right] \right) \times \left( \bigtimes_{\substack{\tilde{w} \text{ of type 1B} \\ i}} S_{p_{\tilde{w},i}} \left[ D_{2i} \right] \right) \times \left( \bigtimes_{\substack{\tilde{w} \text{ of type 2} \\ i}} S_{p_{\tilde{w},i}} \left[ \mathbb{Z}_i \right] \right)
		\label{SO(N) quarter-bps stabilier group}
		\end{equation}
		and
		\begin{align*}
		\widetilde{Z}_{\widetilde{\mathcal{P}}} & = 
		\left(  \prod_{\tilde{w} \text{ of type 1A}} z_{2 p_{\tilde{w}}} \right) 
		\left(  \prod_{\tilde{w} \text{ of type 1B}} z_{4 p_{\tilde{w}}} \right) 
		\left(  \prod_{\tilde{w} \text{ of type 2}} z_{p_{\tilde{w}}} \right)
		\end{align*}
		In this formula the $2 p_{\tilde{w}}$ in the first factor has come from the dihedral group replacing the cyclic group in the first factor of \eqref{SO(N) quarter-bps stabilier group}, while the $4 p_{\tilde{w}}$ in the type 1B factor has come from the dihedral group combined with the doubled number of periods for even double cosets (since traces with an odd number of periods vanish).
		
		By the orbit-stabiliser theorem, the size of an even double coset is
		\begin{equation}
		\frac{\left| S_{n+m}[S_2] \times \left( S_n[S_2] \times S_m[S_2] \right) \right|}{\left| \text{stabiliser} \right|} = \frac{2^{2n+2m} n! m! (n+m)!}{\widetilde{Z}_{\widetilde{\mathcal{P}}}}
		\label{size of even double cosets}
		\end{equation}		
		As well as the abstract interpretation of $\widetilde{ Z }_{ \widetilde{ \mathcal{ P } } }$ as the size of an orbit, it has a physical interpretation. Using the quarter-BPS correlators in section \ref{section: correlators}, one can show that $N^{n+m} \widetilde{ Z }_{ \widetilde{ \mathcal{ P } } }$ is the large $N$ normalisation of the two point function for multi-traces.
		
		In this section we have described the equivalence classes in $S_{2n+2m}$ that lead, via the contraction \eqref{mesonic operator}, to the different $SO(N)$ traces. These classes were orbits under the group action \eqref{SO(N) invariance without minus sign}, and we separated the orbits into two types (odd/even) depending on whether they produced non-vanishing traces. 
		
		The $U(N)$-type contraction, \eqref{general U(N) operator}, also produces $SO(N)$ traces if we treat $X$ and $Y$ as antisymmetric matrices, and therefore we can give an equivalent description using equivalence classes in $S_{n+m}$. Explicitly, given $\sigma \in S_{n+m}$, we have
		\begin{equation}
		\sigma \sim \alpha \sigma \alpha^{-1} \qquad \qquad \alpha \in S_n \times S_m
		\label{equivalence part 1}
		\end{equation}
		and in addition, $\sigma$ is related to any permutation that can be obtained by inverting some subset of the cycles of $\sigma$. Explicitly, if the cycle decomposition of $\sigma $ is $\sigma = c_1 c_2 \ldots c_r$ then 
		\begin{equation}
		\sigma \sim c_1^{i_1} c_2^{i_2} \ldots c_r^{i_r} \qquad \qquad i_j \in \{ -1, 1 \}
		\label{equivalence part 2}
		\end{equation}
		As before, we can split these equivalence classes into those that produce non-zero traces and those whose contraction vanishes. If $\sigma$ contains a cycle $c$ of odd length such that $c$ is conjugate (under $S_n \times S_m$) to $c^{-1}$, then the contraction vanishes. If $\sigma$ contains no such cycle, then it and the corresponding equivalence class produce a non-vanishing trace.
		
		The combination of \eqref{equivalence part 1} and \eqref{equivalence part 2} in $S_{n+m}$ is equivalent to \eqref{SO(N) invariance without minus sign} in $S_{2n+2m}$. We see that the $S_{n+m}$ version is more complicated, and explicitly depends on the cycle structure of $\sigma$. It therefore cannot be described as a group action on $S_{n+m}$, unlike \eqref{SO(N) invariance without minus sign}.

		\subsection{Projection coefficients}
		
		The quarter-BPS operators are given in \eqref{mesonic operators definition} and \eqref{U(N) operators} for $SO(N)$ and $U(N)$ respectively. Using the results of the previous section we can turn these into sums over traces
		\begin{align}
		\mathcal{O}^{U(N)}_{T,R,S,\mu, \nu} & = \frac{d_T n! m!}{(n+m)!} \sum_{\mathcal{P} \Vdash (n,m)} \frac{1}{Z_{\mathcal{P}}} \text{Tr}_T00 \left[ P^T_{R,S;\mu \rightarrow \nu} D^T ( \sigma^{ \ }_{\mathcal{P}} ) \right] T^{U(N)}_{\mathcal{P}} 
		\label{U(N) quarter-BPS operators from traces} \\
		\mathcal{O}^{SO(N)}_{T,R,S,\lambda} & = \frac{d_T 2^{2n+2m} n! m! (n+m)!}{(2n+2m)!} \sum_{\widetilde{P} \Vdash (n,m) } \! \frac{1}{\widetilde{Z}_{\widetilde{\mathcal{P}}}} \left\langle T, [S] \right| D^T \left( \sigma^{ \ }_{\widetilde{\mathcal{P}}} \right) \left| R,S,\lambda, [A] \right\rangle T^{SO(N)}_{\widetilde{\mathcal{P}}}
		\label{mesonic operators from traces}
		\end{align}
		where $\sigma_{\mathcal{P}} \in S_{n+m}$ is a representative member of the conjugacy class labelled by $\mathcal{P}$ and $\sigma_{\widetilde{\mathcal{P}}} \in S_{2n+2m}$ is a positive representative member of the even double coset labelled by $\widetilde{\mathcal{P}}$. Note that for convenience we have introduced $|R,S,\lambda, [A] \rangle = |R, \lambda, [A] \rangle \otimes |S, \lambda, [A] \rangle$.
		
		To evaluate the projection coefficients, we would like to invert \eqref{mesonic operators from traces} and write the traces in terms of the operators. We do this by finding an orthogonality relation for the coefficients
		\begin{equation}
		C^{\widetilde{\mathcal{P}}}_{T,R,S,\lambda} = \frac{d_T}{(2n+2m)!} \frac{2^{2n+2m} n! m! (n+m)!}{\widetilde{Z}_{\widetilde{\mathcal{P}}}} \langle T, [S] | D^T \left( \sigma^{ \ }_{\widetilde{\mathcal{P}}} \right) | R,S,\lambda, [A] \rangle
		\nonumber
		\end{equation}
		These are just change of basis coefficients taking us from the basis of multi-traces to the orthogonal Young diagram basis, so there must be an inverse set of coefficients $D^{T,R,S,\lambda}_{\widetilde{\mathcal{P}}}$ satisfying
		\begin{align}
		\sum_{T,R,S,\lambda} C^{\widetilde{\mathcal{P}}}_{T,R,S,\lambda} D^{T,R,S,\lambda}_{\widetilde{\mathcal{Q}}} = \delta^{\widetilde{\mathcal{P}}}_{\widetilde{\mathcal{Q}}}
		\qquad \qquad 
		\sum_{\widetilde{\mathcal{P}}} C^{\widetilde{\mathcal{P}}}_{T,R,S,\lambda} D^{T',R',S',\lambda'}_{\widetilde{\mathcal{P}}} = \delta^{T'}_{T} \delta^{R'}_{R} \delta^{S'}_{S} \delta^{\lambda'}_{\lambda}
		\label{inverse coefficients}
		\end{align}
		Note that proving either of these two relations is sufficient. We define the inverse coefficients to be
		\begin{equation}
		D^{T,R,S,\lambda}_{\widetilde{\mathcal{P}}} =  \langle R,S,\lambda, [A] | D^T \left( \sigma_{\widetilde{\mathcal{P}}}^{-1} \right) | T, [S] \rangle
		\end{equation}
		and prove the second relation in \eqref{inverse coefficients}.
		
		Plugging $C$ and $D$ in gives
		\begin{align*}
		& \sum_{\widetilde{\mathcal{P}}} C^{\widetilde{\mathcal{P}}}_{T,R,S,\lambda} D^{T',R',S',\lambda'}_{\widetilde{\mathcal{P}}} \\
		& = \frac{d_T 2^{2n+2m} n! m! (n+m)!}{(2n+2m)!} \sum_{\widetilde{\mathcal{P}}} \hspace{-2pt} \frac{1}{\widetilde{Z}_{\widetilde{\mathcal{P}}}}  \langle R',S',\lambda', [A] | D^{T'} \left( \sigma_{\widetilde{\mathcal{P}}}^{-1} \right) | T', [S] \rangle \\
		& \hspace{9cm} \langle T, [S] | D^T \left( \sigma^{ \ }_{\widetilde{\mathcal{P}}} \right) | R,S,\lambda, [A] \rangle \\
		& = \frac{d_T}{(2n+2m)!} \sum_{\sigma \in S_{2n+2m}} \langle R',S',\lambda', [A] | D^{T'} \left( \sigma^{-1} \right) | T', [S] \rangle \langle T, [S] | D^T \left( \sigma \right) | R,S,\lambda, [A] \rangle \\
		& = \delta_{T T'} \langle T, [S] | T', [S] \rangle \langle R',S',\lambda' | R,S,\lambda \rangle \\
		& = \delta_T^{T'} \delta^{R'}_R \delta_S^{S'} \delta_{\lambda}^{\lambda'}
		\end{align*}
		where to augment the sum to one over $S_{2n+2m}$ we have used the invariance of the summand under \eqref{SO(N) invariance without minus sign}, the size of the even double cosets (given in \eqref{size of even double cosets}) and the fact that these matrix elements vanish on odd double cosets. To evaluate the sum over $S_{2n+2m}$, we have then used the orthogonality relation for matrix elements of an irreducible representation, \eqref{orthogonality of matrix elements}.
		
		Using \eqref{inverse coefficients} to invert \eqref{mesonic operators from traces} gives
		\begin{equation}
		T^{SO(N)}_{\widetilde{\mathcal{P}}} = \sum_{T,R,S,\lambda} \langle R,S,\lambda, [A] | D^T \left( \sigma_{\widetilde{\mathcal{P}}}^{-1} \right) | T, [S] \rangle O^{SO(N)}_{T,R,S,\lambda}
		\nonumber
		\end{equation}
		We can now use \eqref{quarter-BPS trace projection} to project the $U(N)$ operators \eqref{U(N) quarter-BPS operators from traces} to the $SO(N)$ theory, and then use the above inversion to express this in terms of the $SO(N)$ operators 
		\begin{align*}
		& \mathcal{O}^{U(N)}_{T,R,S,\mu, \nu} \\ 
		& \quad \overset{\mathbb{Z}_2}{\longrightarrow} \frac{d_T n! m!}{(n+m)!} \sum_{\widetilde{\mathcal{Q}}} \left( \sum_{\mathcal{P}: \text{Orth}(\mathcal{P}) = \widetilde{\mathcal{Q}}} \frac{1}{Z_{\mathcal{P}}} \text{OrthSign}(\mathcal{P}) \text{Tr}_T \left[ P^T_{R,S;\mu \rightarrow \nu} D^T \left( \sigma^{ \ }_{\mathcal{P}} \right) \right] \right) T^{SO(N)}_{\widetilde{\mathcal{Q}}} \\
		& \quad = \frac{d_T n! m!}{(n+m)!} \sum_{\widetilde{\mathcal{Q}}} \sum_{T', R', S', \lambda'} \left( \sum_{\mathcal{P}: \text{Orth}(\mathcal{P}) = \widetilde{\mathcal{Q}}} \frac{1}{Z_{\mathcal{P}}} \text{OrthSign}(\mathcal{P}) \text{Tr}_T \left[ P^T_{R,S;\mu \rightarrow \nu} D^T \left( \sigma^{ \ }_{\mathcal{P}} \right) \right] \right) \\ & \qquad \qquad \qquad \qquad \qquad \qquad \qquad \qquad \langle R',S',\lambda', [A] | D^{T'} \left( \sigma_{\widetilde{\mathcal{Q}}}^{-1} \right) | T', [S] \rangle O^{SO(N)}_{T',R',S',\lambda'} \\
		& \quad = \sum_{T',R',S',\lambda'} \alpha_{T,R,S,\mu,\nu}^{T',R',S',\lambda'} O^{SO(N)}_{T',R',S',\lambda'}
		\end{align*}
		where the projection coefficients are
		\begin{gather}
		\alpha_{T,R,S,\mu,\nu}^{T',R',S',\lambda'} = \frac{d_T n! m!}{(n+m)!} \sum_{\widetilde{\mathcal{Q}}} \left( \sum_{\mathcal{P}: \text{Orth}(\mathcal{P}) = \widetilde{\mathcal{Q}}} \frac{1}{Z_{\mathcal{P}}} \text{OrthSign}(\mathcal{P}) \text{Tr}_T \left[ P^T_{R,S;\mu \rightarrow \nu} D^T \left( \sigma^{ \ }_{\mathcal{P}} \right) \right] \right) 
		\nonumber \\ 
		\qquad \qquad \qquad \qquad \qquad \qquad \qquad \qquad \qquad \qquad
		\langle R',S',\lambda', [A] | D^{T'} \left( \sigma_{\widetilde{\mathcal{Q}}}^{-1} \right) | T', [S] \rangle
		\label{quarter-BPS projection coefficients}
		\end{gather}

		\section{Correlators}
		\label{section: correlators}

		The mesonic operators \eqref{mesonic operators definition} were presented in \cite{Kemp2014}, where the author calculated their correlators using techniques from \cite{Caputa2013}. The conventions here will differ in two important ways, but these do not affect the calculation of the correlator and we can quote the result in order to find the mesonic two-point function.
		
		In contrast, the baryonic operators \eqref{baryonic operators definition} have not been studied in detail before, and their correlator (in the half-BPS sector) was left unevaluated in \cite{Caputa2013}. The methods of \cite{Kemp2014} can be simply extended to prove that the mesonic and baryonic operators are orthogonal to each other, as well as giving an expression for the baryonic correlator in terms of index contractions over the tensor space. We then proceed in two ways. In appendix \ref{section: alternative baryonic correlator} we evaluate this intermediate expression directly by generalising the methods of \cite{Caputa2013}. 
		
		Alternatively, in this section, we partially contract this expression while leaving some indices free. We can then use Schur-Weyl duality on the free indices to relate it to the equivalent expression for the mesonic correlator, and then use the mesonic result to give the baryonic version. Before we begin, we first define the complex conjugate of the operators \eqref{mesonic operators definition} and \eqref{baryonic operators definition} and explain how the conventions here differ from those in \cite{Kemp2014} and \cite{Caputa2013}.
		
		The complex conjugate of $X^{ij}$ is $\left( X^* \right)^{ij} = X_{ij}$, where this defines $X_{ij}$, and similarly for $Y^{ij}$ and $Y_{ij}$. Therefore the complex conjugate of $ \left( X^{\otimes n} Y^{\otimes m} \right)^I$ is $\left( X^{\otimes n} Y^{\otimes m} \right)_I$. Noting that $\sigma^I_J = \left( \sigma^{-1} \right)^J_I$ and $C_I^{(\delta)} = C^{(\delta) \, I}$ (and both are real), we see that
		\begin{equation}
		\left[ C_I^{(\delta)} \sigma^I_J \left( X^{\otimes n} Y^{\otimes m} \right)^J \right]^* = \left( X^{\otimes n} Y^{\otimes m} \right)_I \left( \sigma^{-1} \right)^I_J C^{(\delta) \, J}
		\nonumber
		\end{equation}
		Therefore the complex conjugate of the mesonic operator $\mathcal{O}^\delta_{T,R,S,\lambda}$ (defined in \eqref{mesonic operators definition}) is
		\begin{align}
		\overline{\mathcal{O}}^\delta_{T,R,S,\lambda} = \frac{d_T}{(2n+2m)!} \sum_{\sigma \in S_{2n+2m}} \langle T, [S] | D^T (\sigma ) \Big( | R, \lambda, [A] \rangle \otimes | S, \lambda, [A] \rangle \Big) \nonumber \\
		\qquad \qquad \qquad \qquad \qquad \qquad \qquad \qquad \qquad \qquad \qquad \qquad \left( X^{\otimes n} Y^{\otimes m} \right)_I \left( \sigma^{-1} \right)^I_J C^{(\delta) \, J}
		\label{conjugate mesonic operators}
		\end{align}
		and similarly, the complex conjugate of the baryonic operator $\mathcal{O}^{\varepsilon}_{T,R,S,\lambda}$ (defined in \eqref{baryonic operators definition}) is
		\begin{align}
		\overline{\mathcal{O}}^{\varepsilon}_{T,R,S,\lambda}  = \frac{d_{1^N+T}}{(2n+2m)!} \sum_{\sigma \in S_{2n+2m}} \! \! \Big( \langle 1^N | \otimes \langle T, [S] | \Big) D^{1^N+T}(\sigma) \Big( | R, \lambda, [A] \rangle \otimes | S, \lambda, [A] \rangle \Big) \nonumber \\
		\qquad \qquad \qquad \qquad \qquad \qquad \qquad \qquad \qquad \qquad \left( X^{\otimes n} Y^{\otimes m} \right)_I \left( \sigma^{-1} \right)^I_J C^{(\varepsilon) \, J}
		\label{conjugate baryonic operators}
		\end{align}
		Note that the relations between $T$ and $n+m$ are different in these two formulae. In the mesonic operators, $T \vdash 2n+2m$, whereas in the baryonic operators $T \vdash 2q = 2n + 2m - N$.
		
		The two point function of $X^{ij}$ with $X_{kl}$ and $Y^{ij}$ with $Y_{kl}$ is
		\begin{equation}
		\langle X^{ij} X_{kl} \rangle = \delta^i_k \delta^j_l - \delta^i_l \delta^j_k = \langle Y^{ij} Y_{kl} \rangle
		\label{two point function}
		\end{equation}
		Initially, this appears to be the same as the two point function used in \cite{Caputa2013} and \cite{Kemp2014}, however they have used the definition $X_{ij} = \left( X^\dagger \right)^{ij} = - \left( X^* \right)^{ij}$, so there is actually a relative minus sign. The convention \eqref{two point function} ensures the positivity of the two point function between $X^{ij}$ and its conjugate $X_{ij}$.
		
		We can follow the arguments in \cite{Caputa2013} and \cite{Kemp2014} (keeping track of the implicit minus sign), to show that
		\begin{align}
		\left\langle \left( X^{\otimes n} Y^{\otimes m} \right)^I \left( X^{\otimes n} Y^{\otimes m} \right)_J \right\rangle 
		& = \sum_{\sigma \in S_n[S_2] \times S_m[S_2]} (-1)^{\sigma} \sigma^I_J \nonumber  \\
		& = 2^{n+m} n! m! \left( P_{[A]_n \otimes [A]_m} \right)^I_J
		\label{correlator as projector}
		\end{align}
		This differs from the answer in \cite{Caputa2013} and \cite{Kemp2014} by a factor of $(-1)^{n+m}$.
		
		Both \cite{Caputa2013} and \cite{Kemp2014} define their conjugate operators, \eqref{conjugate mesonic operators} and \eqref{conjugate baryonic operators}, in terms of $\left( X^\dagger \right)^{ij}$ rather than $\left( X^* \right)^{ij}$, and hence there is also a $(-1)^{n+m}$ factor difference in their definition of conjugate operators compared with \eqref{conjugate mesonic operators} and \eqref{conjugate baryonic operators}. This cancels with the minus sign in \eqref{correlator as projector}, and we can use their results to give the mesonic correlators directly.
		
		In \cite{Kemp2014}, Kemp presented the mesonic operators and calculated their correlators. The normalisation in \eqref{mesonic operators definition} differs from his by a factor of $\frac{d_T(2n)!(2m)!}{(2n+2m)!}$, so rescaling his result gives
		\begin{equation}
		\langle \mathcal{O}^\delta_{T,R,S,\lambda} \overline{\mathcal{O}}^\delta_{T',R', S', \lambda'} \rangle = \delta_{T T'} \delta_{R R'} \delta_{S S'} \delta_{\lambda, \lambda'} \frac{d_T 2^{2n+2m}  (n+m)! n! m!}{(2n+2m)!} \prod_{\substack{i \in \text{ odd}\\ \text{columns of }T}} (N+c_i)
		\label{mesonic correlator}
		\end{equation}
		Where $c_i$ is the content of a box, as defined in \eqref{content definition}, and $i$ runs over the boxes in the Young diagram $T$ that are in the odd numbered columns of $T$. So for example, $i$ would run over the starred boxes in the following
		\begin{equation*}
		\begin{ytableau}
		\ast & & \ast & \\
		\ast & & \ast & \\
		\ast & \\
		\ast &
		\end{ytableau}
		\qquad
		\begin{ytableau}
		\ast & & \ast & & \ast & & \ast & \\
		\ast & & \ast & \\
		\ast &
		\end{ytableau}
		\qquad
		\begin{ytableau}
		\ast & \\
		\ast & \\
		\ast & \\
		\ast & \\
		\ast & \\
		\ast &
		\end{ytableau}
		\end{equation*}
		The calculation of the mesonic correlators goes via the intermediate result
		\begin{align}
		\langle \mathcal{O}^\delta_{T,R,S,\lambda} \overline{\mathcal{O}}^\delta_{T', R', S', \lambda'} \rangle & = \delta_{T T'} \delta_{R R'} \delta_{S S'} \delta_{\lambda \lambda'} \frac{d_{T} 2^{n+m} n! m! }{(2n+2m)!} \hspace{-5pt} \sum_{\sigma \in S_{2n+2m} } \hspace{-10pt} \langle T | D^{T}(\sigma) | T \rangle C^{(\delta)}_I \sigma^I_J C^{(\delta) \, J} \nonumber \\
		& = \delta_{T T'} \delta_{R R'} \delta_{S S'} \delta_{\lambda \lambda'} 2^{n+m} n! m! C_I^{(\delta)} (A_T)^I_J C^{(\delta) \, J}
		\label{mesonic correlator intermediary step}
		\end{align}
		where for notational simplicity we have used $| T \rangle = \left| T, [S] \right\rangle$, and $A_T \in \mathbb{C}(S_{2n+2m})$, given by
		\begin{equation}
		A_T = \frac{d_T}{(2n+2m)!} \sum_{\sigma \in S_{2n+2m}} \langle T | D^{T}(\sigma) | T \rangle \sigma
		\nonumber
		\end{equation}
		The method used to acquire this intermediate result works just as well for the correlators of baryonic operators, as well as the two-point function of a baryonic operator with a mesonic one. As expected, the mesonic and baryonic operators are orthogonal
		\begin{equation}
		\langle \mathcal{O}^\varepsilon_{T,R,S,\lambda} \overline{\mathcal{O}}^\delta_{T', R', S', \lambda'} \rangle = 0
		\nonumber
		\end{equation}
		while in the purely baryonic case, we find 
		\begin{align}
		&\langle \mathcal{O}^\varepsilon_{T,R,S,\lambda} \overline{\mathcal{O}}^\varepsilon_{T', R', S', \lambda'} \rangle \nonumber \\ &  \qquad = \delta_{T T'} \delta_{R R'} \delta_{S S'} \delta_{\lambda \lambda'} \frac{d_{1^N+T} 2^{n+m} n! m! }{(2n+2m)!} \sum_{\sigma \in S_{2n+2m} } \langle 1^N + T | D^{1^N+T}(\sigma) | 1^N + T \rangle \nonumber \\ & \hspace{12cm} C^{(\varepsilon)}_I \sigma^I_J C^{(\varepsilon) \, J}
		\label{baryonic correlator intermediary step 1}
		\end{align}
		where we have used $|1^N+T\rangle = \left| 1^N \right\rangle \otimes \left| T, [S] \right\rangle$.
		
		In appendix \ref{section: alternative baryonic correlator} we evaluate \eqref{baryonic correlator intermediary step 1} explicitly. Here, we relate the baryonic correlator to the mesonic case by introducing
		\begin{align}
		\left( B_{1^N + T} \right)^I_J & = \frac{d_{1^N+T}}{(2n+2m)!} \sum_{\sigma \in S_{2n+2m}} \langle 1^N + T | D^{1^N+T}(\sigma) | 1^N + T \rangle \sigma^I_J
		\label{B_N+T definition} \\
		\left( B_T \right)^{i_1 i_2 \ldots i_{2q}}_{j_1 j_2 \ldots j_{2q}} & = \frac{1}{N!} \varepsilon_{k_1 k_2 \ldots k_N} \varepsilon^{l_1 l_2 \ldots l_N} \left( B_{1^N + T} \right)^{k_1 k_2 \ldots k_N \, i_1 i_2 \ldots i_{2q}}_{l_1 l_2 \ldots l_N \, j_1 j_2 \ldots j_{2q}} 
		\label{B_T definition}
		\end{align}
		so that
		\begin{align}
		\langle \mathcal{O}^\varepsilon_{T,R,S,\lambda} \overline{\mathcal{O}}^\varepsilon_{T', R', S', \lambda'} \rangle
		& = \delta_{T T'} \delta_{R R'} \delta_{S S'} \delta_{\lambda \lambda'} 2^{n+m} n! m! C^{(\varepsilon)}_I \left( B_{1^N + T} \right)^I_J C^{(\varepsilon) \, J} \nonumber \\
		& = \delta_{T T'} \delta_{R R'} \delta_{S S'} \delta_{\lambda \lambda'}  2^{n+m} n! m! N! C^{(\delta)}_I \left(B_T\right)^I_J C^{(\delta) \, J}
		\label{baryonic correlator intermediary step 2}
		\end{align}
		Note that $B_T$ has fewer indices than the permutations from which it is constructed. Since the permutations are partially contracted, we cannot (immediately) state that $B_T$ is in $\mathbb{C}(S_{2q})$. Instead it is just an endomorphism on $V^{\otimes 2q}$.
		
		We now want to compare $A_T$ and $B_T$ for the same Young diagram $T$. As noted earlier, the relations between $T$ and $n,m$ are different for mesonic and baryonic operators, so to avoid confusion, for the rest of the section we will use the baryonic relations. Explicitly, $T \vdash 2q = 2n + 2m - N$. This means the coefficient in front of $A_T$ (and the operators in \eqref{mesonic operators definition} and \eqref{conjugate mesonic operators}) is $\frac{d_T}{(2q)!}$ rather than $\frac{d_T}{(2n+2m)!}$.
		
		In \eqref{Schur-Weyl duality} we saw one statement of Schur-Weyl duality. We now use a different version. The group algebra of $S_{2q}$ and the diagonal action of $U(N)$ are sub-algebras of the endomorphism algebra of $V^{\otimes 2q}$. Schur-Weyl duality states that these two sub-algebras are each other's centraliser within the larger endomorphism algebra. Therefore proving that $B_T$ commutes with the diagonal action of $U(N)$ is sufficient to show that it is in $\mathbb{C}(S_{2q})$.
		
\begin{figure}
			\centering
			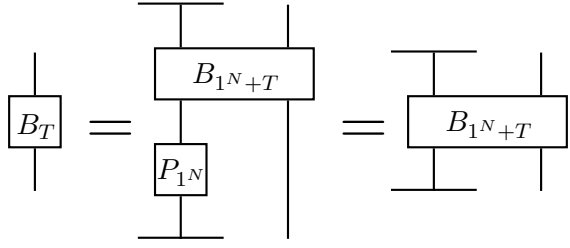
			\caption{$B_T$ is a partial trace of $B_{1^N+T}$ multiplied by the projector $P_{1^N}$. $B_{1^N+T}$ has $N+2q$ indices, which are split into a set of $N$ and a set of $2q$. The two lines represent these two sets respectively. The horizontal lines then indicate we trace over the first $N$ indices, while the open line indicates the remaining $2q$ indices are left free.}
			\label{B_T diagrammatic definition}
		\end{figure}

		Firstly, note that
		\begin{equation}
		\varepsilon_{k_1 k_2 \ldots k_N} \varepsilon^{l_1 l_2 \ldots l_N} = \sum_{\tau \in S_N} (-1)^\tau \tau^{l_1 l_2 \ldots l_N}_{k_1 k_2 \ldots k_N} = N! \left( P_{1^N} \right)^L_K
		\label{epsilons as a projector}
		\end{equation}
		From the definition of $|1^N+T \rangle$, given after \eqref{baryonic correlator intermediary step 1}, we know that $P_{1^N}$ leaves it invariant. Using the definition \eqref{B_N+T definition}, it follows that $B_{1^N + T}$ is invariant under multiplication by $P_{1^N}$. This simplifies the definition \eqref{B_T definition}, which is given diagrammatically in figure \ref{B_T diagrammatic definition}.

\begin{figure}
			\centering
			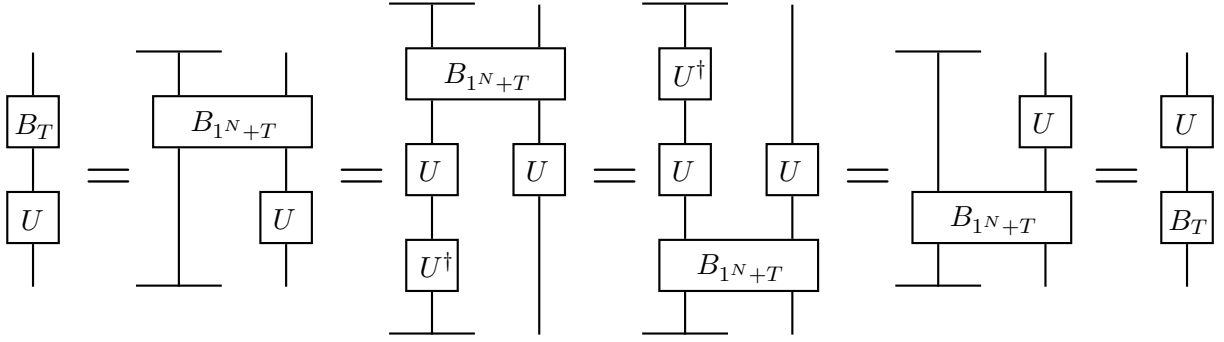
			\caption{Diagrammatic proof that $B_T$ commutes with $U(N)$. $B_{1^N+T}$ has $N+2q$ indices, which are split into a set of $N$ and a set of $2q$. The two lines represent these two sets respectively. Here $U$ stands for the diagonal action of $U$ on the tensor space, so more properly we should write $U^{\otimes N}$ on the left hand side and $U^{\otimes 2q}$ on the right hand side. The central equality follows from Schur-Weyl duality, which implies $U^{\otimes N+2q}$ commutes with $B_{1^N+T}$, and the cyclicity of trace.}
			\label{B_T commutes with U}
		\end{figure}

		Figure \ref{B_T commutes with U} then uses this definition and the statement of Schur-Weyl duality to prove that $B_T$ commutes with $U(N)$, and hence $B_T \in \mathbb{C}(S_{2q})$.
		
		Note that $A_T$ is invariant under pre- or post- multiplication by $S_q[S_2]$. This follows from the vector $|T\rangle$ being invariant under multiplication by $S_q[S_2]$ permutations. In fact, the set $\{ A_T : T \vdash 2q \text{ with even row lengths} \}$ forms a basis of the sub-algebra of $\mathbb{C}(S_{2q})$ defined by this invariance. One can derive this result by following a similar argument to that in section \ref{section: Mesonic operators}.
		
		Similarly, $|1^N+T \rangle$ is invariant under $S_q[S_2]$, and it follows that pre- or post- multiplication of $B_T$ leaves it unchanged. Therefore $B_T$ must be a linear combination of the $A_T$.
		
		To determine which $A_T$ contribute to $B_T$, consider the projector $P_R$ for $R \vdash 2q$. This acts identically on $|T\rangle$ if $R=T$ and annihilates it if $R \neq T$. It follows that $P_R A_T = \delta_{R T} A_T$. Similarly, $D^{1^N+T}(P_R) |1^N+T \rangle = \delta_{R,T} |1^N+T \rangle$ (where $P_R \in S_{2q}$ is embedded into $S_{2n+2m}$ by acting on $\{ N+1, N+2, \ldots , N+2q = 2n+2m \}$), and therefore $P_R B_T =  \delta_{R T} B_T$. We conclude that $B_T$ is proportional to $A_T$.
		
		To find the constant of proportionality, we look at the traces of $A_T$ and $B_T$.
		\begin{align*}
		\text{Tr} \left( A_T \right) & = \frac{d_T}{(2q)!} \sum_{\sigma \in S_{2q}} \langle T | D^T (\sigma ) | T \rangle \text{Tr} (\sigma)  \\
		& = \frac{d_T}{(2q)!} \sum_{\sigma \in S_{2q}} \langle T | D^T (\sigma ) | T \rangle N^{c(\sigma)}  \\
		& = \frac{d_T}{(2q)!} \langle T | D^T (\Omega) | T \rangle  \\
		& = \frac{d_T}{(2q)!} \prod_{\substack{i \in \text{ boxes}\\ \text{of }T}} (N+c_i)
		\end{align*}
		where $c(\sigma)$ is the number of cycles in $\sigma$, $\Omega$ is as defined in \eqref{omega definition} and we have used \eqref{omega in an irrep} to evaluate the matrix element $\langle T | D^T (\Omega) | T \rangle$.
		
		Since $B_T$ is just the partial trace of $B_{1^N + T}$, the trace of $B_T$ is just the full trace of $B_{1^N+T}$
		\begin{align*}
		\text{Tr} \left( B_T \right) = \text{Tr} \left( B_{1^N + T} \right) & = \frac{d_{1^N+T}}{(2n+2m)!} \sum_{\sigma \in S_{2n+2m}} \langle 1^N+T | D^{1^N+T}(\sigma) |1^N+T\rangle N^{c(\sigma)} \\
		& = \frac{d_{1^N+T}}{(2n+2m)!} \langle 1^N+T | D^{1^N+T} (\Omega) | 1^N+T \rangle \\
		& = \frac{d_{1^N+T}}{(2n+2m)!} \prod_{\substack{i \in \text{ boxes}\\ \text{of }1^N+T}} (N+c_i)
		\end{align*}
		To compare the traces of $A_T$ and $B_T$, we now find the ratio of the dimensions $d_T$ and $d_{1^N + T}$ and the ratio of the products of $N+c_i$. The dimensions can be found using the hook length formula \eqref{hook length formula}.
		
		Now $1^N+T$ is the diagram $T$ with a single column of $N$ boxes set to the left of it. Clearly the $T$ part of the diagram has exactly	the same hook lengths as $T$ itself. Denote the components (row lengths) of $T$ by $T_1, T_2, \ldots ,T_N$ (some of the $T_i$ may vanish). Then the hook length of the $j$th box in the first column of $1^N+T$ is $N + T_j - j + 1$. Therefore
		\begin{equation}
		H_{1^N+T} = H_T \prod_{j=1}^N (N + T_j - j + 1)
		\label{ratio of hook lengths}
		\end{equation}
		We can also think of $1^N+T$ as the diagram made by adding a single box to the end of each row of $T$ (including empty rows if $l(T)<N$). The content of the extra box added to the $j$th row is $T_j - j + 1$. Therefore
		\begin{equation}
		\prod_{\substack{i \in \text{ boxes}\\ \text{of }1^N+T}} (N+c_i) = \left[ \prod_{\substack{i \in \text{ boxes}\\ \text{of }T}} (N+c_i) \right] \left[ \prod_{j=1}^N (N + T_j - j + 1) \right]
		\label{ratio of content products}
		\end{equation}
		The latter factors of \eqref{ratio of hook lengths} and \eqref{ratio of content products} are the same, so using \eqref{hook length formula},  we can simplify the ratio between $B_T$ and $A_T$
		\begin{align*}
		B_T & = \frac{d_{1^N+T}}{(2n+2m)!} \frac{(2q)!}{d_T} \left[ \prod_{\substack{i \in \text{ boxes}\\ \text{of }1^N+T}} (N+c_i) \right] \left[ \prod_{\substack{i \in \text{ boxes}\\ \text{of }T}} \frac{1}{N+c_i} \right] A_T \\
		& = \frac{H_T}{H_{1^N+T}} \left[ \prod_{j=1}^N (N + T_j - j + 1) \right] A_T \\
		& = A_T
		\end{align*}
		Comparing \eqref{mesonic correlator} with \eqref{mesonic correlator intermediary step}, and recalling the change in conventions, we see that
		\begin{equation}
		C^{(\delta)}_I \left( A_T \right)^I_J C^{(\delta) \, J} = 2^q q! \frac{d_T}{(2q)!} \prod_{\substack{i \in \text{ odd}\\ \text{columns of }T}} (N+c_i)
		\nonumber
		\end{equation}
		We can now evaluate \eqref{baryonic correlator intermediary step 2} to get the full baryonic correlator
		\begin{align}
		\langle \mathcal{O}^\varepsilon_{T,R,S,\lambda} \overline{\mathcal{O}}^\varepsilon_{T', R', S', \lambda'} \rangle & = \delta_{T T'} \delta_{R R'} \delta_{S S'} \delta_{\lambda \lambda'} 2^{n+m+q} n! m! q! N! \frac{d_T}{(2q)!} \prod_{\substack{i \in \text{ odd}\\ \text{columns of }T}} (N+c_i)
		\nonumber
		\end{align}
		In the description of baryonic operators we have chosen to use $T$ as the label. We could instead have used $1^N+T$, in which case we would want to express the correlator in terms of this Young diagram.  By performing similar manipulations on the hook lengths and contents of $T$ and $1^N + T$, one can show
		\begin{align}
		& \langle \mathcal{O}^\varepsilon_{T,R,S,\lambda} \overline{\mathcal{O}}^\varepsilon_{T', R', S', \lambda'} \rangle = \delta_{T T'} \delta_{R R'} \delta_{S S'} \delta_{\lambda \lambda'} 2^{n+m+q} n! m! q! N! \frac{d_{1^N+T}}{(2n+2m)!} \nonumber \\ & \hspace{10cm} \prod_{\substack{i \in \text{ odd}\\ \text{columns of }1^N+T}} (N+c_i)
		\label{baryonic correlator}
		\end{align}

		\section{Symplectic gauge group} 
		\label{section: Symplectic gauge group}
		
		There are many connections between the orthogonal group and the symplectic group. It was proved in \cite{King1971} that dimensions of $SO(N)$ and $Sp(N)$ irreps (both labelled by Young diagrams) are related by conjugation of the Young diagram and $N \rightarrow -N$. This general pattern of anti-symmetrisation (conjugation) and $N \rightarrow -N$ was also found in \cite{Caputa2013a,Kemp1406}, and will occur repeatedly in this section.
		
		Since the orthogonal group and the symplectic group are so closely related, one would expect that repeating the working of the previous sections for the symplectic case would involve only minor changes. This expectation is correct, and the majority of the work is either identical or directly analogous.
		
		We start by constructing a basis of symplectic gauge-invariant operators. As before, we use the group invariances of the index contractions to guide us. These are similar to the orthogonal versions encountered in \eqref{SO(N) group algebra invariance}, but the invariance and anti-invariance have switched places. This time there is no distinct sector of baryonic operators. While we can use $\varepsilon$ to contract indices, the operators this produces are within the mesonic sector. The symplectic operators are labelled by a trio of Young diagrams conjugate to those labelling the orthogonal mesonic operators. As conjugation of Young diagrams corresponds to anti-symmetrisation, we find that the two sets of operators are related by anti-symmetrisation.
		
		In the large $N$ limit, we prove that symplectic and orthogonal multi-traces have exactly the same form, and therefore the structures given in figure \ref{figure: SO(N) diagram} and \ref{figure: SO(N) periodicity diagram} apply to the space of $Sp(N)$ two-matrix multi-traces.
		
		Replacing the unconstrained $U(N)$ matrices $X$ and $Y$ with matrices satisfying the symplectic condition \eqref{symplectic matrix condition}, we find that the half-BPS symplectic projection coefficients are exactly the same as their orthogonal equivalents. For the quarter-BPS sector they are related by a change of sign inside the sum.
		
		Finally we review symplectic correlators. These were calculated already in \cite{Kemp1406}, and as the symplectic theory has no baryonic sector, this completed the story for symplectic operators.

		\subsection{Symplectic operators}
		
		The Lie algebra $\mathfrak{sp}(N)$ is composed of $N$ by $N$ ($N$ even) matrices $X$ satisfying
		\begin{equation}
		X^T = \Omega X \Omega
		\label{symplectic matrix condition}
		\end{equation}
		where
		\begin{equation}
		\Omega = \begin{pmatrix}
		0 & I \\
		- I & 0
		\end{pmatrix}
		\nonumber
		\end{equation}
		and $I$ is the $\frac{N}{2}$ by $\frac{N}{2}$ identity matrix.
		
		Note that the condition \eqref{symplectic matrix condition} is equivalent to saying that $\Omega X$ (or $X \Omega$) is a symmetric matrix.
		
		In the $Sp(N)$ gauge theory, the quarter-BPS sector is made up of two scalar fields $X$ and $Y$ lying in the adjoint of $\mathfrak{sp}(N)$. Gauge invariant operators can then be constructed in much the same way as in section \ref{section: Construction and counting of operators}. Rather than using the matrices $X$ and $Y$ directly, we use the symmetric combinations $\Omega X$ and $\Omega Y$. Clearly this is an equivalent approach, as $\Omega$ is an invertible matrix, but it has the advantage that the symmetry properties of $\Omega X$ and $\Omega Y$ allow us to use the same techniques as we employed for the special orthogonal group.
		
		In analogy to \eqref{general SO(N) operator}, the most general gauge-invariant operator in the $Sp(N)$ theory can be written as
		\begin{equation}
		\mathcal{O} = C_I \left[ \left( \Omega X \right)^{\otimes n} \left( \Omega Y \right)^{\otimes m} \right]^I
		\nonumber
		\end{equation}
		where $C_I$ is a contractor constructed from $Sp(N)$ invariant tensors. For the orthogonal group we had two such tensors, which split operators into mesonic and baryonic sectors. Here, there is only one independent invariant tensor, namely $\Omega_{ij}$. We could also consider the $\varepsilon$ tensor, but it is not independent of $\Omega$. At $N=2,4$ we have
		\begin{align*}
		\varepsilon_{ij} & = \Omega _{ij} \\
		\varepsilon_{ijkl} & = \Omega_{ij} \Omega_{kl} + \Omega_{ik} \Omega_{lj} + \Omega_{il} \Omega_{jk}
		\end{align*}
		while more generally we have
		\begin{equation}
		\varepsilon_{i_1 i_2 \ldots i_N} = \frac{1}{2^{\frac{N}{2}} \left( \frac{N}{2} \right) !} \sum_{\sigma \in S_N} \Omega_{i_{\sigma(1)} i_{\sigma(2)}} \Omega_{i_{\sigma(3)} i_{\sigma(4)}} \ldots \Omega_{i_{\sigma(N-1)} i_{\sigma(N)}}
		\end{equation}
		The normalisation factor of $2^{\frac{N}{2}} \left( \frac{N}{2} \right)!$ comes from the $S_{\frac{N}{2}} [S_2]$ stabiliser group of the index structure $\Omega_{i_1 i_2} \Omega_{i_3 i_4} \ldots \Omega_{i_{N-1} i_N}$. After removing this redundancy, we are left with $(N - 1)!!$ terms, corresponding to the different ways of splitting $\{1,2, \ldots ,N\}$ up into $\frac{N}{2}$ pairs.
		
\begin{figure}
			\centering
			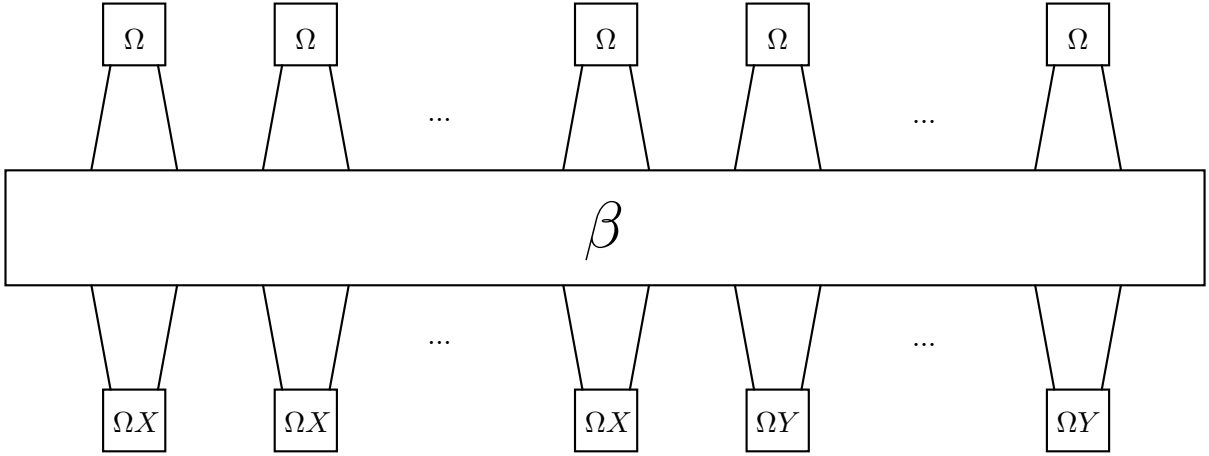
			\caption{Diagrammatic representation of the contraction pattern for symplectic mesonic operators.}
			\label{figure: Sp(N) mesonic contraction}
		\end{figure}

		So, unlike $SO(N)$, we need only consider one type of contractor, the mesonic ones containing $n+m$ $\Omega$s. As argued in section \ref{section: Set-up}, different index arrangements can be absorbed into an element $\beta$ of $\mathbb{C}\left( S_{2n+2m} \right)$. Therefore the most general mesonic operator is
		\begin{align}
		\mathcal{O}^{(\Omega)}_{\beta} & = C^{(\Omega)}_I \beta^I_J \left[ \left( \Omega X \right)^{\otimes n} \left( \Omega Y \right)^{\otimes m} \right]^J 
		\label{Sp(N) contraction}
		\end{align}
		where the standard index arrangement on the contractor is
		\begin{align*}
		C^{(\Omega)}_I & = \Omega_{i_1 i_2} \Omega_{i_3 i_4} \ldots \Omega_{i_{2n+2m-1} i_{2n+2m}}
		\end{align*}
		The contraction \eqref{Sp(N) contraction} is given diagrammatically in figure \ref{figure: Sp(N) mesonic contraction}.
		
		This time, $\beta$ is invariant under the transformation
		\begin{align}
		\beta & \mapsto (-1)^\alpha \alpha \beta \gamma^{-1} & \alpha \in S_{n+m}[S_2] \ \ , \ \ \gamma \in S_n[S_2] \times S_m[S_2] 
		\label{symplectic mesonic invariance}
		\end{align}
		which has the same group structure as the \eqref{SO(N) group algebra invariance}, but the invariance and anti-invariance have swapped sides. This is the first example of the orthogonal and symplectic groups being related by anti-symmetrisation.
		
		The invariance \eqref{symplectic mesonic invariance} defines a subspace $\mathcal{A}^{Sp}_{n,m}$ of $\mathbb{C}(S_{2n+2m})$. As in section \ref{section: Mesonic operators}, we can find a basis for this space. Imposing the finite $N$ restriction, that all Young diagrams must have at most $N$ rows, reduces this basis (and the space it spans) to that relevant for constructing operators. Explicitly, we have
		\begin{align}
		\beta_{T,R,S,\lambda} & = \frac{d_T}{(2n+2m)!} \sum_{\sigma \in S_{2n+2m}} \langle T, [A] | D^T  (\sigma) \Big( | R,\lambda, [S] \rangle \otimes | S, \lambda , [S] \rangle  \Big) \sigma
		\label{symplectic mesonic basis}
		\end{align}
		where the constraints on $T,R,S,\lambda$ are
		\begin{gather}
		\begin{gathered}
		T \vdash 2n+2m \text{ with even column lengths}  \\
		l(T) \leq N \\
		R \vdash 2n \text{ with even row lengths} \\
		S \vdash 2m \text{ with even row lengths} \\
		1 \leq \lambda \leq g_{R,S;T}
		\end{gathered}
		\label{symplectic label conditions}
		\end{gather}
		Just as in the $SO(N)$ case, the space $\mathcal{A}^{Sp}_{n,m}$ forms a right module of $A_{2n,2m}^{U}$. By following the same argument as \eqref{module multiplication}, we have the multiplication relation
		\begin{equation*}
		\beta^{Sp(N)}_{T,R,S,\lambda} \beta^{U(N)}_{T',R',S',\mu,\nu} = \delta_{TT'} \delta_{RR'} \delta_{SS'} \delta_{\mu \lambda} \beta^{Sp(N)}_{T,R,S,\nu}
		\end{equation*}
		From the basis \eqref{symplectic mesonic basis} we proceed to find the bases of gauge-invariant operators
		\begin{align}
		\mathcal{O}^{\Omega}_{T,R,S,\lambda} & = \frac{d_T}{(2n+2m)!} \sum_{\sigma \in S_{2n+2m}} \langle T, [A] | D^T  (\sigma) \Big( | R,\lambda, [S] \rangle \otimes | S, \lambda , [S] \rangle  \Big) 
		\nonumber \\ 
		& \qquad \qquad \qquad \qquad \qquad \qquad \qquad \qquad 
		C^{(\Omega)}_I \sigma^I_J \left[ \left( \Omega X \right)^{\otimes n} \left( \Omega Y \right)^{\otimes m} \right]^J 
		\label{symplectic mesonic operator}
		\end{align}
		The labelling for this basis allows us to give the number of gauge-invariant operators of order $n,m$
		\begin{gather}
		N^\Omega_{n,m} = \sum_{\substack{
				R \vdash 2n \text{ with even row lengths} \\
				S \vdash 2m \text{ with even row lengths} \\
				T \vdash 2n+2m \text{ with even column lengths} \\ 
				l(T) \leq N}}
		g_{R,S;T}
		\label{symplectic mesonic counting}
		\end{gather}
		We can make \eqref{symplectic mesonic operator} and \eqref{symplectic mesonic counting} look more similar to their $SO(N)$ equivalents \eqref{mesonic operators definition} and \eqref{2-matrix delta counting} by making use of the conjugate partitions $R^c, S^c$ and $T^c$.
		
		Let $V_T$ be the representation space for $T$. Then since $T^c = $sgn$\otimes T$, we have an orthogonal map $\rho$ from $V_T$ to $V_{T^c}$ satisfying
		\begin{equation}
		D^{T^c} (\sigma) = (-1)^\sigma \rho D^T (\sigma) \rho^{-1}
		\label{relation between action of T and T'}
		\end{equation}
		Now $|T,[A] \rangle$ and is defined by its behaviour under $S_{n+m}[S_2]$ in the $T$ representation. Similarly $|R, S, \lambda, [S] \rangle = | R,\lambda, [S] \rangle \otimes | S, \lambda , [S] \rangle$ is defined by its behaviour under $P_{R \otimes S}$ and $S_n[S_2] \times S_m[S_2]$ (as well as a choice of basis vector in the Littlewood-Richardson multiplicity space). Explicitly, these behaviours are
		\begin{gather}
		D^{T} (\sigma) |T,[A] \rangle = (-1)^\sigma  | T,[A] \rangle 
		\nonumber \\
		D^T \left( \tau \right) |R, S, \lambda, [S] \rangle = D^T \left( P_{R \otimes S} \right) |R, S, \lambda, [S] \rangle = |R, S, \lambda, [S] \rangle 
		\nonumber
		\end{gather}
		for $\sigma \in S_{n+m}[S_2]$ and $\tau \in S_n[S_2] \times S_m[S_2]$. So from \eqref{relation between action of T and T'}, the equivalent actions in the $T^c$ representation are
		\begin{align}
		D^{T^c}(\sigma) \rho | T,[A] \rangle & = \rho | T , [A] \rangle
		\nonumber \\
		D^{T^c} \left( \tau \right) \rho |R, S, \lambda, [S] \rangle & = (-1)^\tau \rho |R, S, \lambda, [S] \rangle
		\nonumber \\
		D^{T^c} \left( P_{R^c \otimes S^c} \right) \rho |R, S, \lambda, [S] \rangle & = \rho |R, S, \lambda, [S] \rangle 
		\nonumber
		\end{align}
		Therefore $\rho |T,[A] \rangle$ lies in the $[S]$ subspace of $T^c$, and $ \rho |R, S, \lambda, [S] \rangle $ lies in the $[A]$ subspace of $R^c \otimes S^c$, which is itself a subspace of $T^c$. Note that $g_{R,S;T} = g_{R^c, S^c; T^c}$, and therefore $R^c \otimes S^c$ is indeed a subspace of $T^c$. This means we can choose the multiplicity index $\lambda$ to be the same before and after conjugation.  Therefore we have
		\begin{equation}
		| T^c , [S] \rangle \propto \rho | T , [A] \rangle
		\qquad \qquad \qquad \qquad
		| R^c , S^c , \lambda, [A] \rangle \propto \rho | R, S, \lambda , [S] \rangle
		\label{conjugate vectors}
		\end{equation}
		Since $\rho$ is orthogonal and $S_{2n+2m}$ representations are real, the constants of proportionality must be $\pm 1$. It is only the relative sign of the two which is important, but without a positivity condition (similar to \eqref{special case of Ivanov's formula}, but for the quarter-BPS sector) we cannot determine what this should be. We discussed such conditions in section \ref{section: resolving ambiguity}, and suggested a possible candidate for the orthogonal case.
		
		Using \eqref{conjugate vectors} and the orthogonality of $\rho$, we can rewrite \eqref{symplectic mesonic operator} in terms of $T^c, R^c$ and $S^c$
		\begin{align}
		\mathcal{O}^{\Omega}_{T,R,S,\lambda} & = \frac{d_{T^c}}{(2n+2m)!} \sum_{\sigma \in S_{2n+2m}} (-1)^\sigma \langle T^c, [S] | D^{T^c}  ( \sigma )| R^c, S^c, \lambda, [A] \rangle \nonumber \\ & \hspace{8cm}
		C^{(\Omega)}_I \sigma^I_J \left[ \left( \Omega X \right)^{\otimes n} \left( \Omega Y \right)^{\otimes m} \right]^J 
		\end{align}
		Clearly a partition with even row lengths is conjugate to a partition with even column lengths and vice versa.	Therefore, rather than labelling the symplectic operators with the conditions \eqref{symplectic label conditions} we can instead use the conditions
		\begin{gather}
		\begin{gathered}
		T \vdash 2n+2m \text{ with even row lengths}  \\
		l(T^c) \leq N \\
		R \vdash 2n \text{ with even column lengths} \\
		S \vdash 2m \text{ with even column lengths} \\
		1 \leq \lambda \leq g_{R,S;T}
		\end{gathered}
		\label{symplectic conjugate label conditions}
		\end{gather}
		where the corresponding operator is
		\begin{align}
		\mathcal{O}^{\Omega}_{T,R,S,\lambda} & = \frac{d_T}{(2n+2m)!} \sum_{\sigma \in S_{2n+2m}} (-1)^\sigma \langle T, [S] | D^T  ( \sigma )| R, S, \lambda, [A] \rangle \nonumber \\
		& \hspace{8cm} C^{(\Omega)}_I \sigma^I_J \left[ \left( \Omega X \right)^{\otimes n} \left( \Omega Y \right)^{\otimes m} \right]^J 
		\label{symplectic conjugate operators}
		\end{align}
		Therefore the labelling set for symplectic operators is exactly the same as the orthogonal mesonic ones, except for the finite $N$ condition imposes a limit on the length of the rows instead of the length of the columns.
		
		Note that the matrix element in \eqref{symplectic conjugate operators} is the same as that in \eqref{mesonic operators definition}. The only differences are the factor of $(-1)^\sigma$ and the different contractions.
		
		In section \ref{section: symplectic projection} we prove that the $Sp(N)$ contraction produces exactly the same multi-trace as the $SO(N)$ equivalent when using the same double coset representative (compare equations \eqref{Sp(N) to U(N) contraction} and \eqref{SO(N) to U(N) contraction}). So by restricting the sums in \eqref{symplectic conjugate operators} and \eqref{mesonic operators definition} to run over traces (or equivalently even double cosets) rather than permutations (as done in sections \ref{section: Z2 quotient} and \ref{section: quarter-BPS projection}), we see that symplectic operators and orthogonal mesonic operators are just anti-symmetrisations of each other. By this we mean that if a symplectic operator contains a term of the form $c \text{Tr} W_1 \text{Tr} W_2 \ldots \text{Tr} W_k$ for some constant $c$ and matrix words $W_1,W_2, \ldots W_k$, then the orthogonal mesonic operator with the same labels (or conjugate labels, depending on the labelling set being used) will contain a term of the form $(-1)^{l(W)} c \text{Tr} W_1 \text{Tr} W_2 \ldots \text{Tr} W_k$. Note that the matrices $X$ and $Y$ that make up the matrix words satisfy different conditions in the orthogonal and symplectic theories, so one cannot compare these two operators directly, only their form.
		
		The result of this conjugation argument for the counting of operators is that we may rewrite \eqref{symplectic mesonic counting} as
		\[
		N^\Omega_{n,m} = \sum_{\substack{
				R \vdash 2n \text{ with even column lengths} \\
				S \vdash 2m \text{ with even column lengths} \\
				T \vdash 2n+2m \text{ with even row lengths} \\ 
				l(T^c) \leq N}}
		g_{R,S;T}
		\]
		This is now identical to the $SO(N)$ formula \eqref{2-matrix delta counting} except for the restriction $l(T^c) \leq N$ instead of $l(T) \leq N$. Note that in the large $N$ limit, both of these restrictions disappear, so the counting is the same in either group. In particular we have the same generating function \eqref{SO(N) multi trace generating function} for $N^\delta_{n,m}$ and $N^\Omega_{n,m}$.
		
		For the remainder of this section we will use the labelling set \eqref{symplectic label conditions} as standard, and will refer to \eqref{symplectic conjugate label conditions} as the conjugate labels.

		\subsection{Symplectic projection}
		\label{section: symplectic projection}
		
		In section \ref{section: Z2 quotient} we looked at projecting from the $U(N)$ theory into the $SO(N)$ theory in the half-BPS sector by replacing the generic matrix $X$ with an anti-symmetric one. We now study the equivalent in the $Sp(N)$ setting. This replaces the generic $X$ with one satisfying \eqref{symplectic matrix condition}. This implies
		\begin{align}
		\text{Tr} X^n & = \text{Tr} (X^T)^n \nonumber \\
		& = \text{Tr} \left( \Omega X \Omega \right)^n \nonumber \\
		& = \text{Tr} \left( \Omega^2 X \right)^n \nonumber \\
		& = \text{Tr} \left( - X \right)^n \nonumber \\
		& = (-1)^n \text{Tr} X^n
		\label{symplectic odd traces vanish}
		\end{align}
		where we have used $\Omega^2  = -1$. So, just as in the $SO(N)$ case, the odd order single traces vanish while the even ones remain unchanged.
		
		Applying the same logic to the quarter-BPS case, we again find the $Sp(N)$ relations between traces are the same as those for $SO(N)$. For a trace with $k$ periods and aperiodic matrix word $W$, we have
		\begin{equation}
		\text{Tr} W^k = (-1)^{k \, l(W)} \text{Tr} \left( W^{(r)} \right)^k
		\label{relation between traces from symplectic condition}
		\end{equation}
		As claimed, this is identical to \eqref{relation between traces from antisymmetry}.
		
		Returning to the half-BPS sector, \eqref{symplectic odd traces vanish} means the symplectic quotient of a half-BPS $U(N)$ operator (defined in \eqref{U(N) half-BPS operators}) vanishes if $n$ odd, and if $n$ even we have
		\begin{align}
		\mathcal{O}_R^{U(N)} & \overset{\mathbb{Z}_2}{\longrightarrow} d_R \sum_{p \vdash \frac{n}{2}} \frac{1}{z_{2p}} \chi_R (2p) \prod_i \left( \text{Tr} X^{2i} \right)^{p_i} \nonumber \\
		& = \sum_T \alpha^{Sp(N);T}_R \mathcal{O}^{\Omega}_T
		\label{symplectic quotient}
		\end{align}
		where the second line defines the projection coefficients $\alpha^{Sp(N);T}_R$. While the first line appears identical to \eqref{quotient of U(N) operator}, $X$ satisfies different conditions here.
		
		We now study how these multi-traces relate to the $Sp(N)$ operators \eqref{symplectic mesonic operator}. In the half-BPS sector ($m=0$), these reduce to
		\begin{equation}
		\mathcal{O}_{T}^\Omega = \frac{d_T}{(2n)!} \sum_{\sigma \in S_{2n}} \langle T, [A] | D^T ( \sigma ) | T, [S] \rangle C_I^{(\Omega)} \sigma^I_J \left[ \left( \Omega X \right)^{\otimes n} \right]^J
		\label{symplectic half-bps operator}
		\end{equation}
		Since the contractor $C^{(\Omega)}$ is constructed from $\Omega$s, the contraction $C^{(\Omega)}_I \sigma^I_J \left[ \left( \Omega X \right)^{\otimes n} \right]^J$ will be some multi-trace of $\Omega$ and $X$. We know the contraction is invariant under
		\begin{equation}
		\sigma \mapsto (-1)^\alpha \alpha  \sigma \gamma^{-1} \qquad \qquad \alpha, \gamma \in S_n[S_2]
		\label{symplectic half-bps invariance}
		\end{equation}
		(this is just the $m=0$ version of \eqref{symplectic mesonic invariance}). If we ignore the minus sign for a moment, we have exactly the same action that we saw in \eqref{hal-bps invariance without minus sign}. We studied the orbits of this action, called double cosets, in detail. In particular, they were labelled by partitions $p \vdash n$, with representatives $\sigma \in S_n^{(odd)}$, where $\sigma$ acts with cycle type $p$ on the odd numbers. We split the double cosets into two categories, even and odd, and gave two different characterisations of this split. Firstly, a double coset was odd if one or more components of the corresponding partition were odd. Secondly, a double coset was odd if the stabiliser of a representative element contained at least one odd permutation in either factor of the direct product ($Stab(\sigma) \leq S_n[S_2] \times S_n[S_2]$). This second condition is equivalent to saying that it is possible to use the action \eqref{symplectic half-bps invariance} to take $\sigma$ to $-\sigma$. Therefore we can repeat the arguments in \eqref{vanishing contraction} and \eqref{vanishing matrix element} to show the $Sp(N)$ contraction and matrix elements vanish for odd double cosets.
		
\begin{figure}
			\centering
			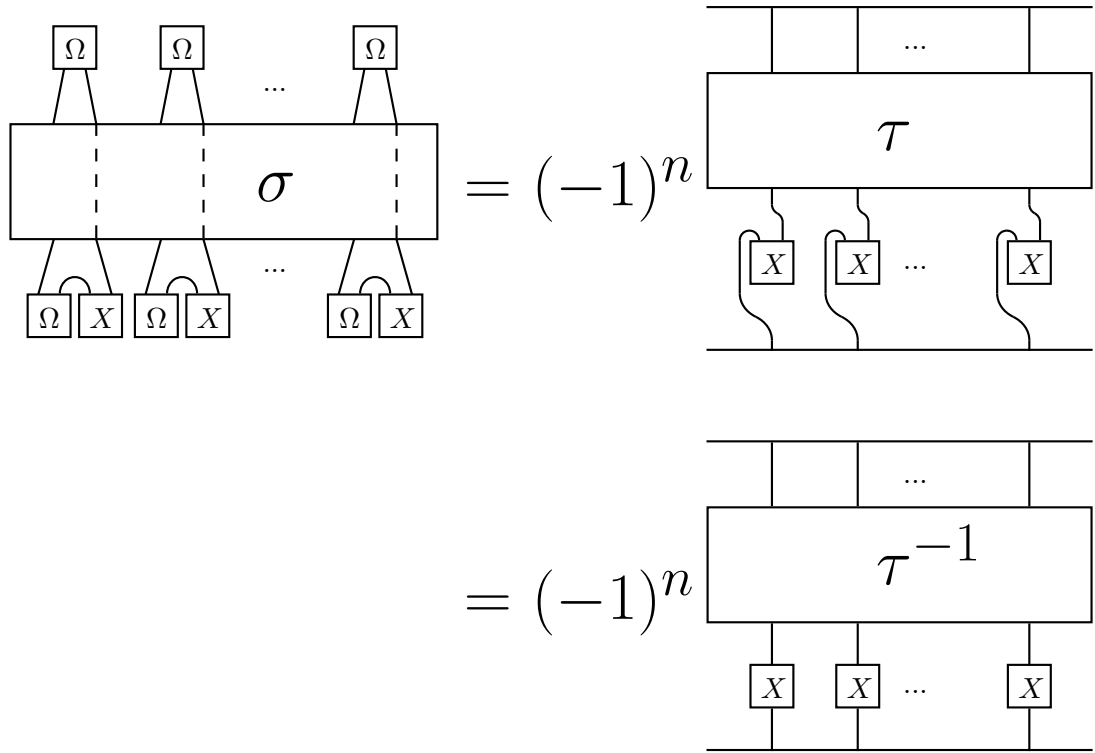
			\caption{A diagrammatic version of \eqref{Sp(N) contraction simplification}. The dotted lines represent the fact that $\sigma$ fixes all even numbers. By following the index contractions on the left, we see that $\tau$ should be contracted with $n$ copies of the matrix $\Omega X \Omega^T$. Using the condition \eqref{symplectic matrix condition}, this is just $- X^T$. We have pulled out the factors of $-1$ and the transpose means the $X$ indices switch roles (compare with figure \ref{figure: SO(N) simplified contraction}). In the second row, we convert this result into a $U(N)$ type contraction by breaking our index conventions and setting $X^i_{\ j} = X^{ij}$. Note that the role switch of the $X$ indices on the first line means $\tau$ is inverted on the second line.}
			\label{figure: Sp(N) simplified contraction}
		\end{figure}

		Consider $\sigma \in S_n^{(odd)}$ the representative of some double coset labelled by $p \vdash n$. Let $\tau \in S_n$ be the equivalent permutation in $S_n$ (so $\tau$ has cycle type $p$). Then we have
		\begin{align}
		C^{(\Omega)}_I \sigma^I_J \left[ \left( \Omega X \right)^{\otimes n} \right]^J & = \Omega_{i_1 j_1} \Omega_{i_2 j_2} \ldots  \Omega_{i_n j_n} \sigma^{i_1 j_1 i_2 j_2 \ldots  i_n j_n}_{k_1 l_1 k_2 l_2 \ldots  k_n l_n} \left( \Omega X \right)^{k_1 l_1} \left( \Omega X \right)^{k_2 l_2} \ldots \left( \Omega X \right)^{k_n l_n} \nonumber \\
		& = \Omega_{i_1 j_1} \Omega_{i_2 j_2} \ldots  \Omega_{i_n j_n} \tau^{i_1 i_2 \ldots  i_n}_{k_1 k_2 \ldots  k_n} \delta^{j_1}_{l_1} \delta^{j_2}_{l_2} \ldots  \delta^{j_n}_{l_n} \nonumber \\ & \hspace{5.9cm} \left( \Omega X \right)^{k_1 l_1} \left( \Omega X \right)^{k_2 l_2} \ldots \left( \Omega X \right)^{k_n l_n} \nonumber \\
		& = \Omega_{i_1 j_1} \left( \Omega X \right)^{j_1 k_1} \Omega_{i_2 j_2} \left( \Omega X \right)^{j_2 k_2} \ldots \Omega_{i_n j_n} \left( \Omega X \right)^{j_n k_n} \tau^{i_1 i_2 \ldots i_n}_{k_1 k_2 \ldots k_n} \nonumber \\
		& = \tau^{i_1 i_2 \ldots i_n}_{k_1 k_2 \ldots k_n} \left( \Omega^2 X \right)^{i_1 k_1} \left( \Omega^2 X \right)^{i_2 k_2} \ldots \left( \Omega^2 X \right)^{i_n k_n} \nonumber \\
		& = (-1)^n \tau^{i_1 i_2 \ldots i_n}_{k_1 k_2 \ldots k_n} X^{i_1 k_1} X^{i_2 k_2} \ldots X^{i_n k_n} \nonumber \\
		& = (-1)^n X^{i_1 i_{\tau^{-1} (1)}} X^{i_2 i_{\tau^{-1} (2)}} \ldots X^{i_n i_{\tau^{-1} (n)}} \nonumber \\
		& = \prod_i \left( \text{Tr} X^i \right)^{p_i}
		\label{Sp(N) contraction simplification}
		\end{align}
		where for the last line we have used \eqref{SO(N) permutations to traces}, noting that $\tau$ and $\tau^{-1}$ have the same cycle type. The same calculation (excluding the last line) is shown diagrammatically in figure \ref{figure: Sp(N) simplified contraction}. Therefore, just as with $SO(N)$, the double coset labelled by $p$ leads to the expected trace. 
		
		So using results from section \ref{section: SO(N) permutations to traces} we can re-express \eqref{symplectic half-bps operator} in terms of multi-traces
		\begin{equation}
		\mathcal{O}^{\Omega}_T = \frac{d_T 2^{2n} (n!)^2}{(2n)!} \sum_{p \vdash \frac{n}{2}} \frac{1}{z_{4p}} \langle T, [A] | D^T ( \sigma_{2p} ) | T, [S] \rangle \prod_i \left( \text{Tr}X^{2i} \right)^{p_i}
		\label{symplectic half-bps operator from traces 1}
		\end{equation}
		where $\sigma_{2p} \in S_n^{(odd)}$ is a representative of the even double coset with partition $2p$. Now since representations of $S_{2n}$ are orthogonal we have
		\begin{equation}
		\langle T, [A] | D^T ( \sigma_{2p} ) | T, [S] \rangle = \langle T, [S] | D^T ( \sigma_{2p}^{-1} ) | T, [A] \rangle
		\nonumber
		\end{equation}
		Then as $\sigma_{2p}^{-1}$ is also in $S_n^{(odd)}$ with cycle type $2p$, we could equally well have chosen this to be our double coset representative. So we can apply the formula \eqref{Ivanov's matrix element} to the symplectic matrix element in exactly the same way as we could for the special orthogonal case. Plugging this into \eqref{symplectic half-bps operator from traces 1} we have
		\begin{equation}
		\mathcal{O}^{\Omega}_T = 2^n n! \sqrt{\frac{d_T}{(2n)!}} \sum_{p \vdash \frac{n}{2}} \frac{1}{z_{2p}} \chi_t (p)  \prod_i \left( \text{Tr}X^{2i} \right)^{p_i}
		\label{symplectic half-bps operator from traces 2}
		\end{equation}
		We see that \eqref{symplectic quotient} and \eqref{symplectic half-bps operator from traces 2} are exactly the same as \eqref{quotient of U(N) operator} and \eqref{SO(N) half-bps operator} respectively, so the projection coefficients $\alpha^{Sp(N);T}_R$ are exactly the same as the $\alpha^T_R$ we found in section \ref{section: projection coefficients}. 
		
		Note that the equality of \eqref{symplectic half-bps operator from traces 2} and \eqref{SO(N) half-bps operator} does not contradict the earlier statement that orthogonal and symplectic operators are anti-symmetrisations of each other. This result stated that $O^\Omega_T$ was an anti-symmetrisation of $O^\delta_{T^c}$, while the above states that $O^\Omega_T \approx O^\delta_T$ (where $\approx$ denotes the fact that the symplectic and orthogonal operators have the exact same form in terms of traces, but the matrix $X$ satisfies different conditions, so they are not strictly equal). One can check that $O^\Omega_T$ is the anti-symmetrisation of $O^\Omega_{T^c}$, and therefore the two statements agree with each other.
		
		We could also consider the projection to the symplectic theory in the quarter-BPS sector. The logic in \eqref{Sp(N) contraction simplification} can be extended, and gives
		\begin{align*}
		C^{(\Omega)}_I \sigma^I_J \left[ \left( \Omega X \right)^{\otimes n} \left( \Omega Y \right)^{\otimes m} \right]^J & = (-1)^{n+m} \left( X^{\otimes n} Y^{\otimes m} \right)^{i_1 i_{\tau^{-1}(1)} i_2 i_{\tau^{-1}(2)} \ldots i_{n+m} i_{\tau^{-1}(n+m)}} \\ & = (-1)^{n+m} X^{i_1 i_{\tau^{-1}(1)}} \ldots X^{i_n i_{\tau^{-1}(n)}} \\ & \hspace{4cm} Y^{i_{n+1} i_{\tau^{-1}(n+1)}} \ldots Y^{i_{n+m} i_{\tau^{-1}(n+m)}}
		\end{align*}
		From \eqref{relation between traces from symplectic condition} we know that we can reverse a single trace at the expense of a minus sign corresponding to the length of the trace. Since reversing a single trace corresponds to inverting a cycle in the permutation $\tau$, we can invert $\tau$ at the expense of a factor of $(-1)^{n+m}$. Therefore we have
		\begin{equation}
		C^{(\Omega)}_I \sigma^I_J \left[ \left( \Omega X \right)^{\otimes n} \left( \Omega Y \right)^{\otimes m} \right]^J = X^{k_1 k_{\tau(1)}} \ldots X^{k_n k_{\tau(n)}} Y^{k_{n+1} k_{\tau(n+1)}} \ldots Y^{k_{n+m} k_{\tau(n+m)}}
		\label{Sp(N) to U(N) contraction}
		\end{equation}
		which is the $Sp(N)$ equivalent of \eqref{SO(N) to U(N) contraction}.
		
		Following the procedures in section \ref{section: quarter-BPS projection}, we can use \eqref{Sp(N) to U(N) contraction} to find the symplectic quarter-BPS projection coefficients. The expression is exactly as in \eqref{quarter-BPS projection coefficients} but with the $[A]$ and $[S]$ swapped (and the implicit difference in constraints on Young diagrams). Using the conjugate labels for symplectic operators swaps the $[A]$ and $[S]$ and removes the difference in constraints at the cost of introducing a factor of $(-1)^{\sigma_{\widetilde{Q}}}$. So using these labels, the symplectic quarter-BPS projection coefficients differ from the orthogonal ones only by a sign in the summation over $\widetilde{Q}$.

		\subsection{Structure of symplectic space of gauge-invariant operators}
		
		In section \ref{section: Generating function at infinite N}, we showed that the large $N$ generating function for $N^{\delta}_{n,m}$ had a lot of structure, associated with corresponding structures in the space of $SO(N)$ multi-traces. Since $N^\Omega_{n,m} = N^\delta_{n,m}$, the generating function for $Sp(N)$ is the same, and we therefore expect the vector space of $Sp(N)$ traces at large $N$ to share this structure.
		
		For $SO(N)$, the structures shown in figure \ref{figure: SO(N) diagram} and \ref{figure: SO(N) periodicity diagram} are derived from the relation \eqref{relation between traces from antisymmetry}, and we have the exact same statement for $Sp(N)$ in \eqref{relation between traces from symplectic condition}. Therefore, at large $N$, the structures of the $Sp(N)$ and $SO(N)$ quarter-BPS sectors are identical.

		\subsection{Symplectic correlators}

		As for $SO(N)$, we define the complex conjugate of $X^{ij}$ to be $\left( X^* \right)^{ij} = X_{ij}$, and similarly for $Y^{ij}$. This leads to the conjugate operators
		\begin{align*}
		\overline{\mathcal{O}}^{\Omega}_{T,R,S,\lambda} & = \frac{d_T}{(2n+2m)!} \sum_{\sigma \in S_{2n+2m}} \langle T, [A] | D^T  (\sigma) \Big( | R,\lambda, [S] \rangle \otimes | S, \lambda , [S] \rangle  \Big) \\ 
		& \qquad \qquad \qquad \qquad \qquad \qquad \qquad \qquad \left[ \left( \Omega X \right)^{\otimes n} \left( \Omega Y \right)^{\otimes m} \right]_I \left( \sigma^{-1} \right)^I_J C^{(\Omega) \, J}
		\end{align*}
		The two point function for symplectic matrices is 
		\begin{equation}
		\langle X^{ij} X_{kl} \rangle = \delta^i_k \delta^j_l - \Omega^i_{\ l} \Omega^j_{\ k} = \langle Y^{ij} Y_{kl} \rangle
		\label{symplectic two point function}
		\end{equation}
		which is equivalent to
		\begin{equation}
		\left\langle \left( \Omega X \right)^{ij} \left( \Omega X \right)_{kl} \right\rangle = \delta^i_k \delta^j_l + \delta^i_l \delta^j_k = \left\langle \left( \Omega Y \right)^{ij} \left( \Omega Y \right)_{kl} \right\rangle
		\label{symplectic two point function 2}
		\end{equation}
		Again the definition \eqref{symplectic two point function} looks the same as that in \cite{Caputa2013a}, but they used the definition $X_{ij} = \left( X^\dagger \right)^{ij}$, so there is a distinction. The convention used here ensures the positivity of the two point function of $X^{ij}$ with its conjugate $X_{ij}$.
		
		The definition \eqref{symplectic two point function 2} leads to
		\begin{align}
		\left\langle \left[ \left( \Omega X \right)^{\otimes n} \left( \Omega Y \right)^{\otimes m} \right]^I \left[ \left( \Omega X \right)^{\otimes n} \left( \Omega Y \right)^{\otimes m} \right]_J \right\rangle & = \sum_{\sigma \in S_n[S_2] \times S_m[S_2]} \sigma^I_J \nonumber \\
		& = 2^{n+m} n! m! \left( P_{[S]_n \otimes [S]_m} \right)^I_J
		\label{symplectic two point function 3}
		\end{align}
		In \cite{Kemp1406}, Kemp presented the symplectic operators and calculated their two point functions. He used the same conventions, \eqref{symplectic two point function 3}, for the two point function, so we can directly quote his result, taking into account the normalisation difference of $\frac{d_T}{(2n+2m)} 2^{2n+2m} \sqrt{n! m! (n+m)!}$ relative to \eqref{symplectic mesonic operator}.
		\begin{align}
		\left\langle \mathcal{O}^\Omega_{T,R,S,\lambda} \overline{\mathcal{O}}^{\Omega}_{T',R',S',\lambda'} \right\rangle & = \delta_{T T'} \delta_{R R'} \delta_{S S'} \delta_{\lambda \lambda'} 2^{2n+2m} n!m!(n+m)! \frac{d_T}{(2n+2m)!} \nonumber \\ & \hspace{7.8cm} \prod_{\substack{i \in \text{ odd}\\ \text{rows of }T}} (N+c_i)
		\label{symplectic correlator}
		\end{align}
		This is very similar to the correlator \eqref{mesonic correlator}, except for the product running over rows rather than columns (and the implicit differences due to different conditions on $T,R,S$).
		
		It is not difficult to show that
		\begin{equation}
		\prod_{\substack{i \in \text{ odd}\\ \text{rows of }T}} (N+c_i) = (-1)^{n+m} \prod_{\substack{i \in \text{ odd}\\ \text{columns of } T^c }} (-N+c_i)
		\nonumber
		\end{equation}
		therefore the symplectic correlator \eqref{symplectic correlator} evaluated with $N \rightarrow -N$ is the same (up to a factor of $(-1)^{n+m}$) as the orthogonal mesonic correlator \eqref{mesonic correlator} with conjugate labels.

		\clearpage
		
	\section{Discussion  } \label{section: Outlook} 
	
	We discuss here a selection of interesting questions raised  by the results 
	of this paper. 
	
	\subsection{ Giant graviton branes and plethyms } 
	
	The classification of half BPS operators in terms of Young diagrams allows an elegant map between these operators and quantum states obtained from  semi-classical giant gravitons \cite{Corley2002,BBNS01}. Young diagrams with order $1$ long columns, and column lengths of order $N$, were proposed to be dual to giant gravitons which are three-spheres expanded in $S^5$ of $ AdS_5 \times S^5$. Single giant states are dual to single column Young diagrams, and multiple giants are dual to multiple-column Young diagrams. A similar picture holds for giant gravitons which are three-spheres in  the $ AdS_5$ directions. This map receives confirmation from a number of directions: holographic comparison of correlators of two Young diagrams with a trace  \cite{BKYZ11,Lin12,CDZ12,KMY15}, moduli space quantisation \cite{BGLM06,PR12} and strings attached to giants \cite{BHLN02,DSS07I,GGO2011,BBFH04,DSS07II,BDS08,DR12}. 
	

	In AdS/CFT, the AdS background dual to $SO(N)$ gauge theories is obtained from the AdS dual of $U(N)$ by an orientifold operation, which acts as a $\mathbb{Z}_2$ in space-time accompanied by an orientation reversal on the string worldsheet. Analogously to the map between branes and states in $U(N)$ theories, we expect, for the $SO(N)$ theories,  a similar detailed map between Young diagram states and branes in the dual $ AdS_5 \times \mathbb{R} P^5$ background. This motivated us to  conduct a  study of the orientifold projection operation on the Young diagram bases. 
	
	The projection is captured by integer coefficients $ \bar{\alpha}_R^T$, which were found to be related to a plethystic refinement of Littlewood-Richardson coefficients
	\begin{align}
	\bar{\alpha}_R^T & = \text{Mult} [ R , S^2 (t)  ] - \text{Mult} [ R , \Lambda^2 (t)  ]
	\nonumber \\
	& = \mathcal{P}(t,[2],R) - \mathcal{P}(t,[1,1],R)
	\label{signcrt} 
	\end{align}
	The sum of these is the LR coefficient 
	\begin{align} 
	g_{t,t;R} & =  \text{Mult} [ R , S^2 (t)  ] + \text{Mult} [ R , \Lambda^2 (t)  ] 
	\nonumber \\
	& = \mathcal{P}(t,[2],R) + \mathcal{P}(t,[1,1],R)
	\end{align}
	As we have seen in section \ref{section: Z2 quotient}, $\bar \alpha_R^T$ has an interpretation in terms of the combinatorics of domino tilings, and we have discussed a brane interpretation of this combinatorics in section \ref{sec:branesdominoes}. 
	Here we will discuss another approach to physically understand the nature of 
	the projection coefficient. 
	
	Interestingly, the LR coefficient $g_{t,t;R}$ appears in the extremal correlator in the $U(N ) $ theory $ \langle \chi_t \chi_t \chi_R^{\dagger } \rangle $ \cite{Corley2002}. Given the correspondence between Young diagrams and branes, this extremal correlator is naturally  interpreted as the amplitude for the overlap between the composite system consisting of the pair of  branes $(t,t)$ and the brane $R$. The effect of the orientifold operation is to change the amplitude of interaction $ t \otimes t \rightarrow R $ by introducing the sign in (\ref{signcrt}).
	
	
	A very interesting problem is to derive  this relative sign from the point of view of strings propagating in $ AdS_5 \times S^5$ and the orientifold of this background. The argument above, which says that the effect of the orientifold is to change the sign of the anti-symmetric part of the interaction, is based on assuming AdS/CFT and using the relation  about gauge invariant operators corresponding to branes. The problem is to explain this sign without using facts about the dual CFT. This is not straightforward. The  consistency of brane physics in spacetime with the formula in terms of LR coefficients has been tested in various limits e.g. \cite{BKYZ11,Lin12,CDZ12,KMY15,HS18}. However a general  understanding,  directly from the spacetime perspective, of why the interaction of branes is given by the Littlewood-Richardson coeffients is not currently available. Understanding the sign from the physics of orientifolds would probably also shed light on this question of why, based purely on the physics of strings in the AdS spacetime without assuming AdS/CFT duality, LR coefficients appear in the interactions of branes. Insights from discussions of signs in orientifolds, such as those in   \cite{HoriWalcher} may be useful. We will leave this as a very interesting question for the future.

	\subsection{ Weak coupling }

	The quarter BPS sector undergoes a step change when going from zero to weak coupling: for a review with extensive references to the previous litereture see  \cite{Pasukonis2010}.  In the $U(N)$ theory this change is equivalent to allowing the matrices to commute inside the trace. These states are related to the quantisation 
	of moduli spaces of giant gravitons \cite{Biswas:2006tj}.   
	An analogous  discussion for the  $SO(N)$ theory, and  the relation of these states to the quantisation of giant graviton moduli spaces  would be interesting to develop. 
	
	\subsection{ General quivers }

	The generating function \eqref{U(N) multi trace generating function} has been generalised to arbitrary $U(N)$ quivers \cite{Pasukonis2013,Mattioli2014}. The structure of the function, with its infinite product of a root function, was found to be very general, and the root function had an interpretation in terms of counting words made from loops in the quiver. Is there an analogous  generalization of \eqref{SO(N) multi trace generating function} for $SO(N)$ (or $ Sp(N)$) gauge theory?

	
	\subsection{ Permutations as background independent structures  in string theory }
	
	In this paper, we have made extensive use of permutations as tools for understanding gauge invariant operators. The formula for the orientifold projection map in the half-BPS sector, which made contact with domino combinatorics, was given as a sum of permutation group characters.  The projection map relates different backgrounds of string theory. We also explored (section \ref{section: U(N) correlators of SO(N) operators}) the $U(N)$ inner product of gauge invariant operators which survive the orientifold projection to $SO(N)$, and observed a  connection to an appropriately defined inverse of the plethysm coefficients. The $U(N)$ and $SO(N)$ inner products for the same operators can be viewed as different (background-dependent) pairings on permutations which are background independent characterizations of gauge invariants. Other diverse applications of permutations in gauge invariant operators (for a short review see \cite{Ramgoolam:2016ciq}) have seen applications of Littlewood-Richardson coefficients as well as Kronecker coefficients in multi-matrix bases and correlators. An interesting exercise is to revisit these applications and disentangle the aspects of permutations and associated representation  theory which contain information about specific backgrounds, and those that are common to different backgrounds, or relate different backgrounds. The integrally-graded word combinatorics (involving Lyndon words and their orthogonal generalisations) which we have here identified as key structures in understanding the space of gauge invariant operators in both $U(N)$ and $ SO/Sp$ gauge theories	may well be structures which contain background independent information. In this connection, it is interesting that another recent physics application of Lyndon words is in connection with knot invariants associated with intersections of M2-branes and M5-branes \cite{KS1608,LMOV}. It would be interesting to seek a background-independent characterization of how word combinatorics appears in the physics of BPS states in string theory.  
		
\vskip2cm 
\begin{center} 
{ \bf Acknowledgements} 
\end{center} 
SR is supported by the STFC consolidated grant ST/L000415/1 “String Theory, Gauge Theory \& Duality” and  a Visiting Professorship at the Mandelstam Institute for Theoretical Physics,  University of the Witwatersrand, funded by a Simons Foundation  grant. SR  thanks the Galileo Galilei Institute for Theoretical Physics for hospitality and the INFN for partial support during the completion of this work. We thank for useful discussions: Robert de Mello Koch, Shinji Hirano, Vishnu Jejjalla, Tarun Sharma, Costas Zoubos.

		\newpage

		\appendix

		\section{M{\"o}bius inversion}
		\label{section: mobius inversion}
		
		\textbf{Proposition: The M{\"o}bius Inversion Formula} \\
		Let $\{ a_n \}$ and $\{ b_n \}$ be two sequences indexed by the positive integers. If $a_n$ can be expressed as
		\begin{equation}
		a_n = \sum_{d | n} b_d = \sum_{d|n} b_{\frac{n}{d}}
		\label{single variate mobius inversion formula 1}
		\end{equation}
		where $d$ runs over all divisors of $n$, denoted by $d | n$, then
		\begin{equation}
		b_n = \sum_{d | n} \mu \left( \frac{n}{d} \right) a_d = \sum_{d|n} \mu(d) a_{\frac{n}{d}}
		\label{single variate mobius inversion formula 2}
		\end{equation}
		where $\mu$ is the M{\"o}bius function defined by
		\begin{equation}
		\mu(d) = \begin{cases}
		1 & d = 1 \\
		(-1)^n & d \text{ a product of } n \text{ distinct prime factors} \\
		0 & d \text{ has a repeated prime in its prime factorisation}
		\end{cases} \label{mobius mu function}
		\end{equation}						
		The proof of this proposition relies on\\
		\textbf{Lemma}
		\begin{equation}
		\sum_{d | n} \mu (d) = \begin{cases}
		1 & n = 1 \\
		0 & n>1
		\end{cases}
		\label{mobius function identity}
		\end{equation}
		\textbf{Proof of Lemma} \\
		This is obvious for $n=1$, so we will only prove the case $n>1$. Writing $n$ in terms of its prime factors, we have
		\begin{equation*}
		n=p_1^{r_1} p_2^{r_2} \ldots p_k^{r_k}
		\end{equation*}
		where $r_i \geq 1$ for each $i$. The divisors of $n$ which contribute to the sum \eqref{mobius function identity} are those which are square free. Explicitly, they can be written
		\begin{equation}
		d=p_1^{s_1} p_2^{s_2} \ldots p_k^{s_k}
		\nonumber
		\end{equation}
		where $s_i \in \{ 0,1 \}$ for each $i$.
		
		We define $S$ to be the set of distinct prime factors of $n$: $S = \{ p_1, p_2, \ldots p_k \}$. Then subsets of $S$ correspond exactly to the divisors $d$ defined above
		\begin{equation}
		d = p_1^{s_1} p_2^{s_2} \ldots p_k^{s_k} \qquad \longleftrightarrow \qquad \{ p_i : s_i = 1\} \leq S
		\end{equation}
		From the definition \eqref{mobius mu function}, we see that
		\begin{equation}
		\mu (d) = (-1)^{|\text{subset of } S \text{ corresponding to }d|}\\
		\nonumber
		\end{equation}
		So
		\begin{equation}
		\sum_{d|n} \mu(d) = \# \text{ of subsets of } S \text{ with even size} - \# \text{ of subsets of } S \text{ with odd size}
		\nonumber
		\end{equation}
		But we have a bijective map between even subsets and odd subsets given by
		\begin{equation}
		A \longrightarrow \begin{cases}
		A \cup \{ p_1 \} & p_1 \not \in A \\
		A / \{ p_1 \} & p_1 \in A
		\nonumber
		\end{cases}
		\end{equation}
		and therefore
		\begin{equation}
		\sum_{d|n} \mu(d) = 0
		\nonumber
		\end{equation}
		$\square$
		\newline
		
		\noindent \textbf{Proof of Proposition} \\
		The first step in the proof is to note that the $a_n$ determine the $b_n$ uniquely via the relation \eqref{single variate mobius inversion formula 1}. Indeed, we have $b_1 = a_1$, $b_2 = a_2 - a_1$, $b_3 = a_3 - a_1$. To prove it in general, we use strong induction with these three as the base cases. Assuming $b_n$ is determined by the sequences of $a$s for all $n \leq k$, we can rearrange \eqref{single variate mobius inversion formula 1} to get
		\begin{equation}
		b_{k+1} = a_{k+1} - \sum_{\substack{d|(k+1) \\ d \neq k+1}} b_d
		\nonumber
		\end{equation}
		Then since the sum over $d$ only includes $d \leq k$, we know inductively that $b_d$ is determined by the $a$s, and hence $b_{k+1}$ is also determined by the $a$s.
		
		We now notice that the $b_n$, as defined in \eqref{single variate mobius inversion formula 2}, satisfy \eqref{single variate mobius inversion formula 1}:
		\begin{align*}
		\sum_{d | n} b_d & = \sum_{d|n} \sum_{e | d} \mu \left( \frac{d}{e} \right) a_e \\
		& = \sum_{e | n} a_e \sum_{f| \frac{n}{e}} \mu (f) \\
		& = a_n
		\end{align*}
		In going from the 1st to the 2nd line we have reordered the sums and reparameterised by $f = \frac{d}{e}$, and in going from the 2nd to the 3rd we have used the lemma \eqref{mobius function identity}.
		
		Since the $b_n$ have a unique solution, \eqref{single variate mobius inversion formula 2} must therefore be the correct formula for the $b_n$, as claimed. $\square$
		\newline
		
		\noindent Note that in this proposition, there was nothing special about addition, the result and proof follow exactly the same way if we replace the addition by multiplication. Explicitly, given
		\begin{equation}
		b_n = \prod_{d | n} a_{d} = \prod_{d | n} a_{\frac{n}{d}}
		\nonumber
		\end{equation}
		we can invert uniquely to get
		\begin{equation}
		a_n = \prod_{d | n} b_d^{\mu \left(\frac{n}{d} \right)} = \prod_{d | n} b_{\frac{n}{d}}^{ \mu (d)}
		\label{multiplicative mobius inversion}
		\end{equation}
		In this paper, we come across relations of the form
		\begin{equation}
		a_{n,m} = \sum_{d | n,m} b_{\frac{n}{d}, \frac{m}{d}}
		\label{multi variate mobius inversion formula 2}
		\end{equation}
		so we would like a generalisation of the M{\"o}bius inversion formula for two variables. This generalisation is \\
		\newline
		\textbf{Lemma} \\
		The $b_{n,m}$ are determined uniquely by \eqref{multi variate mobius inversion formula 2}, with
		\begin{equation}
		b_{n,m} = \sum_{d|n,m} \mu (d) a_{\frac{n}{d},\frac{m}{d}}
		\label{multi variate mobius inversion formula}
		\end{equation}
		\textbf{Proof} \\
		To prove this, consider fixing $\bar{n}, \bar{m}$ to be coprime. We then define
		\begin{equation}
		\bar{a}_k = a_{k \bar{n}, k \bar{m}} \qquad \qquad \bar{b}_k = b_{k \bar{n}, k \bar{m}}
		\nonumber
		\end{equation}
		In terms of these sequences \eqref{multi variate mobius inversion formula 2} reads
		\begin{align*}
		\bar{a}_k & = \sum_{d | k \bar{n}, k\bar{m}} b_{\frac{k}{d} \bar{n}, \frac{k}{d} \bar{m}}\\
		& = \sum_{d|k} \bar{b}_{\frac{k}{d}}
		\end{align*}
		where we have used the fact that $\bar{n}, \bar{m}$ are coprime to conclude that $d|k\bar{n},k\bar{m}$ is equivalent to $d|k$. Then by the standard M{\"o}bius inversion formula, we have
		\begin{equation*}
		\bar{b}_k = \sum_{d|k} \mu(d) \bar{a}_{\frac{k}{d}}
		\end{equation*}
		or in terms of $a$s and $b$s
		\begin{equation*}
		b_{k\bar{n}, k\bar{m}} = \sum_{d| k \bar{n}, k\bar{m}} \mu(d) a_{\frac{k \bar{n}}{d}, \frac{k \bar{m}}{d}}
		\end{equation*}
		This is true for all $k$, and coprime $\bar{n}$, $\bar{m}$. So to prove \eqref{multi variate mobius inversion formula} for an arbitrary $n,m$ we pick $k = \gcd (n,m)$, $\bar{n} = \frac{n}{k}$, $\bar{m} = \frac{m}{k}$. \\ $\square$ \\
		
		\noindent The M{\"o}bius inversion formula can be used to prove some useful identities. We start with the well known identity
		\begin{equation}
		\sum_{d | n} \phi(d) = n
		\label{phi identity}
		\end{equation}
		where $\phi (n)$ is the Euler totient function that counts the number of numbers less than $n$ that are coprime to $n$. Applying the M{\"o}bius inversion formula gives
		\begin{equation}
		\frac{\phi(n)}{n} = \sum_{d|n} \frac{\mu(d)}{d}
		\label{phi mu identity}
		\end{equation}
		and applying it again gives
		\begin{equation}
		\mu (n) = \sum_{d|n} d \mu(d) \phi \left( \frac{n}{d} \right) = \sum_{d | n} \frac{n}{d} \mu \left( \frac{n}{d} \right) \phi (d)
		\label{phi mu identity 2}
		\end{equation}

		\section{Alternative derivation of $SO(N)$ infinite $N$ generating function}
		\label{section: alternative SO(N) generating function}
			
			We now derive the generating function \eqref{SO(N) multi trace generating function} directly from \eqref{2-matrix delta counting}, ignoring the finite $N$ constraint $l(T) \leq N$.
			
			The first step is to find an alternative formula for \eqref{2-matrix delta counting} that lends itself more easily to explicit calculation of the generating function. This is done using results from the theory of symmetric functions, and gives an expression involving the coefficients of the cycle index polynomial of $S_n[S_2]$.
			
			Using this alternative formula we can express the generating function as a product of integrals, each of which can be explicitly evaluated.

			\subsection{An alternative counting formula}
			
			Expanding $g_{R,S;T}$ in terms of characters gives
			\[
			N_{n,m}^{\delta} = \sum_{\substack{
					R \vdash 2n \text{ with even column lengths} \\
					S \vdash 2m \text{ with even column lengths} \\
					T \vdash 2n+2m \text{ with even row lengths}}} \frac{1}{(2n)! (2m)!}
			\sum_{\substack{\sigma \in S_{2n} \\ \tau \in S_{2m}}} \chi_R (\sigma) \chi_S (\tau) \chi_T ( \sigma \circ \tau)
			\]
			where $\sigma \circ \tau$ means the permutation in $S_{2n+2m}$ that acts as $\sigma$ on the first $2n$ objects and $\tau$ on the last $2m$. Since the characters only depend on the cycle type of $\sigma$ and $\tau$, we can rewrite this as 
			\begin{equation}
			N_{n,m}^{\delta} = \sum_{\substack{
					R \vdash 2n \text{ with even column lengths} \\
					S \vdash 2m \text{ with even column lengths} \\
					T \vdash 2n+2m \text{ with even row lengths}}}
			\sum_{\substack{p \vdash 2n \\ q \vdash 2m}} \frac{\chi_R (p) \chi_S (q) \chi_T ( p \cup q)}{z_p z_q}
			\label{2-matrix counting from characters}
			\end{equation}
			where $p \cup q$ was defined in section \ref{section: notation}, and $z_p$ and $z_q$ arise because the number of permutations in $S_{2n}$ with cycle type $p$ is given by $\frac{(2n)!}{z_p}$.
					
			Now we'd like to evaluate
			\begin{equation}
			\sum_{\substack{R \vdash 2n \text{ with even} \\ \text{row/column lengths}}} \chi_R (q)
			\label{row/column sum}
			\end{equation}
			To do this we need to review some facts from the theory of symmetric functions. These are defined as formal polynomials in an infinite number of variables $t_1,t_2, \ldots$ which are completely symmetric under permutations of the $t_i$. We will use two different bases for the order $n$ symmetric functions. The power sum polynomials are defined for integer $r$ by
			\[
			P_r (t_1, t_2, \ldots ) = \sum_i t_i^r
			\]
			and for a partition $q = [\lambda_1, \lambda_2, \ldots ]$ by
			\[
			P_q = P_{\lambda_1} P_{\lambda_2} \ldots
			\]
			The $P_\lambda$ for $\lambda \vdash n$ are a basis for the order $n$ symmetric functions.
			
			Schur polynomials, also indexed by partitions (Young diagrams) $R \vdash n$, are defined by
			\begin{equation}
			s_R (t_1,t_2, \ldots ) = \sum_{q \vdash n} \frac{1}{z_q} \chi_R (q) P_q(t_1, t_2, \ldots )
			\nonumber
			\end{equation}
			From the orthogonality of characters, \eqref{character orthogonality}, we can invert this definition to write the power sum polynomials in terms of the Schur polynomials. Therefore the Schur polynomials also form a basis.
			
			We now introduce the Hall inner product on the space of symmetric functions. It is defined by
			\[
			\left\langle P_p , P_q \right\rangle = \delta_{p q} z_p
			\]
			This enables us to extract the coefficient of the power sum polynomials from $s_R$. Explicitly
			\[
			\chi_R (q) = \langle s_R , P_q \rangle
			\]
			So 
			\[
			\sum_{\substack{R \vdash 2n \text{ with even} \\ \text{row lengths}}} \chi_R (q) = \left\langle \sum_{\substack{R \text{ with even} \\ \text{row lengths}}} s_R \ , \ P_{q} \right\rangle
			\]
			Note that on the right hand side, $R$ can range over all partitions with even row length, not just those with $|R| = 2n$, since the inner product with $P_q$ is non-zero only for those $R$ with $|R| = 2n$. 
			
			In MacDonald's book \cite[Chapter I.5]{Macdonald1995} he shows that
			\[
			s(t_1,t_2, \ldots ) = \sum_{\substack{R \text{ with even} \\ \text{row lengths}}} s_R = \prod_i \frac{1}{1-t_i^2} \prod_{i < j} \frac{1}{1-t_i t_j}
			\]
			To find the inner product of $s$ with $P_q$ we need to express $s$ in terms of the $P_q$. It turns out to be easier to look at log$s$
			\begin{align*}
			\log s & = - \sum_i \log(1-t_i^2) - \sum_{i < j} \log(1-t_i t_j) \\
			& = \sum_{r=1}^{\infty} \frac{1}{2r} \left( \sum_{i,j} t_i^r t_j^r + \sum_i t_i^{2r} \right) \\
			& = \sum_{r=1}^{\infty} \frac{1}{2r} \left( P_r^2 + P_{2r} \right) \\
			& = \sum_{r=1}^{\infty} \frac{1}{r} Z^{S_2} (P_r, P_{2r})
			\end{align*}
			where $Z^{S_2}$ is the cycle index polynomial of the group $S_2$ as defined in \eqref{cycle polynomial definition}. Therefore
			\begin{equation}
			s= \text{exp} \left[ \sum_{r=1}^\infty \frac{1}{r} Z^{S_2}(P_r, P_{2r}) \right]
			\label{sum of schur polynomials over even young diagrams}
			\end{equation}
			Before we proceed further, we recall two useful facts. Firstly, the generating function for the cycle index polynomials of $S_n$ is \cite[Chapter 5.13]{cameron1999permutation}
			\begin{equation}
			\sum_{n=0}^\infty x^n Z^{S_n}(t_1,t_2, \ldots ) = \text{exp} \left[ \sum_{m=1}^\infty \frac{1}{m} x^m t_m \right]
			\label{S_n cycle polynomial generating function}
			\end{equation}
			and secondly, the cycle index polynomial of a wreath product group is \cite[Chapter 15.5]{cameron1994combinatorics}
			\begin{equation}
			Z^{G[H]} (t_1,t_2, \ldots ) = Z^G(r_1, r_2, \ldots )
			\label{wreath product cycle index}
			\end{equation}
			where
			\[
			r_i = Z^H(t_i,t_{2i},t_{3i}, \ldots )
			\]
			Combining \eqref{S_n cycle polynomial generating function} and \eqref{wreath product cycle index} tells us that the generating function for the cycle index polynomials of $S_n[S_2]$ is
			\begin{equation}
			\sum_{n=0}^\infty x^n Z^{S_n[S_2]} (t_1,t_2, \ldots ) = \sum_{n=0}^\infty x^n \sum_{q \vdash 2n} Z^{S_n[S_2]}_q \prod_i t_i^{q_i} = \text{exp} \left[ \sum_{r=1}^\infty \frac{1}{r} x^r Z^{S_2}(t_r,t_{2r}) \right]
			\label{S_n[S_2] cycle polynomial generating function}
			\end{equation}
			Putting together \eqref{S_n[S_2] cycle polynomial generating function} and \eqref{sum of schur polynomials over even young diagrams}
			\begin{align*}
			s& = \sum_{n=0}^\infty Z^{S_n[S_2]}(P_1,P_2, \ldots ) \\
			& = \sum_{n=0}^\infty \sum_{q \vdash 2n} Z^{S_n[S_2]}_q P_q
			\end{align*}
			Therefore the inner product with $P_q$ gives
			\begin{equation}
			\sum_{\substack{R \vdash 2n \text{ with even} \\ \text{row lengths}}} \chi_R (q) = \langle s, P_q \rangle = Z_q^{S_n[S_2]} z_q
			\label{sum of characters over partitions with even row lengths}
			\end{equation}
			Clearly a Young diagram has even row lengths if and only if its conjugate has even column lengths, so to evaluate the column version of \eqref{row/column sum}, we just conjugate the summation variable $R$. Now $R^c =$sgn$\otimes R$, so the characters are related by
			\[
			\chi_{R^c}(q) = (-1)^{q} \chi_R (q)
			\]
			Therefore
			\begin{equation}
			\sum_{\substack{R \vdash 2n \text{ with even} \\ \text{column lengths}}} \chi_R (q) = (-1)^{q} Z_q^{S_n[S_2]} z_q
			\label{sum of characters over partitions with even column lengths}
			\end{equation}
			Plugging \eqref{sum of characters over partitions with even row lengths} and \eqref{sum of characters over partitions with even column lengths} into \eqref{2-matrix counting from characters} gives
			\begin{equation}
			N_{n,m}^{\delta} = \sum_{\substack{p \vdash 2n \\ q \vdash 2m}} (-1)^{p \cup q} Z_p^{S_n[S_2]} Z_q^{S_m[S_2]} Z_{p \cup q}^{S_{n+m}[S_2]} z_{p \cup q}
			\label{infinite N counting}
			\end{equation}
			
			\subsection{The generating function}
			
			We now want to find the function
			\begin{equation}
			F (x,y) = \sum_{n,m}x^n y^m N_{n,m}^\delta = \sum_{n,m} x^n y^m \sum_{\substack{p \vdash 2n \\ q \vdash 2m}} (-1)^{p \cup q} Z_p^{S_n[S_2]} Z_q^{S_m[S_2]} Z_{p \cup q}^{S_{n+m}[S_2]} z_{p \cup q}
			\label{target generating function}
			\end{equation}
			Our approach is to build candidate generating functions by introducing the terms on the right hand side one by one. We begin by using \eqref{S_n[S_2] cycle polynomial generating function} twice
			\begin{equation*}
			\text{exp} \left[ \sum_{k=1}^\infty \frac{1}{2k}( x^k+y^k) (t_{k}^2+t_{2k}) \right] = \sum_{n,m} x^n y^m \sum_{\substack{p \vdash 2n \\ q \vdash 2m}} Z_p^{S_n[S_2]} Z_q^{S_n[S_2]} \prod_i t_i^{p_i+q_i}
			\end{equation*}
			The third cycle index in \eqref{infinite N counting} comes with a factor of $(-1)^{p \cup q}$. To introduce this into \eqref{S_n[S_2] cycle polynomial generating function}, we just replace $t_k$ with $-t_k$ for $n$ even. Multiplying through by this modified version with a new set of variables $s_k$ and no overall level (no equivalent to $x,y$) gives
			\begin{align*}
			&  \text{exp} \left[ \sum_{k=1}^\infty \frac{1}{2k} (s_k^2 - s_{2k}) + \sum_{k=1}^\infty \frac{1}{2k}(x^k + y^k)(t_k^2 + t_{2k}) \right] \\
			& \qquad \qquad \qquad \qquad = \sum_{n,m,o} x^n y^m \sum_{\substack{p \vdash 2n \\ q \vdash 2m \\ r \vdash 2o}} (-1)^r Z_p^{S_n[S_2]} Z_q^{S_m[S_2]} Z_r^{S_o[S_2]} \prod_i t_i^{p_i + q_i} s_i^{r_i}
			\end{align*}
			This looks similar to \eqref{target generating function}, but we need to introduce a factor of $z_{p \cup q}$ and enforce $r=p \cup q$ (and hence $o = n+m$). We do this in two steps, corresponding to the two parts of 
			\[
			z_{p \cup q} = \prod_i i^{p_i + q_i} (p_i + q_i)!
			\]
			To get the powers of $i$, we can just replace $t_k$ and $s_k$ with $\sqrt{k} s_k$ and $\sqrt{k} t_k$.
			\begin{align*}
			& \exp \left[ \sum_{k=1}^{\infty} \left( \frac{1}{2} s_k^2 - \frac{1}{\sqrt{2k}} s_{2k} \right)\right] \exp \left[ \sum_{k=1}^{\infty} (x^k + y^k) \left( \frac{1}{2} t_k^2 + \frac{1}{\sqrt{2k}} t_{2k} \right) \right] \\
			& \qquad = \sum_{n,m,o} x^n y^m \sum_{\substack{p \vdash 2n \\ q \vdash 2m \\ r \vdash 2o}} (-1)^p (-1)^q Z_p^{S_n[S_2]} Z_q^{S_m[S_2]} Z_r^{S_o[S_2]} \prod_i i^{\frac{1}{2}(p_i + q_i + r_i)} t_i^{p_i + q_i} s_i^{r_i}
			\end{align*}
			Now getting the powers of $i$ in $z_{p \cup q}$ just reduces to the same condition we already needed, $r = p \cup q$. So now we just need to replace $\prod_i t_i^{p_i + q_i} s_i^{r_i}$ with $\delta_{r_i, p_i + q_i} (p_i + q_i)!$. This can be done via the integral
			\[
			\int_{\mathbb{C}} \frac{\text{d}z \text{d}\bar{z}}{2 \pi} e^{- z \bar{z}} z^p \bar{z}^r = \delta_{p,r} p!
			\]
			So replacing $t_k$ with $z_k$, $s_k$ with $\bar{z}_k$, multiplying by $e^{-\sum_k z_k \bar{z}_k}$, and integrating over a copy of $\mathbb{C}$ for each $k$ gives us
			\begin{align}
			F(x,y) & = \sum_{n,m} x^n y^m \sum_{\substack{p \vdash 2n \\ q \vdash 2m}} (-1)^{p \cup q} Z_p^{S_n[S_2]} Z_q^{S_m[S_2]} Z_{p \cup q}^{S_{n+m}[S_2]} z_{p \cup q} \nonumber \\
			& = \int \left( \prod_{k=1}^{\infty} \frac{\text{d}z_k \text{d}\bar{z}_k}{2\pi} \right) \exp \left[ \sum_{k=1}^{\infty} \left( \frac{1}{2} \bar{z}_k^2 - \frac{1}{\sqrt{2k}} \bar{z}_{2k} \right) \right] \nonumber \\
			& \qquad \qquad \exp \left[ \sum_{k=1}^{\infty} (x^k + y^k) \left( \frac{1}{2}z_k^2 + \frac{1}{\sqrt{2k}}z_{2k} \right) \right] \exp \left[ - \sum_{k=1}^{\infty} z_k \bar{z}_k \right] \nonumber	\\
			& = \prod_{ k \text{ odd}} \int \frac{\text{d}z \text{d}\bar{z}}{2 \pi} \exp \left[ \frac{1}{2} (\bar{z}^2 - 2 z \bar{z} + (x^k + y^k) z^2) \right] \nonumber \\
			& \prod_{k \text{ even}} \int \frac{\text{d}z \text{d}\bar{z}}{2 \pi} \exp \left[ \frac{1}{2} \left( \bar{z}^2 - 2 z \bar{z} + (x^k + y^k) z^2 - \frac{2}{\sqrt{k}} \left( \bar{z} - \left( x^{\frac{k}{2}} + y^{\frac{k}{2}} \right) z \right)  \right) \right]
			\label{integral formula for generating function}
			\end{align}
			So we have two integrals to compute. To do them we split $z$ into its real and imaginary parts. Using $z = u + iv$, $\bar{z} = u - iv$, and for simplicity writing $\lambda = x^k + y^k, \mu = x^{\frac{k}{2}} + y^{\frac{k}{2}}$, we have
			\begin{align*}
			\bar{z}^2 - 2 z \bar{z} + \lambda z^2 & = - (1- \lambda) (u + i v)^2 - 4 v^2 \\
			\bar{z}^2 - 2 z \bar{z} + \lambda z^2 - \frac{2}{\sqrt{k}} \left( \bar{z} - \mu z \right)  & = - (1- \lambda) \left( u + i v + \frac{1 - \mu}{\sqrt{k}(1 - \lambda)} \right)^2 \\ & \qquad - 4 \left( v - \frac{i}{2 \sqrt{k}} \right)^2 + \frac{\lambda - 2 \mu + \mu^2}{k(1-\lambda)}
			\end{align*}
			So by changing variables from $(z,\bar{z})$ to $(u,v)$ (and remembering that $\text{d} z \text{d} \bar{z} = 2 \text{d} u \text{d} v$), both odd and even integrals can be evaluated using the standard Gaussian integral
			\begin{equation}
			\int_{- \infty}^{\infty}  \text{d}u \, e^{-a (u + b)^2} = \sqrt{ \frac{\pi}{a} }
			\end{equation}
			where $a,b$ are complex numbers with $\text{Re}(a)>0$. Explicitly, the integrals are
			\begin{align*}
			\int \frac{\text{d}z \text{d}\bar{z}}{2 \pi} \exp \left[ \frac{1}{2} (\bar{z}^2 - 2 z \bar{z} + \lambda z^2) \right]  = \frac{1}{\sqrt{1- \lambda}} 
			\end{align*}
			and
			\begin{align*}
			\int \frac{\text{d}z \text{d}\bar{z}}{2 \pi} \exp  \left[ \frac{1}{2}  \left( \bar{z}^2 - 2 z \bar{z} + \lambda z^2 - \sqrt{\frac{2}{k}} \left( \bar{z} - \mu z \right)  \right) \right] = \frac{1}{\sqrt{1- \lambda}} \text{exp} \left[ \frac{\lambda - 2 \mu + \mu^2}{2k(1- \lambda)} \right]
			\end{align*}
			Plugging these into \eqref{integral formula for generating function} gives
			\begin{align*}
			F(x,y) & = \left( \prod_{k \text{ odd}} \frac{1}{\sqrt{1 - x^k - y^k}} \right) \\ & \qquad \left( \prod_{k \text{ even}} \frac{1}{\sqrt{1- x^k - y^k}} \, \text{exp} \left[ \frac{x^k + x^{\frac{k}{2}} y^{\frac{k}{2}} +y^k - x^{\frac{k}{2}} - y^{\frac{k}{2}} }{k(1-x^k-y^k)} \right] \right) \\
			& = \prod_{k=1}^{\infty} \frac{1}{\sqrt{1- x^k - y^k}} \, \text{exp} \left[ \frac{x^{2k} + x^k y^k + y^{2k} - x^k - y^k}{2k(1-x^{2k} - y^{2k})} \right]
			\end{align*}
			which matches the result \eqref{SO(N) multi trace generating function}, as expected.

			\section{List of sequences and generating functions}
			\label{section: sequences}
			
			We introduce a lot of different single and multi-trace counting sequences in this paper. Here we present all of them in one place. For each sequence we give the definition of the $(n,m)$th term, the first few terms, the generating function and (for the single trace sequences) the plethystic exponential of the generating function. We also give the vector spaces which have these functions as Hilbert series.
			
			Many of the results here can be found together with their derivations in sections \ref{section:U(N) results} and \ref{section: Generating function at infinite N}. The single trace sequences are only considered at infinite $N$, while the multi-trace sequences are defined for finite $N$, but we have only found their generating functions at infinite $N$.
			
			After listing the sequences, we give the relations between them and their generating functions.
			
			\subsection{Single trace sequences}
			
			All of the following definitions are valid provided we have one of $n,m \neq 0$. For all single-trace sequences, we implicitly set the $n=m=0$ term to 0.
			
			\subsubsection{$A_{n,m}$}
			
			The $A_{n,m}$ count single traces of generic matrices ($U(N)$ single traces). They are defined by
			\begin{equation}
			A_{n,m} = \frac{1}{n+m} \sum_{d|n,m} \phi(d) \binom{\frac{n+m}{d}}{\frac{n}{d}}
			\nonumber
			\end{equation}
			Their generating function is
			\begin{equation}
			f_{U(N)} (x,y) = - \sum_{d=1}^\infty \frac{\phi(d)}{d} \log (1-x^d-y^d)
			\nonumber
			\end{equation}
			which is the Hilbert series for the vector space $T_{ST}$. The plethystic exponential is
			\begin{equation}
			F_{U(N)}(x,y) = \prod_{n,m} \frac{1}{(1-x^n y^m)^{A_{n,m}}} = \prod_{k=1}^\infty \frac{1}{1-x^k -y^k}
			\label{U(N) multi-trace counting}
			\end{equation}
			which is the Hilbert series for the vector space $T = \operatorname{Sym}\left( T_{ST} \right)$.
			
			The values of $A_{n,m}$ for $n,m \leq 10$ are shown below
			
			\begin{equation}		
			\begin{array}{c|ccccccccccc}
				& 0 & 1 & 2 & 3 & 4 & 5 & 6 & 7 & 8 & 9 & 10  \\ \hline
				0 & 0 & 1 & 1 & 1 & 1 & 1 & 1 & 1 & 1 & 1 & 1 \\
				1 & 1 & 1 & 1 & 1 & 1 & 1 & 1 & 1 & 1 & 1 & 1 \\
				2 & 1 & 1 & 2 & 2 & 3 & 3 & 4 & 4 & 5 & 5 & 6 \\
				3 & 1 & 1 & 2 & 4 & 5 & 7 & 10 & 12 & 15 & 19 & 22 \\
				4 & 1 & 1 & 3 & 5 & 10 & 14 & 22 & 30 & 43 & 55 & 73 \\
				5 & 1 & 1 & 3 & 7 & 14 & 26 & 42 & 66 & 99 & 143 & 201 \\
				6 & 1 & 1 & 4 & 10 & 22 & 42 & 80 & 132 & 217 & 335 & 504 \\
				7 & 1 & 1 & 4 & 12 & 30 & 66 & 132 & 246 & 429 & 715 & 1144 \\
				8 & 1 & 1 & 5 & 15 & 43 & 99 & 217 & 429 & 810 & 1430 & 2438 \\
				9 & 1 & 1 & 5 & 19 & 55 & 143 & 335 & 715 & 1430 & 2704 & 4862 \\
				10 & 1 & 1 & 6 & 22 & 73 & 201 & 504 & 1144 & 2438 & 4862 & 9252 \\
			\end{array}
			\nonumber
			\end{equation}
			
			\subsubsection{$a_{n,m}$}
			
			The $a_{n,m}$ count aperiodic single traces of generic matrices ($U(N)$ aperiodic single traces), or equivalently Lyndon words. They are defined by
			\begin{equation}
			a_{n,m} = \frac{1}{n+m} \sum_{d|n,m} \mu(d) \binom{\frac{n+m}{d}}{\frac{n}{d}}
			\nonumber
			\end{equation}
			Their generating function is
			\begin{equation}
			\bar{f}_{U(N)} (x,y) = - \sum_{d=1}^\infty \frac{\mu(d)}{d} \log (1-x^d-y^d)
			\nonumber
			\end{equation}
			which is the Hilbert series for the vector space $T_{ST}^{(1)}$. The plethystic exponential is
			\begin{equation}
			F_{U(N)}(x,y) = \prod_{n,m} \frac{1}{(1-x^n y^m)^{a_{n,m}}} = \frac{1}{1-x -y}
			\nonumber
			\end{equation}
			which is the Hilbert series for the vector space $T^{(1)} = \operatorname{Sym}\left( T_{ST}^{(1)} \right)$.
			
			The values of $a_{n,m}$ for $n,m \leq 10$ are shown below

			\begin{equation}
			\begin{array}{c|ccccccccccc}
				& 0 & 1 & 2 & 3 & 4 & 5 & 6 & 7 & 8 & 9 & 10  \\ \hline
				0 & 0 & 1 & 0 & 0 & 0 & 0 & 0 & 0 & 0 & 0 & 0 \\
				1 & 1 & 1 & 1 & 1 & 1 & 1 & 1 & 1 & 1 & 1 & 1 \\
				2 & 0 & 1 & 1 & 2 & 2 & 3 & 3 & 4 & 4 & 5 & 5 \\
				3 & 0 & 1 & 2 & 3 & 5 & 7 & 9 & 12 & 15 & 18 & 22 \\
				4 & 0 & 1 & 2 & 5 & 8 & 14 & 20 & 30 & 40 & 55 & 70 \\
				5 & 0 & 1 & 3 & 7 & 14 & 25 & 42 & 66 & 99 & 143 & 200 \\
				6 & 0 & 1 & 3 & 9 & 20 & 42 & 75 & 132 & 212 & 333 & 497 \\
				7 & 0 & 1 & 4 & 12 & 30 & 66 & 132 & 245 & 429 & 715 & 1144 \\
				8 & 0 & 1 & 4 & 15 & 40 & 99 & 212 & 429 & 800 & 1430 & 2424 \\
				9 & 0 & 1 & 5 & 18 & 55 & 143 & 333 & 715 & 1430 & 2700 & 4862 \\
				10 & 0 & 1 & 5 & 22 & 70 & 200 & 497 & 1144 & 2424 & 4862 & 9225 \\
			\end{array}
			\nonumber
			\end{equation}
			
			\subsubsection{$A_{n,m}^{inv}$}
			
			The $A_{n,m}^{inv}$ count matrix words (up to cyclic rotations) which don't change when reversed (up to cyclic rotations). They are defined by
			\begin{equation}
			A_{n,m}^{inv} = \binom{\lfloor \frac{n}{2} \rfloor + \lfloor \frac{m}{2} \rfloor}{\lfloor \frac{n}{2} \rfloor}
			\nonumber
			\end{equation}
			Their generating function is
			\begin{equation}
			f_{inv} (x,y) = \frac{x^2+xy+y^2+x+y}{1-x^2-y^2}
			\nonumber
			\end{equation}
			which is the Hilbert series for the vector space $T_{ST;inv}$. The plethystic exponential is
			\begin{equation}
			F_{inv}(x,y) = \prod_{n,m} \frac{1}{(1-x^n y^m)^{A_{n,m}^{inv}}} = \prod_{k=1}^\infty \text{exp} \left[ \frac{x^{2k}+x^k y^k + y^{2k} + x^k + y^k}{k(1-x^{2k}-y^{2k})} \right]
			\nonumber
			\end{equation}
			which is the Hilbert series for the vector space $T_{inv} = \operatorname{Sym}\left( T_{ST;inv} \right)$
			
			The values of $A^{inv}_{n,m}$ for $n,m \leq 10$ are shown below
			
			\begin{equation}
			\begin{array}{c|ccccccccccc}
				& 0 & 1 & 2 & 3 & 4 & 5 & 6 & 7 & 8 & 9 & 10  \\ \hline
				0 & 1 & 1 & 1 & 1 & 1 & 1 & 1 & 1 & 1 & 1 & 1 \\
				1 & 1 & 1 & 1 & 1 & 1 & 1 & 1 & 1 & 1 & 1 & 1 \\
				2 & 1 & 1 & 2 & 2 & 3 & 3 & 4 & 4 & 5 & 5 & 6 \\
				3 & 1 & 1 & 2 & 2 & 3 & 3 & 4 & 4 & 5 & 5 & 6 \\
				4 & 1 & 1 & 3 & 3 & 6 & 6 & 10 & 10 & 15 & 15 & 21 \\
				5 & 1 & 1 & 3 & 3 & 6 & 6 & 10 & 10 & 15 & 15 & 21 \\
				6 & 1 & 1 & 4 & 4 & 10 & 10 & 20 & 20 & 35 & 35 & 56 \\
				7 & 1 & 1 & 4 & 4 & 10 & 10 & 20 & 20 & 35 & 35 & 56 \\
				8 & 1 & 1 & 5 & 5 & 15 & 15 & 35 & 35 & 70 & 70 & 126 \\
				9 & 1 & 1 & 5 & 5 & 15 & 15 & 35 & 35 & 70 & 70 & 126 \\
				10 & 1 & 1 & 6 & 6 & 21 & 21 & 56 & 56 & 126 & 126 & 252 \\
			\end{array}
			\nonumber
			\end{equation}
			
			\subsubsection{$a_{n,m}^{inv}$}
			
			The $a_{n,m}^{inv}$ count aperiodic matrix words (up to cyclic rotations) which don't change (up to cyclic rotations) when reversed. They are defined by
			\begin{equation}
			a_{n,m}^{inv} = \sum_{d|n,m} \mu (d) \binom{\lfloor \frac{n}{2d} \rfloor + \lfloor \frac{m}{2d} \rfloor}{\lfloor \frac{n}{2d} \rfloor}
			\nonumber
			\end{equation}
			Their generating function is
			\begin{equation}
			\bar{f}_{inv} (x,y) = \sum_{d=1}^\infty \mu(d) \frac{x^{2d}+x^d y^d + y^{2d} + x^d + y^d}{1 - x^{2d} - y^{2d}}
			\nonumber
			\end{equation}
			which is the Hilbert series for the vector space $T_{ST;inv}^{(1)}$. The plethystic exponential is
			\begin{equation}
			\bar{F}_{inv}(x,y) = \prod_{n,m} \frac{1}{(1-x^n y^m)^{a_{n,m}^{inv}}} = \prod_{k=1}^\infty \text{exp} \left[ \frac{x^{2k}+x^k y^k + y^{2k} + x^k + y^k}{k(1-x^{2k}-y^{2k})} \sum_{d|k} d \mu(d) \right]
			\nonumber
			\end{equation}
			which is the Hilbert series for the vector space $T_{inv}^{(1)} = \operatorname{Sym}\left( T_{ST;inv}^{(1)} \right)$
			
			The values of $a^{inv}_{n,m}$ for $n,m \leq 10$ are shown below
			
			\begin{equation}
			\begin{array}{c|ccccccccccc}
			& 0 & 1 & 2 & 3 & 4 & 5 & 6 & 7 & 8 & 9 & 10  \\ \hline
			0 & 0 & 1 & 0 & 0 & 0 & 0 & 0 & 0 & 0 & 0 & 0 \\
			1 & 1 & 1 & 1 & 1 & 1 & 1 & 1 & 1 & 1 & 1 & 1 \\
			2 & 0 & 1 & 1 & 2 & 2 & 3 & 3 & 4 & 4 & 5 & 5 \\
			3 & 0 & 1 & 2 & 1 & 3 & 3 & 3 & 4 & 5 & 4 & 6 \\
			4 & 0 & 1 & 2 & 3 & 4 & 6 & 8 & 10 & 12 & 15 & 18 \\
			5 & 0 & 1 & 3 & 3 & 6 & 5 & 10 & 10 & 15 & 15 & 20 \\
			6 & 0 & 1 & 3 & 3 & 8 & 10 & 17 & 20 & 32 & 33 & 53 \\
			7 & 0 & 1 & 4 & 4 & 10 & 10 & 20 & 19 & 35 & 35 & 56 \\
			8 & 0 & 1 & 4 & 5 & 12 & 15 & 32 & 35 & 64 & 70 & 120 \\
			9 & 0 & 1 & 5 & 4 & 15 & 15 & 33 & 35 & 70 & 68 & 126 \\
			10 & 0 & 1 & 5 & 6 & 18 & 20 & 53 & 56 & 120 & 126 & 245 \\
			\end{array}
			\nonumber
			\end{equation}
			
			\subsubsection{$B_{n,m}$}
			
			The $B_{n,m}$ count single traces of anti-symmetric matrices ($SO(N)$ single traces). They are defined by
			\begin{equation}
			B_{n,m} = \frac{1}{2n+2m} \sum_{d|n,m} \phi(d) \binom{\frac{n+m}{d}}{\frac{n}{d}} + \frac{(-1)^{n+m}}{2}   \binom{\lfloor \frac{n}{2} \rfloor + \lfloor \frac{m}{2} \rfloor}{\lfloor \frac{n}{2} \rfloor}
			\nonumber
			\end{equation}
			Their generating function is
			\begin{equation}
			f_{SO(N)} (x,y) = \frac{1}{2} \left[ - \sum_{d=1}^\infty \frac{\phi(d)}{d}  \log(1-x^d-y^d) + \frac{x^2+xy+y^2-x-y}{1-x^2-y^2} \right]
			\nonumber
			\end{equation}
			which is the Hilbert series for the vector space $\widetilde{T}_{ST} = T_{ST;inv;even} \oplus \widetilde{T}_{ST;var}$. The plethystic exponential is
			\begin{align}
			F_{SO(N)}(x,y) & = \prod_{n,m} \frac{1}{(1-x^n y^m)^{B_{n,m}}} \nonumber \\
			& = \prod_{k=1}^\infty \frac{1}{\sqrt{1-x^k-y^k}} \text{exp} \left[ \frac{x^{2k} + x^k y^k + y^{2k} - x^k -y^k}{2k (1 - x^{2k} - y^{2k})}  \right]
			\label{SO(N) multi-trace counting}
			\end{align}
			which is the Hilbert series for the vector space $\widetilde{T} = \operatorname{Sym} \left( \widetilde{T}_{ST} \right)$.
			
			The values of $B_{n,m}$ for $n,m \leq 10$ are shown below
			
			\begin{equation}
			\begin{array}{c|ccccccccccc}
				& 0 & 1 & 2 & 3 & 4 & 5 & 6 & 7 & 8 & 9 & 10  \\ \hline
				0 & 0 & 0 & 1 & 0 & 1 & 0 & 1 & 0 & 1 & 0 & 1 \\
				1 & 0 & 1 & 0 & 1 & 0 & 1 & 0 & 1 & 0 & 1 & 0 \\
				2 & 1 & 0 & 2 & 0 & 3 & 0 & 4 & 0 & 5 & 0 & 6 \\
				3 & 0 & 1 & 0 & 3 & 1 & 5 & 3 & 8 & 5 & 12 & 8 \\
				4 & 1 & 0 & 3 & 1 & 8 & 4 & 16 & 10 & 29 & 20 & 47 \\
				5 & 0 & 1 & 0 & 5 & 4 & 16 & 16 & 38 & 42 & 79 & 90 \\
				6 & 1 & 0 & 4 & 3 & 16 & 16 & 50 & 56 & 126 & 150 & 280 \\
				7 & 0 & 1 & 0 & 8 & 10 & 38 & 56 & 133 & 197 & 375 & 544 \\
				8 & 1 & 0 & 5 & 5 & 29 & 42 & 126 & 197 & 440 & 680 & 1282 \\
				9 & 0 & 1 & 0 & 12 & 20 & 79 & 150 & 375 & 680 & 1387 & 2368 \\
				10 & 1 & 0 & 6 & 8 & 47 & 90 & 280 & 544 & 1282 & 2368 & 4752 \\
			\end{array}
			\nonumber
			\end{equation}
			
			\subsubsection{$b_{n,m}$}			
			The $b_{n,m}$ count minimally periodic single traces of anti-symmetric matrices, or equivalently orthogonal Lyndon words. They are defined by
			\begin{equation}
			b_{n,m} = \frac{1}{2} \sum_{d | n,m} \mu (d) \left[ \frac{1}{n+m}  \binom{\frac{n+m}{d}}{\frac{n}{d}} + (-1)^{\frac{n+m}{d}} \binom{ \lfloor \frac{n}{2d} \rfloor + \lfloor \frac{m}{2d} \rfloor}{\lfloor \frac{n}{2d} \rfloor} \right]
			\nonumber
			\end{equation}
			Their generating function is
			\begin{equation}
			\bar{f}_{SO(N)} (x,y) = = \frac{1}{2} \sum_{d=1}^\infty \mu(d) \left[ -\frac{1}{d} \log (1-x^d - y^d) + \frac{x^{2d} +x^d y^d + y^{2d} - x^d - y^d}{1 - x^{2d} - y^{2d}} \right]
			\nonumber
			\end{equation}
			which is the Hilbert series for the vector space $\widetilde{T}_{ST}^{(min)} = T_{ST;inv;even}^{(1)} \oplus T_{ST;inv;odd}^{(2)} \oplus \widetilde{T}_{ST;var}^{(1)}$. The plethystic exponential is
			\begin{align*}
			\bar{F}_{SO(N)}(x,y) &  = \prod_{n,m} \frac{1}{(1-x^n y^m)^{b_{n,m}}} \\
			& = \frac{1}{\sqrt{1-x-y}} \prod_{k=1}^\infty \text{exp} \left[ \frac{1}{2k} \frac{x^{2k} + x^k y^k + y^{2k} - x^k - y^k}{1 - x^{2k} - y^{2k}} \sum_{d|k} d \mu(d) \right]
			\end{align*}
			which is the Hilbert series for the vector space $\widetilde{T}^{(min)} = \operatorname{Sym}\left( \widetilde{T}_{ST}^{(min)} \right)$.
			
			The values of $b_{n,m}$ for $n,m \leq 10$ are shown below
			
			\begin{equation}
			\begin{array}{c|ccccccccccc}
			& 0 & 1 & 2 & 3 & 4 & 5 & 6 & 7 & 8 & 9 & 10  \\ \hline
			0 & 0 & 0 & 1 & 0 & 0 & 0 & 0 & 0 & 0 & 0 & 0 \\
			1 & 0 & 1 & 0 & 1 & 0 & 1 & 0 & 1 & 0 & 1 & 0 \\
			2 & 1 & 0 & 1 & 0 & 3 & 0 & 3 & 0 & 5 & 0 & 5 \\
			3 & 0 & 1 & 0 & 2 & 1 & 5 & 3 & 8 & 5 & 11 & 8 \\
			4 & 0 & 0 & 3 & 1 & 6 & 4 & 16 & 10 & 26 & 20 & 47 \\
			5 & 0 & 1 & 0 & 5 & 4 & 15 & 16 & 38 & 42 & 79 & 90 \\
			6 & 0 & 0 & 3 & 3 & 16 & 16 & 46 & 56 & 125 & 150 & 275 \\
			7 & 0 & 1 & 0 & 8 & 10 & 38 & 56 & 132 & 197 & 375 & 544 \\
			8 & 0 & 0 & 5 & 5 & 26 & 42 & 125 & 197 & 432 & 680 & 1278 \\
			9 & 0 & 1 & 0 & 11 & 20 & 79 & 150 & 375 & 680 & 1384 & 2368 \\
			10 & 0 & 0 & 5 & 8 & 47 & 90 & 275 & 544 & 1278 & 2368 & 4735 \\
			\end{array}
			\nonumber
			\end{equation}
			
			\subsubsection{$b_{n,m}^{(odd)}$}
			
			The $b_{n,m}^{(odd)}$ count single traces of anti-symmetric matrices with a specified odd number of periods. Note that $n,m$ refer to the number of $X$s and $Y$s contained in the aperiodic root of the trace, rather than in the whole trace. They are defined by
			\begin{equation}
			b_{n,m}^{(odd)} = \frac{1}{2} \sum_{d|n,m} \mu(d) \left[ \frac{1}{n+m} \binom{\frac{n+m}{d}}{\frac{n}{d}} + (-1)^{n+m} \binom{\lfloor \frac{n}{2d} \rfloor + \lfloor \frac{m}{2d} \rfloor}{\lfloor \frac{n}{2d} \rfloor} \right]
			\nonumber
			\end{equation}
			Their generating function is
			\begin{align*}
			\bar{f}_{SO(N)}^{(odd)} (x,y) & = \frac{1}{2} \sum_{d=1}^\infty \mu(d)  \left[- \frac{1}{d} \log (1-x^d - y^d) \right. \\ & \hspace{5cm} \left. + \frac{x^{2d}+x^d y^d + y^{2d} + (-x)^d + (-y)^d}{1 - x^{2d} - y^{2d}}   \right]
			\end{align*}
			which is the Hilbert series for the vector space $\widetilde{T}_{ST}^{(odd)} = T_{ST;inv;even}^{(1)} \oplus \widetilde{T}_{ST;var}^{(1)}$. The plethystic exponential is
			\begin{align*}
			\bar{F}_{SO(N)}^{(odd)}(x,y) & = \prod_{n,m} \frac{1}{(1-x^n y^m)^{b_{n,m}^{(odd)}}} \\
			& = \frac{1}{\sqrt{1-x-y}} \prod_{k=1}^\infty \text{exp} \left[ \sum_{d|k} \frac{d\mu(d)}{2k} \frac{x^{2k} + x^k y^k + y^{2k} + (-1)^d (x^k + y^k)}{1- x^{2k} - y^{2k}} \right]
			\end{align*}
			which is the Hilbert series for the vector space $\widetilde{T}^{(odd)} = \operatorname{Sym} \left( \widetilde{T}_{ST}^{(odd)} \right)$.
			
			The values of $b_{n,m}^{(odd)}$ for $n,m \leq 10$ are shown below
			
			\begin{equation}
			\begin{array}{c|ccccccccccc}
			& 0 & 1 & 2 & 3 & 4 & 5 & 6 & 7 & 8 & 9 & 10  \\ \hline
			0 & 0 & 0 & 0 & 0 & 0 & 0 & 0 & 0 & 0 & 0 & 0 \\
			1 & 0 & 1 & 0 & 1 & 0 & 1 & 0 & 1 & 0 & 1 & 0 \\
			2 & 0 & 0 & 1 & 0 & 2 & 0 & 3 & 0 & 4 & 0 & 5 \\
			3 & 0 & 1 & 0 & 2 & 1 & 5 & 3 & 8 & 5 & 11 & 8 \\
			4 & 0 & 0 & 2 & 1 & 6 & 4 & 14 & 10 & 26 & 20 & 44 \\
			5 & 0 & 1 & 0 & 5 & 4 & 15 & 16 & 38 & 42 & 79 & 90 \\
			6 & 0 & 0 & 3 & 3 & 14 & 16 & 46 & 56 & 122 & 150 & 275 \\
			7 & 0 & 1 & 0 & 8 & 10 & 38 & 56 & 132 & 197 & 375 & 544 \\
			8 & 0 & 0 & 4 & 5 & 26 & 42 & 122 & 197 & 432 & 680 & 1272 \\
			9 & 0 & 1 & 0 & 11 & 20 & 79 & 150 & 375 & 680 & 1384 & 2368 \\
			10 & 0 & 0 & 5 & 8 & 44 & 90 & 275 & 544 & 1272 & 2368 & 4735 \\
			\end{array}
			\nonumber
			\end{equation}
			
			\subsubsection{$b_{n,m}^{(even)}$}
			
			The $b_{n,m}^{(even)}$ count single traces of anti-symmetric matrices with a specified even number of periods. Note that $n,m$ refer to the number of $X$s and $Y$s contained in the aperiodic root of the trace, rather than in the whole trace. They are defined by
			\begin{equation}
			b_{n,m}^{(even)} = \frac{1}{2} \sum_{d|n,m} \mu(d) \left[ \frac{1}{n+m} \binom{\frac{n+m}{d}}{\frac{n}{d}} + \binom{\lfloor \frac{n}{2d} \rfloor + \lfloor \frac{m}{2d} \rfloor}{\lfloor \frac{n}{2d} \rfloor} \right]
			\nonumber
			\end{equation}
			Their generating function is
			\begin{equation}
			\bar{f}_{SO(N)}^{(even)} (x,y) = \frac{1}{2} \sum_{d=1}^\infty \mu(d) \left[ -\frac{1}{d} \log (1-x^d - y^d) + \frac{x^{2d}+x^d y^d + y^{2d} + x^d + y^d}{1 - x^{2d} - y^{2d}}   \right]
			\nonumber
			\end{equation}
			which is the Hilbert series for the vector space $\widetilde{T}_{ST}^{(even)} = T_{ST;inv;even}^{(1)} \oplus T_{ST;inv;odd}^{(1)} \oplus \widetilde{T}_{ST;var}^{(1)}$. The plethystic exponential is
			\begin{align*}
			\bar{F}_{SO(N)}^{(even)} (x,y) & = \prod_{n,m} \frac{1}{(1-x^n y^m)^{ b_{n,m}^{(even)} }} \\
			& = \frac{1}{\sqrt{1-x-y}} \prod_{k=1}^\infty \text{exp} \left[ \frac{1}{2k} \frac{x^{2k} + x^k y^k + y^{2n} + x^k + y^k}{1-x^{2k}-y^{2k}} \sum_{d|k} d \mu(d) \right]
			\end{align*}
			which is the Hilbert series for the vector space $\widetilde{T}^{(even)} = \operatorname{Sym} \left( \widetilde{T}_{ST}^{(even)} \right)$.
			
			The values of $b_{n,m}^{(even)}$ for $n,m \leq 10$ are shown below
			
			\begin{equation}
			\begin{array}{c|ccccccccccc}
			& 0 & 1 & 2 & 3 & 4 & 5 & 6 & 7 & 8 & 9 & 10  \\ \hline
			0 & 0 & 1 & 0 & 0 & 0 & 0 & 0 & 0 & 0 & 0 & 0 \\
			1 & 1 & 1 & 1 & 1 & 1 & 1 & 1 & 1 & 1 & 1 & 1 \\
			2 & 0 & 1 & 1 & 2 & 2 & 3 & 3 & 4 & 4 & 5 & 5 \\
			3 & 0 & 1 & 2 & 2 & 4 & 5 & 6 & 8 & 10 & 11 & 14 \\
			4 & 0 & 1 & 2 & 4 & 6 & 10 & 14 & 20 & 26 & 35 & 44 \\
			5 & 0 & 1 & 3 & 5 & 10 & 15 & 26 & 38 & 57 & 79 & 110 \\
			6 & 0 & 1 & 3 & 6 & 14 & 26 & 46 & 76 & 122 & 183 & 275 \\
			7 & 0 & 1 & 4 & 8 & 20 & 38 & 76 & 132 & 232 & 375 & 600 \\
			8 & 0 & 1 & 4 & 10 & 26 & 57 & 122 & 232 & 432 & 750 & 1272 \\
			9 & 0 & 1 & 5 & 11 & 35 & 79 & 183 & 375 & 750 & 1384 & 2494 \\
			10 & 0 & 1 & 5 & 14 & 44 & 110 & 275 & 600 & 1272 & 2494 & 4735 \\
			\end{array}
			\nonumber
			\end{equation}

			\subsection{Multi-trace sequences}
			
			\subsubsection{$N^{U(N)}_{n,m}$}

			The $N^{U(N)}_{n,m}$ count the multi-traces of generic $N \times N$ matrices, where $N$ can be finite or infinite. They are defined by			
			\begin{equation}
			N^{U(N)}_{n,m} = \sum_{\substack{
					R \vdash n \\
					S \vdash m \\
					T \vdash n+m \\
					l(T) \leq N}}
			g_{R,S;T}^2
			\nonumber
			\end{equation}
			At infinite $N$, $A_{n,m}$ and $N^{U(N)}_{n,m}$ are related by the plethystic exponential, so the generating function is given by \eqref{U(N) multi-trace counting}, which is the Hilbert series for $T$.
			
			The values of $N^{U(\infty)}_{n,m}$ for $n,m \leq 10$ are shown below
			
			\fontsize{11pt}{12pt}
			
			\begin{equation}
			\begin{array}{c|ccccccccccc}
			& 0 & 1 & 2 & 3 & 4 & 5 & 6 & 7 & 8 & 9 & 10  \\ \hline
			0 & 1 & 1 & 2 & 3 & 5 & 7 & 11 & 15 & 22 & 30 & 42 \\
			1 & 1 & 2 & 4 & 7 & 12 & 19 & 30 & 45 & 67 & 97 & 139 \\
			2 & 2 & 4 & 10 & 18 & 34 & 56 & 94 & 146 & 228 & 340 & 506 \\
			3 & 3 & 7 & 18 & 38 & 74 & 133 & 233 & 385 & 623 & 977 & 1501 \\
			4 & 5 & 12 & 34 & 74 & 158 & 297 & 550 & 951 & 1614 & 2627 & 4202 \\
			5 & 7 & 19 & 56 & 133 & 297 & 602 & 1166 & 2133 & 3775 & 6437 & 10692 \\
			6 & 11 & 30 & 94 & 233 & 550 & 1166 & 2382 & 4551 & 8424 & 14953 & 25835 \\
			7 & 15 & 45 & 146 & 385 & 951 & 2133 & 4551 & 9142 & 17639 & 32680 & 58659 \\
			8 & 22 & 67 & 228 & 623 & 1614 & 3775 & 8424 & 17639 & 35492 & 68356 & 127443 \\
			9 & 30 & 97 & 340 & 977 & 2627 & 6437 & 14953 & 32680 & 68356 & 136936 & 264747 \\
			10 & 42 & 139 & 506 & 1501 & 4202 & 10692 & 25835 & 58659 & 127443 & 264747 & 530404 \\
			\end{array}
			\nonumber
			\end{equation}
			
			\normalsize

			\subsubsection{$N^{SO(N); \delta}_{n,m}$}

			The $N^{SO(N); \delta}_{n,m}$ count the multi-traces of anti-symmetric $N \times N$ matrices, where $N$ can be finite or infinite. They are defined by			
			\begin{equation}
			N^{SO(N); \delta}_{n,m} = \sum_{ \substack{
					R \vdash 2n \text{ with even column lengths} \\
					S \vdash 2m \text{ with even column lengths} \\
					T \vdash 2n+2m \text{ with even row lengths} \\
					l(T) \leq N } }
			g_{R,S;T}
			\nonumber
			\end{equation}
			At infinite $N$, $B_{n,m}$ and $N^{SO(N); \delta}_{n,m}$ are related by the plethystic exponential, so the generating function is given by \eqref{SO(N) multi-trace counting}, which is the Hilbert series for $\widetilde{T}$.
			
			The values of $N^{SO(\infty); \delta}_{n,m}$ for $n,m \leq 10$ are shown below

			\begin{equation}
			\begin{array}{c|ccccccccccc}
			& 0 & 1 & 2 & 3 & 4 & 5 & 6 & 7 & 8 & 9 & 10  \\ \hline
			0 & 1 & 0 & 1 & 0 & 2 & 0 & 3 & 0 & 5 & 0 & 7 \\
			1 & 0 & 1 & 0 & 2 & 0 & 4 & 0 & 7 & 0 & 12 & 0 \\
			2 & 1 & 0 & 4 & 0 & 9 & 0 & 19 & 0 & 35 & 0 & 62 \\
			3 & 0 & 2 & 0 & 9 & 1 & 23 & 4 & 52 & 10 & 105 & 22 \\
			4 & 2 & 0 & 9 & 1 & 33 & 6 & 85 & 21 & 198 & 56 & 410 \\
			5 & 0 & 4 & 0 & 23 & 6 & 86 & 33 & 243 & 114 & 600 & 313 \\
			6 & 3 & 0 & 19 & 4 & 85 & 33 & 297 & 152 & 845 & 512 & 2137 \\
			7 & 0 & 7 & 0 & 52 & 21 & 243 & 152 & 879 & 664 & 2646 & 2227 \\
			8 & 5 & 0 & 35 & 10 & 198 & 114 & 845 & 664 & 3003 & 2742 & 9168 \\
			9 & 0 & 12 & 0 & 105 & 56 & 600 & 512 & 2646 & 2742 & 9702 & 11033 \\
			10 & 7 & 0 & 62 & 22 & 410 & 313 & 2137 & 2227 & 9168 & 11033 & 33704 \\
			\end{array}
			\nonumber
			\end{equation}

			\subsection{Relations between different sequences}
			
			The $a_{n,m}$ are the M{\"o}bius transform of the $A_{n,m}$.
			\begin{align*}
			A_{n,m} & = \sum_{d|n,m} a_{\frac{n}{d}, \frac{m}{d}} & a_{n,m} & = \sum_{d|n,m} \mu(d) A_{\frac{n}{d}, \frac{m}{d}} \\
			f_{U(N)}(x,y) & = \sum_{k=1}^\infty \bar{f}_{U(N)} (x^k, y^k) & \bar{f}_{U(N)} (x,y) & = \sum_{k=1}^\infty \mu(k) f_{U(N)} (x^k,y^k) \\
			F_{U(N)}(x,y) & = \prod_{k=1}^\infty \bar{F}_{U(N)} (x^k, y^k) & \bar{F}_{U(N)} (x,y) & = \prod_{k=1}^\infty  F_{U(N)} (x^k,y^k)^{\mu(k)}
			\end{align*}
			The $a_{n,m}^{inv}$ are the M{\"o}bius transform of the $A_{n,m}^{inv}$.
			\begin{align*}
			A_{n,m}^{inv} & = \sum_{d|n,m} a_{\frac{n}{d}, \frac{m}{d}}^{inv} & a_{n,m}^{inv} & = \sum_{d|n,m} \mu(d) A_{\frac{n}{d}, \frac{m}{d}}^{inv} \\
			f_{inv}(x,y) & = \sum_{k=1}^\infty \bar{f}_{inv} (x^k, y^k) & \bar{f}_{inv} (x,y) & = \sum_{k=1}^\infty \mu(k) f_{inv} (x^k,y^k) \\
			F_{inv}(x,y) & = \prod_{k=1}^\infty \bar{F}_{inv} (x^k, y^k) & \bar{F}_{inv} (x,y) & = \prod_{k=1}^\infty  F_{inv} (x^k,y^k)^{\mu(k)}
			\end{align*}
			The $B_{n,m}$ can be expressed in terms of the $A_{n,m}$ and the $A_{n,m}^{inv}$.
			\begin{align*}
			B_{n,m} & = \frac{1}{2} \left[ A_{n,m} + (-1)^{n+m} A_{n,m}^{inv} \right] \\
			f_{SO(N)}(x,y) & = \frac{1}{2} \left[ f_{U(N)}(x,y) + f_{inv}(-x,-y) \right]
			\end{align*}
			The $b_{n,m}$ are the M{\"o}bius transform of the $B_{n,m}$.
			\begin{align*}
			B_{n,m} & = \sum_{d|n,m} b_{\frac{n}{d}, \frac{m}{d}} & b_{n,m} & = \sum_{d|n,m} \mu(d) B_{\frac{n}{d}, \frac{m}{d}} \\
			f_{SO(N)}(x,y) & = \sum_{k=1}^\infty \bar{f}_{SO(N)} (x^k, y^k) & \bar{f}_{SO(N)} (x,y) & = \sum_{k=1}^\infty \mu(k) f_{SO(N)} (x^k,y^k) \\
			F_{SO(N)}(x,y) & = \prod_{k=1}^\infty \bar{F}_{SO(N)} (x^k, y^k) & \bar{F}_{SO(N)} (x,y) & = \prod_{k=1}^\infty  F_{SO(N)} (x^k,y^k)^{\mu(k)}
			\end{align*}
			The $b_{n,m}^{(odd)}$ and $b_{n,m}^{(even)}$ can be expressed in terms of the $a_{n,m}$ and the $a_{n,m}^{inv}$.
			\begin{align*}
			b_{n,m}^{(odd)} & = \frac{1}{2} \left[ a_{n,m} + (-1)^{n+m} a_{n,m}^{inv} \right] & b_{n,m}^{(even)} & = \frac{1}{2} \left[ a_{n,m} + a_{n,m}^{inv} \right] \\
			\bar{f}_{SO(N)}^{(odd)} (x,y) & = \frac{1}{2} \left[ \bar{f}_{U(N)}(x,y) + \bar{f}_{inv}(-x,-y) \right] & & \\ & & 
			\bar{f}_{SO(N)}^{(even)} (x,y) & = \frac{1}{2} \left[ \bar{f}_{U(N)}(x,y) + \bar{f}_{inv}(x,y) \right]
			\end{align*}

	\section{Jucys-Murphy elements}
	\label{section: Jucys-Murphy elements}
	
	Take a Young diagram $R \vdash n$, and label each box by their row and column number, where the top left box is $(1,1)$ and numbers increase to the right and downwards. Then for the box $i = (r,c)$, we define the content of that box to be $c_i = c-r$. For example, the contents of $R = [4,4,2,2], [8,4,2]$ and $[2,2,2,2,2,2]$ are shown below
	\begin{equation}
	\begin{ytableau}
	0 & 1 & 2 & 3 \\
	-1 & 0 & 1 & 2 \\
	-2 & -1 \\
	-3 & -2
	\end{ytableau}
	\qquad
	\begin{ytableau}
	0 & 1 & 2 & 3 & 4 & 5 & 6 & 7 \\
	-1 & 0 & 1 & 2 \\
	-2 & -1
	\end{ytableau}
	\qquad
	\begin{ytableau}
	0 & 1 \\
	-1 & 0 \\
	-2 & -1 \\
	-3 & -2 \\
	-4 & -3 \\
	-5 & -4
	\end{ytableau}
	\label{content definition}		
	\end{equation}	
	The contents of a box relate to the eigenvalues of a certain set elements of $\mathbb{C}(S_n)$ called the Jucys-Murphy elements, defined by
	\begin{equation}
	J_k = \sum_{i=1}^{k-1} \left( i, k \right)
	\label{Jucys-Murphy elements}
	\end{equation}
	These span a maximal commuting sub-algebra of $\mathbb{C}(S_n)$, and therefore one can choose a basis of any irreducible representation to be eigenvectors of the $J_k$. To describe this basis, we first recall the definition of a standard Young tableaux.
	
	For a Young diagram $R \vdash n$, a Young tableaux of shape $R$ is produced by placing a positive integer into each box of $R$. The tableaux is called semi-standard if the numbers increase weakly along the rows and strictly down the columns. It is called standard if in addition the $n$ integers are the numbers $1$ to $n$. For example, the possible standard Young tableaux of shape $R = [3,2]$ are
	\begin{gather*}
	\begin{ytableau} 1 & 2 & 3 \\ 4 & 5 \end{ytableau} 
	\qquad 
	\begin{ytableau} 1 & 2 & 4 \\ 3 & 5 \end{ytableau}
	\qquad 
	\begin{ytableau} 1 & 2 & 5 \\ 3 & 4 \end{ytableau} 
	\qquad 
	\begin{ytableau} 1 & 3 & 4 \\ 2 & 5 \end{ytableau} 
	\qquad 
	\begin{ytableau} 1 & 3 & 5 \\ 2 & 4 \end{ytableau} 
	\end{gather*}	
	In an irrep $R \vdash n$ of $S_n$, the basis of eigenvectors for the Jucys-Murphy elements are labelled by the standard Young tableaux of shape $R$. Consider such a tableau $r$. Then the eigenvalue of $|r \rangle$ under $J_k$ is the content of the box containing $k$ in $r$. So for example if we have $R = [3,2,1]$
	the contents of the cells are
	\begin{equation}
	\begin{ytableau}
	0 & 1 & 2 \\
	-1 & 0 \\
	-2
	\end{ytableau}
	\nonumber
	\end{equation}
	so the eigenvalues of the Jucys-Murphy elements on 4 of the 16 different standard Young tableaux are
	\begin{center}
		\begin{tabular}{c | c | c | c | c}
			& \begin{ytableau} 1 & 2 & 3 \\ 4 & 5 \\ 6	\end{ytableau}
			& \begin{ytableau} 1 & 3 & 5 \\ 2 & 6 \\ 4	\end{ytableau} 
			& \begin{ytableau} 1 & 2 & 6 \\ 3 & 4 \\ 5	\end{ytableau} 
			& \begin{ytableau} 1 & 4 & 6 \\ 2 & 5 \\ 3  \end{ytableau} \\[5ex] \hline
			$J_2$ & 1 & -1 & 1 & -1 \\
			$J_3$ & 2 & 1 & -1 & -2 \\
			$J_4$ & -1 & -2 & 0 & 1 \\
			$J_5$ & 0 & 2 & -2 & 0 \\
			$J_6$ & -2 & 0 & 2 & 2
		\end{tabular}
	\end{center}		
	We will be particularly interested in the product
	\begin{equation}
	\Omega = \prod_{i=1}^n \left( N + J_i \right)
	\label{omega from Jucys-Murphy elements}
	\end{equation}
	It is a standard result, see for example \cite{Zinn-Justin2010}, that this can also be written
	\begin{equation}
	\Omega = \prod_{i=1}^n \left( N + J_i \right) = \sum_{\sigma \in S_n} N^{c(\sigma)} \sigma
	\label{omega definition}
	\end{equation}
	where $c(\sigma)$ is the number of cycles in $\sigma$. From this second expression we can see that $\Omega$ is in the centre of the $\mathbb{C}(S_n)$.
	\begin{align*}
	\alpha \Omega \alpha^{-1} & = \sum_{\sigma \in S_n} N^{c(\sigma)} \alpha \sigma \alpha^{-1} \\
	& = \sum_{\sigma \in S_n} N^{c( \alpha^{-1} \sigma \alpha)} \sigma \\
	& = \Omega
	\end{align*}
	Therefore, by Schur's lemma, in any irrep of $S_n$ the representative of $\Omega$ is proportional to the identity. To find the constant of proportionality, consider $\Omega$ acting on a standard Young tableaux $r$ of shape $R \vdash n$. Since the product in \eqref{omega from Jucys-Murphy elements} includes all the Jucys-Murphy elements, the content of every box will be picked up. Therefore
	\begin{equation}
	D^R \left( \Omega \right) | r \rangle = \left[ \prod_{\substack{i \in \text{ boxes}\\ \text{of }R}} (N+c_i) \right] | r \rangle
	\nonumber
	\end{equation}
	As expected, the eigenvalue of $\Omega$ on $r$ does not depend on $r$, only on the irrep $R$. So we have 
	\begin{equation}
	D^R \left( \Omega \right) = \prod_{\substack{i \in \text{ boxes}\\ \text{of }R}} (N+c_i) 
	\label{omega in an irrep}
	\end{equation}
	Another important result, similar in spirit to \eqref{omega definition}, is as follows: consider $S_n[S_2]$ as a subgroup of $S_{2n}$. Then one can choose the left coset representatives of this subgroup such that
	\begin{equation}
	\sum_{\substack{\beta \text{ left coset} \\ \text{representatives}}} C^{(\delta)}_I \beta^I_J C^{(\delta) \, J} \beta = \prod_{i=1}^n \left( N + J_{2i-1} \right)
	\label{mesonic Jucys-Murphy result}
	\end{equation}
	This is the key result behind the evaluation of the mesonic correlator in \cite{Caputa2013,Kemp2014}, and is proved inductively in \cite{Zinn-Justin2010}.

	\section{Alternative derivation of baryonic correlator}
	\label{section: alternative baryonic correlator}

	To evaluate \eqref{baryonic correlator intermediary step 1} explicitly, we first look at how the equivalent mesonic calculation was performed in \cite{Kemp2014}. The mesonic starting point is \eqref{mesonic correlator intermediary step}, and the first step is to split the sum over $S_{2n+2m}$ into two, one over the subgroup $S_{n+m}[S_2]$ and the other over the (left) coset representatives of said subgroup. The invariance properties of $|T \rangle$ and $C_I^{(\delta)}$ mean the subgroup sum becomes trivial and merely contributes a normalisation factor. The sum over the coset representatives is evaluated using \eqref{mesonic Jucys-Murphy result}. Finally the correlator is found by evaluating the product of Jucys-Murphy elements on the vector $| T \rangle$.
	
	This process works for the baryonic case, but with subgroup $S_N \times S_q[S_2]$ (where $q= n+m - \frac{N}{2}$) instead of $S_{n+m}[S_2]$. Splitting the sum as before, we see that both $|1^N+T\rangle$ and $C^{(\varepsilon)}$ are invariant (up to minus signs, which cancel) under $S_N \times S_q[S_2]$, so the subgroup sum becomes trivial and just produces a factor of $N! 2^q q!$. Writing $C(\beta)$ for $C^{(\varepsilon)}_I \beta^I_J C^{(\varepsilon)J}$, the sum we are left with is
	\begin{align}
	&\langle \mathcal{O}^\varepsilon_{T,R,S,\lambda} \overline{\mathcal{O}}^\varepsilon_{T', R', S', \lambda'} \rangle \nonumber \\ & = \delta_{T T'} \delta_{R R'} \delta_{S S'} \delta_{\lambda \lambda'} \frac{d_{1^N+T} 2^{n+m+q} n! m! N! q! }{(2n+2m)!} \langle 1^N+T | D^{1^N+T}\left( \sum_{\substack{\beta \text{ left coset} \\ \text{representatives}}} \hspace{-18pt} C(\beta) \beta \right) |1^N+T\rangle
	\label{sum over coset reps}
	\end{align}
	Since $\beta$ are the coset representatives of $S_N \times S_q[S_2]$, we cannot use \eqref{mesonic Jucys-Murphy result}. Instead we prove a generalisation, \eqref{intermediary claim}, that performs the same role, expressing the sum over coset representatives in terms of Jucys-Murphy elements. The majority of this section is taken up by proving this result, following the methods used by \cite{Zinn-Justin2010} in proving \eqref{omega definition} and \eqref{mesonic Jucys-Murphy result}. After the proof, we then determine how the product of Jucys-Murphy elements acts on the vector $|1^N+T\rangle$.

	Since we are using left coset representatives, a generic element $\sigma \in S_{2n+2m}$ can be written $\sigma = \beta \tau$ where $\beta$ is the coset rep and $\tau \in S_N \times S_q[S_2]$. The cosets are labelled by a choice of $q$ pairs from $\{ 1,2, \ldots ,N+2q \}$. Let the set of all such choices be $P_q$. An element of $p \in P_q$ then has the form
	\begin{equation}
	p = \Big\{ \{ i_{1,1}, i_{1,2} \} , \{ i_{2,1} , i_{2,2} \}, \ldots ,\{ i_{q,1}, i_{q,2} \} \Big\}
	\nonumber
	\end{equation} 
	The coset representative for $p$ could be any permutation $\beta_{p}$ satisfying
	\begin{align}
	\beta_{p} (p) = \Big\{ \{ N+1, N+2 \} , \{ N+3 , N+4 \}, \ldots ,\{ N+2q-1, N+2q \} \Big\}
	\label{coset rep conditions}
	\end{align}
	Using this notation for the cosets, we propose \\
	\ \\
	\textbf{Proposition} \\
	\nopagebreak \\
	\nopagebreak
	The coset representatives $\beta_{p}$ can be chosen such that
	\begin{align}
	\sum_{p \in P_q} C(\beta_{p}) \beta_{p} & = N! \prod_{i=1}^q \left[ N + \sum_{j=1}^N (j, N+2i-1) + \sum_{j=1}^N (j,N+2i) \right. \nonumber \\ & \hspace{7.5cm} \left. + \sum_{j=1}^{2i-2} (N+j, N+2i-1) \right] \nonumber \\
	& = N! \prod_{i=1}^q \left[ N + J_{N+2i-1} + (N+1,N+2i) J_{N+1} (N+1,N+2i) \right]
	\label{intermediary claim}
	\end{align}
	where the product is ordered $[i=1][i=2] \ldots [i=q]$. \\ \ \\
	\textbf{Proof} \\
	\nopagebreak \\
	\nopagebreak	
	We prove this by induction on $q$ at fixed $N$, following the example of \cite{Zinn-Justin2010}. First we consider the base case with $q=1$. The possible $p$, along with the associated $\beta_{p}$ and $C(\beta_{p})$ are
	
\begin{figure}
		\centering
		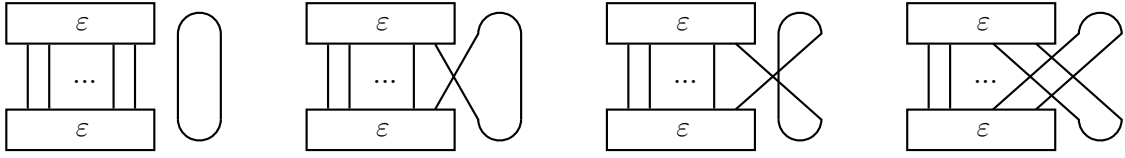
		\caption{Diagrammatic calculation of $C(\beta)$ for $\beta = 1, (N,N+1) , (N,N+2) $ and $ (N-1,N+1)(N,N+2)$ respectively. Two $\varepsilon$s fully contracted contribute $\varepsilon_{i_1 \ldots i_N} \varepsilon^{i_1 \ldots i_N} = N!$ while a loop gives $\delta_{i j} \delta^{i j} = N$. Since $\varepsilon$ is anti-symmetric and $\delta$ is symmetric, a contraction between the two gives 0.}
		\label{figure: q=1 contractions}
	\end{figure}

	\begin{center}
	\begin{tabular}{c | c c c c }
	$ p $ & $ \{ N+1, N+2 \} $ & $ \{ k, N+1 \} $ & $ \{ k, N+2 \} $ & $ \{ l_1, l_2 \} $  \\ \hline
	$ \beta_{p} $ & $1$ & $(k,N+2)$ & $(k,N+1)$ & $(l_1,N+1)(l_2,N+2)$ \\
	$C(\beta_{p})$ & $ N! N$ & $N!$ & $N!$ & 0
	\end{tabular} 
	\end{center}
	where $1 \leq k, l_1,l_2 \leq N$ and $l_1,l_2$ are distinct. It is simple to check that these $\beta_{p}$ satisfy the conditions in \eqref{coset rep conditions} and therefore serve as coset representatives. The calculations for $C(\beta_p)$ are shown diagrammatically in figure \ref{figure: q=1 contractions}. For simplicity the figure shows $k=N$ in the pairings $p = \{ k, N+1 \} , \{ k, N+2 \}$ and $l_1=N-1, l_2=N$ in the pairing $p = \{ k_1 , k_2 \}$, but it is clear that the results hold for all $k,l_1,l_2$.
	
	So we have
	\begin{equation}
	\sum_{p \in P_1} C(\beta_{p}) \beta_{p} = N! \left[ N + \sum_{j=1}^N (j,N+1) + \sum_{j=1}^N (j,N+2) \right]
	\nonumber
	\end{equation}
	as claimed in \eqref{intermediary claim}.

	Now assume the claim is true for $q-1$. In particular this means that there is a map from $P_{q-1} \rightarrow S_{N+2q-2}$, namely $p \rightarrow \beta_p$, such that for each $p$, $\beta_p$ satisfies \eqref{coset rep conditions}, and the $\beta_p$ combine so as to satisfy \eqref{intermediary claim}.
	
	Now we consider the case at $q$. The pairings $p \in P_q$ fall into 5 categories depending on how $N+2q-1$ and $N+2q$ pair (or don't pair) up with the first $N+2q-2$ numbers.
	
	\begin{enumerate}
	\item $\{ N+2q-1, N+2q \}$ is a pair
	
	\item $\{k_1,N+2q-1\}$ and $\{k_2,N+2q\}$ are pairs, for some $k_1,k_2 < N+2q-1, k_1 \neq k_2$
	
	\item $N+2q$ is unpaired and $\{k, N+2q-1 \}$ is a pair, for some $k < N+2q-1$
	
	\item $N+2q-1$ is unpaired and $\{k, N+2q \}$ is a pair, for some $k < N+2q-1$
	
	\item $N+2q-1$ and $N+2q$ are both unpaired
	\end{enumerate}
	We now split up the sum over $P_q$ into five sums, one for each type of pairing.

	\ \\
	\noindent \textbf{Type 1} \\
	\nopagebreak \\
	\nopagebreak
	\noindent Let $P_{q;1}$ be the set of pairings that are of type 1. Given $p \in P_{q,1}$, first note that $p$ reduces uniquely to a $\bar{p} \in P_{q-1}$ given by $\bar{p} = p \backslash \{ N+2q-1, N+2q \}$. Using this $\bar{p}$, we choose the coset representative of $p$ to be
	\begin{equation}
	\beta_p = \beta_{\bar{p}}
	\nonumber
	\end{equation}
	By which we mean that $\beta_p$ acts as $\beta_{\bar{p}}$ on $\{ 1,2, \ldots ,N+2q-2 \}$ and as the identity on $\{ N+2q-1,N+2q\}$. It is simple to check that this satisfies the conditions \eqref{coset rep conditions}.
	
\begin{figure}
		\centering
		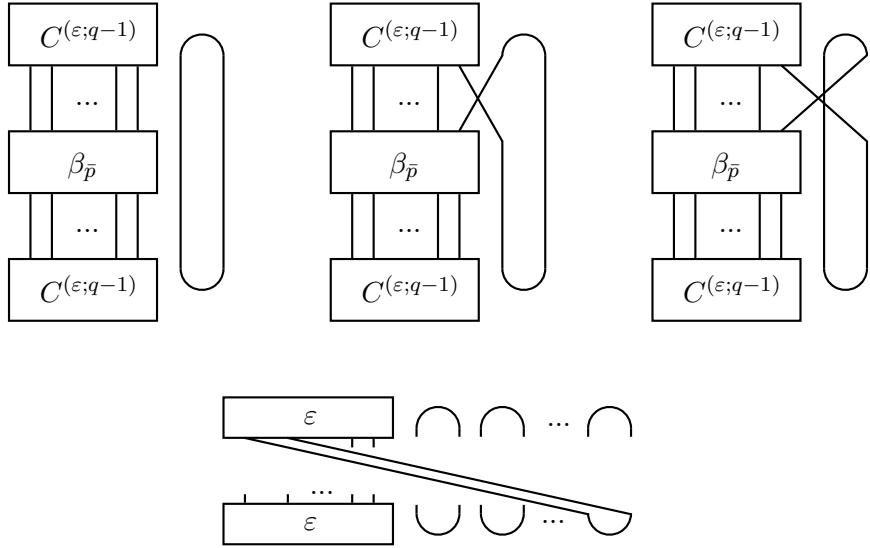
		\caption{Diagrammatic calculation of $C(\beta)$ for various $\beta \in S_{N+2q}$. The top row shows $\beta = \beta_{\bar{p}} , (N+2q-2,N+2q-1) \beta_{\bar{p}}$ and $ (N+2q-2,N+2q) \beta_{\bar{p}}$ respectively, where $\beta_{\bar{p}} \in S_{N+2q-2}$. The bottom row shows a $\beta$ with $\beta (N+2q-1) = 1$ and $\beta (N+2q) = 2$. These two values of $\beta$ are enough to ensure $C(\beta)=0$, so the remaining parts of $\beta$ are not included in the diagram.}
		\label{figure: inductive contractions}
	\end{figure}

	To calculate $C( \beta_p )$, add an extra label $q$ onto the contractor $C^{(\varepsilon)}$ to record how many indices it has. So $C^{(\varepsilon ; q)}$ will have $N+2q$ indices, the first $N$ in an $\varepsilon$ and the remaining $2q$ in $q$ $\delta$s. Then we can write $C^{(\varepsilon ; q)}_{i_1 \ldots i_{N+2q}} = C^{(\varepsilon ; q-1)}_{i_1 \ldots i_{N+2q-2}} \delta_{i_{N+2q-1} i_{N+2q}}$. Since $C^{(\varepsilon ; q)}_{i_1 \ldots i_{N+2q}}$ is the contractor used in $C ( \beta_p )$, while $C^{(\varepsilon ; q-1 )}_{i_1 \ldots i_{N+2q}}$ is used in $C ( \beta_{\bar{p}})$, this allows us to relate $C( \beta_p )$ and $C( \beta_{\bar{p}} )$. This calculation is shown diagrammatically at the top left of figure \ref{figure: inductive contractions}. In particular we find
	\begin{equation}
	C ( \beta_p ) = N C ( \beta_{\bar{p}} )
	\nonumber
	\end{equation}
	Given a $\bar{p} \in P_{q-1}$, there is a unique $p \in P_{q;1}$ which reduces to $\bar{p}$, namely $p = \bar{p} \cup \{N+2q-1, N+2q\}$. Therefore
	\begin{align}
	\sum_{p \in P_{q;1}} C(\beta_p) \beta_p & = N \sum_{\bar{p} \in P_{q-1}} C(\beta_{\bar{p}}) \beta_{\bar{p}}
	\label{type 1}
	\end{align}

	\noindent \textbf{Type 2} \\
	\nopagebreak \\
	\nopagebreak
	\noindent We follow the same route as for type 1. Let $P_{q;2}$ be the set of pairings that are of type 2. Given $p \in P_{q,2}$, we define $\bar{p} \in P_{q-1}$ by $\bar{p} = \left( p \cup \{k_1,k_2\} \right) \backslash \{ \{k_1,N+2q-1\} , \{ k_2, N+2q \} \}$. We then choose the coset representative of $p$ to be
	\begin{equation}
	\beta_p = \beta_{\bar{p}} \left( \beta_{\bar{p}} (k_2) , N+2q-1 \right) = \left( k_2 , N+2q-1 \right) \beta_{\bar{p}} 
	\nonumber
	\end{equation}
	Again, one can check that this satisfies the conditions \eqref{coset rep conditions}. 
	
	The calculation for $C( \beta_p )$ is shown diagrammatically in figure \ref{figure: inductive contractions} in the middle of the top row. For simplicity, the calculation shown has $k_2 = N+2q-2$, but it is clear that for any $k_2$ we arrive at the relation
	\begin{equation}
	C ( \beta_p ) = C ( \beta_{\bar{p}} )
	\nonumber
	\end{equation}
	Consider $\bar{p} \in P_{q-1}$. We can explicitly write this out as
	\begin{equation}
	\bar{p} = \Big\{ \{ l_{1,1}, l_{1,2} \} , \{ l_{2,1} , l_{2,2} \}, \ldots ,\{ l_{q-1,1}, l_{q-1,2} \} \Big\}
	\nonumber
	\end{equation}
	Now consider the different $p$ which reduce to $\bar{p}$. For each pair $\{ l_{i,1}, l_{i,2} \}$, we obtain 2 possible $p$ by setting $k_1 = l_{i,1}$ and $k_2 = l_{i,2}$ or $k_1 = l_{i,2}$ and $k_2 = l_{i,1}$. Therefore we can specify a $p \in P_{q;2}$ by the trio $(\bar{p}, i, j)$, where the two cases above correspond to $j=2$ and $j=1$ respectively (so $p$ has $k_2 = l_{i,j}$). Changing variables in this way, we have
	\begin{equation*}
	\sum_{p \in P_{q;2}} C(\beta_p) \beta_p = \sum_{\bar{p} \in P_{q-1}} C(\beta_{\bar{p}}) \beta_{\bar{p}} \sum_{i=1}^{q-1} \sum_{j=1}^2  \left( \beta_{\bar{p}}(l_{i,j}) , N+2q-1 \right)
	\end{equation*}
	From \eqref{coset rep conditions} we know that $\beta_{\bar{p}} ( \{ l_{i,j} \} ) = \{ N+1, N+2, \ldots , N+2q-2 \}$, so we can simplify this to		
	\begin{align}
	\sum_{p \in P_{q;2}} C(\beta_p) \beta_p = \sum_{\bar{p} \in P_{q-1}} C(\beta_{\bar{p}}) \beta_{\bar{p}} \sum_{j=1}^{2q-2} (N+j, N+2q-1)
	\label{type 2}
	\end{align}
	
	\noindent \textbf{Types 3 and 4} \\
	\nopagebreak \\
	\nopagebreak
	\noindent Let $P_{q;3}$ be the set of pairings that are of type 3. Given $p \in P_{q,3}$, we define $\bar{p} \in P_{q-1}$ by $\bar{p} = p \backslash \{ k, N+2q-1 \}$. We then choose the coset representative of $p$ to be
	\begin{equation}
	\beta_p = \beta_{\bar{p}} \left( \beta_{\bar{p}}(k) , N+2q \right)  = \left( k, N+2q \right) \beta_{\bar{p}}
	\nonumber
	\end{equation}
	The calculation for $C( \beta_p )$ is shown diagrammatically at the top right of figure \ref{figure: inductive contractions}, and demonstrates that 
	\begin{equation}
	C ( \beta_p ) = C ( \beta_{\bar{p}} )
	\nonumber
	\end{equation}
	For simplicity, the calculation shown has $k = N+2q-2$, but clearly $k$ can be arbitrary and we still arrive at the same result.
	
	Take $\bar{p} \in P_{q-1}$. This contains $q-1$ pairs from the set $\{ 1,2, \ldots N+2q-2\}$, so there are $N$ numbers that are omitted. Let these be $\{ l_1, \ldots ,l_N \}$. The different $p$ which reduce to $\bar{p}$ are then given by $\bar{p} \cup \{ l_i, N+2q-1 \}$ for $i=1,2, \ldots ,N$. Therefore a $p \in P_{q;3}$ can be specified by the pair $(\bar{p}, i)$. Changing to these variables, we have
	\begin{equation*}
	\sum_{p \in P_{q;3}} C(\beta_p) \beta_p = \sum_{\bar{p} \in P_{q-1}} C(\beta_{\bar{p}}) \beta_{\bar{p}} \sum_{i=1}^{N}  \left( \beta_{\bar{p}}(l_{i}) , N+2q \right)
	\end{equation*}
	From \eqref{coset rep conditions} we know that $\beta_{\bar{p}}(\{ l_i \} ) = \{ 1,2, \ldots ,N\}$, therefore this simplifies to
	\begin{align}
	\sum_{p \in P_{q;3}} C(\beta_p) \beta_p  = \sum_{\bar{p} \in P_{q-1}} C( \beta_{ \bar{p} } ) \beta_{\bar{p}} \sum_{j=1}^{N} (j, N+2q)
	\label{type 3}
	\end{align}
	We can repeat the above process with $N+2q-1$ and $N+2q$ swapped to give the sum over type 4 pairings
	\begin{equation}
	\sum_{p \in P_{q;4}} C(\beta_p) \beta_p = \sum_{\bar{p} \in P_{q-1}} C(\beta_{\bar{p}}) \beta_{\bar{p}} \sum_{j=1}^{N} (j, N+2q-1)
	\label{type 4}
	\end{equation}
	
	\noindent \textbf{Type 5} \\
	\nopagebreak \\
	\nopagebreak
	\noindent Let $P_{q;5}$ be the set of pairings that are of type 5. Given $p \in P_{q,5}$, we can choose the coset representative $\beta_p$ such that
	\begin{equation}
	\beta_p (N+2q-1) = 1 , \qquad \beta_p (N+2q) = 2
	\nonumber
	\end{equation}
	We do not need to specify the remaining values of $\beta_p$ as this is enough to show that $C(\beta_p)$ vanishes. The calculation is shown diagrammatically on the bottom row of figure \ref{figure: inductive contractions}. This means
	\begin{equation}
	\sum_{p \in P_{q;5}} C(\beta_p) \beta_p = 0
	\label{type 5}
	\end{equation}	
	Adding together \eqref{type 1}, \eqref{type 2}, \eqref{type 3}, \eqref{type 4} and \eqref{type 5}, we get
	\begin{align*}
	\sum_{p \in P_q} C(\beta_p) \beta_p = \sum_{\bar{p} \in P_{q-1}} C(\beta_{\bar{p}}) \beta_{\bar{p}} \left[ N + \sum_{j=1}^N (j, N+2q-1) \right. &  + \sum_{j=1}^N (j,N+2q) \\ & \left. + \sum_{j=1}^{2q-2} (N+j, N+2q-1) \right]
	\end{align*}
	The factor on the right is just the $i=q$ factor in \eqref{intermediary claim}, so plugging in the inductive assumption proves the proposition.\\ $\square$ \\

	\noindent We now study how the two Jucys-Murphy elements in \eqref{intermediary claim}, $J_{N+2i-1}$ and the conjugate of $J_{N+1}$, act on $| 1^N+T \rangle$ and its constituents. We start with $J_{N+1}$.
	
	The vector $| 1^N+T \rangle$ is a linear combination of various standard Young tableaux of shape $1^N+T$. Let $r$ be one of these tableaux. Since $|1^N+T \rangle$ is completely anti-symmetric under the $S_N$ acting on $\{ 1,2, \ldots ,N \}$, the first column of $r$ must consist of the numbers $1,2, \ldots ,N$. Then as $|1^N+T \rangle$ is invariant under the $S_q[S_2]$ acting on the pairs $\{ \{ N+1,N+2 \} , \ldots , \{ N+2q-1, N+2q\} \}$, the numbers $\{ N+1,N+2, \ldots ,N+2q-1,N+2q\}$ must appear in pairs, with each even number appearing directly to the right of the preceding odd number (this is proved in \cite{Zinn-Justin2010}). This means that the odd numbers greater than $N$ occupy the 2nd, 4th, 6th, \ldots columns of $r$ while the even numbers greater than $N$ occupy the 3rd, 5th, 7th, \ldots columns. So for example, given $T = [4,4,2]$, the possible $r$ are displayed in figure \ref{possible tableaux for T=[4,4,2]}.
	
	\begin{figure}
	\ytableausetup{boxsize=35pt}
	\begin{gather*}
	\begin{ytableau} 1 & N+1 & N+2 & N+3 & N+4 \\ 2 & N+5 & N+6 & N+7 & N+8 \\ 3 & N+9 & N+10 \\ 4 \\ \vdots \\ N \end{ytableau}
	\quad \quad
	\begin{ytableau} 1 & N+1 & N+2 & N+3 & N+4 \\ 2 & N+5 & N+6 & N+9 & N+10 \\ 3 & N+7 & N+8 \\ 4 \\ \vdots \\ N \end{ytableau} \\
	\begin{ytableau} 1 & N+1 & N+2 & N+5 & N+6 \\ 2 & N+3 & N+4 & N+7 & N+8 \\ 3 & N+9 & N+10 \\ 4 \\ \vdots \\ N \end{ytableau}
	\quad \quad
	\begin{ytableau} 1 & N+1 & N+2 & N+5 & N+6 \\ 2 & N+3 & N+4 & N+9 & N+10 \\ 3 & N+7 & N+8 \\ 4 \\ \vdots \\ N \end{ytableau} \\
	\begin{ytableau} 1 & N+1 & N+2 & N+7 & N+8 \\ 2 & N+3 & N+4 & N+9 & N+10 \\ 3 & N+5 & N+6 \\ 4 \\ \vdots \\ N \end{ytableau}
	\end{gather*}
	\caption{The standard Young tableaux that contribute to the vector $|1^N+T \rangle$ for $T = [4,4,2]$}
	\label{possible tableaux for T=[4,4,2]}
	\ytableausetup{boxsize=normal}
	\end{figure}
	
	Since the numbers $ 1,2, \ldots ,N$ take up the first column of $r$, the number $N+1$ must in the first box of the second column (one can see that this is the case for all the tableaux shown in figure \ref{possible tableaux for T=[4,4,2]}). The content of this box is 1, and therefore $J_{N+1}$ has eigenvalue 1 when acting on $r$:
	\begin{equation}
	D^{1^N+T}\left( J_{N+1} \right) | r \rangle = | r \rangle
	\nonumber
	\end{equation}
	We are interested in the conjugate of $J_{N+1}$, where the conjugating element is $(N+1,N+2i)$, $1 \leq i \leq q$. Since $S_N$ commutes with $(N+1,N+2i)$, we know that $D^{1^N+T}[(N+1,N+2i)] | r \rangle$ is still anti-symmetric under $S_N$, and hence, by the same argument as above, it is made up of standard Young tableaux with $N+1$ in the first box of the second column. Therefore
	\begin{equation}
	D^{1^N+T} \left[ J_{N+1} (N+1,N+2i) \right] | r \rangle = D^{1^N+T} [(N+1,N+2i)] | r \rangle
	\nonumber
	\end{equation}
	Multiplying by $D^{1^N+T} [(N+1,N+2i)]$ gives
	\begin{equation}
	D^{1^N+T} \left[ \sum_{j=1}^N (j,N+2i) \right] | r \rangle = D^{1^N+T} \left[ (N+1,N+2i) J_{N+1} (N+1,N+2i) \right] | r \rangle = | r \rangle
	\nonumber
	\end{equation}
	This gives the behaviour of the second term in each factor of \eqref{intermediary claim}. We now look at the other term, of the form $J_{N+2i-1}$.
	
	Denote the contents of the cell labelled by $k$ in $r$ by $c(r,k)$. Then
	\begin{equation}
	D^{1^N+T} \left( J_{N+2i-1} \right) | r \rangle = c(r, N+2i-1) | r \rangle
	\nonumber
	\end{equation}
	Decomposing $| 1^N+T \rangle$ into its constituent standard Young tableaux, we have
	\begin{equation}
	|1^N+T \rangle = \sum_r \alpha_r | r \rangle
	\nonumber
	\end{equation}
	for some coefficients $\alpha_r$. Therefore
	\begin{align}
	& \langle 1^N+T | D^{1^N+T}\left( \sum_{\substack{\beta \text{ left coset} \\ \text{representatives}}} C(\beta) \beta \right) |1^N+T\rangle \nonumber \\
	& \qquad  = N! \langle 1^N+T | \prod_{i=1}^q \left[ N + J_{N+2i-1} + (N+1,N+2i) J_{N+1} (N+1,N+2i) \right] \sum_r \alpha_r | r \rangle \nonumber \\
	& \qquad = N! \langle 1^N+T | \sum_r \alpha_r \prod_{i=1}^q \left[ N+1 + c(r,N+2i-1) \right] | r \rangle
	\label{intermediary claim applied to |1^N+T>}
	\end{align}
	As we noted earlier, the odd numbers greater than $N$ occupy the even numbered columns of $r$, so the set $\{c(r,N+2i-1):1 \leq i \leq q \}$ is just the contents of these columns. Therefore the product in \eqref{intermediary claim applied to |1^N+T>} does not depend on $r$, just on the shape of $1^N+T$, and we can pull it out of the sum. Using the same notation as for the mesonic correlator, we have
	\begin{align*}
	& \langle 1^N+T | D^{1^N+T}\left( \sum_{\substack{\beta \text{ left coset} \\ \text{representatives}}} C(\beta) \beta \right) |1^N+T\rangle \\
	& \qquad \qquad \qquad = N! \prod_{\substack{i \in \text{ even}\\ \text{columns of }1^N+T}} (N+1+c_i) \langle 1^N+T | 1^N + T \rangle \\
	& \qquad \qquad \qquad = N! \prod_{\substack{i \in \text{ even}\\ \text{columns of }1^N+T}} (N+1+c_i)
	\end{align*}
	Now we note that
	\begin{equation}
	\prod_{\substack{i \in \text{ even}\\ \text{columns of }1^N+T}} (N+1+c_i) = \prod_{\substack{i \in \text{ odd}\\ \text{columns of }1^N+T \\ \text{excluding first column}}} (N+c_i)
	\nonumber
	\end{equation}
	and
	\begin{equation}
	N! = \prod_{\substack{i \in \text{ first}\\ \text{column of }1^N+T}} (N+c_i)
	\nonumber
	\end{equation}
	Therefore
	\begin{equation}
	\langle 1^N+T | D^{1^N+T}\left( \sum_{\substack{\beta \text{ left coset} \\ \text{representatives}}} C(\beta) \beta \right) |1^N+T\rangle = \prod_{\substack{i \in \text{ odd}\\ \text{columns of }1^N+T}} (N+c_i)
	\nonumber
	\end{equation}
	Reinstating the normalisation factor from \eqref{sum over coset reps}, the full baryonic correlator is
	\begin{align*}
	\langle \mathcal{O}^\varepsilon_{T,R,S,\lambda} \overline{\mathcal{O}}^\varepsilon_{T', R', S', \lambda'} \rangle =  \delta_{T T'} \delta_{R R'} \delta_{S S'} \delta_{\lambda \lambda'} 2^{n+m +q} n! m! N! q! & \frac{d_{1^N+T}}{(2n+2m)!} \\ & \prod_{\substack{i \in \text{ odd}\\ \text{columns of }1^N+T}} (N+c_i)
	\end{align*} 
	which, as expected, agrees with \eqref{baryonic correlator}.

	\bibliography{library}
	\bibliographystyle{unsrt}

\end{document}

%% file: SnS2.pdf_tex
\begingroup%
  \makeatletter%
  \providecommand\color[2][]{%
    \errmessage{(Inkscape) Color is used for the text in Inkscape, but the package 'color.sty' is not loaded}%
    \renewcommand\color[2][]{}%
  }%
  \providecommand\transparent[1]{%
    \errmessage{(Inkscape) Transparency is used (non-zero) for the text in Inkscape, but the package 'transparent.sty' is not loaded}%
    \renewcommand\transparent[1]{}%
  }%
  \providecommand\rotatebox[2]{#2}%
  \ifx\svgwidth\undefined%
    \setlength{\unitlength}{160.8bp}%
    \ifx\svgscale\undefined%
      \relax%
    \else%
      \setlength{\unitlength}{\unitlength * \real{\svgscale}}%
    \fi%
  \else%
    \setlength{\unitlength}{\svgwidth}%
  \fi%
  \global\let\svgwidth\undefined%
  \global\let\svgscale\undefined%
  \makeatother%
  \begin{picture}(1,0.82721159)%
    \put(0,0){\includegraphics[width=\unitlength,page=1]{SnS2.pdf}}%
    \put(-0.02186334,0.6744209){\color[rgb]{0,0,0}\makebox(0,0)[lb]{\smash{1}}}%
    \put(-0.01457556,0.02765465){\color[rgb]{0,0,0}\makebox(0,0)[lb]{\smash{2}}}%
    \put(0.23359764,0.67723884){\color[rgb]{0,0,0}\makebox(0,0)[lb]{\smash{3}}}%
    \put(0.23903918,0.02765465){\color[rgb]{0,0,0}\makebox(0,0)[lb]{\smash{4}}}%
    \put(0.48215951,0.67723884){\color[rgb]{0,0,0}\makebox(0,0)[lb]{\smash{5}}}%
    \put(0.48361707,0.03047259){\color[rgb]{0,0,0}\makebox(0,0)[lb]{\smash{6}}}%
    \put(0.70508782,0.3759135){\color[rgb]{0,0,0}\makebox(0,0)[lb]{\smash{\ldots}}}%
    \put(0.8947562,0.67601444){\color[rgb]{0,0,0}\makebox(0,0)[lb]{\smash{$2n-1$}}}%
    \put(0.95893954,0.02924829){\color[rgb]{0,0,0}\makebox(0,0)[lb]{\smash{$2n$}}}%
  \end{picture}%
\endgroup%

%% file: Mesonic_Contraction.pdf_tex
\begingroup%
  \makeatletter%
  \providecommand\color[2][]{%
    \errmessage{(Inkscape) Color is used for the text in Inkscape, but the package 'color.sty' is not loaded}%
    \renewcommand\color[2][]{}%
  }%
  \providecommand\transparent[1]{%
    \errmessage{(Inkscape) Transparency is used (non-zero) for the text in Inkscape, but the package 'transparent.sty' is not loaded}%
    \renewcommand\transparent[1]{}%
  }%
  \providecommand\rotatebox[2]{#2}%
  \ifx\svgwidth\undefined%
    \setlength{\unitlength}{400.78259773bp}%
    \ifx\svgscale\undefined%
      \relax%
    \else%
      \setlength{\unitlength}{\unitlength * \real{\svgscale}}%
    \fi%
  \else%
    \setlength{\unitlength}{\svgwidth}%
  \fi%
  \global\let\svgwidth\undefined%
  \global\let\svgscale\undefined%
  \makeatother%
  \begin{picture}(1,0.44167437)%
    \put(0,0){\includegraphics[width=\unitlength,page=1]{Mesonic_Contraction.pdf}}%
    \put(0.47906271,0.19221711){\color[rgb]{0,0,0}\makebox(0,0)[lb]{\smash{\Huge$\beta$}}}%
    \put(0,0){\includegraphics[width=\unitlength,page=2]{Mesonic_Contraction.pdf}}%
    \put(0.37526581,0.30881773){\color[rgb]{0,0,0}\makebox(0,0)[lb]{\smash{\ldots}}}%
    \put(0.75452386,0.30881773){\color[rgb]{0,0,0}\makebox(0,0)[lb]{\smash{\ldots}}}%
    \put(0,0){\includegraphics[width=\unitlength,page=3]{Mesonic_Contraction.pdf}}%
    \put(0.37526581,0.09922776){\color[rgb]{0,0,0}\makebox(0,0)[lb]{\smash{\ldots}}}%
    \put(0.75452381,0.09922778){\color[rgb]{0,0,0}\makebox(0,0)[lb]{\smash{\ldots}}}%
    \put(0,0){\includegraphics[width=\unitlength,page=4]{Mesonic_Contraction.pdf}}%
    \put(0.11578311,0.3810118){\color[rgb]{0,0,0}\makebox(0,0)[lb]{\smash{$\delta$}}}%
    \put(0.2761252,0.38101178){\color[rgb]{0,0,0}\makebox(0,0)[lb]{\smash{$\delta$}}}%
    \put(0.49486258,0.38101179){\color[rgb]{0,0,0}\makebox(0,0)[lb]{\smash{$\delta$}}}%
    \put(0.65399979,0.3810118){\color[rgb]{0,0,0}\makebox(0,0)[lb]{\smash{$\delta$}}}%
    \put(0.87370408,0.38101176){\color[rgb]{0,0,0}\makebox(0,0)[lb]{\smash{$\delta$}}}%
    \put(0,0){\includegraphics[width=\unitlength,page=5]{Mesonic_Contraction.pdf}}%
    \put(0.10744925,0.01872117){\color[rgb]{0,0,0}\makebox(0,0)[lb]{\smash{$X$}}}%
    \put(0,0){\includegraphics[width=\unitlength,page=6]{Mesonic_Contraction.pdf}}%
    \put(0.26713678,0.01872117){\color[rgb]{0,0,0}\makebox(0,0)[lb]{\smash{$X$}}}%
    \put(0,0){\includegraphics[width=\unitlength,page=7]{Mesonic_Contraction.pdf}}%
    \put(0.48670715,0.01872117){\color[rgb]{0,0,0}\makebox(0,0)[lb]{\smash{$X$}}}%
    \put(0,0){\includegraphics[width=\unitlength,page=8]{Mesonic_Contraction.pdf}}%
    \put(0.6463948,0.01872127){\color[rgb]{0,0,0}\makebox(0,0)[lb]{\smash{$Y$}}}%
    \put(0,0){\includegraphics[width=\unitlength,page=9]{Mesonic_Contraction.pdf}}%
    \put(0.86596524,0.01872132){\color[rgb]{0,0,0}\makebox(0,0)[lb]{\smash{$Y$}}}%
    \put(0,0){\includegraphics[width=\unitlength,page=10]{Mesonic_Contraction.pdf}}%
  \end{picture}%
\endgroup%

%% file: Baryonic_Contraction.pdf_tex
\begingroup%
  \makeatletter%
  \providecommand\color[2][]{%
    \errmessage{(Inkscape) Color is used for the text in Inkscape, but the package 'color.sty' is not loaded}%
    \renewcommand\color[2][]{}%
  }%
  \providecommand\transparent[1]{%
    \errmessage{(Inkscape) Transparency is used (non-zero) for the text in Inkscape, but the package 'transparent.sty' is not loaded}%
    \renewcommand\transparent[1]{}%
  }%
  \providecommand\rotatebox[2]{#2}%
  \ifx\svgwidth\undefined%
    \setlength{\unitlength}{399.99999995bp}%
    \ifx\svgscale\undefined%
      \relax%
    \else%
      \setlength{\unitlength}{\unitlength * \real{\svgscale}}%
    \fi%
  \else%
    \setlength{\unitlength}{\svgwidth}%
  \fi%
  \global\let\svgwidth\undefined%
  \global\let\svgscale\undefined%
  \makeatother%
  \begin{picture}(1,0.44437479)%
    \put(0,0){\includegraphics[width=\unitlength,page=1]{Baryonic_Contraction.pdf}}%
    \put(0.50000001,0.19259319){\color[rgb]{0,0,0}\makebox(0,0)[lb]{\smash{\Huge$\beta$}}}%
    \put(0.77599996,0.30942187){\color[rgb]{0,0,0}\makebox(0,0)[lb]{\smash{\ldots}}}%
    \put(0,0){\includegraphics[width=\unitlength,page=2]{Baryonic_Contraction.pdf}}%
    \put(0.67615861,0.38175721){\color[rgb]{0,0,0}\makebox(0,0)[lb]{\smash{$\delta$}}}%
    \put(0.89628985,0.38175722){\color[rgb]{0,0,0}\makebox(0,0)[lb]{\smash{$\delta$}}}%
    \put(0,0){\includegraphics[width=\unitlength,page=3]{Baryonic_Contraction.pdf}}%
    \put(0.3644107,0.0994219){\color[rgb]{0,0,0}\makebox(0,0)[lb]{\smash{\ldots}}}%
    \put(0.77600004,0.09942187){\color[rgb]{0,0,0}\makebox(0,0)[lb]{\smash{\ldots}}}%
    \put(0,0){\includegraphics[width=\unitlength,page=4]{Baryonic_Contraction.pdf}}%
    \put(0.08765945,0.01875785){\color[rgb]{0,0,0}\makebox(0,0)[lb]{\smash{$X$}}}%
    \put(0,0){\includegraphics[width=\unitlength,page=5]{Baryonic_Contraction.pdf}}%
    \put(0.24765943,0.01875785){\color[rgb]{0,0,0}\makebox(0,0)[lb]{\smash{$X$}}}%
    \put(0,0){\includegraphics[width=\unitlength,page=6]{Baryonic_Contraction.pdf}}%
    \put(0.50765943,0.01875785){\color[rgb]{0,0,0}\makebox(0,0)[lb]{\smash{$X$}}}%
    \put(0,0){\includegraphics[width=\unitlength,page=7]{Baryonic_Contraction.pdf}}%
    \put(0.66765957,0.01875795){\color[rgb]{0,0,0}\makebox(0,0)[lb]{\smash{$Y$}}}%
    \put(0,0){\includegraphics[width=\unitlength,page=8]{Baryonic_Contraction.pdf}}%
    \put(0.88765946,0.018758){\color[rgb]{0,0,0}\makebox(0,0)[lb]{\smash{$Y$}}}%
    \put(0,0){\includegraphics[width=\unitlength,page=9]{Baryonic_Contraction.pdf}}%
    \put(0.35149202,0.30942191){\color[rgb]{0,0,0}\makebox(0,0)[lb]{\smash{\ldots}}}%
    \put(0.31452136,0.38359354){\color[rgb]{0,0,0}\makebox(0,0)[lb]{\smash{$\varepsilon$}}}%
    \put(0,0){\includegraphics[width=\unitlength,page=10]{Baryonic_Contraction.pdf}}%
  \end{picture}%
\endgroup%

%% file: UN_Contraction.pdf_tex
\begingroup%
  \makeatletter%
  \providecommand\color[2][]{%
    \errmessage{(Inkscape) Color is used for the text in Inkscape, but the package 'color.sty' is not loaded}%
    \renewcommand\color[2][]{}%
  }%
  \providecommand\transparent[1]{%
    \errmessage{(Inkscape) Transparency is used (non-zero) for the text in Inkscape, but the package 'transparent.sty' is not loaded}%
    \renewcommand\transparent[1]{}%
  }%
  \providecommand\rotatebox[2]{#2}%
  \ifx\svgwidth\undefined%
    \setlength{\unitlength}{384.05725098bp}%
    \ifx\svgscale\undefined%
      \relax%
    \else%
      \setlength{\unitlength}{\unitlength * \real{\svgscale}}%
    \fi%
  \else%
    \setlength{\unitlength}{\svgwidth}%
  \fi%
  \global\let\svgwidth\undefined%
  \global\let\svgscale\undefined%
  \makeatother%
  \begin{picture}(1,0.40827251)%
    \put(0,0){\includegraphics[width=\unitlength,page=1]{UN_Contraction.pdf}}%
    \put(0.47916975,0.14226333){\color[rgb]{0,0,0}\makebox(0,0)[lb]{\smash{\Huge$\beta$}}}%
    \put(0.25628634,0.30976828){\color[rgb]{0,0,0}\makebox(0,0)[lb]{\smash{\ldots}}}%
    \put(0.71455134,0.29935315){\color[rgb]{0,0,0}\makebox(0,0)[lb]{\smash{\ldots}}}%
    \put(0,0){\includegraphics[width=\unitlength,page=2]{UN_Contraction.pdf}}%
    \put(0.25628632,0.04522437){\color[rgb]{0,0,0}\makebox(0,0)[lb]{\smash{\ldots}}}%
    \put(0.71455133,0.04522439){\color[rgb]{0,0,0}\makebox(0,0)[lb]{\smash{\ldots}}}%
    \put(0,0){\includegraphics[width=\unitlength,page=3]{UN_Contraction.pdf}}%
    \put(0.07095606,0.30074454){\color[rgb]{0,0,0}\makebox(0,0)[lb]{\smash{$X$}}}%
    \put(0.15427699,0.30074454){\color[rgb]{0,0,0}\makebox(0,0)[lb]{\smash{$X$}}}%
    \put(0.36216586,0.30074455){\color[rgb]{0,0,0}\makebox(0,0)[lb]{\smash{$X$}}}%
    \put(0.44590017,0.30074457){\color[rgb]{0,0,0}\makebox(0,0)[lb]{\smash{$X$}}}%
    \put(0.53338715,0.30074454){\color[rgb]{0,0,0}\makebox(0,0)[lb]{\smash{$Y$}}}%
    \put(0.61670805,0.30074454){\color[rgb]{0,0,0}\makebox(0,0)[lb]{\smash{$Y$}}}%
    \put(0.82501029,0.30074454){\color[rgb]{0,0,0}\makebox(0,0)[lb]{\smash{$Y$}}}%
    \put(0.90833119,0.30074454){\color[rgb]{0,0,0}\makebox(0,0)[lb]{\smash{$Y$}}}%
  \end{picture}%
\endgroup%

%% file: SON_Simplified_Contraction2.pdf_tex
\begingroup%
  \makeatletter%
  \providecommand\color[2][]{%
    \errmessage{(Inkscape) Color is used for the text in Inkscape, but the package 'color.sty' is not loaded}%
    \renewcommand\color[2][]{}%
  }%
  \providecommand\transparent[1]{%
    \errmessage{(Inkscape) Transparency is used (non-zero) for the text in Inkscape, but the package 'transparent.sty' is not loaded}%
    \renewcommand\transparent[1]{}%
  }%
  \providecommand\rotatebox[2]{#2}%
  \ifx\svgwidth\undefined%
    \setlength{\unitlength}{351.99999998bp}%
    \ifx\svgscale\undefined%
      \relax%
    \else%
      \setlength{\unitlength}{\unitlength * \real{\svgscale}}%
    \fi%
  \else%
    \setlength{\unitlength}{\svgwidth}%
  \fi%
  \global\let\svgwidth\undefined%
  \global\let\svgscale\undefined%
  \makeatother%
  \begin{picture}(1,0.84207273)%
    \put(0,0){\includegraphics[width=\unitlength,page=1]{SON_Simplified_Contraction2.pdf}}%
    \put(0.23863636,0.64390985){\color[rgb]{0,0,0}\makebox(0,0)[lb]{\smash{\Huge$\sigma$}}}%
    \put(0.2318182,0.76136368){\color[rgb]{0,0,0}\makebox(0,0)[lb]{\smash{\ldots}}}%
    \put(0,0){\includegraphics[width=\unitlength,page=2]{SON_Simplified_Contraction2.pdf}}%
    \put(0.23181819,0.49820669){\color[rgb]{0,0,0}\makebox(0,0)[lb]{\smash{\ldots}}}%
    \put(0,0){\includegraphics[width=\unitlength,page=3]{SON_Simplified_Contraction2.pdf}}%
    \put(0.06476517,0.82475271){\color[rgb]{0,0,0}\makebox(0,0)[lb]{\smash{$\delta$}}}%
    \put(0.05302215,0.48722482){\color[rgb]{0,0,0}\makebox(0,0)[lb]{\smash{$X$}}}%
    \put(0,0){\includegraphics[width=\unitlength,page=4]{SON_Simplified_Contraction2.pdf}}%
    \put(0.14393126,0.48722482){\color[rgb]{0,0,0}\makebox(0,0)[lb]{\smash{$X$}}}%
    \put(0,0){\includegraphics[width=\unitlength,page=5]{SON_Simplified_Contraction2.pdf}}%
    \put(0.15567426,0.82475271){\color[rgb]{0,0,0}\makebox(0,0)[lb]{\smash{$\delta$}}}%
    \put(0,0){\includegraphics[width=\unitlength,page=6]{SON_Simplified_Contraction2.pdf}}%
    \put(0.32574942,0.48722482){\color[rgb]{0,0,0}\makebox(0,0)[lb]{\smash{$X$}}}%
    \put(0,0){\includegraphics[width=\unitlength,page=7]{SON_Simplified_Contraction2.pdf}}%
    \put(0.33749244,0.82475274){\color[rgb]{0,0,0}\makebox(0,0)[lb]{\smash{$\delta$}}}%
    \put(0.46477272,0.64346587){\color[rgb]{0,0,0}\makebox(0,0)[lb]{\smash{\Huge=}}}%
    \put(0,0){\includegraphics[width=\unitlength,page=8]{SON_Simplified_Contraction2.pdf}}%
    \put(0.77045485,0.68825531){\color[rgb]{0,0,0}\makebox(0,0)[lb]{\smash{\Huge$\tau$}}}%
    \put(0.81818186,0.79704038){\color[rgb]{0,0,0}\makebox(0,0)[lb]{\smash{\ldots}}}%
    \put(0,0){\includegraphics[width=\unitlength,page=9]{SON_Simplified_Contraction2.pdf}}%
    \put(0.82954543,0.54252483){\color[rgb]{0,0,0}\makebox(0,0)[lb]{\smash{\ldots}}}%
    \put(0,0){\includegraphics[width=\unitlength,page=10]{SON_Simplified_Contraction2.pdf}}%
    \put(0.73484036,0.53157022){\color[rgb]{0,0,0}\makebox(0,0)[lb]{\smash{$X$}}}%
    \put(0,0){\includegraphics[width=\unitlength,page=11]{SON_Simplified_Contraction2.pdf}}%
    \put(0.64393119,0.53157028){\color[rgb]{0,0,0}\makebox(0,0)[lb]{\smash{$X$}}}%
    \put(0,0){\includegraphics[width=\unitlength,page=12]{SON_Simplified_Contraction2.pdf}}%
    \put(0.91665854,0.53157022){\color[rgb]{0,0,0}\makebox(0,0)[lb]{\smash{$X$}}}%
    \put(0,0){\includegraphics[width=\unitlength,page=13]{SON_Simplified_Contraction2.pdf}}%
    \put(0.46477272,0.17866663){\color[rgb]{0,0,0}\makebox(0,0)[lb]{\smash{\Huge=}}}%
    \put(0,0){\includegraphics[width=\unitlength,page=14]{SON_Simplified_Contraction2.pdf}}%
    \put(0.77045485,0.20072808){\color[rgb]{0,0,0}\makebox(0,0)[lb]{\smash{\Huge$\tau$}}}%
    \put(0.8181818,0.31297943){\color[rgb]{0,0,0}\makebox(0,0)[lb]{\smash{\ldots}}}%
    \put(0,0){\includegraphics[width=\unitlength,page=15]{SON_Simplified_Contraction2.pdf}}%
    \put(0.81818183,0.07775215){\color[rgb]{0,0,0}\makebox(0,0)[lb]{\smash{\ldots}}}%
    \put(0,0){\includegraphics[width=\unitlength,page=16]{SON_Simplified_Contraction2.pdf}}%
    \put(0.73484036,0.06677013){\color[rgb]{0,0,0}\makebox(0,0)[lb]{\smash{$X$}}}%
    \put(0,0){\includegraphics[width=\unitlength,page=17]{SON_Simplified_Contraction2.pdf}}%
    \put(0.64393119,0.06677013){\color[rgb]{0,0,0}\makebox(0,0)[lb]{\smash{$X$}}}%
    \put(0,0){\includegraphics[width=\unitlength,page=18]{SON_Simplified_Contraction2.pdf}}%
    \put(0.91665854,0.06677013){\color[rgb]{0,0,0}\makebox(0,0)[lb]{\smash{$X$}}}%
    \put(0,0){\includegraphics[width=\unitlength,page=19]{SON_Simplified_Contraction2.pdf}}%
  \end{picture}%
\endgroup%

%% file: UN_diagram.pdf_tex
\begingroup%
  \makeatletter%
  \providecommand\color[2][]{%
    \errmessage{(Inkscape) Color is used for the text in Inkscape, but the package 'color.sty' is not loaded}%
    \renewcommand\color[2][]{}%
  }%
  \providecommand\transparent[1]{%
    \errmessage{(Inkscape) Transparency is used (non-zero) for the text in Inkscape, but the package 'transparent.sty' is not loaded}%
    \renewcommand\transparent[1]{}%
  }%
  \providecommand\rotatebox[2]{#2}%
  \ifx\svgwidth\undefined%
    \setlength{\unitlength}{400bp}%
    \ifx\svgscale\undefined%
      \relax%
    \else%
      \setlength{\unitlength}{\unitlength * \real{\svgscale}}%
    \fi%
  \else%
    \setlength{\unitlength}{\svgwidth}%
  \fi%
  \global\let\svgwidth\undefined%
  \global\let\svgscale\undefined%
  \makeatother%
  \begin{picture}(1,0.91800002)%
    \put(0,0){\includegraphics[width=\unitlength,page=1]{UN_diagram.pdf}}%
    \put(0.4717324,0.87862246){\color[rgb]{0,0,0}\makebox(0,0)[lb]{\smash{$T_{ST}^{(1)}$}}}%
    \put(0.46661996,0.82338092){\color[rgb]{0,0,0}\makebox(0,0)[lb]{\smash{$H_{T_{ST}^{(1)}}$}}}%
    \put(0.30861731,0.75511069){\color[rgb]{0,0,0}\makebox(0,0)[lb]{\smash{Counts aperiodic single traces}}}%
    \put(0,0){\includegraphics[width=\unitlength,page=2]{UN_diagram.pdf}}%
    \put(0.10496848,0.51862245){\color[rgb]{0,0,0}\makebox(0,0)[lb]{\smash{$T_{ST}=K \otimes T_{ST}^{(1)}$}}}%
    \put(0.06534134,0.45938091){\color[rgb]{0,0,0}\makebox(0,0)[lb]{\smash{$H_{T_{ST}} = \mathcal{M}^{-1} \left( H_{T_{ST}^{(1)}} \right)$}}}%
    \put(0.06850224,0.39652488){\color[rgb]{0,0,0}\makebox(0,0)[lb]{\smash{Counts single traces}}}%
    \put(0,0){\includegraphics[width=\unitlength,page=3]{UN_diagram.pdf}}%
    \put(0.68923732,0.51938088){\color[rgb]{0,0,0}\makebox(0,0)[lb]{\smash{$T^{(1)}=\operatorname{Sym} \left( T_{ST}^{(1)} \right)$}}}%
    \put(0.66337167,0.45938087){\color[rgb]{0,0,0}\makebox(0,0)[lb]{\smash{$H_{T^{(1)}} = \operatorname{PExp }\left( H_{T_{ST}^{(1)}} \right)$}}}%
    \put(0.60861494,0.39511068){\color[rgb]{0,0,0}\makebox(0,0)[lb]{\smash{Counts aperiodic multi-traces}}}%
    \put(0,0){\includegraphics[width=\unitlength,page=4]{UN_diagram.pdf}}%
    \put(0.2591727,0.15338091){\color[rgb]{0,0,0}\makebox(0,0)[lb]{\smash{$T=\operatorname{Sym} \left( K \otimes T_{ST}^{(1)} \right) = \operatorname{Sym} \left( T_{ST}^{(1)} \right)^{\otimes K}$}}}%
    \put(0.15950678,0.09138088){\color[rgb]{0,0,0}\makebox(0,0)[lb]{\smash{$H_{T} = \operatorname{PExp} \left[ \mathcal{M}^{-1} \left( H_{T_{ST}^{(1)}} \right) \right] = \mathcal{M}_{mult}^{-1} \left[ \operatorname{PExp} \left( H_{T_{ST}^{(1)}} \right) \right] $}}}%
    \put(0.34827402,0.02452482){\color[rgb]{0,0,0}\makebox(0,0)[lb]{\smash{Counts all multi-traces}}}%
    \put(0,0){\includegraphics[width=\unitlength,page=5]{UN_diagram.pdf}}%
    \put(0.11004953,0.67619327){\color[rgb]{0,0,0}\makebox(0,0)[lb]{\smash{Tensor with $K$}}}%
    \put(0.30496266,0.61754931){\color[rgb]{0,0,0}\makebox(0,0)[lb]{\smash{$\mathcal{M}^{-1}$}}}%
    \put(0.71214148,0.67921779){\color[rgb]{0,0,0}\makebox(0,0)[lb]{\smash{$\operatorname{Sym}$}}}%
    \put(0.61991084,0.61921779){\color[rgb]{0,0,0}\makebox(0,0)[lb]{\smash{$\operatorname{PExp}$}}}%
    \put(0.22299489,0.25800195){\color[rgb]{0,0,0}\makebox(0,0)[lb]{\smash{$\operatorname{Sym}$}}}%
    \put(0.30058303,0.31557025){\color[rgb]{0,0,0}\makebox(0,0)[lb]{\smash{$\operatorname{PExp}$}}}%
    \put(0.72552105,0.26195346){\color[rgb]{0,0,0}\makebox(0,0)[lb]{\smash{Tensor power of $K$}}}%
    \put(0.61813283,0.32215722){\color[rgb]{0,0,0}\makebox(0,0)[lb]{\smash{$\mathcal{M}_{mult}^{-1}$}}}%
  \end{picture}%
\endgroup%

%% file: ON_diagram.pdf_tex
\begingroup%
  \makeatletter%
  \providecommand\color[2][]{%
    \errmessage{(Inkscape) Color is used for the text in Inkscape, but the package 'color.sty' is not loaded}%
    \renewcommand\color[2][]{}%
  }%
  \providecommand\transparent[1]{%
    \errmessage{(Inkscape) Transparency is used (non-zero) for the text in Inkscape, but the package 'transparent.sty' is not loaded}%
    \renewcommand\transparent[1]{}%
  }%
  \providecommand\rotatebox[2]{#2}%
  \ifx\svgwidth\undefined%
    \setlength{\unitlength}{412bp}%
    \ifx\svgscale\undefined%
      \relax%
    \else%
      \setlength{\unitlength}{\unitlength * \real{\svgscale}}%
    \fi%
  \else%
    \setlength{\unitlength}{\svgwidth}%
  \fi%
  \global\let\svgwidth\undefined%
  \global\let\svgscale\undefined%
  \makeatother%
  \begin{picture}(1,0.89126215)%
    \put(0,0){\includegraphics[width=\unitlength,page=1]{ON_diagram.pdf}}%
    \put(0.44928975,0.85303151){\color[rgb]{0,0,0}\makebox(0,0)[lb]{\smash{$\widetilde{T}_{ST}^{(min)}$}}}%
    \put(0.44387197,0.79939896){\color[rgb]{0,0,0}\makebox(0,0)[lb]{\smash{$H_{\widetilde{T}_{ST}^{(min)}}$}}}%
    \put(0.22972552,0.73311718){\color[rgb]{0,0,0}\makebox(0,0)[lb]{\smash{Counts minimally periodic single traces}}}%
    \put(0,0){\includegraphics[width=\unitlength,page=2]{ON_diagram.pdf}}%
    \put(0.09026064,0.50351694){\color[rgb]{0,0,0}\makebox(0,0)[lb]{\smash{$\widetilde{T}_{ST}=K \otimes \widetilde{T}_{ST}^{(min)}$}}}%
    \put(0.04984598,0.44600089){\color[rgb]{0,0,0}\makebox(0,0)[lb]{\smash{$H_{\widetilde{T}_{ST}} = \mathcal{M}^{-1} \left( H_{\widetilde{T}_{ST}^{(min)}} \right)$}}}%
    \put(0.07233226,0.38497561){\color[rgb]{0,0,0}\makebox(0,0)[lb]{\smash{Counts single traces}}}%
    \put(0,0){\includegraphics[width=\unitlength,page=3]{ON_diagram.pdf}}%
    \put(0.62068204,0.50425331){\color[rgb]{0,0,0}\makebox(0,0)[lb]{\smash{$\widetilde{T}^{(min)}=\operatorname{Sym} \left( \widetilde{T}_{ST}^{(min)} \right)$}}}%
    \put(0.60715694,0.44600085){\color[rgb]{0,0,0}\makebox(0,0)[lb]{\smash{$H_{\widetilde{T}^{(min)}} = \operatorname{PExp} \left( H_{\widetilde{T}_{ST}^{(min)}} \right)$}}}%
    \put(0.50575133,0.38360261){\color[rgb]{0,0,0}\makebox(0,0)[lb]{\smash{Counts minimally periodic multi-traces}}}%
    \put(0,0){\includegraphics[width=\unitlength,page=4]{ON_diagram.pdf}}%
    \put(0.21298725,0.14891354){\color[rgb]{0,0,0}\makebox(0,0)[lb]{\smash{$\widetilde{T}=\operatorname{Sym} \left( K \otimes \widetilde{T}_{ST}^{(min)} \right) = \operatorname{Sym} \left( \widetilde{T}_{ST}^{(min)} \right)^{\otimes K}$}}}%
    \put(0.11602598,0.0887193){\color[rgb]{0,0,0}\makebox(0,0)[lb]{\smash{$H_{\widetilde{T}} = \operatorname{PExp} \left[ \mathcal{M}^{-1} \left( H_{\widetilde{T}_{ST}^{(min)}} \right) \right] = \mathcal{M}_{mult}^{-1} \left[ \operatorname{PExp} \left( H_{\widetilde{T}_{ST}^{(min)}} \right) \right] $}}}%
    \put(0.32842138,0.02381051){\color[rgb]{0,0,0}\makebox(0,0)[lb]{\smash{Counts all multi-traces}}}%
    \put(0,0){\includegraphics[width=\unitlength,page=5]{ON_diagram.pdf}}%
    \put(0.09713546,0.65649832){\color[rgb]{0,0,0}\makebox(0,0)[lb]{\smash{Tensor with $K$}}}%
    \put(0.2819766,0.59956243){\color[rgb]{0,0,0}\makebox(0,0)[lb]{\smash{$\mathcal{M}^{-1}$}}}%
    \put(0.68169075,0.65943474){\color[rgb]{0,0,0}\makebox(0,0)[lb]{\smash{$\operatorname{Sym}$}}}%
    \put(0.59214645,0.60118232){\color[rgb]{0,0,0}\makebox(0,0)[lb]{\smash{$\operatorname{PExp}$}}}%
    \put(0.20679115,0.25048733){\color[rgb]{0,0,0}\makebox(0,0)[lb]{\smash{$\operatorname{Sym}$}}}%
    \put(0.28211945,0.30637888){\color[rgb]{0,0,0}\makebox(0,0)[lb]{\smash{$\operatorname{PExp}$}}}%
    \put(0.69468063,0.25432375){\color[rgb]{0,0,0}\makebox(0,0)[lb]{\smash{Tensor power of $K$}}}%
    \put(0.59042022,0.312774){\color[rgb]{0,0,0}\makebox(0,0)[lb]{\smash{$\mathcal{M}_{mult}^{-1}$}}}%
  \end{picture}%
\endgroup%

%% file: ON_periodicity_diagram.pdf_tex
\begingroup%
  \makeatletter%
  \providecommand\color[2][]{%
    \errmessage{(Inkscape) Color is used for the text in Inkscape, but the package 'color.sty' is not loaded}%
    \renewcommand\color[2][]{}%
  }%
  \providecommand\transparent[1]{%
    \errmessage{(Inkscape) Transparency is used (non-zero) for the text in Inkscape, but the package 'transparent.sty' is not loaded}%
    \renewcommand\transparent[1]{}%
  }%
  \providecommand\rotatebox[2]{#2}%
  \ifx\svgwidth\undefined%
    \setlength{\unitlength}{421.02953128bp}%
    \ifx\svgscale\undefined%
      \relax%
    \else%
      \setlength{\unitlength}{\unitlength * \real{\svgscale}}%
    \fi%
  \else%
    \setlength{\unitlength}{\svgwidth}%
  \fi%
  \global\let\svgwidth\undefined%
  \global\let\svgscale\undefined%
  \makeatother%
  \begin{picture}(1,1.16286379)%
    \put(0,0){\includegraphics[width=\unitlength,page=1]{ON_periodicity_diagram.pdf}}%
    \put(0.35388752,1.12545306){\color[rgb]{0,0,0}\makebox(0,0)[lb]{\smash{$\widetilde{T}_{ST}^{(odd)}$}}}%
    \put(0.35238577,1.07297073){\color[rgb]{0,0,0}\makebox(0,0)[lb]{\smash{$H_{\widetilde{T}_{ST}^{(odd)}}$}}}%
    \put(0.24950015,1.00811045){\color[rgb]{0,0,0}\makebox(0,0)[lb]{\smash{Counts single traces of a specified periodicity}}}%
    \put(0,0){\includegraphics[width=\unitlength,page=2]{ON_periodicity_diagram.pdf}}%
    \put(0.01854211,0.55542194){\color[rgb]{0,0,0}\makebox(0,0)[lb]{\smash{$\widetilde{T}_{ST}=K_{odd} \otimes \widetilde{T}_{ST}^{(odd)} \oplus K_{even} \otimes \widetilde{T}_{ST}^{(even)}$}}}%
    \put(0.07547457,0.49913942){\color[rgb]{0,0,0}\makebox(0,0)[lb]{\smash{$H_{\widetilde{T}_{ST}} = \mathcal{S} \left( H_{\widetilde{T}_{ST}^{(odd)}} , H_{\widetilde{T}_{ST}^{(even)}} \right)$}}}%
    \put(0.1049829,0.43942293){\color[rgb]{0,0,0}\makebox(0,0)[lb]{\smash{Counts all single traces}}}%
    \put(0,0){\includegraphics[width=\unitlength,page=3]{ON_periodicity_diagram.pdf}}%
    \put(0.35100845,0.78415486){\color[rgb]{0,0,0}\makebox(0,0)[lb]{\smash{$\widetilde{T}^{(odd)}=\operatorname{Sym} \left( \widetilde{T}_{ST}^{(odd)} \right)$}}}%
    \put(0.32630078,0.72335158){\color[rgb]{0,0,0}\makebox(0,0)[lb]{\smash{$H_{\widetilde{T}^{(odd)}} = \operatorname{PExp} \left( H_{\widetilde{T}_{ST}^{(odd)}} \right)$}}}%
    \put(0.37870497,0.66609168){\color[rgb]{0,0,0}\makebox(0,0)[lb]{\smash{Counts multi-traces of a specified periodicity}}}%
    \put(0,0){\includegraphics[width=\unitlength,page=4]{ON_periodicity_diagram.pdf}}%
    \put(0.19898424,0.20842326){\color[rgb]{0,0,0}\makebox(0,0)[lb]{\smash{$\widetilde{T}=\operatorname{Sym} \left( K_{odd} \otimes \widetilde{T}_{ST}^{(odd)} \oplus K_{even} \otimes \widetilde{T}_{ST}^{(even)} \right)$ }}}%
    \put(0.01283215,0.08681631){\color[rgb]{0,0,0}\makebox(0,0)[lb]{\smash{$H_{\widetilde{T}} = \operatorname{PExp} \left[ \mathcal{S} \left( H_{\widetilde{T}_{ST}^{(odd)}} , H_{\widetilde{T}_{ST}^{(even)}} \right) \right] = \mathcal{S}_{mult} \left[ \operatorname{PExp} \left( H_{\widetilde{T}_{ST}^{(odd)}} \right) , \operatorname{PExp} \left( H_{\widetilde{T}_{ST}^{(even)}} \right) \right] $}}}%
    \put(0.34987955,0.02330032){\color[rgb]{0,0,0}\makebox(0,0)[lb]{\smash{Counts all multi-traces}}}%
    \put(0,0){\includegraphics[width=\unitlength,page=5]{ON_periodicity_diagram.pdf}}%
    \put(0.0133478,0.93083085){\color[rgb]{0,0,0}\makebox(0,0)[lb]{\smash{Tensor with $K_{odd}$ and}}}%
    \put(0.24012034,0.73301281){\color[rgb]{0,0,0}\makebox(0,0)[lb]{\smash{$\mathcal{S}$}}}%
    \put(0.73357466,0.93600823){\color[rgb]{0,0,0}\makebox(0,0)[lb]{\smash{$\operatorname{Sym}$}}}%
    \put(0.64595075,0.87900511){\color[rgb]{0,0,0}\makebox(0,0)[lb]{\smash{$\operatorname{PExp}$}}}%
    \put(0.26885991,0.30781889){\color[rgb]{0,0,0}\makebox(0,0)[lb]{\smash{$\operatorname{Sym}$}}}%
    \put(0.34257269,0.36251176){\color[rgb]{0,0,0}\makebox(0,0)[lb]{\smash{$\operatorname{PExp}$}}}%
    \put(0.63569213,0.51697805){\color[rgb]{0,0,0}\makebox(0,0)[lb]{\smash{$\mathcal{S}_{mult}$}}}%
    \put(0.62206418,1.12731287){\color[rgb]{0,0,0}\makebox(0,0)[lb]{\smash{$\widetilde{T}_{ST}^{(even)}$}}}%
    \put(0.62402923,1.07030974){\color[rgb]{0,0,0}\makebox(0,0)[lb]{\smash{$H_{\widetilde{T}_{ST}^{(even)}}$}}}%
    \put(0,0){\includegraphics[width=\unitlength,page=6]{ON_periodicity_diagram.pdf}}%
    \put(0.67403146,0.78481909){\color[rgb]{0,0,0}\makebox(0,0)[lb]{\smash{$\widetilde{T}^{(even)}=\operatorname{Sym} \left( \widetilde{T}_{ST}^{(even)} \right)$}}}%
    \put(0.65118474,0.72335157){\color[rgb]{0,0,0}\makebox(0,0)[lb]{\smash{$H_{\widetilde{T}^{(even)}} = \operatorname{PExp} \left( H_{\widetilde{T}_{ST}^{(even)}} \right)$}}}%
    \put(0.04236906,0.90324184){\color[rgb]{0,0,0}\makebox(0,0)[lb]{\smash{$K_{even}$ respectively}}}%
    \put(0.69808948,0.37716271){\color[rgb]{0,0,0}\makebox(0,0)[lb]{\smash{Tensor power of $K_{odd}$ }}}%
    \put(0.69257333,0.34957373){\color[rgb]{0,0,0}\makebox(0,0)[lb]{\smash{and $K_{even}$ respectively}}}%
    \put(0.22742072,0.14820975){\color[rgb]{0,0,0}\makebox(0,0)[lb]{\smash{$= \operatorname{Sym} \left( \widetilde{T}_{ST}^{(odd)} \right)^{\otimes K_{odd}} \otimes \operatorname{Sym} \left( \widetilde{T}_{ST}^{(even)} \right)^{\otimes K_{even}}$}}}%
  \end{picture}%
\endgroup%

%% file: Baryonic_correlator3.pdf_tex
\begingroup%
  \makeatletter%
  \providecommand\color[2][]{%
    \errmessage{(Inkscape) Color is used for the text in Inkscape, but the package 'color.sty' is not loaded}%
    \renewcommand\color[2][]{}%
  }%
  \providecommand\transparent[1]{%
    \errmessage{(Inkscape) Transparency is used (non-zero) for the text in Inkscape, but the package 'transparent.sty' is not loaded}%
    \renewcommand\transparent[1]{}%
  }%
  \providecommand\rotatebox[2]{#2}%
  \ifx\svgwidth\undefined%
    \setlength{\unitlength}{208.79999998bp}%
    \ifx\svgscale\undefined%
      \relax%
    \else%
      \setlength{\unitlength}{\unitlength * \real{\svgscale}}%
    \fi%
  \else%
    \setlength{\unitlength}{\svgwidth}%
  \fi%
  \global\let\svgwidth\undefined%
  \global\let\svgscale\undefined%
  \makeatother%
  \begin{picture}(1,0.42528732)%
    \put(0,0){\includegraphics[width=\unitlength,page=1]{Baryonic_correlator3.pdf}}%
    \put(0.01006946,0.18836243){\color[rgb]{0,0,0}\makebox(0,0)[lb]{\smash{$B_T$}}}%
    \put(0,0){\includegraphics[width=\unitlength,page=2]{Baryonic_correlator3.pdf}}%
    \put(0.32910842,0.28096471){\color[rgb]{0,0,0}\makebox(0,0)[lb]{\smash{$B_{1^N + T}$}}}%
    \put(0.26548635,0.10776966){\color[rgb]{0,0,0}\makebox(0,0)[lb]{\smash{$P_{1^N}$}}}%
    \put(0.13984673,0.17038169){\color[rgb]{0,0,0}\makebox(0,0)[lb]{\smash{\Huge=}}}%
    \put(0,0){\includegraphics[width=\unitlength,page=3]{Baryonic_correlator3.pdf}}%
    \put(0.78121577,0.19475776){\color[rgb]{0,0,0}\makebox(0,0)[lb]{\smash{$B_{1^N + T}$}}}%
    \put(0.591954,0.1703816){\color[rgb]{0,0,0}\makebox(0,0)[lb]{\smash{\Huge=}}}%
    \put(0,0){\includegraphics[width=\unitlength,page=4]{Baryonic_correlator3.pdf}}%
  \end{picture}%
\endgroup%

%% file: Baryonic_correlator2.pdf_tex
\begingroup%
  \makeatletter%
  \providecommand\color[2][]{%
    \errmessage{(Inkscape) Color is used for the text in Inkscape, but the package 'color.sty' is not loaded}%
    \renewcommand\color[2][]{}%
  }%
  \providecommand\transparent[1]{%
    \errmessage{(Inkscape) Transparency is used (non-zero) for the text in Inkscape, but the package 'transparent.sty' is not loaded}%
    \renewcommand\transparent[1]{}%
  }%
  \providecommand\rotatebox[2]{#2}%
  \ifx\svgwidth\undefined%
    \setlength{\unitlength}{410.39999995bp}%
    \ifx\svgscale\undefined%
      \relax%
    \else%
      \setlength{\unitlength}{\unitlength * \real{\svgscale}}%
    \fi%
  \else%
    \setlength{\unitlength}{\svgwidth}%
  \fi%
  \global\let\svgwidth\undefined%
  \global\let\svgscale\undefined%
  \makeatother%
  \begin{picture}(1,0.30409355)%
    \put(0,0){\includegraphics[width=\unitlength,page=1]{Baryonic_correlator2.pdf}}%
    \put(0.00512306,0.18355284){\color[rgb]{0,0,0}\makebox(0,0)[lb]{\smash{$B_T$}}}%
    \put(0.01339004,0.09564777){\color[rgb]{0,0,0}\makebox(0,0)[lb]{\smash{$U$}}}%
    \put(0,0){\includegraphics[width=\unitlength,page=2]{Baryonic_correlator2.pdf}}%
    \put(0.14794815,0.18680661){\color[rgb]{0,0,0}\makebox(0,0)[lb]{\smash{$B_{1^N + T}$}}}%
    \put(0.0614035,0.13054507){\color[rgb]{0,0,0}\makebox(0,0)[lb]{\smash{\Huge=}}}%
    \put(0,0){\includegraphics[width=\unitlength,page=3]{Baryonic_correlator2.pdf}}%
    \put(0.22271047,0.0964721){\color[rgb]{0,0,0}\makebox(0,0)[lb]{\smash{$U$}}}%
    \put(0.27192984,0.13054507){\color[rgb]{0,0,0}\makebox(0,0)[lb]{\smash{\Huge=}}}%
    \put(0,0){\includegraphics[width=\unitlength,page=4]{Baryonic_correlator2.pdf}}%
    \put(0.35847501,0.23066626){\color[rgb]{0,0,0}\makebox(0,0)[lb]{\smash{$B_{1^N + T}$}}}%
    \put(0,0){\includegraphics[width=\unitlength,page=5]{Baryonic_correlator2.pdf}}%
    \put(0.43323728,0.14033173){\color[rgb]{0,0,0}\makebox(0,0)[lb]{\smash{$U$}}}%
    \put(0,0){\includegraphics[width=\unitlength,page=6]{Baryonic_correlator2.pdf}}%
    \put(0.33577139,0.14033173){\color[rgb]{0,0,0}\makebox(0,0)[lb]{\smash{$U$}}}%
    \put(0,0){\includegraphics[width=\unitlength,page=7]{Baryonic_correlator2.pdf}}%
    \put(0.3348141,0.05261241){\color[rgb]{0,0,0}\makebox(0,0)[lb]{\smash{$U^\dagger$}}}%
    \put(0.48245615,0.13054507){\color[rgb]{0,0,0}\makebox(0,0)[lb]{\smash{\Huge=}}}%
    \put(0,0){\includegraphics[width=\unitlength,page=8]{Baryonic_correlator2.pdf}}%
    \put(0.56900149,0.05522764){\color[rgb]{0,0,0}\makebox(0,0)[lb]{\smash{$B_{1^N + T}$}}}%
    \put(0,0){\includegraphics[width=\unitlength,page=9]{Baryonic_correlator2.pdf}}%
    \put(0.64376365,0.14033173){\color[rgb]{0,0,0}\makebox(0,0)[lb]{\smash{$U$}}}%
    \put(0,0){\includegraphics[width=\unitlength,page=10]{Baryonic_correlator2.pdf}}%
    \put(0.54629776,0.14033173){\color[rgb]{0,0,0}\makebox(0,0)[lb]{\smash{$U$}}}%
    \put(0,0){\includegraphics[width=\unitlength,page=11]{Baryonic_correlator2.pdf}}%
    \put(0.54534042,0.22805111){\color[rgb]{0,0,0}\makebox(0,0)[lb]{\smash{$U^\dagger$}}}%
    \put(0,0){\includegraphics[width=\unitlength,page=12]{Baryonic_correlator2.pdf}}%
    \put(0.69298245,0.13054507){\color[rgb]{0,0,0}\makebox(0,0)[lb]{\smash{\Huge=}}}%
    \put(0,0){\includegraphics[width=\unitlength,page=13]{Baryonic_correlator2.pdf}}%
    \put(0.77952778,0.09908726){\color[rgb]{0,0,0}\makebox(0,0)[lb]{\smash{$B_{1^N + T}$}}}%
    \put(0,0){\includegraphics[width=\unitlength,page=14]{Baryonic_correlator2.pdf}}%
    \put(0.85428992,0.18419135){\color[rgb]{0,0,0}\makebox(0,0)[lb]{\smash{$U$}}}%
    \put(0,0){\includegraphics[width=\unitlength,page=15]{Baryonic_correlator2.pdf}}%
    \put(0.95639023,0.09583355){\color[rgb]{0,0,0}\makebox(0,0)[lb]{\smash{$B_T$}}}%
    \put(0.96465741,0.18336705){\color[rgb]{0,0,0}\makebox(0,0)[lb]{\smash{$U$}}}%
    \put(0.89961012,0.13054507){\color[rgb]{0,0,0}\makebox(0,0)[lb]{\smash{\Huge=}}}%
    \put(0,0){\includegraphics[width=\unitlength,page=16]{Baryonic_correlator2.pdf}}%
  \end{picture}%
\endgroup%

%% file: Symplectic_Mesonic_Contraction.pdf_tex
\begingroup%
  \makeatletter%
  \providecommand\color[2][]{%
    \errmessage{(Inkscape) Color is used for the text in Inkscape, but the package 'color.sty' is not loaded}%
    \renewcommand\color[2][]{}%
  }%
  \providecommand\transparent[1]{%
    \errmessage{(Inkscape) Transparency is used (non-zero) for the text in Inkscape, but the package 'transparent.sty' is not loaded}%
    \renewcommand\transparent[1]{}%
  }%
  \providecommand\rotatebox[2]{#2}%
  \ifx\svgwidth\undefined%
    \setlength{\unitlength}{399.9999999bp}%
    \ifx\svgscale\undefined%
      \relax%
    \else%
      \setlength{\unitlength}{\unitlength * \real{\svgscale}}%
    \fi%
  \else%
    \setlength{\unitlength}{\svgwidth}%
  \fi%
  \global\let\svgwidth\undefined%
  \global\let\svgscale\undefined%
  \makeatother%
  \begin{picture}(1,0.42266925)%
    \put(0,0){\includegraphics[width=\unitlength,page=1]{Symplectic_Mesonic_Contraction.pdf}}%
    \put(0.48,0.19259319){\color[rgb]{0,0,0}\makebox(0,0)[lb]{\smash{\Huge$\beta$}}}%
    \put(0,0){\includegraphics[width=\unitlength,page=2]{Symplectic_Mesonic_Contraction.pdf}}%
    \put(0.36600001,0.30942191){\color[rgb]{0,0,0}\makebox(0,0)[lb]{\smash{\ldots}}}%
    \put(0.76599999,0.30942189){\color[rgb]{0,0,0}\makebox(0,0)[lb]{\smash{\ldots}}}%
    \put(0.36600002,0.09942191){\color[rgb]{0,0,0}\makebox(0,0)[lb]{\smash{\ldots}}}%
    \put(0.76600009,0.09942191){\color[rgb]{0,0,0}\makebox(0,0)[lb]{\smash{\ldots}}}%
    \put(0,0){\includegraphics[width=\unitlength,page=3]{Symplectic_Mesonic_Contraction.pdf}}%
    \put(0.07898481,0.0187578){\color[rgb]{0,0,0}\makebox(0,0)[lb]{\smash{$\Omega X$}}}%
    \put(0,0){\includegraphics[width=\unitlength,page=4]{Symplectic_Mesonic_Contraction.pdf}}%
    \put(0.23868109,0.0187578){\color[rgb]{0,0,0}\makebox(0,0)[lb]{\smash{$\Omega X$}}}%
    \put(0,0){\includegraphics[width=\unitlength,page=5]{Symplectic_Mesonic_Contraction.pdf}}%
    \put(0.47904001,0.0187578){\color[rgb]{0,0,0}\makebox(0,0)[lb]{\smash{$\Omega X$}}}%
    \put(0,0){\includegraphics[width=\unitlength,page=6]{Symplectic_Mesonic_Contraction.pdf}}%
    \put(0.63877756,0.01875792){\color[rgb]{0,0,0}\makebox(0,0)[lb]{\smash{$\Omega Y$}}}%
    \put(0,0){\includegraphics[width=\unitlength,page=7]{Symplectic_Mesonic_Contraction.pdf}}%
    \put(0.87891742,0.01875795){\color[rgb]{0,0,0}\makebox(0,0)[lb]{\smash{$\Omega Y$}}}%
    \put(0,0){\includegraphics[width=\unitlength,page=8]{Symplectic_Mesonic_Contraction.pdf}}%
    \put(0.09124731,0.37947578){\color[rgb]{0,0,0}\makebox(0,0)[lb]{\smash{$\Omega$}}}%
    \put(0.25124732,0.37947578){\color[rgb]{0,0,0}\makebox(0,0)[lb]{\smash{$\Omega$}}}%
    \put(0.49124728,0.37947578){\color[rgb]{0,0,0}\makebox(0,0)[lb]{\smash{$\Omega$}}}%
    \put(0.89124725,0.37947578){\color[rgb]{0,0,0}\makebox(0,0)[lb]{\smash{$\Omega$}}}%
    \put(0.65124727,0.37947578){\color[rgb]{0,0,0}\makebox(0,0)[lb]{\smash{$\Omega$}}}%
  \end{picture}%
\endgroup%

%% file: SpN_Simplified_Contraction2.pdf_tex
\begingroup%
  \makeatletter%
  \providecommand\color[2][]{%
    \errmessage{(Inkscape) Color is used for the text in Inkscape, but the package 'color.sty' is not loaded}%
    \renewcommand\color[2][]{}%
  }%
  \providecommand\transparent[1]{%
    \errmessage{(Inkscape) Transparency is used (non-zero) for the text in Inkscape, but the package 'transparent.sty' is not loaded}%
    \renewcommand\transparent[1]{}%
  }%
  \providecommand\rotatebox[2]{#2}%
  \ifx\svgwidth\undefined%
    \setlength{\unitlength}{404.00034785bp}%
    \ifx\svgscale\undefined%
      \relax%
    \else%
      \setlength{\unitlength}{\unitlength * \real{\svgscale}}%
    \fi%
  \else%
    \setlength{\unitlength}{\svgwidth}%
  \fi%
  \global\let\svgwidth\undefined%
  \global\let\svgscale\undefined%
  \makeatother%
  \begin{picture}(1,0.69308647)%
    \put(0,0){\includegraphics[width=\unitlength,page=1]{SpN_Simplified_Contraction2.pdf}}%
    \put(0.22772264,0.51152498){\color[rgb]{0,0,0}\makebox(0,0)[lb]{\smash{\Huge$\sigma$}}}%
    \put(0,0){\includegraphics[width=\unitlength,page=2]{SpN_Simplified_Contraction2.pdf}}%
    \put(0.05107035,0.64350477){\color[rgb]{0,0,0}\makebox(0,0)[lb]{\smash{$\Omega$}}}%
    \put(0,0){\includegraphics[width=\unitlength,page=3]{SpN_Simplified_Contraction2.pdf}}%
    \put(0.02730799,0.39400011){\color[rgb]{0,0,0}\makebox(0,0)[lb]{\smash{$\Omega$}}}%
    \put(0,0){\includegraphics[width=\unitlength,page=4]{SpN_Simplified_Contraction2.pdf}}%
    \put(0.07206347,0.39400006){\color[rgb]{0,0,0}\makebox(0,0)[lb]{\smash{$X$}}}%
    \put(0,0){\includegraphics[width=\unitlength,page=5]{SpN_Simplified_Contraction2.pdf}}%
    \put(0.15008017,0.6435048){\color[rgb]{0,0,0}\makebox(0,0)[lb]{\smash{$\Omega$}}}%
    \put(0,0){\includegraphics[width=\unitlength,page=6]{SpN_Simplified_Contraction2.pdf}}%
    \put(0.12631781,0.39400019){\color[rgb]{0,0,0}\makebox(0,0)[lb]{\smash{$\Omega$}}}%
    \put(0,0){\includegraphics[width=\unitlength,page=7]{SpN_Simplified_Contraction2.pdf}}%
    \put(0.17107329,0.39400006){\color[rgb]{0,0,0}\makebox(0,0)[lb]{\smash{$X$}}}%
    \put(0,0){\includegraphics[width=\unitlength,page=8]{SpN_Simplified_Contraction2.pdf}}%
    \put(0.32829782,0.64350484){\color[rgb]{0,0,0}\makebox(0,0)[lb]{\smash{$\Omega$}}}%
    \put(0,0){\includegraphics[width=\unitlength,page=9]{SpN_Simplified_Contraction2.pdf}}%
    \put(0.30453546,0.39400026){\color[rgb]{0,0,0}\makebox(0,0)[lb]{\smash{$\Omega$}}}%
    \put(0,0){\includegraphics[width=\unitlength,page=10]{SpN_Simplified_Contraction2.pdf}}%
    \put(0.34929094,0.39400011){\color[rgb]{0,0,0}\makebox(0,0)[lb]{\smash{$X$}}}%
    \put(0,0){\includegraphics[width=\unitlength,page=11]{SpN_Simplified_Contraction2.pdf}}%
    \put(0.22970277,0.40200117){\color[rgb]{0,0,0}\makebox(0,0)[lb]{\smash{\ldots}}}%
    \put(0.22970279,0.64843736){\color[rgb]{0,0,0}\makebox(0,0)[lb]{\smash{\ldots}}}%
    \put(0.41955525,0.50778182){\color[rgb]{0,0,0}\makebox(0,0)[lb]{\smash{\Huge=  $(-1)^n$}}}%
    \put(0,0){\includegraphics[width=\unitlength,page=12]{SpN_Simplified_Contraction2.pdf}}%
    \put(0.79999959,0.55906742){\color[rgb]{0,0,0}\makebox(0,0)[lb]{\smash{\Huge$\tau$}}}%
    \put(0.84158345,0.65316289){\color[rgb]{0,0,0}\makebox(0,0)[lb]{\smash{\ldots}}}%
    \put(0,0){\includegraphics[width=\unitlength,page=13]{SpN_Simplified_Contraction2.pdf}}%
    \put(0.83762307,0.45586284){\color[rgb]{0,0,0}\makebox(0,0)[lb]{\smash{\ldots}}}%
    \put(0,0){\includegraphics[width=\unitlength,page=14]{SpN_Simplified_Contraction2.pdf}}%
    \put(0.69186494,0.44352272){\color[rgb]{0,0,0}\makebox(0,0)[lb]{\smash{$X$}}}%
    \put(0,0){\includegraphics[width=\unitlength,page=15]{SpN_Simplified_Contraction2.pdf}}%
    \put(0.77107276,0.44352267){\color[rgb]{0,0,0}\makebox(0,0)[lb]{\smash{$X$}}}%
    \put(0,0){\includegraphics[width=\unitlength,page=16]{SpN_Simplified_Contraction2.pdf}}%
    \put(0.92948848,0.44352272){\color[rgb]{0,0,0}\makebox(0,0)[lb]{\smash{$X$}}}%
    \put(0.41955609,0.12297685){\color[rgb]{0,0,0}\makebox(0,0)[lb]{\smash{\Huge=  $(-1)^n$}}}%
    \put(0,0){\includegraphics[width=\unitlength,page=17]{SpN_Simplified_Contraction2.pdf}}%
    \put(0.80000037,0.15508953){\color[rgb]{0,0,0}\makebox(0,0)[lb]{\smash{\Huge$\tau^{-1}$}}}%
    \put(0.84158348,0.2538828){\color[rgb]{0,0,0}\makebox(0,0)[lb]{\smash{\ldots}}}%
    \put(0,0){\includegraphics[width=\unitlength,page=18]{SpN_Simplified_Contraction2.pdf}}%
    \put(0.84158348,0.06279389){\color[rgb]{0,0,0}\makebox(0,0)[lb]{\smash{\ldots}}}%
    \put(0,0){\includegraphics[width=\unitlength,page=19]{SpN_Simplified_Contraction2.pdf}}%
    \put(0.69186575,0.04944576){\color[rgb]{0,0,0}\makebox(0,0)[lb]{\smash{$X$}}}%
    \put(0,0){\includegraphics[width=\unitlength,page=20]{SpN_Simplified_Contraction2.pdf}}%
    \put(0.77107358,0.04944576){\color[rgb]{0,0,0}\makebox(0,0)[lb]{\smash{$X$}}}%
    \put(0,0){\includegraphics[width=\unitlength,page=21]{SpN_Simplified_Contraction2.pdf}}%
    \put(0.92948936,0.04944576){\color[rgb]{0,0,0}\makebox(0,0)[lb]{\smash{$X$}}}%
  \end{picture}%
\endgroup%

%% file: q=1_Contractions.pdf_tex
\begingroup%
  \makeatletter%
  \providecommand\color[2][]{%
    \errmessage{(Inkscape) Color is used for the text in Inkscape, but the package 'color.sty' is not loaded}%
    \renewcommand\color[2][]{}%
  }%
  \providecommand\transparent[1]{%
    \errmessage{(Inkscape) Transparency is used (non-zero) for the text in Inkscape, but the package 'transparent.sty' is not loaded}%
    \renewcommand\transparent[1]{}%
  }%
  \providecommand\rotatebox[2]{#2}%
  \ifx\svgwidth\undefined%
    \setlength{\unitlength}{416.80456336bp}%
    \ifx\svgscale\undefined%
      \relax%
    \else%
      \setlength{\unitlength}{\unitlength * \real{\svgscale}}%
    \fi%
  \else%
    \setlength{\unitlength}{\svgwidth}%
  \fi%
  \global\let\svgwidth\undefined%
  \global\let\svgscale\undefined%
  \makeatother%
  \begin{picture}(1,0.14659333)%
    \put(0,0){\includegraphics[width=\unitlength,page=1]{q=1_Contractions.pdf}}%
    \put(0.33189748,0.10935822){\color[rgb]{0,0,0}\makebox(0,0)[lb]{\smash{$\varepsilon$}}}%
    \put(0,0){\includegraphics[width=\unitlength,page=2]{q=1_Contractions.pdf}}%
    \put(0.33189748,0.01338989){\color[rgb]{0,0,0}\makebox(0,0)[lb]{\smash{$\varepsilon$}}}%
    \put(0,0){\includegraphics[width=\unitlength,page=3]{q=1_Contractions.pdf}}%
    \put(0.32053395,0.06064213){\color[rgb]{0,0,0}\makebox(0,0)[lb]{\smash{\ldots}}}%
    \put(0,0){\includegraphics[width=\unitlength,page=4]{q=1_Contractions.pdf}}%
    \put(0.06318643,0.10935822){\color[rgb]{0,0,0}\makebox(0,0)[lb]{\smash{$\varepsilon$}}}%
    \put(0,0){\includegraphics[width=\unitlength,page=5]{q=1_Contractions.pdf}}%
    \put(0.06318643,0.01338989){\color[rgb]{0,0,0}\makebox(0,0)[lb]{\smash{$\varepsilon$}}}%
    \put(0,0){\includegraphics[width=\unitlength,page=6]{q=1_Contractions.pdf}}%
    \put(0.05182284,0.06012516){\color[rgb]{0,0,0}\makebox(0,0)[lb]{\smash{\ldots}}}%
    \put(0,0){\includegraphics[width=\unitlength,page=7]{q=1_Contractions.pdf}}%
    \put(0.60060855,0.10935822){\color[rgb]{0,0,0}\makebox(0,0)[lb]{\smash{$\varepsilon$}}}%
    \put(0,0){\includegraphics[width=\unitlength,page=8]{q=1_Contractions.pdf}}%
    \put(0.60060855,0.01338989){\color[rgb]{0,0,0}\makebox(0,0)[lb]{\smash{$\varepsilon$}}}%
    \put(0,0){\includegraphics[width=\unitlength,page=9]{q=1_Contractions.pdf}}%
    \put(0.58924492,0.06056828){\color[rgb]{0,0,0}\makebox(0,0)[lb]{\smash{\ldots}}}%
    \put(0,0){\includegraphics[width=\unitlength,page=10]{q=1_Contractions.pdf}}%
    \put(0.86931967,0.10935819){\color[rgb]{0,0,0}\makebox(0,0)[lb]{\smash{$\varepsilon$}}}%
    \put(0,0){\includegraphics[width=\unitlength,page=11]{q=1_Contractions.pdf}}%
    \put(0.86931967,0.01338986){\color[rgb]{0,0,0}\makebox(0,0)[lb]{\smash{$\varepsilon$}}}%
    \put(0,0){\includegraphics[width=\unitlength,page=12]{q=1_Contractions.pdf}}%
    \put(0.8579554,0.06019901){\color[rgb]{0,0,0}\makebox(0,0)[lb]{\smash{\ldots}}}%
    \put(0,0){\includegraphics[width=\unitlength,page=13]{q=1_Contractions.pdf}}%
  \end{picture}%
\endgroup%

%% file: Inductive_Contractions.pdf_tex
\begingroup%
  \makeatletter%
  \providecommand\color[2][]{%
    \errmessage{(Inkscape) Color is used for the text in Inkscape, but the package 'color.sty' is not loaded}%
    \renewcommand\color[2][]{}%
  }%
  \providecommand\transparent[1]{%
    \errmessage{(Inkscape) Transparency is used (non-zero) for the text in Inkscape, but the package 'transparent.sty' is not loaded}%
    \renewcommand\transparent[1]{}%
  }%
  \providecommand\rotatebox[2]{#2}%
  \ifx\svgwidth\undefined%
    \setlength{\unitlength}{320.80456336bp}%
    \ifx\svgscale\undefined%
      \relax%
    \else%
      \setlength{\unitlength}{\unitlength * \real{\svgscale}}%
    \fi%
  \else%
    \setlength{\unitlength}{\svgwidth}%
  \fi%
  \global\let\svgwidth\undefined%
  \global\let\svgscale\undefined%
  \makeatother%
  \begin{picture}(1,0.6359011)%
    \put(0,0){\includegraphics[width=\unitlength,page=1]{Inductive_Contractions.pdf}}%
    \put(0.03690546,0.5843089){\color[rgb]{0,0,0}\makebox(0,0)[lb]{\smash{$\Large C^{(\varepsilon ; q-1 )}$}}}%
    \put(0,0){\includegraphics[width=\unitlength,page=2]{Inductive_Contractions.pdf}}%
    \put(0.06733071,0.51512328){\color[rgb]{0,0,0}\makebox(0,0)[lb]{\smash{\ldots}}}%
    \put(0,0){\includegraphics[width=\unitlength,page=3]{Inductive_Contractions.pdf}}%
    \put(0.34393646,0.14208324){\color[rgb]{0,0,0}\makebox(0,0)[lb]{\smash{$\varepsilon$}}}%
    \put(0,0){\includegraphics[width=\unitlength,page=4]{Inductive_Contractions.pdf}}%
    \put(0.34393646,0.01739668){\color[rgb]{0,0,0}\makebox(0,0)[lb]{\smash{$\varepsilon$}}}%
    \put(0,0){\includegraphics[width=\unitlength,page=5]{Inductive_Contractions.pdf}}%
    \put(0.18524854,1.28070848){\color[rgb]{0,0,0}\makebox(0,0)[lt]{\begin{minipage}{1.24864628\unitlength}\raggedright \end{minipage}}}%
    \put(0.03690546,0.2850613){\color[rgb]{0,0,0}\makebox(0,0)[lb]{\smash{$\Large C^{(\varepsilon ; q-1 )}$}}}%
    \put(0,0){\includegraphics[width=\unitlength,page=6]{Inductive_Contractions.pdf}}%
    \put(0.06733073,0.36536242){\color[rgb]{0,0,0}\makebox(0,0)[lb]{\smash{\ldots}}}%
    \put(0,0){\includegraphics[width=\unitlength,page=7]{Inductive_Contractions.pdf}}%
    \put(0.06932394,0.43645732){\color[rgb]{0,0,0}\makebox(0,0)[lb]{\smash{$\Large \beta_{\bar{p}}$}}}%
    \put(0,0){\includegraphics[width=\unitlength,page=8]{Inductive_Contractions.pdf}}%
    \put(0.41096501,0.5843089){\color[rgb]{0,0,0}\makebox(0,0)[lb]{\smash{$\Large C^{(\varepsilon ; q-1 )}$}}}%
    \put(0,0){\includegraphics[width=\unitlength,page=9]{Inductive_Contractions.pdf}}%
    \put(0.44139026,0.5151233){\color[rgb]{0,0,0}\makebox(0,0)[lb]{\smash{\ldots}}}%
    \put(0,0){\includegraphics[width=\unitlength,page=10]{Inductive_Contractions.pdf}}%
    \put(0.41096501,0.2850613){\color[rgb]{0,0,0}\makebox(0,0)[lb]{\smash{$\Large C^{(\varepsilon ; q-1 )}$}}}%
    \put(0,0){\includegraphics[width=\unitlength,page=11]{Inductive_Contractions.pdf}}%
    \put(0.44139028,0.36604785){\color[rgb]{0,0,0}\makebox(0,0)[lb]{\smash{\ldots}}}%
    \put(0,0){\includegraphics[width=\unitlength,page=12]{Inductive_Contractions.pdf}}%
    \put(0.44338347,0.43645739){\color[rgb]{0,0,0}\makebox(0,0)[lb]{\smash{$\Large \beta_{\bar{p}}$}}}%
    \put(0,0){\includegraphics[width=\unitlength,page=13]{Inductive_Contractions.pdf}}%
    \put(0.78502449,0.58430896){\color[rgb]{0,0,0}\makebox(0,0)[lb]{\smash{$\Large C^{(\varepsilon ; q-1 )}$}}}%
    \put(0,0){\includegraphics[width=\unitlength,page=14]{Inductive_Contractions.pdf}}%
    \put(0.81544967,0.5149862){\color[rgb]{0,0,0}\makebox(0,0)[lb]{\smash{\ldots}}}%
    \put(0,0){\includegraphics[width=\unitlength,page=15]{Inductive_Contractions.pdf}}%
    \put(0.78502449,0.28506142){\color[rgb]{0,0,0}\makebox(0,0)[lb]{\smash{$\Large C^{(\varepsilon ; q-1 )}$}}}%
    \put(0,0){\includegraphics[width=\unitlength,page=16]{Inductive_Contractions.pdf}}%
    \put(0.81544978,0.36522553){\color[rgb]{0,0,0}\makebox(0,0)[lb]{\smash{\ldots}}}%
    \put(0,0){\includegraphics[width=\unitlength,page=17]{Inductive_Contractions.pdf}}%
    \put(0.81744295,0.43645751){\color[rgb]{0,0,0}\makebox(0,0)[lb]{\smash{$\Large \beta_{\bar{p}}$}}}%
    \put(0,0){\includegraphics[width=\unitlength,page=18]{Inductive_Contractions.pdf}}%
    \put(0.61595135,0.13304636){\color[rgb]{0,0,0}\makebox(0,0)[lb]{\smash{\ldots}}}%
    \put(0,0){\includegraphics[width=\unitlength,page=19]{Inductive_Contractions.pdf}}%
    \put(0.61595129,0.02331022){\color[rgb]{0,0,0}\makebox(0,0)[lb]{\smash{\ldots}}}%
    \put(0,0){\includegraphics[width=\unitlength,page=20]{Inductive_Contractions.pdf}}%
    \put(0.34164102,0.06071609){\color[rgb]{0,0,0}\makebox(0,0)[lb]{\smash{\ldots}}}%
  \end{picture}%
\endgroup%

%% file: PaperJHEPVersion.bbl
\begin{thebibliography}{10}

\bibitem{Maldacena1998}
Juan~Martin Maldacena.
\newblock {The Large N Limit of Superconformal field theories and
  supergravity}.
\newblock {\em Adv. Theor. Math. Phys.}, 2:231--252, 1998.

\bibitem{Gubser1998}
S.~S. Gubser, I.~R. Klebanov, and A.~M. Polyakov.
\newblock {Gauge Theory Correlators from Non-Critical String Theory}.
\newblock {\em Phys. Lett.}, B428:105--114, 1998.

\bibitem{Witten1998}
Edward Witten.
\newblock {Anti de Sitter space and holography}.
\newblock {\em Adv. Theor. Math. Phys.}, 2:253--291, 1998.

\bibitem{BBNS01}
Vijay Balasubramanian, Micha Berkooz, Asad Naqvi, and Matthew~J. Strassler.
\newblock {Giant gravitons in conformal field theory}.
\newblock {\em JHEP}, 04:034, 2002.

\bibitem{Corley2002}
Steve Corley, Antal Jevicki, and Sanjaye Ramgoolam.
\newblock {Exact Correlators of Giant Gravitons from Dual N = 4 SYM theory}.
\newblock {\em Adv. Theor. Math. Phys.}, 5:809--839, 2002.

\bibitem{MST2000}
John McGreevy, Leonard Susskind, and Nicolaos Toumbas.
\newblock {Invasion of the giant gravitons from Anti-de Sitter space}.
\newblock {\em JHEP}, 06:008, 2000.

\bibitem{GMT2000}
Marcus~T. Grisaru, Robert~C. Myers, and Oyvind Tafjord.
\newblock {SUSY and goliath}.
\newblock {\em JHEP}, 08:040, 2000.

\bibitem{HHI2000}
Akikazu Hashimoto, Shinji Hirano, and N.~Itzhaki.
\newblock {Large branes in AdS and their field theory dual}.
\newblock {\em JHEP}, 08:051, 2000.

\bibitem{LLM}
Hai Lin, Oleg Lunin, and Juan~Martin Maldacena.
\newblock {Bubbling AdS space and 1/2 BPS geometries}.
\newblock {\em JHEP}, 10:025, 2004.

\bibitem{Maldacena:1998bw}
Juan~Martin Maldacena and Andrew Strominger.
\newblock {AdS(3) black holes and a stringy exclusion principle}.
\newblock {\em JHEP}, 12:005, 1998.

\bibitem{BHLN02}
Vijay Balasubramanian, Min-xin Huang, Thomas~S. Levi, and Asad Naqvi.
\newblock {Open strings from N=4 superYang-Mills}.
\newblock {\em JHEP}, 08:037, 2002.

\bibitem{BBFH04}
Vijay Balasubramanian, David Berenstein, Bo~Feng, and Min-xin Huang.
\newblock {D-branes in Yang-Mills theory and emergent gauge symmetry}.
\newblock {\em JHEP}, 03:006, 2005.

\bibitem{DSS07I}
Robert de~Mello~Koch, Jelena Smolic, and Milena Smolic.
\newblock {Giant Gravitons - with Strings Attached (I)}.
\newblock {\em JHEP}, 06:074, 2007.

\bibitem{Kimura:2007wy}
Yusuke Kimura and Sanjaye Ramgoolam.
\newblock {Branes, anti-branes and brauer algebras in gauge-gravity duality}.
\newblock {\em JHEP}, 11:078, 2007.

\bibitem{Brown:2007xh}
Thomas~William Brown, P.~J. Heslop, and S.~Ramgoolam.
\newblock {Diagonal multi-matrix correlators and BPS operators in N=4 SYM}.
\newblock {\em JHEP}, 02:030, 2008.

\bibitem{Bhattacharyya:2008rb}
Rajsekhar Bhattacharyya, Storm Collins, and Robert de~Mello~Koch.
\newblock {Exact Multi-Matrix Correlators}.
\newblock {\em JHEP}, 03:044, 2008.

\bibitem{Bhattacharyya:2008xy}
Rajsekhar Bhattacharyya, Robert de~Mello~Koch, and Michael Stephanou.
\newblock {Exact Multi-Restricted Schur Polynomial Correlators}.
\newblock {\em JHEP}, 06:101, 2008.

\bibitem{Brown:2008ij}
Thomas~William Brown, P.~J. Heslop, and S.~Ramgoolam.
\newblock {Diagonal free field matrix correlators, global symmetries and giant
  gravitons}.
\newblock {\em JHEP}, 04:089, 2009.

\bibitem{Pasukonis2013}
Jurgis Pasukonis and Sanjaye Ramgoolam.
\newblock {Quivers as Calculators: Counting, Correlators and Riemann Surfaces}.
\newblock {\em JHEP}, 04:094, 2013.

\bibitem{Mattioli2014}
Paolo Mattioli and Sanjaye Ramgoolam.
\newblock {Quivers, Words and Fundamentals}.
\newblock {\em JHEP}, 03:105, 2015.

\bibitem{Mattioli:2016gyl}
Paolo Mattioli and Sanjaye Ramgoolam.
\newblock {Gauge Invariants and Correlators in Flavoured Quiver Gauge
  Theories}.
\newblock {\em Nucl. Phys.}, B911:638--711, 2016.

\bibitem{Witten1998a}
Edward Witten.
\newblock {Baryons And Branes In Anti de Sitter Space}.
\newblock {\em JHEP}, 07:006, 1998.

\bibitem{AABF02}
Ofer Aharony, Yaron~E. Antebi, Micha Berkooz, and Ram Fishman.
\newblock {`Holey sheets': Pfaffians and subdeterminants as D-brane operators
  in large N gauge theories}.
\newblock {\em JHEP}, 12:069, 2002.

\bibitem{Caputa2013}
Pawel Caputa, Robert {De Mello Koch}, and Pablo Diaz.
\newblock {A basis for large operators in N=4 SYM with orthogonal gauge group}.
\newblock {\em Journal of High Energy Physics}, 2013(3), 2013.

\bibitem{Caputa2013a}
Pawel Caputa, Robert {De Mello Koch}, and Pablo Diaz.
\newblock {Operators, correlators and free fermions for SO(N) and Sp(N)}.
\newblock {\em Journal of High Energy Physics}, 2013(6), 2013.

\bibitem{Kemp2014}
Garreth Kemp.
\newblock {SO(N) restricted Schur polynomials}.
\newblock {\em J. Math. Phys.}, 56(2):022302, 2015.

\bibitem{Kemp1406}
Garreth Kemp.
\newblock {Restricted Schurs and correlators for $SO(N)$ and $Sp(N)$}.
\newblock {\em JHEP}, 08:137, 2014.

\bibitem{PlethysmL}
D.~E. Littlewood.
\newblock {Polynomial concomitants and invariant matrices}.
\newblock {\em Jour. London Math. Soc.}, 11, 1936.

\bibitem{Stanley1999}
R.~P. Stanley.
\newblock {\em {Enumerative Combinatorics}}, volume~2 of {\em Cambridge Studies
  in Advanced Mathematics}.
\newblock Cambridge University Press, 1999.

\bibitem{Carre1995}
Christophe Carr{\'e} and Bernard Leclerc.
\newblock Splitting the square of a schur function into its symmetric and
  antisymmetric parts.
\newblock {\em Journal of Algebraic Combinatorics}, 4(3):201--231, Jul 1995.

\bibitem{Sundborg2000}
Bo~Sundborg.
\newblock {The Hagedorn transition, deconfinement and N=4 SYM theory}.
\newblock {\em Nucl. Phys.}, B573:349--363, 2000.

\bibitem{Polyakov2001}
Alexander~M. Polyakov.
\newblock {Gauge fields and space-time}.
\newblock {\em Int. J. Mod. Phys.}, A17S1:119--136, 2002.

\bibitem{BDHO07}
M.~Bianchi, F.~A. Dolan, P.~J. Heslop, and H.~Osborn.
\newblock {N=4 superconformal characters and partition functions}.
\newblock {\em Nucl. Phys.}, B767:163--226, 2007.

\bibitem{Lyndon54}
R~C Lyndon.
\newblock On burnside's problem.
\newblock {\em Transactions of the American Mathematical Society}, 77:202--215,
  1954.

\bibitem{wiki-words}
{Wikipedia:} combinatorics on words.
\newblock \url{ https://en.wikipedia.org/wiki/Combinatorics_on_words}.
\newblock Accessed: 19-03-2018.

\bibitem{Lothaire1983combinatorics}
M.~Lothaire.
\newblock {\em {Combinatorics on Words}}.
\newblock Encyclopedia of Mathematics and its Applications. Addison-Wesley,
  1983.

\bibitem{Benvenuti2006}
Sergio Benvenuti, Bo~Feng, Amihay Hanany, and Yang-Hui He.
\newblock {Counting BPS Operators in Gauge Theories -Quivers, Syzygies and
  Plethystics}.
\newblock {\em JHEP}, 11:050, 2007.

\bibitem{Willenbring07}
Jeb Willenbring.
\newblock Stable hilbert series of $s(g)^k$ for classical groups.
\newblock {\em Journal of Algebra}, 314:844–871, 08 2007.

\bibitem{Macdonald1995}
I~G Macdonald.
\newblock {\em {Symmetric functions and Hall polynomials}}.
\newblock Oxford Mathematical Monographs. Oxford University Press, 1995.

\bibitem{cameron1999permutation}
P.J. Cameron.
\newblock {\em Permutation Groups}.
\newblock London Mathematical Society St. Cambridge University Press, 1999.

\bibitem{Kimura:2016bzo}
Yusuke Kimura, Sanjaye Ramgoolam, and Ryo Suzuki.
\newblock {Flavour singlets in gauge theory as Permutations}.
\newblock {\em JHEP}, 12:142, 2016.

\bibitem{Ivanov1999}
V.~N. Ivanov.
\newblock {Bispherical functions on the symmetric group associated with the
  hyperoctahedral subgroup}.
\newblock {\em Journal of Mathematical Sciences}, 96(5), 1999.

\bibitem{EHS08}
Yusuke Kimura and Sanjaye Ramgoolam.
\newblock {Enhanced symmetries of gauge theory and resolving the spectrum of
  local operators}.
\newblock {\em Phys. Rev.}, D78:126003, 2008.

\bibitem{Mattioli:2016eyp}
Paolo Mattioli and Sanjaye Ramgoolam.
\newblock {Permutation Centralizer Algebras and Multi-Matrix Invariants}.
\newblock {\em Phys. Rev.}, D93(6):065040, 2016.

\bibitem{Kimura:2017vpp}
Yusuke Kimura.
\newblock {Noncommutative Frobenius algebras and open-closed duality}.
\newblock 2017.

\bibitem{james1984representation}
G.D. James and A.~Kerber.
\newblock {\em The Representation Theory of the Symmetric Group}.
\newblock Encyclopedia of Mathematics and its Applications. Cambridge
  University Press, 1984.

\bibitem{Berenstein2017}
David Berenstein and Alexandra Miller.
\newblock Superposition induced topology changes in quantum gravity.
\newblock {\em Journal of High Energy Physics}, 2017(11):121, Nov 2017.

\bibitem{Mukhi:2005cv}
Sunil Mukhi and Mikael Smedback.
\newblock {Bubbling orientifolds}.
\newblock {\em JHEP}, 08:005, 2005.

\bibitem{Gukov:1999yn}
Sergei Gukov.
\newblock {K theory, reality, and orientifolds}.
\newblock {\em Commun. Math. Phys.}, 210:621--639, 2000.

\bibitem{Hanany:2000fq}
Amihay Hanany and Barak Kol.
\newblock {On orientifolds, discrete torsion, branes and M theory}.
\newblock {\em JHEP}, 06:013, 2000.

\bibitem{GR2017}
P~L Giscard and P~Rochet.
\newblock {Algebraic combinatorics on trace monoids: extending number theory to
  walks on graphs}.
\newblock {\em SIAM J. Discrete Math.}, 31(2):1428--1453, 2017.

\bibitem{Feng2007}
Bo~Feng, Amihay Hanany, and Yang-Hui He.
\newblock {Counting Gauge Invariants: the Plethystic Program}.
\newblock {\em JHEP}, 03:090, 2007.

\bibitem{BGM99}
Per Berglund, Eric~G. Gimon, and Djordje Minic.
\newblock {The AdS / CFT correspondence and spectrum generating algebras}.
\newblock {\em JHEP}, 07:025, 1999.

\bibitem{Bae:2017fcs}
Jin-Beom Bae, Euihun Joung, and Shailesh Lal.
\newblock {Exploring Free Matrix CFT Holographies at One-Loop}.
\newblock {\em Universe}, 3(4):77, 2017.

\bibitem{Yokoyama2016}
Yosuke Imamura and Shuichi Yokoyama.
\newblock Superconformal index of n=3 orientifold theories.
\newblock {\em Journal of Physics A: Mathematical and Theoretical}, 49, 03
  2016.

\bibitem{oeis}
Oeis foundation inc. (2018), the on-line encyclopedia of integer sequences,
  https://oeis.org/a023900.

\bibitem{King1971}
R.C. King.
\newblock The dimensions of irreducible tensor representation of the orthogonal
  and symplectic groups.
\newblock {\em Can. J. Math.}, 23:176--188, 1971.

\bibitem{BKYZ11}
A.~Bissi, C.~Kristjansen, D.~Young, and K.~Zoubos.
\newblock {Holographic three-point functions of giant gravitons}.
\newblock {\em JHEP}, 06:085, 2011.

\bibitem{Lin12}
Hai Lin.
\newblock {Giant gravitons and correlators}.
\newblock {\em JHEP}, 12:011, 2012.

\bibitem{CDZ12}
Pawel Caputa, Robert de~Mello~Koch, and Konstantinos Zoubos.
\newblock {Extremal versus Non-Extremal Correlators with Giant Gravitons}.
\newblock {\em JHEP}, 08:143, 2012.

\bibitem{KMY15}
Charlotte Kristjansen, Stefano Mori, and Donovan Young.
\newblock {On the Regularization of Extremal Three-point Functions Involving
  Giant Gravitons}.
\newblock {\em Phys. Lett.}, B750:379--383, 2015.

\bibitem{BGLM06}
Indranil Biswas, Davide Gaiotto, Subhaneil Lahiri, and Shiraz Minwalla.
\newblock {Supersymmetric states of N=4 Yang-Mills from giant gravitons}.
\newblock {\em JHEP}, 12:006, 2007.

\bibitem{PR12}
Jurgis Pasukonis and Sanjaye Ramgoolam.
\newblock {Quantum states to brane geometries via fuzzy moduli spaces of giant
  gravitons}.
\newblock {\em JHEP}, 04:077, 2012.

\bibitem{GGO2011}
Robert de~Mello~Koch, Matthias Dessein, Dimitrios Giataganas, and Christopher
  Mathwin.
\newblock {Giant Graviton Oscillators}.
\newblock {\em JHEP}, 10:009, 2011.

\bibitem{DSS07II}
Robert de~Mello~Koch, Jelena Smolic, and Milena Smolic.
\newblock {Giant Gravitons - with Strings Attached (II)}.
\newblock {\em JHEP}, 09:049, 2007.

\bibitem{BDS08}
David Bekker, Robert de~Mello~Koch, and Michael Stephanou.
\newblock {Giant Gravitons - with Strings Attached (III)}.
\newblock {\em JHEP}, 02:029, 2008.

\bibitem{DR12}
Robert de~Mello~Koch and Sanjaye Ramgoolam.
\newblock {A double coset ansatz for integrability in AdS/CFT}.
\newblock {\em JHEP}, 06:083, 2012.

\bibitem{HS18}
Shinji Hirano and Yuki Sato.
\newblock {Giant graviton interactions and M2-branes ending on multiple
  M5-branes}.
\newblock 2018.

\bibitem{HoriWalcher}
Kentaro Hori and Johannes Walcher.
\newblock {D-brane Categories for Orientifolds: The Landau-Ginzburg Case}.
\newblock {\em JHEP}, 04:030, 2008.

\bibitem{Pasukonis2010}
Jurgis Pasukonis and Sanjaye Ramgoolam.
\newblock {From counting to construction of BPS states in N = 4 SYM}.
\newblock {\em JHEP}, 02:078, 2011.

\bibitem{Biswas:2006tj}
Indranil Biswas, Davide Gaiotto, Subhaneil Lahiri, and Shiraz Minwalla.
\newblock {Supersymmetric states of N=4 Yang-Mills from giant gravitons}.
\newblock {\em JHEP}, 12:006, 2007.

\bibitem{Ramgoolam:2016ciq}
Sanjaye Ramgoolam.
\newblock {Permutations and the combinatorics of gauge invariants for general
  N}.
\newblock {\em PoS}, CORFU2015:107, 2016.

\bibitem{KS1608}
Piotr Kucharski and Piotr Sułkowski.
\newblock {BPS counting for knots and combinatorics on words}.
\newblock {\em JHEP}, 11:120, 2016.

\bibitem{LMOV}
J.~M.~F. Labastida, Marcos Marino, and Cumrun Vafa.
\newblock {Knots, links and branes at large N}.
\newblock {\em JHEP}, 11:007, 2000.

\bibitem{cameron1994combinatorics}
P.J. Cameron.
\newblock {\em Combinatorics: Topics, Techniques, Algorithms}.
\newblock Cambridge University Press, 1994.

\bibitem{Zinn-Justin2010}
Paul Zinn-Justin.
\newblock Jucys–murphy elements and weingarten matrices.
\newblock {\em Letters in Mathematical Physics}, 91:119--127, 02 2010.

\end{thebibliography}
